\documentclass[aps,prd,10pt,twocolumn,nofootinbib,superscriptaddress,longbibliography,floatfix]{revtex4-2}

\usepackage{amsmath,amssymb,amsfonts}
\usepackage{bm}
\usepackage{graphicx}
\usepackage[dvipsnames,table]{xcolor}
\usepackage{tikz}
\usetikzlibrary{decorations.pathmorphing,arrows.meta,positioning,calc,shapes.misc}
\usepackage{booktabs}
\usepackage{multirow}
\usepackage{microtype}
\definecolor{darkred}{rgb}{0.5,0.0,0.0}
\usepackage[colorlinks=true,linkcolor=darkred,citecolor=darkred,urlcolor=darkred]{hyperref}
\usepackage{orcidlink}

\hypersetup{pdftitle={Far-Zone Gravitational Interactions at the Sixth Post-Newtonian Order},pdfauthor={Giacomo Brunello}}

\numberwithin{equation}{section}
\allowdisplaybreaks
\newcommand{\dd}{\mathrm{d}}
\newcommand{\eps}{\varepsilon}
\newcommand{\Lg}{\log \left(\frac{\omega_1^2}{\bar{\mu}^2}\right)}

\newcommand{\cD}{\mathcal{D}}

\newcommand{\cH}{\mathcal{H}}
\newcommand{\cI}{\mathcal{I}}
\newcommand{\cJ}{\mathcal{J}}
\newcommand{\cL}{\mathcal{L}}
\newcommand{\cM}{\mathcal{M}}
\newcommand{\cN}{\mathcal{N}}
\newcommand{\cO}{\mathcal{O}}

\newcommand{\cS}{\mathcal{S}}
\newcommand{\cR}{\mathcal{R}}

\newcommand{\intp}[1]{\int_{\bm{#1}}}
\newcommand{\form}{\texttt{FORM}}
\newcommand{\litered}{\texttt{LiteRed}}
\newcommand{\PN}{\mathrm{PN}}
\DeclareRobustCommand{\inin}{in\nobreakdash-in}
\DeclareRobustCommand{\inout}{in\nobreakdash-out}
\newcommand{\lo}{\mathrm{(0)}}
\newcommand{\hd}{\mathrm{(1)}}
\newcommand{\PV}{\mathrm{PV}}
\newcommand{\sgn}{\mathrm{sign}}

\begin{document}

\title{Far-Zone Gravitational Interactions at the Sixth Post-Newtonian Order}
\author{Giacomo Brunello\,\orcidlink{0009-0004-4788-738X}}
\email{giacomo.brunello@sns.it}
\affiliation{Scuola Normale Superiore, Piazza dei Cavalieri 7, 56126 Pisa, Italy}
\affiliation{INFN Sezione di Pisa, Largo Pontecorvo 3, 56127 Pisa, Italy}
\date{\today}
\begin{abstract}
We compute the nonlinear far-zone dynamics of compact binaries at the sixth post-Newtonian (6PN) order in the effective field theory of general relativity.
We derive the complete $\cO(G^2)$ far-zone action of a generic multipolar source in both the causal and the conservative prescriptions.
Besides the known tails and failed tails, it contains five new sectors generated by the nonlinear memory of the emitted waves.
On scattering orbits, they contain a boundary term, nonlocal in time, fixed by the balance of angular momentum.
The same construction yields the emitted waveforms, whose energy and angular-momentum fluxes balance the losses of the action.
For the binary, the 6PN dynamics also requires PN corrections to the lower-order sectors, the iterated radiation reaction, and a coupling quadratic in the curvature of the composite source, whose coefficient we fix by matching the gravitational Compton amplitude to that of the constituents.
We compute the two-radiation contributions to the total scattering angle at fourth, fifth and sixth post-Minkowskian (PM) order and to the conservative angle at 4PM and 5PM, retaining the integer-PN terms through 6PN.
For the conservative angle, we give both the local and $\gamma$-3 prescriptions.
Combined with the tail and the known potential contributions, our results reproduce the conservative and total 4PM angles and the second-self-force part of the conservative 5PM angle.
Including the dissipative tail, we recover the total 5PM angle through 5PN and determine its new 6PN coefficient.
At 6PM, the two-radiation contribution to the total angle extends the known 5PN result to 6PN.
\end{abstract}

\maketitle

\section{Introduction}
\label{sec:intro}

Gravitational waves from coalescing compact
objects~\cite{LIGOScientific:2016aoc,KAGRA:2021vkt,LIGOScientific:2025slb} 
carry a direct imprint of the relativistic two-body dynamics.
The sensitivity of the next generation of
detectors~\cite{Punturo:2010zz,Reitze:2019iox,LISA:2017pwj} 
increases the need for accurate
predictions of the dynamics and radiation of compact binaries~\cite{Purrer:2019jcp,Owen:2023mid,Dhani:2024jja}.
During the inspiral phase, the orbital velocity $v$ is small compared to the speed of light $c$, 
and the dynamics admits the post-Newtonian (PN)
expansion~\cite{Einstein:1938yz,Blanchet:2013haa,Porto:2016pyg,Schafer:2018jfw,Levi:2018nxp,Jaranowski:1997ky,Damour:2014jta,Jaranowski:2015lha,Bernard:2015njp,Bernard:2016wrg,Damour:2016abl,Blanchet:2023sbv,Blanchet:2023bwj,Damour:2017ced,Goldberger:2004jt,Gilmore:2008gq,Foffa:2011ub,Porto:2017dgs,Kol:2013ega,Foffa:2012rn,Foffa:2019rdf,Foffa:2016rgu,Foffa:2019yfl,Blumlein:2020pog,Blumlein:2020znm,Goldberger:2012kf,Galley:2015kus,Foffa:2019hrb,Blumlein:2019zku,Foffa:2020nqe,Blumlein:2020pyo,Blumlein:2021txe,Porto:2024cwd,Blumlein:2021txj,Porto:2005ac,Porto:2006bt,Porto:2008tb,Levi:2008nh,Levi:2015msa,Levi:2016ofk,Kim:2021rfj,Levi:2020uwu,Levi:2019kgk,Kim:2022pou,Kim:2022bwv,Levi:2022dqm,Levi:2022rrq,Mandal:2022nty,Mandal:2022ufb,Goldberger:2009qd,Ross:2012fc,Porto:2010zg,Maia:2017gxn,Cho:2021mqw,Cho:2022syn,Amalberti:2023ohj,Amalberti:2024jaa,Mandal:2024iug}. 
An $n$PN correction is suppressed by $(v/c)^{2n}$ relative to the
Newtonian result.
This phase is characterised by a separation of scales between the sizes of the compact objects, the orbital separation, and the
wavelength of the emitted radiation. 
This separation of scales is organised by a tower of effective field theories (EFTs)~\cite{Goldberger:2004jt,Goldberger:2007hy,Foffa:2013qca,Porto:2016pyg,Levi:2018nxp,Goldberger:2022rqf}: potential graviton modes mediate the
near-zone orbital interaction, while radiation graviton modes propagate to infinity.
In the far zone, the binary is described as a composite source with a mass, angular momentum and multipole moments~\cite{Goldberger:2009qd,Ross:2012fc}.
Multi-loop methods from quantum field theory have carried these EFTs to high PN orders~\cite{Kol:2013ega,Foffa:2016rgu}.
The complete binary dynamics has been determined at
4PN~\cite{Damour:2014jta,Jaranowski:2015lha,Damour:2016abl,Bernard:2017bvn,Marchand:2017pir}, also within the EFT approach~\cite{Foffa:2016rgu,Foffa:2019rdf,Foffa:2019yfl,Blumlein:2020pog}. 
The 5PN impulse and scattering angle of the binary have recently been computed~\cite{Porto:2026fsd}, combining the near-zone potential~\cite{Foffa:2019hrb,Blumlein:2019zku,Blumlein:2020pyo,Blumlein:2021txe} with the far-zone \inin\ action~\cite{Porto:2024cwd}.
This also fixes the two coefficients left undetermined in the 5PN Hamiltonian of Ref.~\cite{Bini:2020wpo}.
At 6PN, only partial results are known: the conservative
potential up to $\cO(G^4)$~\cite{Blumlein:2020znm,Blumlein:2021txj}, the local- and nonlocal-in-time
dynamics accessible through the self-force expansion~\cite{Bini:2020nsb,Bini:2020hmy}, 
and
the static $\cO(G^7)$ contribution~\cite{Brunello:2025gpf}, extended to $\cO(G^8)$ at 7PN~\cite{Brunello:2026anu}.
Complementary information comes from scattering orbits, where the dynamics can be expanded in $GM/b$, with $M$ the total mass and $b$ the impact parameter, retaining the full velocity dependence: this is the post-Minkowskian (PM) expansion~\cite{Westpfahl:1979gu,Westpfahl:1985tsl,Damour:2016gwp,Damour:2017zjx}.
This expansion has been pushed to high orders with scattering-amplitude~\cite{Bjerrum-Bohr:2018xdl,Cheung:2018wkq,Kosower:2018adc} and worldline~\cite{Kalin:2020mvi,Mogull:2020sak,Kalin:2022hph,Jakobsen:2022psy} methods.
The conservative scattering angle is known at 3PM~\cite{Bern:2019nnu,Bern:2019crd,Cheung:2020gyp,Parra-Martinez:2020dzs,Kalin:2020fhe} and 4PM~\cite{Bern:2021dqo,Bern:2021yeh,Bern:2022jvn,Dlapa:2021npj,Dlapa:2021vgp,Dlapa:2024cje}.
Radiation-reaction effects on the scattering angle are known at 3PM~\cite{Damour:2020tta,DiVecchia:2021bdo,Herrmann:2021tct,Bini:2021gat,Kalin:2022hph,Manohar:2022dea} and 4PM~\cite{Dlapa:2022lmu,Jakobsen:2023hig,Damgaard:2023ttc}.
The radiated momentum is known at 3PM~\cite{Herrmann:2021lqe,Herrmann:2021tct,Kalin:2022hph} and 4PM~\cite{Dlapa:2022lmu}, and the loss of angular momentum through 3PM~\cite{Damour:2020tta,Manohar:2022dea,Riva:2023xxm,Jakobsen:2026ayw} and at 4PM~\cite{Heissenberg:2025ocy,Georgoudis:2026zzn,Bohnenblust:2026ytj}.
At 5PM, the conservative angle is known at first~\cite{Driesse:2024xad,Bern:2024adl,Driesse:2024feo,Dlapa:2025biy} and second~\cite{Driesse:2026qiz} self-force order, and in the potential region~\cite{Bern:2025wyd}.
Its nonlocal-in-time tail contribution is known through tenth self-force order~\cite{Dlapa:2026oyq}.
The radiation-reacted angle, the radiated energy and the recoil are known at 5PM at first self-force order~\cite{Driesse:2024feo}.
In the PN expansion, radiation-reaction effects start at 2.5PN~\cite{Burke:1970wx,Iyer:1993xi,Iyer:1995rn}.
They were derived in the EFT framework in Ref.~\cite{Galley:2009px} and extended to
3.5PN~\cite{Galley:2012qs,Leibovich:2019cxo} and
4.5PN~\cite{Leibovich:2023xpg}.
At higher orders, radiation interacts with the gravitational field of the source and with itself.
These \emph{hereditary} effects depend on the entire past history of the source. 
Scattering off the mass of the source produces the \emph{tail} terms, which first enter the dynamics at 4PN~\cite{Blanchet:1987wq,Blanchet:1992br}, and their iterations, the tails of tails, at 5.5PN~\cite{Blanchet:1997jj}.
Scattering off the angular momentum produces the \emph{failed tail} terms~\cite{Almeida:2023yia,Henry:2023sdy,Porto:2024cwd,Almeida:2026clf}.
Interactions among the emitted waves generate gravitational-wave \emph{memory} terms~\cite{Christodoulou:1991cr,Wiseman:1991ss,Thorne:1992sdb,Blanchet:1992br,Favata:2008yd,Trestini:2023wwg}.
In the multipolar post-Minkowskian (MPM) formalism, these effects are known in the flux and
phasing of quasi-circular binaries through 4.5PN order and in the waveform modes through
4PN order~\cite{Blanchet:2013haa,Blanchet:2023bwj,Blanchet:2023sbv}.
In the EFT, causal forces and radiative observables are studied with the closed-time-path
(\inin) formalism~\cite{Schwinger:1960qe,Keldysh:1964ud,Calzetta:1986ey,Galley:2009px,Galley:2012hx,Galley:2014wla}.
The 4PN tail was derived in the EFT in
Refs.~\cite{Foffa:2011np,Galley:2015kus}, and the 5PN hereditary terms were studied in
Refs.~\cite{Foffa:2019eeb,Blumlein:2021txe,Brunello:2022zui,Almeida:2022jrv,Henry:2023sdy,Almeida:2023yia,Almeida:2024lbv}.
The complete 5PN \inin\ effective action was obtained in
Ref.~\cite{Porto:2024cwd}, including the nonlocal boundary terms of the memory contribution.
In the far zone at 6PN only the tails and failed tails are known for general multipoles~\cite{Almeida:2021xwn,Almeida:2026clf}.

In this work, we compute the nonlinear far-zone contribution to the dynamics of two non-spinning point masses at 6PN.
We first derive, from a Berends--Giele recursion for the \inin\ one-point function, the complete $\cO(G^2)$ far-zone action of a generic source, which contains eleven sectors: two tails, four failed tails and five memories.
We reproduce the known tail and failed-tail results~\cite{Almeida:2021xwn,Almeida:2026clf} and compute the five memory actions in both the causal and the conservative prescriptions.
On scattering orbits, they contain in addition a boundary term, nonlocal in time, fixed by the balance of angular momentum, in which the nonlinear memory of the emitted waves~\cite{Christodoulou:1991cr,Wiseman:1991ss,Thorne:1992sdb,Blanchet:1992br} acts on the change of the mass multipoles.
From the same generating functional we compute the nonlinear waveforms, whose stress tensor is conserved up to the recoil of the source.
We verify that the radiated energy and angular momentum balance the losses predicted by the action at generic frequencies, while the zero-frequency part of the angular-momentum flux reproduces the coefficients of the boundary term.
The far-zone action of a generic source is not yet the far-zone contribution to the binary dynamics.
The multipoles are matched to the near-zone dynamics~\cite{Goldberger:2009qd,Ross:2012fc}, the lower-order far-zone sectors are evaluated with PN-corrected multipoles and motion~\cite{Foffa:2019eeb,Bini:2020hmy,Henry:2021cek,Larrouturou:2021gqo,Trestini:2022tot,Amalberti:2024jaa}, and the radiation-reaction forces are iterated in the equations of motion, which generates conservative-like contributions at second order in the leading force~\cite{Porto:2024cwd}.
A coupling quadratic in the electric curvature of the composite source also enters, which the covariant multipolar couplings do not generate.
We fix its coefficient by matching the gravitational Compton amplitude of the multipoles to that of the constituents.
We compute the two-radiation contributions to the total scattering angle at 4PM, 5PM and 6PM and to the conservative angle at 4PM and 5PM, retaining the integer-PN terms through 6PN.
The conservative results are given in both the local and $\gamma$-3 prescriptions~\cite{Porto:2026fsd,Driesse:2026qiz}.

At 4PM, combining the two-radiation contributions with the tail evaluated in dimensional regularisation and the known potential~\cite{Blumlein:2021txj,Blumlein:2020pog}, we reproduce both the conservative and the total angle~\cite{Bern:2021yeh,Dlapa:2021vgp,Dlapa:2022lmu} through 6PN.

At 5PM, the conservative angle depends on the prescription used to separate it, while the total angle does not.
After adding the potential region~\cite{Bern:2025wyd} and the conservative tail, the local result agrees with Ref.~\cite{Porto:2026fsd} at 5PN and extends it to 6PN.
The $\gamma$-3 result agrees with Ref.~\cite{Driesse:2026qiz} at second self-force order through 6PN.
In both prescriptions, the cubic terms in the symmetric mass ratio $\nu$ cancel.
Combining the retarded two-radiation contribution with the potential and tail contributions, including the dissipative tail, gives the total 5PM angle at orders $\nu^2$ and $\nu^3$ through 6PN.
The 5PN result agrees with Ref.~\cite{Porto:2026fsd}, while the 6PN $\nu^2$ coefficient is new and the $\nu^3$ coefficient agrees with Ref.~\cite{Driesse:2024feo}.

At 6PM, we give the two-radiation contribution to the total angle.
It reproduces the 5PN result of Ref.~\cite{Porto:2026fsd} and extends it to 6PN.
\paragraph*{Conventions.}
We work in dimensional regularisation with $D=d+1$ space-time dimensions, 
$d=3-2\eps$, in
the mostly-minus signature $\eta_{\mu\nu}=\mathrm{diag}(+,-,\dots,-)$, 
and we set $c=1$.
The gravitational coupling is denoted by
$\kappa=\sqrt{32\pi G}$, with $G$ Newton's constant.
We use shorthand notation for integrals
\begin{equation}
    \int_k\equiv\mu^{2\eps}\int\frac{\dd^Dk}{(2\pi)^D}\,,\qquad \intp{p}\equiv\mu^{2\eps}\int\frac{\dd^dp}{(2\pi)^d}\,,
\label{eq:shortint}
\end{equation}
where bold notation denotes spatial loop momenta, and $\mu$ is the scale introduced by dimensional regularisation.
We define the scale $\bar{\mu}$ as
\begin{equation}
    \bar{\mu}^2=\mu^2\pi e^{-\gamma_E}\,.
\end{equation}
Fourier transforms are defined as
\begin{equation}
  h_{\mu\nu}(x)=\int_k e^{-ik\cdot x}h_{\mu\nu}(k)\,,\quad
  \cI^{L}(t)=\int\!\frac{\dd\omega}{2\pi}e^{-i\omega t}\cI^{L}(\omega)\,.
\label{eq:fourier}
\end{equation}
Multi-index notation $L=i_1\cdots i_l$ is used for the symmetric trace-free (STF) multipoles, and spatial indices are contracted with the Euclidean metric.
Angular brackets on indices, as in $x^{\langle L\rangle}$, or a superscript STF on a product denote the STF projection.
Time derivatives are denoted by dots or by superscripts, $X^{(n)}=\dd^nX/\dd t^n$, and
$\Delta X\equiv X(+\infty)-X(-\infty)$ is the change of a quantity over a scattering history.
\section{Far-zone effective theory and in-in observables}
\label{sec:eft}
The far-zone theory describes the binary as a multipolar source coupled to the radiation field.
We specify its action and Feynman rules, then construct the causal \inin\ effective action and the one-point function using recursion relations.
We introduce power counting relations to identify the diagrams relevant at 6PN order.
\subsection{From the binary to the far-zone source}
\label{sec:scales}
The inspiral phase of a compact binary coalescence is characterised by a hierarchy of length scales,
\begin{equation}
 R\ll r\ll\lambda_{\mathrm{GW}}\,,
\label{eq:hierarchy}
\end{equation}
where $R$ is the characteristic size of each compact object, $r$ the orbital separation and $\lambda_{\mathrm{GW}}$ the radiation wavelength.
For compact objects, $R\sim GM/c^2$, and the virial theorem for bound orbits gives $v^2/c^2\sim GM/(rc^2)$, so that $R/r\sim v^2/c^2$.
Since the radiation is emitted at the orbital frequency, $r/\lambda_{\mathrm{GW}}\sim v/c$.
Hence, all scales are related by the small parameter $v/c\ll1$, which defines the post-Newtonian (PN) expansion.
This hierarchy allows the problem to be studied with a tower of EFTs, one for each scale~\cite{Goldberger:2004jt,Goldberger:2009qd} (see Refs.~\cite{Porto:2016pyg,Levi:2018nxp,Goldberger:2022rqf} for comprehensive reviews).
At distances larger than $R$, each body is represented by a worldline, with its internal structure encoded in finite-size operators.
At the orbital scale $r$, the two worldlines interact through the exchange of gravitational fields, described by the action
\begin{equation}
    S_{\mathrm{near}}=S_{\mathrm{bulk}}+\sum_{a=1}^2 S^{(a)}_{\mathrm{pp}}\,,
\end{equation}
where $S^{(a)}_{\mathrm{pp}}$ is the point-particle action
\begin{equation}
    S^{(a)}_{\mathrm{pp}}=-m_a\int\dd\lambda\,
    \sqrt{g_{\mu\nu}u^\mu_a u^\nu_a}\,,
\label{eq:Spp}
\end{equation}
with $m_a$ the mass of body $a$, $\lambda$ a worldline parameter and $u^\mu_a=\dd x^\mu_a/\dd\lambda$ its four-velocity.
The bulk gravitational dynamics is governed by the
Einstein--Hilbert action with the harmonic gauge-fixing term,
\begin{equation}
 S_{\mathrm{bulk}}=S_{\mathrm{EH}}+S_{\mathrm{gf}}
 =\int\!\dd^Dx\Big[-\frac{2}{\kappa^2}\sqrt{-g}\,R
 +\frac{1}{\kappa^2}\,G_\mu G^\mu\Big]\,,
\label{eq:SEHgf}
\end{equation}
where $G_\mu=\kappa\big(\partial^\alpha h_{\alpha\mu}-\tfrac12\partial_\mu h\big)$ is written
in terms of the graviton field, defined by expanding the metric around flat space-time,
\begin{equation}
  g_{\mu\nu}=\eta_{\mu\nu}+\kappa\,h_{\mu\nu}\,.
\label{eq:metricexp}
\end{equation}

The graviton modes do not scale homogeneously with the velocity $v$.
Potential modes $H_{\mu\nu}$, with momenta $k^\mu\sim(v/r,1/r)$, are off shell and mediate the orbital interaction.
Radiation modes $\bar h_{\mu\nu}$, with momenta $k^\mu\sim(v/r,v/r)$, propagate on shell to infinity,
\begin{equation}
  h_{\mu\nu}=H_{\mu\nu}+\bar h_{\mu\nu}\,.
\label{eq:split}
\end{equation}
This separation implements the method of regions~\cite{Beneke:1997zp} in the two-body
problem~\cite{Goldberger:2004jt}, and is well defined in dimensional regularisation, where the scaleless integrals that would mix the regions vanish.
The two effective theories that result are tied to each other by the cancellation of the ultraviolet poles of the far zone against the infrared poles of the near zone~\cite{Porto:2017dgs,Bernard:2017bvn,Foffa:2019yfl,Blumlein:2020pog}.

We integrate out the potential modes while retaining the orbital variables and treating the radiation field as a background.
Since $k_{\mathrm{rad}}r\sim v\ll1$, the radiation field does not resolve the orbital scale, and its
coupling is organised as a multipole expansion about the centre of mass,
\begin{equation}
\begin{aligned}
  e^{\,iS_{\mathrm{eff}}[x_a,\bar h]}&=\int\cD H\,
  e^{\,iS_{\mathrm{near}}[x_a,H+\bar h]}\,,\\
  S_{\mathrm{eff}}[x_a,\bar h]&=S_{\mathrm{bulk}}[\bar h]+S_{\mathrm{orb}}[x_a]+S_{\mathrm{mult}}[\bar h;Q]+\cdots\,.
\end{aligned}
\label{eq:NZmatching}
\end{equation}
Here $S_{\mathrm{orb}}$ is the conservative orbital action computed in the near zone, while
$S_{\mathrm{mult}}$ couples the long-wavelength field to the charges and multipole moments
$Q=\{M,L,\cI_L,\cJ_L,\dots\}$.
The multipoles are time-dependent Wilson coefficients, fixed by matching to the near-zone dynamics~\cite{Goldberger:2009qd,Ross:2012fc,Goldberger:2012kf}.
The far-zone EFT itself describes any multipole-expanded source, whatever its internal dynamics.
\subsection{Far-zone source action}
\label{sec:action}
\begin{figure*}[t]
\centering
\includegraphics[width=0.24\textwidth]{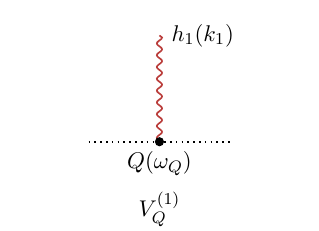}\hspace{0.06\textwidth}
\includegraphics[width=0.24\textwidth]{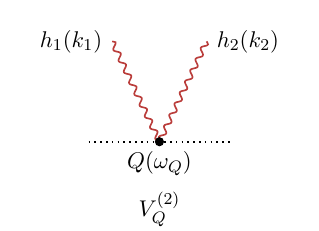}\hspace{0.06\textwidth}
\includegraphics[width=0.24\textwidth]{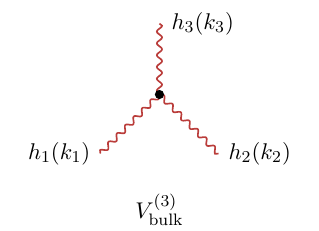}
\caption{The vertices of the far-zone EFT: the one-graviton worldline vertex $V^{(1)}_Q$, the two-graviton worldline vertex $V^{(2)}_Q$ (double emission) and the cubic bulk vertex $V^{(3)}_{\mathrm{bulk}}$. All momenta are outgoing and every worldline vertex conserves the energy, $\omega_Q+\sum_jk_j\cdot u=0$. Dotted lines denote the source worldline, red wavy lines radiation gravitons.}
\label{fig:vertices}
\end{figure*}
The far-zone action of the system is given by
\begin{equation}
  S_{\mathrm{far}}[h,Q]=S_{\mathrm{bulk}}[h]+S_{\mathrm{mult}}[h,Q]+S^{(2R)}_{\mathrm{comp}}[h,Q]+\cdots\,.
\label{eq:Stot}
\end{equation}
Here $S_{\mathrm{bulk}}$ is the bulk action of Eq.~\eqref{eq:SEHgf} restricted to the radiation modes, which we denote by $h_{\mu\nu}$ from now on,
$S_{\mathrm{mult}}$ is the multipolar action linear in the curvature, which contains the mass, angular momentum and multipolar couplings,
while $S^{(2R)}_{\mathrm{comp}}$ contains
independent quadratic-curvature couplings.

Let $X^\mu(\lambda)$ denote the centre-of-mass worldline,
$u^\mu=\dd X^\mu/\dd\lambda$, and $z=\sqrt{g_{\mu\nu}u^\mu u^\nu}$.
The moments are defined in a local orthonormal frame along the worldline, which we take to be the symmetric tetrad,
\begin{equation}
  e^\mu{}_a=\Big[(1+\kappa h)^{-1/2}\Big]^\mu{}_a
  =\delta^\mu_a-\frac{\kappa}{2}h^\mu{}_a+\frac{3\kappa^2}{8}h^\mu{}_\alpha h^\alpha{}_a+\cdots\,.
\label{eq:tetrad}
\end{equation}
The covariant multipolar action is~\cite{Goldberger:2009qd,Ross:2012fc,Porto:2024cwd}
\begin{equation}
\begin{aligned}
  S_{\mathrm{mult}}=\int\!\dd\lambda\bigg[&-M\,z-\frac12\,L_{ab}\,\omega^{ab}{}_{\mu}u^\mu\\
  &+\sum_{l\ge 2}\frac{z}{l!}\,\cI^{a_1\cdots a_l}\,\nabla_{a_3\cdots a_l}E_{a_1a_2}\\
  &-\sum_{l\ge 2}\frac{2l\,z}{(l+1)!}\,\cJ^{c|a_1\cdots a_l}\,\nabla_{a_3\cdots a_l}B_{c|a_1a_2}\bigg]\,.
\end{aligned}
\label{eq:Smult}
\end{equation}
The first two terms couple the source energy $M$ and angular momentum $L_{ab}$ to the metric and spin connection,
\begin{equation}
  \omega^{ab}{}_\mu=e^{b\rho}\big(\partial_\mu e^a{}_\rho-\Gamma^\lambda_{\mu\rho}e^a{}_\lambda\big)\,.
\label{eq:spinconn}
\end{equation}
The mass multipoles $\cI_L$ and the current multipoles $\cJ_L$ couple to the electric and magnetic components of
the curvature,
\begin{equation}
\begin{aligned}
  E_{ab}&=e^\mu{}_a e^\nu{}_b\,R_{\mu\alpha\nu\beta}\hat u^\alpha\hat u^\beta\,,\\
  B_{c|ab}&=e^\alpha{}_c\,e^\mu{}_a e^\nu{}_b\,R_{\mu\alpha\nu\beta}\hat u^\beta\,,
\end{aligned}
\label{eq:EB}
\end{equation}
where $\hat u^\mu=u^\mu/z$ and $\nabla_a=e^\gamma{}_a\nabla_\gamma$.
The mass multipoles are spatial in the local frame, symmetric and trace-free,
$\cI^{a_1\cdots a_l}u_{a_1}=0$, and $L_{ab}u^b=0$.

\begin{figure}[b]
\centering
\includegraphics[width=0.32\columnwidth]{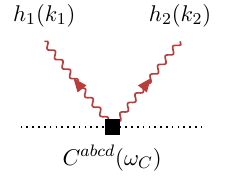}
\caption{Two-graviton vertex of the electric quadratic-curvature operator~\eqref{eq:S2R}, with $\omega_C$ the frequency of the Wilson tensor. All momenta are outgoing.}
\label{fig:e2rule}
\end{figure}

In their $d$-dimensional form, the current moments $\cJ^{c|L}$ have mixed symmetry: they are STF in $L$,
with $\cJ^{(c|L)}=0$.
In $d=3$ they are related to the STF current moments by
\begin{equation}
\begin{aligned}
    \cJ^{L}&=\epsilon^{i_lab}\cJ^{a|bL-1}\,,\qquad L^{ab}=\epsilon^{abc}L^c\,,\\
    \cJ^{a|bL-1}&=\textstyle\frac{1}{l+1}\sum_{i\in L}\epsilon^{aic}\cJ^{cL\setminus i}\,.
\end{aligned}
\label{eq:J3d}
\end{equation}
In $d$ dimensions, the action also admits an evanescent Weyl-type moment~\cite{Amalberti:2023ohj}.
At leading order it is built from the orbital angular momentum, which is conserved at this order, so it does not contribute through 6PN.
\paragraph*{Composite-source operators.}
Expanding the couplings of Eq.~\eqref{eq:Smult} to second order in the field fixes the two-graviton vertices of every multipole.
A composite source also admits independent operators quadratic in the curvature~\cite{Ross:2012fc}.
At the order considered here, the relevant operator is the electric quadratic-curvature operator,
\begin{equation}
\begin{aligned}
    S^{(2R)}_{\mathrm{comp}}&=\int\!\dd\lambda\,z\,
    \Big[\cL_{E^2}+\cdots\Big]\,,\\
    \cL_{E^2}&=C^{abcd}E_{ab}E_{cd}\,.
\end{aligned}
\label{eq:S2R}
\end{equation}
Dots denote magnetic, mixed and higher-derivative operators, entering at higher PN orders.
The Wilson tensor $C^{abcd}=C^{(ab)(cd)}=C^{cdab}$ depends on the orbital variables.
It is distinct from the finite-size response of either constituent~\cite{Goldberger:2004jt,Bini:2012gu,Porto:2016zng}.
Its coefficient is not fixed by covariance, and we match it to the constituents in Sec.~\ref{sec:source_matching}.
\paragraph*{Feynman rules.}
We work in the rest frame of the centre of mass and neglect recoil\footnote{The radiation-reaction force on the centre of mass vanishes at 2.5PN, since it multiplies the mass dipole, and starts at 3.5PN from the flux of linear momentum.
Recoil therefore first enters the dynamics at 7PN.}, so that
$X^\mu(t)=(t,\bm 0)$ and $u^\mu=(1,\bm 0)$.
The background tetrad $\delta^\mu_a$ identifies local-frame and space-time indices at zeroth order in the field.
Expanding the source action $S_{\mathrm{mult}}$ generates the worldline vertices $V_Q^{(n)}$, which describe the emission of $n$ gravitons by the multipole $Q$.
Expanding the bulk action $S_{\mathrm{bulk}}$ gives the graviton propagator and the self-interaction vertices $V_{\mathrm{bulk}}^{(n)}$.
The Feynman rules are illustrated in Fig.~\ref{fig:vertices}.
The linear worldline couplings are given by
\begin{equation}
\begin{aligned}
  V^{(1)}_{M}&=-\frac{i\kappa}{2}\,M\,h_{uu}\,,
  \\
  V^{(1)}_{L}&=-\frac{\kappa}{2}\,L^{ab}\,k_a\,h_{bu}\,,
  \\
  V^{(1)}_{\cI_l}(k)&=\frac{i}{l!}\,\cI^{abL-2}\,(-ik)_{L-2}\,E^{(1)}_{ab}\,,
  \\
  V^{(1)}_{\cJ_l}(k)&=-\frac{2l\,i}{(l+1)!}\,\cJ^{c|abL-2}\,(-ik)_{L-2}\,B^{(1)}_{c|ab}\,.
\end{aligned}
\label{eq:V1}
\end{equation}
Here $h_{u\mu}=u^\nu h_{\nu\mu}$, and $E^{(1)}$ and $B^{(1)}$ are the linearised curvatures in Fourier space,
\begin{equation}
\begin{aligned}
  E^{(1)}_{ab}& =\frac{\kappa}{2}\Big[k_ak_b\,h_{uu}-2(k\cdot u)k_{(a}h_{b)u}+(k\cdot u)^2h_{ab}\Big]\,, 
  \\
   B^{(1)}_{c|ab}&=\frac{\kappa}{2}\Big[k_ak_b h_{uc}-k_ck_a h_{ub}+(k\cdot u) \big(k_c h_{ab}-k_a h_{cb}\big)\Big]\,.
\end{aligned}
\label{eq:linEB}
\end{equation}
The electric operator~\eqref{eq:S2R} gives the two-graviton vertex of Fig.~\ref{fig:e2rule}, with rule
\begin{equation}
    V_C^{(2)}(k_1,k_2)=2i\,C^{abcd}(\omega_C)\,
    E^{(1)}_{ab}(k_1)E^{(1)}_{cd}(k_2)\,.
\label{eq:E2vertexrule}
\end{equation}
The multipolar and bulk Feynman rules required for the present calculation 
are given in the ancillary file \texttt{6pn\_far\_feynman\_rules.m}.
\begin{figure*}[t]
\centering
\includegraphics[width=0.95\textwidth]{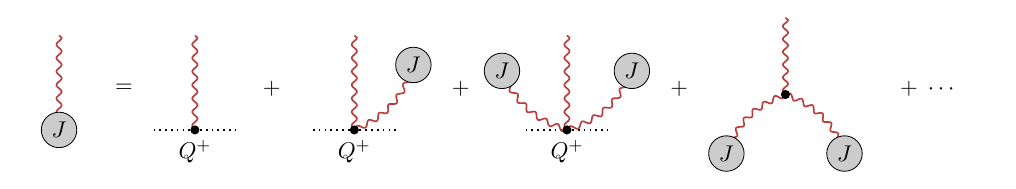}
\caption{The Berends--Giele recursion~\eqref{eq:BG} for the far-zone current $J_{\mu\nu}$: a single worldline emission, a
worldline vertex absorbing $m\ge1$ currents and emitting one graviton (``double
emission''), or a bulk vertex joining $m\ge2$ currents.}
\label{fig:BG}
\end{figure*}
\subsection{In-in effective action and recursion relations}
\label{sec:inin}
Integrating out the radiation field produces dissipative interactions that depend on the past history of the source, which a conventional \inout\ effective action does not generate.
We impose causal conditions using the closed-time-path (Schwinger--Keldysh, or \inin)
formalism~\cite{Schwinger:1960qe,Keldysh:1964ud,Calzetta:1986ey,Galley:2009px,Galley:2012hx,Galley:2014wla} (see also Refs.~\cite{Jakobsen:2022psy,Kalin:2022hph}).
The fields and source coefficients are doubled,
$h_{1,2}$ and $Q_{1,2}$, 
with one copy on each branch of the time contour.
The field histories are identified at the final time and obey vacuum boundary conditions in the far past.
The radiation effective action is defined 
by~\cite{Galley:2012hx,Galley:2015kus}
\begin{equation}
  e^{\,i\Gamma[Q_1,Q_2]}=\int\!\cD h_1\cD h_2\,
  \exp\Big\{i\,S[h_1,Q_1]-i\,S[h_2,Q_2]\Big\}
  \,.
\label{eq:CTP}
\end{equation}
We introduce the Keldysh basis,
\begin{equation}
  Q_+=\tfrac12(Q_1+Q_2)\,,\qquad Q_-=Q_1-Q_2\,,
\label{eq:keldysh}
\end{equation}
and similarly for $h$.
In this representation, the $+$ variables are the physical histories, while the $-$ variables are their variations.
The physical limit (PL) is given by $h_-=Q_-=0$, $h_+=h$, $Q_+=Q$.
The equations of motion follow by varying with respect to the difference variables before taking the physical limit, so that only the part of $\Gamma$ linear in $Q_-$ contributes to them.
In the classical ($\hbar\to0$) limit the path integral is dominated by its saddle point, 
and $\Gamma$ is given by the sum of the connected tree diagrams.
The part linear in the difference variables is generated by connected trees with one $Q_-$ insertion, 
called the \emph{root}.
Their propagators are retarded, 
with frequency flow oriented towards the root,
\begin{equation}
  \vcenter{\hbox{\includegraphics[width=0.42\columnwidth]{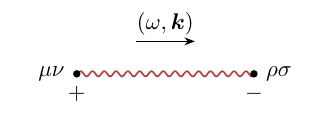}}}
  =\frac{i\,P_{\mu\nu\rho\sigma}}{(\omega+i0)^2-\bm k^2}\,,
\label{eq:retarded}
\end{equation}
with the frequency $\omega$ flowing from the $+$ towards the $-$ insertion, 
and with the $D$-dimensional polarisation sum in the harmonic gauge,
\begin{equation}
P_{\mu\nu\rho\sigma}=\tfrac12\big(\eta_{\mu\rho}\eta_{\nu\sigma}+\eta_{\mu\sigma}\eta_{\nu\rho}\big)
  -\tfrac{1}{D-2}\eta_{\mu\nu}\eta_{\rho\sigma}\,.
\label{eq:Pnum}
\end{equation}
Since every source insertion lies on the same worldline, the energy carried by an
internal line is the sum of the frequencies in the $+$ subtree below that line.
This fixes the retarded prescription of Eq.~\eqref{eq:retarded} throughout the tree.
\paragraph*{Berends--Giele recursion.}
\label{sec:BG}
The trees can be generated by solving the classical field equation iteratively.
To this end, we add to Eq.~\eqref{eq:CTP} an auxiliary source $\Xi_{1,2}^{\mu\nu}$ for the graviton, which defines the generating functional $W[\Xi_+,\Xi_-]$ of the connected \inin\ correlators of the graviton in the presence of the multipoles,
\begin{equation}
\begin{aligned}
  e^{\,iW}=\int\!\cD h_1\cD h_2\,
  \exp\Big\{&i\,S[h_1,Q]-i\,S[h_2,Q]\\
  &+i\!\int\!\dd^Dx\,\big(\Xi_+\!\cdot h_-+\Xi_-\!\cdot h_+\big)\Big\}\,,
\end{aligned}
\label{eq:CTPsource}
\end{equation}
with $\Xi\cdot h=\Xi^{\mu\nu}h_{\mu\nu}$.
Its functional derivatives are the classical response functions of the source~\cite{Bohnenblust:2026ujk},
\begin{equation}
  R^{(n)}_{\mu_1\nu_1\cdots\mu_n\nu_n}
  =\prod_{i=1}^{n-1}\frac{\delta}{\delta\Xi_+^{\mu_i\nu_i}}\,
  \frac{\delta W}{\delta\Xi_-^{\mu_n\nu_n}}\bigg|_{\Xi=0}\,,
\label{eq:responsefn}
\end{equation}
in which $n-1$ incoming legs, generated by $\Xi_+$, are connected to a single outgoing leg, generated by $\Xi_-$, with retarded propagators throughout.

We need only the lowest one, the one-point function
\begin{equation}
\begin{aligned}
  J_{\mu\nu}(x)
  &=\langle h_{+\mu\nu}(x)\rangle
  =\frac{\delta W}{\delta\Xi_-^{\mu\nu}(x)}\Big|_{\Xi=0}\\
  &=\int\!\dd^Dy\;G_{\mathrm{ret}}(x-y)_{\mu\nu}{}^{\rho\sigma}
  \Big(-\frac{\kappa}{2}\Big)T_{\rho\sigma,+}(y)\,,
\end{aligned}
\label{eq:onepointpath}
\end{equation}
which is the classical field generated by the $+$ multipoles.
The effective stress tensor $T_{\rho\sigma,+}$ includes the source and the nonlinear field interactions.
The perturbative expansion of Eq.~\eqref{eq:onepointpath} is the Berends--Giele
recursion~\cite{Berends:1987me,Jakobsen:2023ndj,Bohnenblust:2026ujk,Bautista:2026qse,Brunello:2026rdk,Kalin:2026txf} shown in Fig.~\ref{fig:BG},
\begin{equation}
\begin{aligned}
  J=G_{\mathrm{ret}}\cdot\Big\{&\sum_{Q^+}\Big[V^{(1)}_{Q}+\sum_{m\ge 1}\frac{V^{(m+1)}_{Q}\,J^{m}}{m!}\Big]\\
  &+\sum_{m\ge 2}\frac{V^{(m+1)}_{\mathrm{bulk}}\,J^{m}}{m!}\Big\}\,.
\end{aligned}
\label{eq:BG}
\end{equation}
The first term is a direct source emission.
The remaining terms join previously generated fields at a source $V^{(n)}_{Q}$ or bulk $V^{(n)}_{\mathrm{bulk}}$ vertex.
The explicit $G_{\mathrm{ret}}$ is the propagator of the outgoing graviton, and every current $J$ already contains its own outgoing propagator.
The factors $1/m!$ account for permutations of the fields entering each vertex.
The source sums include the two-graviton vertex~\eqref{eq:E2vertexrule} of the composite-source operator.
\begin{figure*}[t]
\centering
\includegraphics[width=0.22\textwidth]{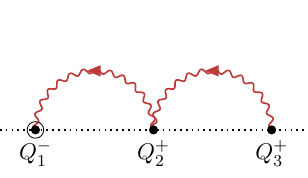}\hspace{0.04\textwidth}
\includegraphics[width=0.22\textwidth]{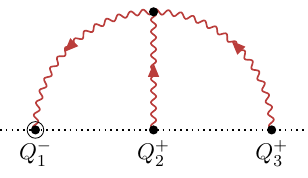}\hspace{0.04\textwidth}
\includegraphics[width=0.22\textwidth]{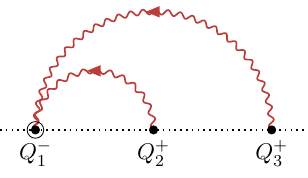}
\caption{The three rooted topologies with three insertions: chain (left), cubic (centre) and seagull (right).
The circled dot denotes the root and
the arrows show the retarded momentum flow towards it.}
\label{fig:topologies}
\end{figure*}
\paragraph*{Effective action.}
The effective action is obtained by attaching 
the currents $J$ of the $+$ insertions to a $-$ source vertex
\begin{equation}
  \Gamma=\sum_{Q^-}\sum_{m\ge1}\frac{1}{m!}\;
  \vcenter{\hbox{\includegraphics[scale=0.62]{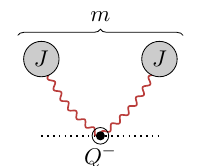}}}\;
  =\sum_{Q^-}\sum_{m\ge1}\frac{V^{(m)}_{Q^-}J^m}{m!}\,.
\label{eq:BGclose}
\end{equation}
Each term is a tree-level diagram with a single root.
With two source insertions, there is a single exchange.
With three, the diagrams have the three topologies shown in Fig.~\ref{fig:topologies}.
Each allowed choice of the root must be included.
A contribution with $n$ source insertions takes the frequency-space form
\begin{equation}
\begin{aligned}
  \Gamma={}&\int\frac{\dd\omega_{s}}{2\pi}\prod_{s'=2}^{n}\frac{\dd\omega_{s'}}{2\pi}\,
  2\pi\delta\Big(\omega_{s}+\sum_{s'=2}^{n}\omega_{s'}\Big)\\
  &\times\Big[Q_{s,-}(\omega_{s})\prod_{s'=2}^{n}Q_{s',+}(\omega_{s'})\Big]\,F(\omega_{s},\omega_2,\dots,\omega_n)\,.
\end{aligned}
\label{eq:Gammadef}
\end{equation}
The tensor contractions are implicit, and $F$ is the kernel obtained after the spatial momentum integrations.
For a radiative root, we write $\omega_{s}=\omega_1$.
For a static $M$ or $L$ root, $\omega_{s}=0$ and $\omega_1$ denotes the first independent radiative frequency.
The charges $M$ and $L$ must be doubled and varied before their conservation is imposed.
Terms proportional to their root frequency contain $\dot M_-$ or $\dot L_-$.
When proportional to the lower-order orbital equations of motion, these terms can be removed by a time reparametrisation or a time-dependent rotation of the orbit.
\paragraph*{Conservative prescription.}
\label{sec:inoutpresc}
Following Refs.~\cite{Porto:2024cwd,Porto:2026fsd}, we define the conservative radiation contribution from the real part of the \inout\ functional,
\begin{equation}
  e^{\,iS^{\mathrm{in\text{-}out}}[Q]}=\int\!\cD h\,e^{\,iS[h,Q]}\,,
\label{eq:inout}
\end{equation}
with Feynman boundary conditions.
Its imaginary part encodes emission through the optical theorem.
This is the prescription used in Refs.~\cite{Foffa:2011np,Galley:2015kus,Almeida:2021xwn,Almeida:2023yia}.
Diagrammatically, the recursion of Eqs.~\eqref{eq:BG} and~\eqref{eq:BGclose} with Feynman propagators generates the unrooted trees, with symmetry factors. 
The conservative part is taken after symmetrising over the source labels.
\subsection{Power counting and the 6PN sectors}
\label{sec:powercounting}
Each diagram can be assigned a definite PN scaling.
The PN order follows from the leading binary values of the multipole moments and the derivatives in their couplings~\cite{Goldberger:2009qd,Ross:2012fc,Porto:2016pyg}.
With $\omega r\sim v$, the vertices of Eq.~\eqref{eq:Smult} scale relative to the mass coupling as
\begin{equation}
  M\sim v^0\,,\qquad L\sim v^2\,,\qquad \cI_l\sim v^{l}\,,\qquad \cJ_l\sim v^{l+1}\,.
\label{eq:multscaling}
\end{equation}
The current moments contain one additional orbital velocity, 
and $L$ counts as a rank-one current moment.
A worldline vertex carries one factor of $\kappa$ per graviton leg, while an $m$-graviton bulk vertex carries $\kappa^{m-2}$.
Counting the legs of a connected tree with $k$ source insertions therefore gives $\kappa^{2(k-1)}\sim G^{k-1}$, independently of its topology.
The dimensionless factor associated with each additional gravitational interaction is $GM\omega\sim v^3$.
For a diagram with $k$ multipole insertions of total rank $s=\sum_{i=1}^kl_i$, $\sigma$ of which
are current moments or the angular momentum, the counting rule is
\begin{equation}
  \vcenter{\hbox{\includegraphics[width=0.5\columnwidth]{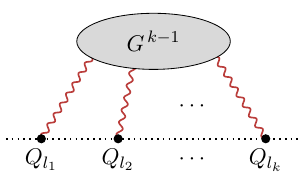}}}
  \;\sim\;v^{\,3(k-1)+s+\sigma}\,.
\label{eq:PNcount}
\end{equation}
Here $M$ has rank zero and $L$ rank one.
Since the Newtonian Lagrangian scales as $Mv^2$, the PN order is $[3(k-1)+s+\sigma-2]/2$.
For example, the radiation reaction $\{\cI_2,\cI_2\}$ scales as $v^7$ and enters at 2.5PN.
The quadrupole memory $\{\cI_2,\cI_2,\cI_2\}$ scales as $v^{12}$ and enters at 5PN.
The failed tail $\{L,\cI_2,\cJ_3\}$ scales as $v^{14}$ and enters at 6PN.
The counting rule must be supplemented by parity and rotational invariance.
A contributing combination must admit a scalar contraction and contain at least two radiative insertions.

We call each allowed combination a \emph{sector}.
Its causal action is the sum of all rooted trees and all allowed assignments of the root.
Table~\ref{tab:sectors} lists the leading sectors through 6PN.
The table contains the $\cO(G)$ radiation-reaction sectors~\cite{Burke:1970wx,Galley:2009px,Galley:2012qs,Leibovich:2023xpg}, the $\cO(G^2)$ mass tails~\cite{Blanchet:1987wq,Foffa:2011np,Galley:2015kus,Almeida:2021xwn}, angular-momentum failed tails~\cite{Porto:2024cwd,Henry:2023sdy,Almeida:2023yia,Almeida:2026clf} and memories~\cite{Porto:2024cwd,Almeida:2024lbv}, and the leading $\cO(G^3)$ tail of tail at 5.5PN~\cite{Blanchet:1997jj,Marchand:2016vox,Edison:2023qvg}.
\begin{table*}[t]
\centering\footnotesize\setlength{\tabcolsep}{4pt}
\begin{ruledtabular}
\begin{tabular}{lccccc}
Order & Radiation reaction & Tail & Failed tail & Memory & Tail of tail\\
 & $\{Q_l,Q_l\}$  & $\{M,Q_l,Q_l\}$  & $\{L,Q_l,Q_{l'}\}$ & $\{Q_l,Q_{l'},Q_{l''}\}$ & $\{M,M,Q_l,Q_l\}$\\\colrule
2.5PN & $\{\cI_2,\cI_2\}$ & & & & \\
3.5PN & $\{\cI_3,\cI_3\}$, $\{\cJ_2,\cJ_2\}$ & & & & \\
4PN & & $\{M,\cI_2,\cI_2\}$ & & & \\
4.5PN & $\{\cI_4,\cI_4\}$, $\{\cJ_3,\cJ_3\}$ & & & & \\
5PN & & $\{M,\cI_3,\cI_3\}$, $\{M,\cJ_2,\cJ_2\}$ & $\{L,\cI_2,\cI_2\}$ & $\{\cI_2,\cI_2,\cI_2\}$ & \\
5.5PN & $\{\cI_5,\cI_5\}$, $\{\cJ_4,\cJ_4\}$& & & & $\{M,M,\cI_2,\cI_2\}$\\
6PN & & $\{M,\cI_4,\cI_4\}$ (7) & $\{L,\cI_3,\cI_3\}$ (7) & \cellcolor{gray!15}$\{\cI_2,\cI_2,\cI_4\}$ (7), $\{\cI_2,\cI_3,\cI_3\}$ (7) & \\
 & & $\{M,\cJ_3,\cJ_3\}$ (7) & $\{L,\cJ_2,\cJ_2\}$ (7) & \cellcolor{gray!15}$\{\cI_2,\cJ_2,\cJ_2\}$ (7), $\{\cI_2,\cI_2,\cJ_3\}$ (7) & \\
 & & & $\{L,\cI_2,\cJ_3\}$ (12) & \cellcolor{gray!15}$\{\cI_2,\cI_3,\cJ_2\}$ (12) & \\
 & & & $\{L,\cI_3,\cJ_2\}$ (12) & & \\
\end{tabular}\end{ruledtabular}
\caption{The far-zone contributions through 6PN from the counting rule~\eqref{eq:PNcount}.
The eleven $\cO(G^2)$ sectors of the
6PN action are listed in the 6PN rows, with 
in parentheses the number of rooted \inin\
trees generated by the recursion~\eqref{eq:BG}, summed over the choices of the root.
The five new memory sectors are shaded.}
\label{tab:sectors}
\end{table*}

\section{Far-zone effective action at 6PN}
\label{sec:results}
We evaluate the eleven $\cO(G^2)$ sectors of Table~\ref{tab:sectors} 
using the recursive expansion~\eqref{eq:BGclose}, together with the action of the quadratic-curvature operator~\eqref{eq:S2R}.
The radiation-reaction, tail and failed-tail actions reproduce known
general-multipole results.
The 6PN memory sectors are new.
For these, we derive the local action at generic frequencies, while the nonlocal-in-time boundary term is derived in Sec.~\ref{sec:soft}.
\subsection{Computational pipeline}
\label{sec:strategy}
For each rooted tree generated by the recursion~\eqref{eq:BG}, we insert the Feynman
rules of Sec.~\ref{sec:action} and perform the tensor contractions in
\form~\cite{Vermaseren:2000nd,Ruijl:2017dtg}.
Energy conservation fixes the internal frequencies.
Momentum conservation at the bulk vertices leaves integrals over $d$-dimensional spatial momenta.
We reduce their tensor numerators to scalar integrals with the projector orthogonal to $u^\mu$.
The resulting scalar integrals are multi-loop vacuum integrals in $d$ dimensions, with
the multipole frequencies entering as effective propagator masses,
\begin{equation}
    D_\omega(\bm p)\equiv\bm p^2-(\omega+i0)^2\,.
\label{eq:Domega}
\end{equation}
They are reduced to master integrals by integration-by-parts (IBP) identities~\cite{Chetyrkin:1981qh,Laporta:2000dsw} as implemented in the public package \litered~\cite{Lee:2012cn,Lee:2013mka}.
All the master integrals with a non-vanishing coefficient are products of one-loop
tadpoles evaluated in $d=3-2\eps$.

In the \inin\ computation, the propagators carry the retarded prescription, and the tadpole reads
\begin{equation}
\begin{aligned}
  T(\omega)&\equiv\intp{p}\frac{1}{\bm p^2-(\omega+i0)^2}\\
  &=\frac{\Gamma(1-\tfrac d2)}{(4\pi)^{d/2}}\,(\omega^2)^{\frac d2-1}\,e^{-i\pi(\frac d2-1)\sgn\omega}
  =\frac{i\,\omega}{4\pi}+\mathcal{O}(\eps)\,.
\end{aligned}
\label{eq:tadpole}
\end{equation}
The Feynman prescription, used in the \inout\ computation, gives
\begin{equation}
  T_F(\omega)=\frac{\Gamma(1-\tfrac d2)}{(4\pi)^{d/2}}\,(\omega^2)^{\frac d2-1}\,e^{-i\pi(\frac d2-1)}
  =\frac{i\,|\omega|}{4\pi}+\mathcal{O}(\eps)\,,
\label{eq:tadpoleF}
\end{equation}
so that $T_F(\omega)=\sgn(\omega)\,T(\omega)$ in $d=3$.
\subsection{Radiation reaction}
\label{sec:RR}
\begin{figure}[!ht]
\centering
\includegraphics[width=0.46\columnwidth]{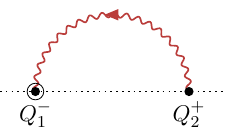}
\caption{The $\cO(G)$ radiation-reaction exchange.}
\label{fig:rr}
\end{figure}
At $\cO(G)$, the action comes from the single retarded exchange of
Fig.~\ref{fig:rr}.
Its frequency is $\omega=\omega_2=-\omega_1$, and its spatial integral is the retarded tadpole~\eqref{eq:tadpole}.
For every multipole, the action takes the closed form~\cite{Blanchet:1993ng,Goldberger:2009qd},
\begin{equation}
\begin{aligned}
  \Gamma_{\mathrm{RR}}=\int\!\frac{\dd\omega_1}{2\pi}\sum_{l\ge2}\Big[&F_l\,\cI^{L}_-(\omega_1)\cI^{L}_+(-\omega_1)\\
  &+F^J_l\,\cJ^{L}_-(\omega_1)\cJ^{L}_+(-\omega_1)\Big]\,,\\
  F_l=-iG\,c_l\,\omega_1^{2l+1}\,,\qquad &F^J_l=-iG\,c^J_l\,\omega_1^{2l+1}\,,
\end{aligned}
\label{eq:RRfreq}
\end{equation}
with
\begin{equation}
\begin{aligned}
    c_l&=\frac{(l+1)(l+2)}
    {(l-1)l\,l!\,(2l+1)!!}\,,\\
    c_l^J&=\frac{4l(l+2)}
    {(l-1)(l+1)!\,(2l+1)!!}\,,
\end{aligned}
\label{eq:cl}
\end{equation}
which coincide with the multipolar flux coefficients of Ref.~\cite{Thorne:1980ru}.
In the time domain, Eq.~\eqref{eq:RRfreq} reads
\begin{equation}
\begin{aligned}
  \Gamma_{\mathrm{RR}}=-G\sum_{l\ge2}(-1)^l\!\int\!\dd t\,\Big[&c_l\,\cI^{L}_-\cI^{L(2l+1)}_+\\[-4pt]
  &+c^{J}_l\,\cJ^{L}_-\cJ^{L(2l+1)}_+\Big]\,.
\end{aligned}
\label{eq:RR}
\end{equation}
The $l=2$ term is the Burke--Thorne action~\cite{Burke:1970wx,Galley:2009px}.
The mass-octupole and current-quadrupole sectors first enter at 3.5PN, and the mass-hexadecapole and current-octupole sectors at 4.5PN.
Together with PN corrections to the lower moments, they give the known 3.5PN~\cite{Iyer:1993xi,Iyer:1995rn,Galley:2012qs} and 4.5PN~\cite{Leibovich:2023xpg} radiation-reaction forces.
\subsection{Tail terms}
\label{sec:tails}
\begin{figure}[!ht]
\centering
\includegraphics[width=0.46\columnwidth]{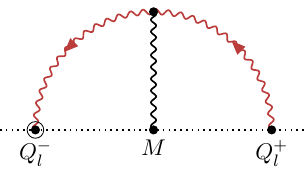}\hfill
\includegraphics[width=0.46\columnwidth]{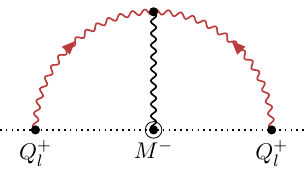}
\caption{The two tail trees: the root is a radiative multipole (left), or the static mass, with both radiative multipoles as $+$ insertions (right). Black wavy lines denote static gravitons.
}
\label{fig:tail}
\end{figure}
The mass tail arises when the emitted radiation scatters from the static gravitational field of the source~\cite{Blanchet:1987wq,Blanchet:1992br}.
In the far-zone EFT, it is the $\cO(G^2)$ interaction of two radiative multipoles with one mass insertion~\cite{Goldberger:2009qd,Foffa:2011np,Galley:2015kus}.
Only the cubic bulk vertex contributes to the action, 
with the two root assignments shown in Fig.~\ref{fig:tail}.
Their loop integrals belong to the two-loop family with one static line,
\begin{equation}
  I^{\mathrm{(rrs)}}_{a_1,a_2,a_3}=\intp{p}\intp{q}\frac{1}{D_{\omega}(\mathbf{p})^{a_1}\,D_{\omega}(\mathbf{p}+\mathbf{q})^{a_2}\,(\mathbf{q}^2)^{a_3}}\,,
\label{eq:tailfamily}
\end{equation}
with $D_\omega$ defined as in Eq.~\eqref{eq:Domega}.
After IBP reduction, there is a single master integral, 
the product of two one-loop tadpoles,
\begin{equation}
  J^{\mathrm{(rrs)}}_1=I^{\mathrm{(rrs)}}_{1,1,0}=T(\omega)^2\,.
\label{eq:tailmaster}
\end{equation}
When the root is the mass insertion, $M^-$, the two radiative lines carry opposite frequencies, $\omega$ and $-\omega$, and the master is $T(\omega)T(-\omega)$, whose phase cancels.
For a mass multipole $Q_l=\cI_l$, the result takes the universal form~\cite{Almeida:2021xwn}
\begin{equation}
\begin{aligned}
\Gamma_{\{M,Q_l,Q_l\}}&=\int\!\frac{\dd\omega_1}{2\pi}\big[F_{Q_l^-}M_+\,Q^{L}_-(\omega_1)Q^{L}_+(-\omega_1)\\
&\qquad+F_{M^-}M_-\,Q^{L}_+(\omega_1)Q^{L}_+(-\omega_1)\big]\,,\\
  F_{Q_l^-}&=+2G^2c_l\,\omega_1^{2l+2}\\
  &\quad\times\Big[\frac{1}{2\eps}+K_l-i\pi\sgn\omega_1-\Lg\Big]\,,\\
  F_{M^-}&=-G^2c_l\,\omega_1^{2l+2}\Big[\frac{1}{2\eps}+K_l-\Lg\Big]\,,
\end{aligned}
\label{eq:tails}
\end{equation}
with $c_l$ of Eq.~\eqref{eq:cl} and $K_l$ a rational constant.
The frequency integrals run over the full real axis.
The retarded radiative-root kernels satisfy $F_{Q_l^-}(-\omega_1)=F_{Q_l^-}(\omega_1)^*$.
For a current multipole, $c_l$ and $K_l$ are replaced by $c_l^J$ and $K_l^J$.
The finite constants are known for arbitrary rank~\cite{Almeida:2021xwn},
\begin{equation}
\begin{aligned}
  K_l&=H_{2l+1}-\tfrac12H_l+H_{l-2}-\tfrac12+\frac{2}{l(l+1)}\,,\\
  K^J_l&=H_{2l+1}-\tfrac12H_l+H_{l+1}-\frac{2l^2+5l+4}{l(l+1)(l+2)}\,,
\end{aligned}
\label{eq:Klclosed}
\end{equation}
where $H_n=\sum_{k=1}^n1/k$ and $H_0=0$.
The frequency logarithm gives the nonlocal time dependence of the tail.
The term proportional to $i\pi\sgn\omega_1$ is its dissipative part.
The sectors that first enter at 6PN are $\{M,\cI_4,\cI_4\}$ and $\{M,\cJ_3,\cJ_3\}$.
Their causal and \inout\ actions are
\begin{widetext}
\begin{align}
\Gamma_{\{M,\cI_4,\cI_4\}}&=\int\!\frac{\dd\omega_1}{2\pi}\Big[F_{\cI_4^-}\,M_+\,\cI^{ijkl}_-(\omega_1)\cI^{ijkl}_+(-\omega_1)
 +F_{M^-}\,M_-\,\cI^{ijkl}_+(\omega_1)\cI^{ijkl}_+(-\omega_1)\Big]\,,\nonumber\\
F_{\cI_4^-}&=\frac{G^2}{4536}\,\omega_1^{10}\Big[\frac1{2\eps}+\frac{1819}{630}-i\pi\sgn\omega_1-\Lg\Big]\,,\qquad
F_{M^-}=-\frac{G^2}{9072}\,\omega_1^{10}\Big[\frac1{2\eps}+\frac{1819}{630}-\Lg\Big]\,,\nonumber\\
S^{\mathrm{in\text{-}out}}_{\{M,\cI_4,\cI_4\}}&=\int\!\frac{\dd\omega_1}{2\pi}\;\frac{G^2}{9072}\,\omega_1^{10}\Big[\frac1{2\eps}+\frac{1819}{630}+i\pi-\Lg\Big]\,
 M\,\cI^{ijkl}(\omega_1)\cI^{ijkl}(-\omega_1)\,,\nonumber\\[1pt]
\Gamma_{\{M,\cJ_3,\cJ_3\}}&=\int\!\frac{\dd\omega_1}{2\pi}\Big[F_{\cJ_3^-}\,M_+\,\cJ^{ijk}_-(\omega_1)\cJ^{ijk}_+(-\omega_1)
 +F_{M^-}\,M_-\,\cJ^{ijk}_+(\omega_1)\cJ^{ijk}_+(-\omega_1)\Big]\,,\nonumber\\
F_{\cJ_3^-}&=\frac{G^2}{42}\,\omega_1^{8}\Big[\frac1{2\eps}+\frac{22}{7}-i\pi\sgn\omega_1-\Lg\Big]\,,\qquad
F_{M^-}=-\frac{G^2}{84}\,\omega_1^{8}\Big[\frac1{2\eps}+\frac{22}{7}-\Lg\Big]\,,\nonumber\\
S^{\mathrm{in\text{-}out}}_{\{M,\cJ_3,\cJ_3\}}&=\int\!\frac{\dd\omega_1}{2\pi}\;\frac{G^2}{84}\,\omega_1^{8}\Big[\frac1{2\eps}+\frac{22}{7}+i\pi-\Lg\Big]\,
 M\,\cJ^{ijk}(\omega_1)\cJ^{ijk}(-\omega_1)\,.
\label{eq:tails6PN}
\end{align}
\end{widetext}
The conservative action is the real part of the \inout\ result, following Sec.~\ref{sec:inoutpresc}.
The ancillary file \texttt{6pn\_actions.m} gives the two 6PN tail sectors for every root and in both prescriptions on the positive-frequency branch.
\subsection{Angular-momentum failed tails}
\label{sec:failedtails}
\begin{figure}[t]
\centering
\includegraphics[width=0.46\columnwidth]{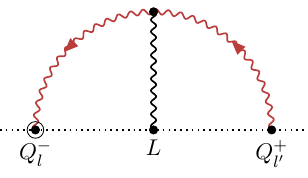}\hfill
\includegraphics[width=0.46\columnwidth]{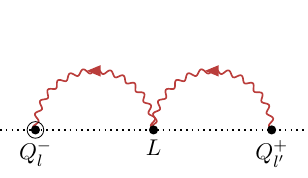}
\caption{Two failed-tail topologies: the cubic vertex with the static angular momentum
attached to it (left) and the chain with the double-emission vertex on the angular momentum
(right). Unlike for the mass tail, the chain and seagull trees also contribute.}
\label{fig:failedtail}
\end{figure}
The angular-momentum tail describes radiation scattering from the gravitomagnetic field of the source~\cite{Almeida:2023yia,Henry:2023sdy,Porto:2024cwd,Almeida:2026clf}.
The calculation includes the cubic bulk interaction and the two-graviton source vertices, with all allowed root assignments.
Two of them are shown in Fig.~\ref{fig:failedtail}.
After IBP reduction, they map to the same family~\eqref{eq:tailfamily}.
The angular-momentum vertex~\eqref{eq:V1} carries one more power of momentum than the mass vertex, which removes the logarithm of the mass tail: after summing the diagrams, the kernels are finite polynomials in frequency.
The 5PN sector and the four 6PN sectors, together with their \inout\ counterparts, are
\begin{widetext}
{\setlength{\jot}{1pt}
\begin{align}
\Gamma_{\{L,\cI_2,\cI_2\}}&=\int\!\frac{\dd\omega_1}{2\pi}\,\epsilon^{abc}\;\frac{iG^2}{30}\,\omega_1^{7}\Big[-L^c_-\,\cI^{ai}_+(\omega_1)\cI^{bi}_+(-\omega_1)
 +2L^c_+\,\cI^{ai}_-(\omega_1)\cI^{bi}_+(-\omega_1)\Big]\,,\nonumber\\
S^{\mathrm{in\text{-}out}}_{\{L,\cI_2,\cI_2\}}&=\int\!\frac{\dd\omega_1}{2\pi}\,\epsilon^{abc}\;\frac{iG^2}{30}\,\omega_1^{7}\,L^c\,\cI^{ai}(\omega_1)\cI^{bi}(-\omega_1)\,,\nonumber\\[1pt]
\Gamma_{\{L,\cI_3,\cI_3\}}&=\int\!\frac{\dd\omega_1}{2\pi}\,\epsilon^{abc}\;\frac{iG^2}{840}\,\omega_1^{9}\Big[-L^c_-\,\cI^{aij}_+(\omega_1)\cI^{bij}_+(-\omega_1)
 +2L^c_+\,\cI^{aij}_-(\omega_1)\cI^{bij}_+(-\omega_1)\Big]\,,\nonumber\\
S^{\mathrm{in\text{-}out}}_{\{L,\cI_3,\cI_3\}}&=\int\!\frac{\dd\omega_1}{2\pi}\,\epsilon^{abc}\;\frac{iG^2}{840}\,\omega_1^{9}\,L^c\,\cI^{aij}(\omega_1)\cI^{bij}(-\omega_1)\,,\nonumber\\
\Gamma_{\{L,\cJ_2,\cJ_2\}}&=\int\!\frac{\dd\omega_1}{2\pi}\,\epsilon^{abc}\;\frac{8iG^2}{45}\,\omega_1^{7}\Big[-L^c_-\,\cJ^{ai}_+(\omega_1)\cJ^{bi}_+(-\omega_1)
 +2L^c_+\,\cJ^{ai}_-(\omega_1)\cJ^{bi}_+(-\omega_1)\Big]\,,\nonumber\\
S^{\mathrm{in\text{-}out}}_{\{L,\cJ_2,\cJ_2\}}&=\int\!\frac{\dd\omega_1}{2\pi}\,\epsilon^{abc}\;\frac{8iG^2}{45}\,\omega_1^{7}\,L^c\,\cJ^{ai}(\omega_1)\cJ^{bi}(-\omega_1)\,, \nonumber\\
\Gamma_{\{L,\cI_2,\cJ_3\}}&=\int\!\frac{\dd\omega_1}{2\pi}\;\frac{G^2}{42}\,\omega_1^{8}\Big[L^c_-\,\cI^{ij}_+(\omega_1)\cJ^{ijc}_+(-\omega_1)
 -L^c_+\,\cI^{ij}_-(\omega_1)\cJ^{ijc}_+(-\omega_1)-L^c_+\,\cJ^{ijc}_-(\omega_1)\cI^{ij}_+(-\omega_1)\Big]\,,\nonumber\\
S^{\mathrm{in\text{-}out}}_{\{L,\cI_2,\cJ_3\}}&=-\int\!\frac{\dd\omega_1}{2\pi}\;\frac{G^2}{42}\,\omega_1^{8}\,L^c\,\cI^{ij}(\omega_1)\cJ^{ijc}(-\omega_1)\,,\nonumber\\[1pt]
\Gamma_{\{L,\cI_3,\cJ_2\}}&=\int\!\frac{\dd\omega_1}{2\pi}\;\frac{4G^2}{315}\,\omega_1^{8}\Big[-L^c_-\,\cI^{ijc}_+(\omega_1)\cJ^{ij}_+(-\omega_1)
 +L^c_+\,\cI^{ijc}_-(\omega_1)\cJ^{ij}_+(-\omega_1)+L^c_+\,\cJ^{ij}_-(\omega_1)\cI^{ijc}_+(-\omega_1)\Big]\,,\nonumber\\
S^{\mathrm{in\text{-}out}}_{\{L,\cI_3,\cJ_2\}}&=\int\!\frac{\dd\omega_1}{2\pi}\;\frac{4G^2}{315}\,\omega_1^{8}\,L^c\,\cI^{ijc}(\omega_1)\cJ^{ij}(-\omega_1)\,.
\label{eq:Ltails}
\end{align}}
\end{widetext}
At 5PN, the causal action reproduces Ref.~\cite{Porto:2024cwd}, and the conservative action agrees with the Fokker-Lagrangian result of Ref.~\cite{Henry:2023sdy}.
The four 6PN sectors agree with the general-multipole expressions and explicit 6PN action of Ref.~\cite{Almeida:2026clf}, for every root.
They are given in the ancillary file \texttt{6pn\_actions.m}.
\subsection{Nonlinear memory: the local action}
\label{sec:memory}
In the memory sectors, the waves emitted by two radiative multipoles interact nonlinearly, and the field they generate acts on the third multipole.
In the waveform, the same interaction produces the nonlinear memory~\cite{Christodoulou:1991cr,Wiseman:1991ss,Thorne:1992sdb,Blanchet:1992br}.
The quadrupolar sector at 5PN was computed in Refs.~\cite{Porto:2024cwd,Almeida:2024lbv}.
Here we evaluate the five sectors required at 6PN, to which all three rooted topologies of Fig.~\ref{fig:topologies} contribute.
Their spatial integrals belong to the family
\begin{equation}
  I^{\mathrm{(rrr)}}_{a_1,a_2,a_3}=\intp{p}\intp{q}\frac{1}{D_{\omega_a}(\bm p)^{a_1}\,D_{\omega_b}(\bm q)^{a_2}\,D_{\omega_c}(\bm p+\bm q)^{a_3}}\,.
\label{eq:memfamily}
\end{equation}
We perform the IBP reduction with $\omega_a$, $\omega_b$ and $\omega_c$ independent,
finding four master integrals
\begin{equation}
\begin{aligned}
  J^{\mathrm{(rrr)}}_1&=I^{\mathrm{(rrr)}}_{1,1,0}=T(\omega_a)\,T(\omega_b)\,,\\
  J^{\mathrm{(rrr)}}_2&=I^{\mathrm{(rrr)}}_{1,0,1}=T(\omega_a)\,T(\omega_c)\,,\\
  J^{\mathrm{(rrr)}}_3&=I^{\mathrm{(rrr)}}_{0,1,1}=T(\omega_b)\,T(\omega_c)\,,\\
  J^{\mathrm{(rrr)}}_4&=I^{\mathrm{(rrr)}}_{1,1,1}\,, \\ 
\end{aligned}
\label{eq:memmasters}
\end{equation}
\\
with $T$ the tadpole integral of Eq.~\eqref{eq:tadpole}.
After IBP reduction, energy conservation imposes $\omega_c=\omega_a+\omega_b$.
The coefficient of the sunrise master $J^{\mathrm{(rrr)}}_4$ then vanishes at generic frequencies.
The kernels are therefore homogeneous polynomials of degree $l_1+l_2+l_3+2$ in the frequencies, that is, local in time.
Each sector contains one term for every multipole that can carry the root, eleven in all.
The five 6PN local actions are,
\begin{widetext}
\begin{align}
\Gamma_{\{\cI_2,\cI_2,\cI_4\}}&=\int\!\frac{\dd\omega_1\dd\omega_2}{(2\pi)^2}\Big[F_{\cI_2^-}\,\cI^{ij}_-(\omega_1)\cI^{kl}_+(\omega_2)\cI^{ijkl}_+(\omega_3)
 +F_{\cI_4^-}\,\cI^{ijkl}_-(\omega_1)\cI^{ij}_+(\omega_2)\cI^{kl}_+(\omega_3)\Big]\,,\nonumber\\
F_{\cI_2^-}&=\frac{G^2}{45360}\Big[-21\,\omega_1^{10}-168\,\omega_1^{9}\omega_2-597\,\omega_1^{8}\omega_2^{2}-1228\,\omega_1^{7}\omega_2^{3}-1597\,\omega_1^{6}\omega_2^{4}+672\,\omega_1^{5}\omega_2^{5}\nonumber\\
&\hspace{1.6cm}+167\,\omega_1^{4}\omega_2^{6}+788\,\omega_1^{3}\omega_2^{7}+537\,\omega_1^{2}\omega_2^{8}+168\,\omega_1\omega_2^{9}+21\,\omega_2^{10}\Big]\,,\nonumber\\
F_{\cI_4^-}&=\frac{G^2}{90720}\Big[-21\,\omega_1^{10}-42\,\omega_1^{9}\omega_2-30\,\omega_1^{8}\omega_2^{2}-20\,\omega_1^{7}\omega_2^{3}-15\,\omega_1^{6}\omega_2^{4}+2010\,\omega_1^{5}\omega_2^{5}\nonumber\\
&\hspace{1.6cm}+9196\,\omega_1^{4}\omega_2^{6}+17640\,\omega_1^{3}\omega_2^{7}+17325\,\omega_1^{2}\omega_2^{8}+8610\,\omega_1\omega_2^{9}+1722\,\omega_2^{10}\Big]\,,\nonumber\\[4pt]
\Gamma_{\{\cI_2,\cI_3,\cI_3\}}&=\int\!\frac{\dd\omega_1\dd\omega_2}{(2\pi)^2}\Big[F_{\cI_2^-}\,\cI^{ij}_-(\omega_1)\cI^{ikl}_+(\omega_2)\cI^{jkl}_+(\omega_3)
 +F_{\cI_3^-}\,\cI^{ikl}_-(\omega_1)\cI^{ij}_+(\omega_2)\cI^{jkl}_+(\omega_3)\Big]\,,\nonumber\\
F_{\cI_2^-}&=\frac{G^2}{22680}\Big[93\,\omega_1^{10}+375\,\omega_1^{9}\omega_2+285\,\omega_1^{8}\omega_2^{2}-940\,\omega_1^{7}\omega_2^{3}-2355\,\omega_1^{6}\omega_2^{4}-2289\,\omega_1^{5}\omega_2^{5}\nonumber\\
&\hspace{1.6cm}-1435\,\omega_1^{4}\omega_2^{6}-150\,\omega_1^{3}\omega_2^{7}+165\,\omega_1^{2}\omega_2^{8}+75\,\omega_1\omega_2^{9}+15\,\omega_2^{10}\Big]\,,\nonumber\\
F_{\cI_3^-}&=\frac{G^2}{2268}\Big[-3\,\omega_1^{10}-15\,\omega_1^{9}\omega_2-33\,\omega_1^{8}\omega_2^{2}-42\,\omega_1^{7}\omega_2^{3}-35\,\omega_1^{6}\omega_2^{4}+147\,\omega_1^{5}\omega_2^{5}\nonumber\\
&\hspace{1.6cm}+411\,\omega_1^{4}\omega_2^{6}+520\,\omega_1^{3}\omega_2^{7}+363\,\omega_1^{2}\omega_2^{8}+135\,\omega_1\omega_2^{9}+21\,\omega_2^{10}\Big]\,,\nonumber \\
\Gamma_{\{\cI_2,\cJ_2,\cJ_2\}}&=\int\!\frac{\dd\omega_1\dd\omega_2}{(2\pi)^2}\Big[F_{\cI_2^-}\,\cI^{ij}_-(\omega_1)\cJ^{ik}_+(\omega_2)\cJ^{jk}_+(\omega_3)
 +F_{\cJ_2^-}\,\cJ^{ik}_-(\omega_1)\cI^{ij}_+(\omega_2)\cJ^{jk}_+(\omega_3)\Big]\,,\nonumber\\
F_{\cI_2^-}&=\frac{G^2}{315}\Big[-180\,\omega_1^{8}-752\,\omega_1^{7}\omega_2-1176\,\omega_1^{6}\omega_2^{2}-832\,\omega_1^{5}\omega_2^{3}-120\,\omega_1^{4}\omega_2^{4}\nonumber\\
&\hspace{1.6cm}-48\,\omega_1^{3}\omega_2^{5}-128\,\omega_1^{2}\omega_2^{6}-96\,\omega_1\omega_2^{7}-24\,\omega_2^{8}\Big]\,,\nonumber\\
F_{\cJ_2^-}&=\frac{G^2}{315}\Big[48\,\omega_1^{8}+192\,\omega_1^{7}\omega_2+192\,\omega_1^{6}\omega_2^{2}-96\,\omega_1^{5}\omega_2^{3}-240\,\omega_1^{4}\omega_2^{4}\nonumber\\
&\hspace{1.6cm}+352\,\omega_1^{3}\omega_2^{5}+672\,\omega_1^{2}\omega_2^{6}+384\,\omega_1\omega_2^{7}+80\,\omega_2^{8}\Big]\,,\nonumber\\[4pt]
\Gamma_{\{\cI_2,\cI_2,\cJ_3\}}&=\int\!\frac{\dd\omega_1\dd\omega_2}{(2\pi)^2}\,\epsilon^{jlm}\Big[F_{\cI_2^-}\,\cI^{ij}_-(\omega_1)\cI^{kl}_+(\omega_2)\cJ^{ikm}_+(\omega_3)
 +F_{\cJ_3^-}\,\cJ^{ikm}_-(\omega_1)\cI^{ij}_+(\omega_2)\cI^{kl}_+(\omega_3)\Big]\,,\nonumber\\
F_{\cI_2^-}&=\frac{iG^2}{840}\Big[-\omega_1^{9}+\omega_1^{8}\omega_2+24\,\omega_1^{7}\omega_2^{2}+76\,\omega_1^{6}\omega_2^{3}+119\,\omega_1^{5}\omega_2^{4}\nonumber\\
&\hspace{1.6cm}+329\,\omega_1^{4}\omega_2^{5}+176\,\omega_1^{3}\omega_2^{6}+44\,\omega_1^{2}\omega_2^{7}+\omega_1\omega_2^{8}-\omega_2^{9}\Big]\,,\nonumber\\
F_{\cJ_3^-}&=\frac{iG^2}{1680}\Big[-\omega_1^{9}-10\,\omega_1^{8}\omega_2-20\,\omega_1^{7}\omega_2^{2}-20\,\omega_1^{6}\omega_2^{3}-15\,\omega_1^{5}\omega_2^{4}\nonumber\\
&\hspace{1.6cm}+218\,\omega_1^{4}\omega_2^{5}+770\,\omega_1^{3}\omega_2^{6}+1000\,\omega_1^{2}\omega_2^{7}+585\,\omega_1\omega_2^{8}+130\,\omega_2^{9}\Big]\,,\nonumber\\[4pt]
\Gamma_{\{\cI_2,\cI_3,\cJ_2\}}&=\int\!\frac{\dd\omega_1\dd\omega_2}{(2\pi)^2}\,\epsilon^{jlm}\Big[F_{\cI_2^-}\,\cI^{ij}_-(\omega_1)\cI^{ikl}_+(\omega_2)\cJ^{km}_+(\omega_3)
 +F_{\cI_3^-}\,\cI^{ikl}_-(\omega_1)\cI^{ij}_+(\omega_2)\cJ^{km}_+(\omega_3)\nonumber\\
&\hspace{2.6cm}+F_{\cJ_2^-}\,\cJ^{km}_-(\omega_1)\cI^{ij}_+(\omega_2)\cI^{ikl}_+(\omega_3)\Big]\,,\nonumber\\
F_{\cI_2^-}&=\frac{iG^2}{945}\Big[-150\,\omega_1^{9}-655\,\omega_1^{8}\omega_2-1080\,\omega_1^{7}\omega_2^{2}-800\,\omega_1^{6}\omega_2^{3}-230\,\omega_1^{5}\omega_2^{4}\nonumber\\
&\hspace{1.6cm}-15\,\omega_1^{4}\omega_2^{5}-20\,\omega_1^{3}\omega_2^{6}-20\,\omega_1^{2}\omega_2^{7}-18\,\omega_1\omega_2^{8}-9\,\omega_2^{9}\Big]\,,\nonumber\\
F_{\cI_3^-}&=\frac{iG^2}{945}\Big[9\,\omega_1^{9}+18\,\omega_1^{8}\omega_2-20\,\omega_1^{6}\omega_2^{3}-15\,\omega_1^{5}\omega_2^{4}+218\,\omega_1^{4}\omega_2^{5}\nonumber\\
&\hspace{1.6cm}+796\,\omega_1^{3}\omega_2^{6}+1080\,\omega_1^{2}\omega_2^{7}+655\,\omega_1\omega_2^{8}+150\,\omega_2^{9}\Big]\,,\nonumber\\
F_{\cJ_2^-}&=\frac{iG^2}{945}\Big[-9\,\omega_1^{9}-63\,\omega_1^{8}\omega_2-180\,\omega_1^{7}\omega_2^{2}-272\,\omega_1^{6}\omega_2^{3}-231\,\omega_1^{5}\omega_2^{4}\nonumber\\
&\hspace{1.6cm}-329\,\omega_1^{4}\omega_2^{5}-118\,\omega_1^{3}\omega_2^{6}-30\,\omega_1^{2}\omega_2^{7}-7\,\omega_1\omega_2^{8}-\omega_2^{9}\Big]\,.
\label{eq:mem6PN}
\end{align}
\end{widetext}
Together with the 5PN sector, they form the local memory action $\Gamma^{\mathrm{loc}}_{\mathrm{mem}}$.
\paragraph*{Conservative memory action.}
\label{sec:inout}
Adopting the Feynman prescription of Sec.~\ref{sec:inoutpresc}, each tadpole in the master integrals~\eqref{eq:memmasters} is replaced by $T_F$ of Eq.~\eqref{eq:tadpoleF},
\begin{equation}
\begin{aligned}
  J^{\mathrm{(rrr)}}_{1,F}&=\sgn(\omega_a)\,\sgn(\omega_b)\,T(\omega_a)\,T(\omega_b)\,,\\
  J^{\mathrm{(rrr)}}_{2,F}&=\sgn(\omega_a)\,\sgn(\omega_c)\,T(\omega_a)\,T(\omega_c)\,,\\
  J^{\mathrm{(rrr)}}_{3,F}&=\sgn(\omega_b)\,\sgn(\omega_c)\,T(\omega_b)\,T(\omega_c)\,.
\end{aligned}
\label{eq:Finout}
\end{equation}
The coefficient of the sunrise vanishes as in the \inin\ case, so each \inout\ kernel is a sum of three terms, in which a homogeneous polynomial multiplies the product of the sign functions of two insertion frequencies. 
The conservative action is the real part of the \inout\ action after symmetrisation over the six relabellings of the three insertions, frequencies and tensor indices included.
Each product of two sign functions splits as
\begin{equation}
\begin{aligned}
    \sgn(\omega_1)\,\sgn(\omega_2)&=-1+2\,\theta(\omega_1)\,\theta(\omega_2)\\
    &\quad+2\,\theta(-\omega_1)\,\theta(-\omega_2)\,,
\end{aligned}
\label{eq:signPB}
\end{equation}
with $\theta$ the Heaviside function.
In the time domain, the sign function is the kernel of the Hilbert transform,
\begin{equation}
\begin{aligned}
    \int\!\frac{\dd\omega}{2\pi}\,e^{-i\omega t}\,\sgn(\omega)\,f(\omega)&=i\,\cH[f](t)\,,\\
    \cH[f](t)&=\frac{1}{\pi}\,\PV\!\int\!\dd t'\,\frac{f(t')}{t'-t}\,,
\end{aligned}
\label{eq:Hilbert}
\end{equation}
and a product of two sign functions acting on two functions $f$ and $g$ of time gives
\begin{equation}
    -\cH[f]\,\cH[g]=-f\,g-\cH\big[f\,\cH[g]+g\,\cH[f]\big]\,.
\label{eq:HHsplit}
\end{equation}
The first term, the constant of Eq.~\eqref{eq:signPB}, is local, while the second, a double principal-value integral, is nonlocal in time~\cite{Porto:2024cwd,Porto:2026fsd}.
Eq.~\eqref{eq:HHsplit} follows from the Poincaré--Bertrand theorem~\cite{Poincare:1910,Bertrand:1923,Muskhelishvili:1953}, and Eq.~\eqref{eq:signPB} is its form in frequency space.
The two nonlocal terms of Eq.~\eqref{eq:signPB} are supported on frequencies of equal sign and vanish when the third frequency, $\omega_3=-\omega_1-\omega_2$, is zero.
The kernels of Eq.~\eqref{eq:mem6PN} and their \inout\ counterparts are given in the ancillary file \texttt{6pn\_actions.m}.
\subsection{Quadratic-curvature source coupling}
\label{sec:e2action}
\begin{figure}[!ht]
\centering
\includegraphics[width=0.46\columnwidth]{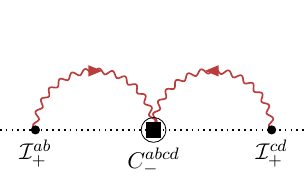}\hfill
\includegraphics[width=0.46\columnwidth]{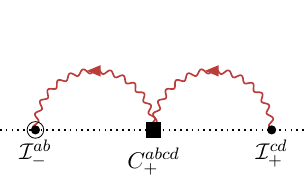}
\caption{Electric quadratic-curvature diagrams with a coefficient root (left) and a quadrupole root (right).
The square denotes the vertex of Fig.~\ref{fig:e2rule}, the root is circled and the arrows point towards it.}
\label{fig:e2vertex}
\end{figure}

The electric vertex~\eqref{eq:E2vertexrule} gives the diagrams of Fig.~\ref{fig:e2vertex}.
For a binary, the Wilson tensor contains no $G$ at leading order and scales as a mass hexadecapole, $C^{abcd}\sim Mr^4$, while its vertex carries four derivatives.
Since each graviton leg carries one factor of $\kappa$, the counting rule~\eqref{eq:PNcount} applies with insertions $\{C,\cI_2,\cI_2\}$, which scale as $v^{14}$ and enter at 6PN.
Summing the coefficient and quadrupole roots gives the \inin\ action
\begin{equation}
\begin{aligned}
    \Gamma_{E^2}=\frac{4G^2}{25}\int\!\dd t\,\Big[&
    C_-^{abcd}\cI_{+,ab}^{(5)}\cI_{+,cd}^{(5)}-2C_+^{abcd}\cI_{-,ab}^{(5)}\cI_{+,cd}^{(5)}\Big]\,.
\end{aligned}
\label{eq:E2causal}
\end{equation}
With Feynman boundary conditions, the conservative action is
\begin{equation}
\begin{aligned}
    S^{\mathrm{cons}}_{E^2}={}&-\frac{4G^2}{25}\int\!\frac{\dd\omega_1\dd\omega_2}{(2\pi)^2}\,
    \sgn(\omega_1)\,\sgn(\omega_2)\,\omega_1^5\,\omega_2^5\\
    &\times C^{abcd}(-\omega_1-\omega_2)\,
    \cI_{ab}(\omega_1)\cI_{cd}(\omega_2)\,.
\end{aligned}
\label{eq:E2generalcons}
\end{equation}

\section{Radiation, balance equations and nonlinear memory}
\label{sec:onepoint}
The one-point function $J_{\mu\nu}$ of Eq.~\eqref{eq:onepointpath}, which generates the far-zone action of Sec.~\ref{sec:results}, also gives the nonlinear gravitational radiation~\cite{Goldberger:2009qd,Ross:2012fc,Leibovich:2019cxo}.
At generic frequencies, the energy and angular momentum carried by the waveform must balance 
the corresponding losses of the \inin\ action, which tests the action.
On scattering orbits, the hereditary part of the angular-momentum flux determines the boundary terms of the memory action, which the diagrams at generic frequencies cannot supply, and its zero-frequency part checks them.
We establish these relations for the sectors relevant through 6PN, and identify the memory of every radiative moment.
\subsection{Radiative one-point function}
\label{sec:Teff}
\begin{figure*}[t]
\centering
\includegraphics[width=0.15\textwidth]{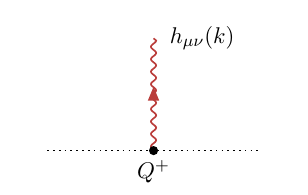}\hfill
\includegraphics[width=0.18\textwidth]{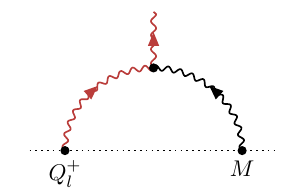}\hfill
\includegraphics[width=0.18\textwidth]{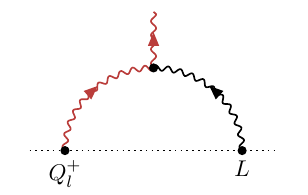}\hfill
\includegraphics[width=0.18\textwidth]{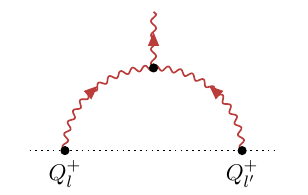}\hfill
\includegraphics[width=0.18\textwidth]{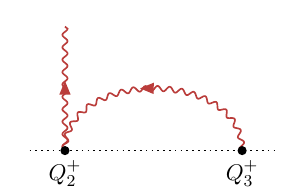}
\caption{
The one-point function of the radiation field.
From left to right, the diagrams show the leading-order current of a single multipole, the cubic bulk interactions of a radiative multipole with the mass, with the angular momentum and with a second radiative multipole, and the two-graviton source vertex.
The arrows point towards the emitted field.
 }
\label{fig:onepoint}
\end{figure*}
Amputating the outgoing propagator of the one-point function~\eqref{eq:onepointpath} gives the effective stress tensor,
\begin{equation}
  T^{\mu\nu}_+(k)=-\frac{2}{\kappa}\,\big[G_{\mathrm{ret}}^{-1}J\big]^{\mu\nu}(k)\,,
\label{eq:Tdef}
\end{equation}
whose diagrams at leading and first nonlinear order are shown in Fig.~\ref{fig:onepoint}.
The emitted graviton has on-shell momentum
\begin{equation}
    k^\mu=\omega(u^\mu+n^\mu)\,,
    \qquad n^\mu=(0,\bm n)\,,
    \qquad \bm n^2=1\,.
\label{eq:onshellk}
\end{equation}
Here $\bm n$ is the direction of propagation.
The transverse-traceless (TT) projection of the stress tensor gives
\begin{equation}
\begin{aligned}
  f_{ij}(\omega,\omega\bm n)&=\Lambda_{ijab}\,T_{ab,+}(\omega,\omega\bm n)\,,\\
\Lambda_{ijab}&=\tfrac12\big(P_{ia}P_{jb}+P_{ib}P_{ja}\big)-\tfrac{1}{d-1}P_{ij}P_{ab}\,,
\end{aligned}
\label{eq:LambdaTT}
\end{equation}
with $P_{ij}=\delta_{ij}-n_in_j$.
In the time domain the physical waveform is
\begin{equation}
    h^{\mathrm{TT}}_{ij}(t,\bm x)=\frac{4G}{|\bm x|}\int\frac{\dd\omega}{2\pi}e^{-i\omega t_{\mathrm{ret}}}f_{ij}(\omega,\omega\bm n)\,,
\end{equation}
with $\bm n=\bm x/|\bm x|$ and $t_{\mathrm{ret}}=t-|\bm x|$ the retarded time.
The leading-order waveform $f^{\lo}$ is generated from the current of a single multipole $T^{\lo}_+$,
represented by the first diagram of Fig.~\ref{fig:onepoint}.
In $d=3$, the result is~\cite{Thorne:1980ru,Blanchet:2013haa}
\begin{equation}
\begin{aligned}
f^{\lo}_{ij}=\Lambda_{ijab}\sum_{l\ge2}\frac{(-i\omega)^l}{l!}\Big[&n_{L-2}\,\cI_{abL-2}(\omega)\\
  &-\tfrac{2l}{l+1}\,n_{cL-2}\,\epsilon_{cd(a}\cJ_{b)dL-2}(\omega)\Big]\,.
\end{aligned}
\label{eq:wavecheck}
\end{equation}
Here $n_{L-2}$ is the product of $l-2$ components of $\bm n$, contracted with the corresponding multipole indices, and $n_{cL-2}=n_cn_{L-2}$.

The nonlinear wave $f^{\hd}$ is generated by a pair of $+$ insertions,
corresponding to the remaining diagrams of Fig.~\ref{fig:onepoint}.
The pairs needed for the 5PN and 6PN action checks are listed in Table~\ref{tab:waves}.
For two radiative insertions $Q_2(\omega_2)$ and $Q_3(\omega_3)$, the emitted wave has frequency $\omega=\omega_2+\omega_3$.
The cubic graph contains the one-loop integral family,
\begin{equation}
  I^{\mathrm{(rr)}}_{a_1,a_2}=\intp{p}\frac{1}{D_{\omega_2}(\mathbf{p})^{a_1}\,D_{\omega_3}(\mathbf{k}-\mathbf{p})^{a_2}}\,,
\label{eq:BubOS}
\end{equation}
with $D_\omega$ as in Eq.~\eqref{eq:Domega} and $\bm k^2=(\omega_2+\omega_3)^2$.
IBP identities reduce the family to two master integrals,
\begin{equation}
\begin{aligned}
    J^{\mathrm{(rr)}}_1 &= I^{\mathrm{(rr)}}_{1,0}=T(\omega_2)\,,\ \ 
    J^{\mathrm{(rr)}}_2 &= I^{\mathrm{(rr)}}_{0,1}=T(\omega_3)\,,
\end{aligned}
\label{eq:BubOSmasters}
\end{equation}
with $T$ defined in Eq.~\eqref{eq:tadpole}.
If one insertion is static, $\omega_3=0$ and only $T(\omega_2)$ survives.
The analytic expressions are collected in the ancillary file \texttt{6pn\_waveforms.m}.
\begin{table}[t]\centering\footnotesize\setlength{\tabcolsep}{3pt}
\begin{ruledtabular}\begin{tabular}{lll}
type & pair & sectors\\
\colrule
tail & $\{M,\cI_4\}$ & $\{M,\cI_4,\cI_4\}$\\
 & $\{M,\cJ_3\}$ & $\{M,\cJ_3,\cJ_3\}$\\
failed tail & $\{L,\cI_2\}$ & $\{L,\cI_2,\cI_2\}$, $\{L,\cI_2,\cJ_3\}$\\
 & $\{L,\cI_3\}$ & $\{L,\cI_3,\cI_3\}$, $\{L,\cI_3,\cJ_2\}$\\
 & $\{L,\cJ_2\}$ & $\{L,\cJ_2,\cJ_2\}$, $\{L,\cI_3,\cJ_2\}$\\
 & $\{L,\cJ_3\}$ & $\{L,\cI_2,\cJ_3\}$\\
memory & $\{\cI_2,\cI_2\}$ & $\{\cI_2,\cI_2,\cI_2\}$, $\{\cI_2,\cI_2,\cI_4\}$,\\
 & & $\{\cI_2,\cI_2,\cJ_3\}$\\
 & $\{\cI_2,\cI_3\}$ & $\{\cI_2,\cI_3,\cI_3\}$, $\{\cI_2,\cI_3,\cJ_2\}$\\
 & $\{\cI_2,\cI_4\}$ & $\{\cI_2,\cI_2,\cI_4\}$\\
 & $\{\cI_3,\cI_3\}$ & $\{\cI_2,\cI_3,\cI_3\}$\\
 & $\{\cJ_2,\cJ_2\}$ & $\{\cI_2,\cJ_2,\cJ_2\}$\\
 & $\{\cI_2,\cJ_2\}$ & $\{\cI_2,\cJ_2,\cJ_2\}$, $\{\cI_2,\cI_3,\cJ_2\}$\\
 & $\{\cI_2,\cJ_3\}$ & $\{\cI_2,\cI_2,\cJ_3\}$\\
 & $\{\cI_3,\cJ_2\}$ & $\{\cI_2,\cI_3,\cJ_2\}$\\
\end{tabular}\end{ruledtabular}
\caption{Next-to-leading-order pair currents used in the 5PN and 6PN
consistency checks. The third column gives the source-level trilinear sectors in
which each pair interferes with the leading-order wave of the remaining
multipole.
}
\label{tab:waves}
\end{table}

\begin{figure}[t]
\centering
\includegraphics[width=0.42\columnwidth]{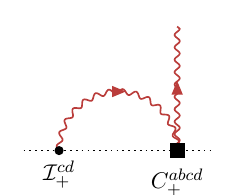}
\caption{Composite-source contribution to the radiative one-point function.
The incoming field is sourced by a mass quadrupole and the arrows point towards the emitted graviton.}
\label{fig:e2onepoint}
\end{figure}

\paragraph*{Composite-source radiation.}
\label{sec:e2wave}
The quadratic-curvature coupling also radiates.
The radiation-reaction field of the quadrupole acts on the composite source through $C^{abcd}$, and the diagram of Fig.~\ref{fig:e2onepoint} couples the emitted graviton to an induced mass quadrupole,
\begin{equation}
    \cI^{ij}_C=-\frac{8G}{5}
    \big(C^{ijcd}\cI_{cd}^{(5)}\big)_{\mathrm{STF}}\,.
\label{eq:E2induced}
\end{equation}
Its waveform contribution is
\begin{equation}
\begin{aligned}
    f^{\hd}_{ij,C}(\omega,\omega\bm n)
    &=\tfrac12(-i\omega)^2\Lambda_{ijab}
    \cI^{ab}_C(\omega)\,,\\
    \delta h^{\mathrm{TT}}_{ij}(t,\bm x)
    &=\frac{2G}{|\bm x|}\Lambda_{ijab}\,
    \cI^{(2)ab}_C(t_{\mathrm{ret}})\,.
\end{aligned}
\label{eq:E2waveform}
\end{equation}
\paragraph*{Ward identities.}
\label{sec:ward}
The waveform~\eqref{eq:LambdaTT} is the TT projection of the effective stress tensor.
The latter has to be conserved.
We therefore compute $k_\mu T^{\mu\nu}_+(k)$ for every pair current of Table~\ref{tab:waves}, at generic off-shell momentum and before the TT projection, including the two-graviton worldline vertices.
At leading order, each radiative multipole current is conserved separately.
At next-to-leading order the source loses energy and angular momentum through the leading-order radiation, so $M$ and $\bm L$ become time dependent.
The current of a pair of identical multipoles, which carries these losses, is then conserved only together with the leading-order currents of $M$ and $\bm L$.
These are the one-point functions of the vertices $V^{(1)}_M$ and $V^{(1)}_L$ of Eq.~\eqref{eq:V1}, the stress tensors of a point mass and a point spin at rest at the origin, with $M(t)$ and $\bm L(t)$ time dependent.
The rates of $M$ and $\bm L$ follow from the radiation-reaction action~\eqref{eq:RR} of that pair~\cite{Porto:2024cwd},
\begin{equation}
\begin{aligned}
    \dot M&=-G(-1)^l c_l\,Q^{(1)}_LQ^{(2l+1)}_L\,,\\
    \dot L^i&=-G(-1)^l\,l\,c_l\,\epsilon^{ijk}Q_{jL-1}Q^{(2l+1)}_{kL-1}\,,
\end{aligned}
\label{eq:wardlosses}
\end{equation}
with $Q_l$ the multipole of the pair, and $c_l\to c_l^J$ for a current multipole.
They differ from the fluxes of Ref.~\cite{Thorne:1980ru} by total time derivatives, the Schott terms, which vanish on scattering histories and upon averaging over a period.
With these currents, every pair current of Table~\ref{tab:waves} except $\{L,\cJ_2\}$, $\{\cI_2,\cI_3\}$ and $\{\cI_2,\cJ_2\}$ is conserved,
\begin{equation}
    k_\mu T^{\mu\nu}_+(k)=0\,.
\label{eq:ward}
\end{equation}
For $\{L,\cJ_2\}$, the radiation-reaction field of $\cJ_2$ exerts a force on the spinning source,
\begin{equation}
    \partial_\mu T^{\mu i}_+(x)=\frac{8G}{15}\,\cJ^{(5)}_{ij}(t)\,L^j\,\delta^3(\bm x)\,,
\label{eq:wardLJ2}
\end{equation}
while the time component is conserved.
Since $\bm L$ is constant at this order, this is a total time derivative.
It changes the momentum of the source, which we keep at the origin, but not its energy or angular momentum.
For $\{\cI_2,\cI_3\}$ and $\{\cI_2,\cJ_2\}$, the divergence is local and is the flux of linear momentum carried by the waves~\cite{Thorne:1980ru}, with the corresponding change of the mass dipole in the time component.
It is cancelled by the leading-order currents of the linear momentum and the mass dipole of the source, which we omit since we neglect recoil~\cite{Porto:2024cwd}.
\subsection{Fluxes and balance relations}
\label{sec:fluxes}
The waveform of Sec.~\ref{sec:Teff} determines the energy and angular momentum radiated to infinity, which are the mass and angular momentum lost by the source~\cite{Thorne:1980ru}.
\paragraph*{Fluxes from the waveform.}
At leading order the integrated losses are quadratic in $f^{\lo}$,
\begin{align}
  \Delta M^{\lo}&=\frac{\kappa^2}{64\pi^2}\int\!\dd\Omega\int\!\frac{\dd\omega}{2\pi}\,
  \omega\,(-\omega)\,f^{\lo}_{ij}(\omega)\,f^{\lo}_{ij}(-\omega)\,,
\label{eq:MdotfluxLO}\\
  \Delta L^{ij,\lo}&=\frac{i\kappa^2}{64\pi^2}\int\!\dd\Omega\int\!\frac{\dd\omega}{2\pi}\,
  (-2\omega)\Big[2f^{\lo}_{l[i}(\omega)\,f^{\lo}_{j]l}(-\omega)\nonumber\\
  &\qquad\qquad-f^{\lo}_{kl}(-\omega)\,n_{[i}\tilde\partial_{j]}f^{\lo}_{kl}(\omega)\Big]\,,
\label{eq:LdotfluxLO}
\end{align}
where $\tilde\partial_j=\partial/\partial n^j$ acts on the angular dependence of the wave.
At next-to-leading order they are given by the interference of $f^{\lo}$ with $f^{\hd}$,
\begin{align}
  \Delta M^{\hd}&=\frac{\kappa^2}{32\pi^2}\int\!\dd\Omega\int\!\frac{\dd\omega}{2\pi}\,
  \omega\,(-\omega)\,f^{\lo}_{ij}(\omega)\,f^{\hd}_{ij}(-\omega)\,,
\label{eq:MdotfluxNLO}\\
  \Delta L^{ij,\hd}&=\frac{i\kappa^2}{32\pi^2}\int\!\dd\Omega\int\!\frac{\dd\omega}{2\pi}\,
  (-2\omega)\Big[2f^{\lo}_{l[i}(\omega)\,f^{\hd}_{j]l}(-\omega)\nonumber\\
  &\qquad\qquad-f^{\hd}_{kl}(-\omega)\,n_{[i}\tilde\partial_{j]}f^{\lo}_{kl}(\omega)\Big]\,.
\label{eq:LdotfluxNLO}
\end{align}
The angular integrals are performed in $d$ dimensions.
We express the angular-momentum loss along an arbitrary constant vector $\bm a$,
\begin{equation}
  \Delta\bm L\cdot\bm a=\tfrac12\,\epsilon_{ijc}\,a^c\,\Delta L^{ij}\,.
\label{eq:Lproj}
\end{equation}
This permits a direct comparison with the rotations of the source multipoles.
The next-to-leading-order losses take the form
\begin{equation}
\begin{aligned}
    \Delta X^{\hd}&=\int\!\frac{\dd\omega_1\dd\omega_2}{(2\pi)^2}\,
    K^X[Q_1,Q_2,Q_3]\,,
\end{aligned}
\label{eq:fluxkernels}
\end{equation}
where $X=M,L$, $\omega_3=-\omega_1-\omega_2$, $Q_s=Q_s(\omega_s)$.
The kernel $K^X$ is a scalar linear in each of the three multipoles, 
with frequency-dependent coefficients.
For $X=L$, $\Delta X$ denotes $\Delta\bm L\cdot\bm a$, and $K^L$ is also linear in $\bm a$.
At generic frequencies on the surface $\omega_1+\omega_2+\omega_3=0$, the failed-tail and memory kernels are finite polynomials.
The unsymmetrised tail kernels also contain $1/\eps$ poles and frequency logarithms.
For identical multipoles, the physical kernel is obtained by permuting their frequency labels and tensor indices together.
We denote this symmetrisation by $\cS$.
\paragraph*{Flux kernels.}
\label{sec:fluxresults}
At leading order, the losses are the multipolar fluxes of Ref.~\cite{Thorne:1980ru}, 
which the waves~\eqref{eq:wavecheck} reproduce for every multipole used here,
\begin{equation}
\begin{aligned}
    \Delta M^{\lo}&=-G\sum_{l\ge2}c_l\int\!\dd t\,Q^{(l+1)}_{L}Q^{(l+1)}_{L}\,,\\
    \Delta L^{i,\lo}&=-G\sum_{l\ge2}l\,c_l\,\epsilon^{ijk}\int\!\dd t\,Q^{(l)}_{jL-1}Q^{(l+1)}_{kL-1}\,,
\end{aligned}
\label{eq:Thornefluxes}
\end{equation}
with $c_l$ of Eq.~\eqref{eq:cl}, replaced by $c_l^J$ for a current multipole.
In frequency space, each multipole contributes the kernel
\begin{equation}
\begin{aligned}
    \Delta X^{\lo}&=\int\!\frac{\dd\omega}{2\pi}\,K^X_{\lo}[Q]\,,\\
    K^M_{\lo}[Q]&=-Gc_l\,\omega^{2l+2}\,Q_L(\omega)Q_L(-\omega)\,,\\
    K^L_{\lo}[Q]&=-iG\,l\,c_l\,\omega^{2l+1}\,\epsilon^{njk}a^n Q_{jL-1}(\omega)Q_{kL-1}(-\omega)\,.
\end{aligned}
\label{eq:LOkernels}
\end{equation}
The nonlinear kernels come in three classes.
In the tail sectors $\{M,Q_l,Q_l\}$, the symmetrised kernel is the leading-order one multiplied by a universal frequency factor~\cite{Blanchet:1992br,Goldberger:2009qd},
\begin{equation}
    \Delta X^{\hd}_{\mathrm{tail}}=\int\!\frac{\dd\omega}{2\pi}\,
    2\pi GM|\omega|\,K^X_{\lo}[Q_l]\,.
\label{eq:tailcheck}
\end{equation}
In the failed-tail sectors, the energy kernel integrates to zero after the sum over roots and the symmetrisation, 
while the angular-momentum kernel does not~\cite{Porto:2024cwd,Almeida:2026clf}.
In the memory sectors, both kernels are nonzero: at 5PN they agree with those of Ref.~\cite{Porto:2024cwd},
and the five 6PN sectors give new ones.
The kernels of all eleven 6PN sectors are given in the ancillary file \texttt{6pn\_fluxes.m}.

\paragraph*{Balance with the effective action.}
\label{sec:balance}
The same losses follow from the \inin\ effective action of Sec.~\ref{sec:results} by varying it with respect to the $-$ multipoles.
For a contribution rooted on the multipole $Q_{s}$, the force conjugate to it is
\begin{equation}
    \mathcal F_{s}=\left.\frac{\delta\Gamma}{\delta Q_{s,-}}\right|_{\mathrm{PL}}\,.
\label{eq:Kr}
\end{equation}
After integration by parts, time translations and rotations give the source losses~\cite{Galley:2012hx},
\begin{equation}
\begin{aligned}
    \dot M_{\mathrm{near}}&\simeq\sum_{s}
    \dot Q_{s}\cdot\mathcal F_{s}\,,\\
    \dot L^n_{\mathrm{near}}&\simeq\sum_{s}
    (\cR_nQ_{s})\cdot\mathcal F_{s}\,,
\end{aligned}
\label{eq:nearlosses}
\end{equation}
where $\simeq$ denotes equality up to local total time derivatives and $\cR_n$ rotates every index of the multipole,
\begin{equation}
    (\cR_nQ)_{i_1\cdots i_l}=\sum_{k=1}^{l}\epsilon^{i_{k}nm}\,Q_{i_1\cdots m\cdots i_l}\,.
\label{eq:rotationop}
\end{equation}
Integrating Eq.~\eqref{eq:nearlosses} over time, and using that the action~\eqref{eq:Gammadef} is linear in its root, the losses are the action with the root replaced by its time derivative or by its rotation,
\begin{equation}
\begin{aligned}
    \Delta M_{\mathrm{near}}&=\sum_{s}\int\!\dd t\,\dot Q_{s}\cdot\mathcal F_{s}
    =\Gamma\big|_{Q_{s,-}\to\dot Q_{s}}\,,\\
    \Delta\bm L_{\mathrm{near}}\cdot\bm a&=\sum_{s}\int\!\dd t\,(\cR_{\bm a}Q_{s})\cdot\mathcal F_{s}
    =\Gamma\big|_{Q_{s,-}\to\cR_{\bm a}Q_{s}}\,,
\end{aligned}
\label{eq:nearlossint}
\end{equation}
with $\cR_{\bm a}=a^n\cR_n$.
In frequency space, this gives the losses in the form of Eq.~\eqref{eq:fluxkernels}, with kernels
\begin{equation}
\begin{aligned}
    K^M_{\mathrm{near}}&=\sum_{s}(-i\omega_{s})\Big[Q_{s}(\omega_{s})\prod_{s'\neq s}Q_{s'}(\omega_{s'})\Big]F_{s}\,,\\
    K^L_{\mathrm{near}}&=\sum_{s}\Big[(\cR_{\bm a}Q_{s})(\omega_{s})\prod_{s'\neq s}Q_{s'}(\omega_{s'})\Big]F_{s}\,,
\end{aligned}
\label{eq:nearkernels}
\end{equation}
where $F_{s}$ is the kernel of Eq.~\eqref{eq:Gammadef} for the terms rooted in slot $s$.
After symmetrisation of the identical insertions, balance requires
\begin{equation}
    \cS\big[K^X-K^X_{\mathrm{near}}\big]=0\,,\qquad X=M,L\,.
\label{eq:balancecheck}
\end{equation}
The two sides agree, for energy and for angular momentum, in every $\cO(G^2)$ sector of Table~\ref{tab:sectors} through 6PN, at generic frequencies on the surface $\omega_1+\omega_2+\omega_3=0$, where local total time derivatives do not contribute.
\paragraph*{Composite source.}
The induced quadrupole~\eqref{eq:E2induced} interferes with the leading quadrupolar wave, giving
\begin{equation}
\begin{aligned}
    \Delta M_C&=-\frac{2G}{5}\int\!\dd t\,
    \cI^{(3)}_{ij}\cI^{(3)}_{C,ij}\,,\\
    \Delta L_C^n&=-\frac{2G}{5}\int\!\dd t\,\epsilon^{njk}
    \Big[\cI^{(2)}_{j\ell}\cI^{(3)}_{C,k\ell}+\cI^{(2)}_{C,j\ell}\cI^{(3)}_{k\ell}\Big]\,.
\end{aligned}
\label{eq:E2fluxes}
\end{equation}
These losses balance those generated by Eq.~\eqref{eq:E2causal}, up to local Schott terms.
Their kernel is polynomial and has no pole at zero frequency, so the composite source gives no zero-frequency flux.
\subsection{Nonlinear memory: boundary terms}
\label{sec:soft}
The balance relations of Sec.~\ref{sec:fluxes} hold at generic frequencies, where the memory action is the local action $\Gamma^{\mathrm{loc}}_{\mathrm{mem}}$ of Sec.~\ref{sec:memory}.
On scattering orbits the moments do not return to their initial values.
The memory action acquires a boundary term $\Gamma^{\mathrm{bdy}}$, nonlocal in time,
\begin{equation}
    \Gamma_{\mathrm{mem}}=\Gamma^{\mathrm{loc}}_{\mathrm{mem}}+\Gamma^{\mathrm{bdy}}\,.
\label{eq:memsplit}
\end{equation}
We derive it here from the radiated angular momentum, for the five 6PN memory sectors, extending the 5PN analysis of Ref.~\cite{Porto:2024cwd}.
\paragraph*{Boundary term.}
The memory kernels of Sec.~\ref{sec:memory} are computed on the surface $\omega_1+\omega_2+\omega_3=0$, where two polynomials that differ by a multiple of $\omega_1+\omega_2+\omega_3$ coincide.
In the time domain, such a multiple is the total time derivative of a product of the moments.
The kernels therefore determine the local action only up to total time derivatives.
The integral of a total time derivative is a local expression in the moments, evaluated at the initial and final times, and vanishes upon averaging over a period.
A boundary term that depends on the whole history of the moments, such as the memory, is not of this form, so the kernels do not fix it.
To parametrise this boundary term, consider a functional $U^{\mathrm{bdy}}_{ij}[Q_2,Q_3]$, quadratic in the multipoles $Q_2$ and $Q_3$, whose time derivative is local and which satisfies $U^{\mathrm{bdy}}_{ij}(-\infty)=0$.
Integration by parts gives~\cite{Porto:2024cwd}
\begin{equation}
\begin{aligned}
    \int\!\dd t\,\ddot{\cI}^{ij}_-\,\dot U^{\mathrm{bdy}}_{ij}
    ={}&-\int\!\dd t\,\cI^{(3)ij}_-\,U^{\mathrm{bdy}}_{ij}+\Delta\ddot{\cI}^{ij}_-\,U^{\mathrm{bdy}}_{ij}(+\infty)\,,
\end{aligned}
\label{eq:memoryboundary}
\end{equation}
with $\Delta\ddot{\cI}^{ij}_-=\ddot{\cI}^{ij}_-(+\infty)-\ddot{\cI}^{ij}_-(-\infty)$.
The last term is the boundary term~\cite{Manohar:2022dea,DiVecchia:2022owy,Riva:2023xxm}: it leaves the bulk equations of motion unchanged but contributes to the canonical charges, the momentum and angular momentum of the binary at late times.
They follow by varying the action with respect to the final values of the difference coordinates $\bm x_{a,-}$, whose initial values and velocities vanish, so that $\Delta\ddot{\cI}^{ij}_-=\ddot{\cI}^{ij}_-(+\infty)$.
In the physical limit, the boundary term contributes to the angular momentum~\cite{Porto:2024cwd}
\begin{equation}
    L^\ell_{\mathrm{nloc}}=\epsilon^{\ell mn}\sum_a x^m_{a,+}\,
    \frac{\partial\Gamma^{\mathrm{bdy}}}{\partial x^n_{a,-}}\Big|_{\mathrm{PL}}\,.
\label{eq:Lnlocvariation}
\end{equation}
For free asymptotic motion, $\bm x_a(T)\to\bm v_aT$.
The derivative in Eq.~\eqref{eq:Lnlocvariation} then rotates the final difference coordinates, under which $\ddot{\cI}^{ij}_-$ changes by $(\cR_\ell\ddot{\cI})^{ij}$, with $\cR_\ell$ of Eq.~\eqref{eq:rotationop}.
This gives
\begin{equation}
    L^{\ell}_{\mathrm{nloc}}
    =2\,\epsilon^{\ell mn}\,\ddot{\cI}^{mj}(+\infty)\,
    U^{\mathrm{bdy}}_{nj}(+\infty)\,.
\label{eq:Lnlocgeneral}
\end{equation}
We fix $U^{\mathrm{bdy}}_{ij}$ by requiring that this charge balance the angular momentum carried away by the nonlinear memory of the radiation.
\paragraph*{Radiative memory.}
The radiative mass quadrupole $U_{ij}$, which equals $\ddot{\cI}_{ij}$ at leading order, receives a memory contribution from the waves emitted by two multipoles $Q_2$ and $Q_3$.
Their interference makes the radiated energy direction dependent, and its quadrupolar component gives~\cite{Thorne:1992sdb,Favata:2008yd}
\begin{equation}
    U^{\mathrm{mem}}_{ij}(t)=\frac52\int_{-\infty}^{t}\dd t'\int\dd\Omega_{\bm n'}\,
    n'_{\langle ij\rangle}\,\frac{\dd E}{\dd t'\dd\Omega_{\bm n'}}\,,
\label{eq:quadrupolememory}
\end{equation}
with $\dd E/(\dd t\,\dd\Omega_{\bm n})$ the angular distribution of the radiated energy.
For two mass quadrupoles, it is the Christodoulou memory~\cite{Christodoulou:1991cr,Wiseman:1991ss},
\begin{equation}
    U^{\mathrm{mem}}_{ij}[\cI_2,\cI_2]=-\frac{2G}{7}\int_{-\infty}^{t}\dd t'\,
    \cI^{(3)}_{a\langle i}(t')\,\cI^{(3)}_{j\rangle a}(t')\,.
\label{eq:quadmemI2I2}
\end{equation}
For two generic multipoles, the same projection gives $U^{\mathrm{mem}}_{ij}[Q_2,Q_3]$ in closed form for every $l$, as discussed in App.~\ref{sec:angproj}.
The higher-rank radiative moments $U_L$ receive analogous contributions $U^{\mathrm{mem}}_L$, which we discuss in Sec.~\ref{sec:DC}.
\paragraph*{Hereditary flux.}
The angular momentum lost to the memory follows from the flux of Eq.~\eqref{eq:Thornefluxes}, which in terms of the radiative mass moments $U_L$ reads
\begin{equation}
    \dot L^i=-G\sum_{l\ge2}l\,c_l\,\epsilon^{ijk}\,U_{jL-1}\,\dot U_{kL-1}\,.
\label{eq:fluxradiative}
\end{equation}
Expanding $U_L=\cI_L^{(l)}+U_L^{\mathrm{mem}}$ to first order in the memory we get,
\begin{equation}
\begin{aligned}
    \dot L^i_{\mathrm{mem}}=-G\sum_{l\ge2}l\,c_l\,
    \epsilon^{iab}\Big[&
    \cI_{aL-1}^{(l)}\dot U_{bL-1}^{\mathrm{mem}}+U_{aL-1}^{\mathrm{mem}}\cI_{bL-1}^{(l+1)}
    \Big]\,.
\end{aligned}
\label{eq:memoryfluxdictionary}
\end{equation}
Its hereditary part is a total time derivative,
\begin{equation}
\begin{aligned}
    \dot L^i_{\mathrm{mem}}&=-2G\sum_{l\ge2}l\,c_l\,\epsilon^{iab}\,\cI_{aL-1}^{(l)}\dot U_{bL-1}^{\mathrm{mem}}+\frac{\dd}{\dd t}L^i_{\mathrm{nloc}}\,,\\
    L^i_{\mathrm{nloc}}&=G\sum_{l\ge2}l\,c_l\,\epsilon^{iab}\,\cI_{aL-1}^{(l)}\,U_{bL-1}^{\mathrm{mem}}\,.
\end{aligned}
\label{eq:memoryendpointbalance}
\end{equation}
The first term is local, since $\dot U^{\mathrm{mem}}_L$ is a product of differentiated source moments.
Together with the other local terms of the nonlinear flux, it is the loss balanced by the local action discussed in Sec.~\ref{sec:balance}.
The second term is the time derivative of $L^i_{\mathrm{nloc}}$, which depends on the whole history through $U^{\mathrm{mem}}_L$ and cannot come from the local action.
Since $U^{\mathrm{mem}}_L(-\infty)=0$, it integrates to $L^i_{\mathrm{nloc}}(+\infty)$.
For the quadrupole, this endpoint charge reads
\begin{equation}
    L^{\ell}_{\mathrm{nloc}}=\frac{2G}{5}\,\epsilon^{\ell mn}\,\ddot\cI^{mj}(+\infty)\,
    U^{\mathrm{mem}}_{nj}[Q_2,Q_3](+\infty)\,.
\label{eq:Lnloc}
\end{equation}
\paragraph*{Memory coefficients.}
Equating Eqs.~\eqref{eq:Lnlocgeneral} and~\eqref{eq:Lnloc} fixes the late-time value of $U^{\mathrm{bdy}}_{ij}$, the only value that enters $\Gamma^{\mathrm{bdy}}$.
We can therefore take
\begin{equation}
    U^{\mathrm{bdy}}_{ij}(t)
    =\frac{G}{5}\,U^{\mathrm{mem}}_{ij}[Q_2,Q_3](t)\,,
\label{eq:Ufix}
\end{equation}
so the boundary term of Eq.~\eqref{eq:memoryboundary} becomes
\begin{equation}
    \Gamma^{\mathrm{bdy}}_{\{\cI_2,Q_2,Q_3\}}
    =\frac{G}{5}\,\Delta\ddot{\cI}^{ij}_-\,
    U^{\mathrm{mem}}_{ij}[Q_2,Q_3](+\infty)\,.
\label{eq:bdymemory}
\end{equation}
This is the source quadrupole coupled to the memory part of the radiative quadrupole at the end of the scattering.
Since the memory requires only the leading-order waves, Eq.~\eqref{eq:bdymemory} gives the quadrupole-root boundary term of each memory sector considered here.
The equality of Eqs.~\eqref{eq:Lnlocgeneral} and~\eqref{eq:Lnloc} holds for every scattering history, along which $\ddot{\cI}_{ij}(+\infty)$ and $U^{\mathrm{bdy}}_{ij}(+\infty)$ vary independently.
It therefore fixes the trace-free tensor $U^{\mathrm{bdy}}_{ij}(+\infty)$, and with it the boundary term, uniquely.
On scattering orbits, $L^\ell_{\mathrm{nloc}}$ vanishes at $\cO(G^4)$, in agreement with Refs.~\cite{Georgoudis:2026zzn,Bohnenblust:2026ytj}, and at $\cO(G^5)$ it equals minus the memory contribution to the radiated angular momentum of Ref.~\cite{Bini:2022enm}.
The construction does not depend on the rank of the root.
For a mass-type root of rank $l$, $\cI^{(l)}_{L,-}$ tends to a constant at late times, and the rank-$l$ term of Eq.~\eqref{eq:memoryendpointbalance} requires
\begin{equation}
    \Gamma^{\mathrm{bdy}}_{\{\cI_l,Q_2,Q_3\}}
    =G\,c_l\,\Delta\cI^{(l)}_{L,-}\,
    U^{\mathrm{mem}}_{L}[Q_2,Q_3](+\infty)\,,
\label{eq:bdymemoryl}
\end{equation}
which reduces to Eq.~\eqref{eq:bdymemory} for $l=2$.
At 6PN, the octupole roots of $\{\cI_2,\cI_3,\cI_3\}$ and $\{\cI_2,\cI_3,\cJ_2\}$ and the hexadecapole root of $\{\cI_2,\cI_2,\cI_4\}$ also contribute, with the memories $U^{\mathrm{mem}}_{ijk}$ and $U^{\mathrm{mem}}_{ijkl}$.
In frequency space, the projections of App.~\ref{sec:angproj} give
\begin{equation}
\begin{aligned}
    \Gamma^{\mathrm{bdy}}_{\{\cI_2,Q_2,Q_3\}}
    &=-\frac{1}{2}\,\varrho\,\Delta\ddot{\cI}^{ij}_-
    \int\frac{\dd\omega}{2\pi}\,\omega^{\,\beta}\,\big[Q_2(\omega)Q_3(-\omega)\big]_{ij}\,,\\
    \beta&=l_2+l_3+2\,,
\end{aligned}
\label{eq:bdyfreq}
\end{equation}
where $[Q_2Q_3]_{ij}$ is the rank-two contraction of the two radiative multipoles of each sector, and $\varrho$ collects the numerical factors and phases.
For the quadrupole roots of the 5PN memory sector and of the five 6PN sectors,
\begin{widetext}
\begin{align}
\Gamma^{\mathrm{bdy}}_{\{\cI_2,\cI_2,\cI_2\}}&=-\frac{2G^2}{35}\,\Delta\ddot{\cI}^{ij}_-
 \int\!\frac{\dd\omega}{2\pi}\,\omega^6\,\cI^{jk}_+(\omega)\cI^{ki}_+(-\omega)\,,\nonumber\\
\Gamma^{\mathrm{bdy}}_{\{\cI_2,\cI_2,\cI_4\}}&=-\frac{G^2}{756}\,\Delta\ddot{\cI}^{ij}_-
 \int\!\frac{\dd\omega}{2\pi}\,\omega^8\,\cI^{kl}_+(\omega)\cI^{ijkl}_+(-\omega)\,,\nonumber\\
\Gamma^{\mathrm{bdy}}_{\{\cI_2,\cI_3,\cI_3\}}&=0\,,\nonumber\\
\Gamma^{\mathrm{bdy}}_{\{\cI_2,\cJ_2,\cJ_2\}}&=-\frac{32G^2}{315}\,\Delta\ddot{\cI}^{ij}_-
 \int\!\frac{\dd\omega}{2\pi}\,\omega^6\,\cJ^{ik}_+(\omega)\cJ^{jk}_+(-\omega)\,,\nonumber\\
\Gamma^{\mathrm{bdy}}_{\{\cI_2,\cI_2,\cJ_3\}}&=\frac{i\,G^2}{42}\,\Delta\ddot{\cI}^{ij}_-
 \int\!\frac{\dd\omega}{2\pi}\,\omega^7\,\epsilon^{jlm}\,\cI^{kl}_+(\omega)\cJ^{ikm}_+(-\omega)\,,\nonumber\\
\Gamma^{\mathrm{bdy}}_{\{\cI_2,\cI_3,\cJ_2\}}&=-\frac{4i\,G^2}{189}\,\Delta\ddot{\cI}^{ij}_-
 \int\!\frac{\dd\omega}{2\pi}\,\omega^7\,\epsilon^{jlm}\,\cI^{ikl}_+(\omega)\cJ^{km}_+(-\omega)\,.
\label{eq:SFL6PN}
\end{align}
\end{widetext}
For $\{\cI_2,\cI_2,\cI_2\}$, Eq.~\eqref{eq:SFL6PN} reproduces the 5PN boundary term of Ref.~\cite{Porto:2024cwd}.
\subsection{Zero-frequency fluxes and memory of the radiative moments}
\label{sec:DC}
The one-point function of Sec.~\ref{sec:Teff} gives an independent computation of the zero-frequency part of the flux.
It allows us to recompute the coefficients of Eq.~\eqref{eq:SFL6PN}, and to identify the memory of every radiative moment through 6PN.
\paragraph*{Zero-frequency flux.}
In each memory sector, the flux at next-to-leading order is the interference of two waves, Eqs.~\eqref{eq:MdotfluxNLO} and~\eqref{eq:LdotfluxNLO}.
One is the leading-order wave of a single multipole $Q_1(\omega_1)$, the root of the action, and the other is the nonlinear wave generated by $Q_2(\omega_2)$ and $Q_3(\omega_3)$, with frequency $\varpi=\omega_2+\omega_3$, equal to $-\omega_1$ in the integrated flux.
Unlike the action, the flux is singular as $\varpi\to0$.
The nonlinear wave contains the memory, whose retarded time integral, Eq.~\eqref{eq:quadrupolememory}, gives a pole in frequency space,
\begin{equation}
    U^{\mathrm{mem}}_{ij}(\varpi)=\frac{i}{\varpi+i0}\,\frac52\int\dd\Omega_{\bm n'}\,n'_{\langle ij\rangle}\,\frac{\dd E}{\dd t\,\dd\Omega_{\bm n'}}(\varpi)\,.
\label{eq:memorypole}
\end{equation}
The flux kernel $K^X$ of Eq.~\eqref{eq:fluxkernels} inherits this pole, and we define its residue $R^{\mathrm{flux}}_X$ by
\begin{equation}
    K^X[Q_1,Q_2,Q_3]=\frac{R^{\mathrm{flux}}_X[Q_1,Q_2,Q_3]}{\varpi+i0}
    +\text{regular}\,.
\label{eq:fluxpole}
\end{equation}
For a mass-quadrupole root, $Q_1=\cI$.
On a scattering orbit $\cI^{(3)}$ falls off as $1/t^2$ at early and late times, and its Fourier transform at zero frequency is its time integral, $\Delta\ddot{\cI}$.
Then $\cI(\omega_1)$ is singular as $\omega_1\to0$, with
\begin{equation}
    \lim_{\omega\to0}\omega^3\cI^{ij}(\omega)=i^3\,\Delta\ddot{\cI}^{ij}\,,
\label{eq:softjump}
\end{equation}
and the pole does not cancel.
We can use the distributional identity
\begin{equation}
    \frac{1}{\varpi+i0}=\PV\frac{1}{\varpi}-i\pi\delta(\varpi)
\label{eq:PVdecomposition}
\end{equation}
to separate the principal-value part from the delta-function term.
The latter gives the zero-frequency, or DC, part of the flux,
\begin{equation}
\begin{aligned}
    \Delta X_{\mathrm{DC}}={}&-\frac{i}{2}\int\frac{\dd\omega}{2\pi}\\
    &\quad\times\lim_{\omega_1\to0}R^{\mathrm{flux}}_X\big[Q_1(\omega_1),Q_2(\omega),Q_3(-\omega)\big]\,.
\end{aligned}
\label{eq:DCflux}
\end{equation}
In the energy flux~\eqref{eq:MdotfluxLO}, a time derivative acts on each wave.
The resulting factor $\omega_1\varpi$ cancels the pole, so the zero-frequency field carries no energy, 
\begin{equation}
    R^{\mathrm{flux}}_M=0\,.
\end{equation}
\paragraph*{Mass-quadrupole root.}
For a mass-quadrupole root, the angular-momentum flux residue is the coefficient of the boundary term~\eqref{eq:bdyfreq}, with the root rotated,
\begin{equation}
\begin{aligned}
    R^{\mathrm{flux}}_L[Q_1,Q_2,Q_3]&=\tfrac12\,\varrho\,\omega_1^3\,\omega_2^{\,\beta}\,
    (\cR_{\bm a}Q_1)^{ij}\big[Q_2Q_3\big]_{ij}\,.
\end{aligned}
\label{eq:DCcheck}
\end{equation}
This holds for the 5PN sector and for the five 6PN sectors.
Inserting this residue in Eq.~\eqref{eq:DCflux} gives
\begin{equation}
    \Delta L^\ell_{\mathrm{DC}}
    =\frac{G}{5}\,\epsilon^{\ell mn}\,
    \Delta\ddot{\cI}^{mj}\,
    U^{\mathrm{mem}}_{nj}[Q_2,Q_3](+\infty)\,.
\label{eq:DCendpointcheck}
\end{equation}
The zero-frequency angular-momentum flux of the one-point function therefore reproduces the coefficients of the boundary terms, independently of the angular projections of App.~\ref{sec:angproj}.
It equals the charge~\eqref{eq:Lnloc} with $\ddot{\cI}(+\infty)$ replaced by $\tfrac12\Delta\ddot{\cI}$, and the principal-value part of Eq.~\eqref{eq:PVdecomposition} supplies the rest of the charge together with the local term of Eq.~\eqref{eq:memoryendpointbalance}.
\paragraph*{Higher-rank roots.}
The waveform calculation also determines the residues when the leading-order wave is a mass octupole or hexadecapole.
For a mass-type root of rank $l$, the residue scales as $\omega_1^{l+1}$.
On a scattering orbit $Q_L^{(l+1)}$ falls off as $1/t^2$ at early and late times, and the Fourier transform gives
\begin{equation}
\begin{aligned}
    \lim_{\omega\to0}\omega^{l+1}Q_L(\omega)
    &=i^{l+1}\int_{-\infty}^{+\infty}\dd t\,Q_L^{(l+1)}(t)=i^{l+1}\Delta Q_L^{(l)}\,,
\end{aligned}
\label{eq:softjumpgeneral}
\end{equation}
which generalises Eq.~\eqref{eq:softjump} and makes the limit in Eq.~\eqref{eq:DCflux} finite.
The zero-frequency loss is then the product of the asymptotic change $\Delta Q_L^{(l)}$ of the leading-order wave and the nonlinear memory generated by the other two moments,
\begin{equation}
    \Delta L^\ell_{\mathrm{DC}}
    =\frac{l}{2}\,G\,c_l\,\epsilon^{\ell mn}\,
    \Delta Q^{(l)}_{mL-1}\,
    U^{\mathrm{mem}}_{nL-1}[Q_2,Q_3](+\infty)\,,
\label{eq:DCendpointcheckl}
\end{equation}
which reduces to Eq.~\eqref{eq:DCendpointcheck} for $l=2$.
Matching it to the zero-frequency part of Eq.~\eqref{eq:memoryfluxdictionary} identifies the memory contribution of each pair.
The pairs of the 5PN and 6PN memory sectors contribute to $U_{ij}$, the pairs $\{\cI_2,\cI_3\}$ and $\{\cI_2,\cJ_2\}$ also to $U_{ijk}$, and the pair $\{\cI_2,\cI_2\}$ also to $U_{ijkl}$.
All of them agree with the radiative moments of Refs.~\cite{Blanchet:2013haa,Faye:2014fra}.
Inserted in Eq.~\eqref{eq:bdymemoryl}, they give the octupole and hexadecapole boundary terms of Sec.~\ref{sec:soft}.
All current-type roots have zero residue, since the current-type radiative moments receive no nonlinear memory at this order~\cite{Faye:2014fra}.
The flux residues are given in the ancillary file \texttt{6pn\_fluxes.m}.

\section{Binary matching at 6PN}
\label{sec:forces}
We express the far-zone action in terms of the binary's orbital variables.
At 6PN, the failed tails and memory contribute through the leading 6PN sectors and through the 1PN corrections to the 5PN sectors, evaluated with the 1PN moments and orbit.
Two further contributions appear.
Inserting the radiation-reaction force back into the orbit gives at second order the iterated radiation reaction, which has a part even under time reversal.
The source action also contains a coupling quadratic in the curvature, which the multipolar couplings do not fix, and whose coefficient we match to the constituents.
We evaluate the effects of all far-zone contributions on the scattering angles through 6PN.
\subsection{Matching and far-zone forces}
\label{sec:moments}
\paragraph*{Source moments.}
The source multipoles are matched to the near-zone dynamics at the PN accuracy required by each far-zone sector.
We use relative harmonic coordinates,
\begin{equation}\begin{aligned}
  \bm r&=\bm x_1-\bm x_2\,,&\qquad \bm v&=\dot{\bm r}\,,\\
  \hat{\bm r}&=\bm r/r\,,&\qquad \dot r&=\hat{\bm r}\cdot\bm v\,,
\end{aligned}\label{eq:relcoords}\end{equation}
and the mass parameters
\begin{equation}
  M=m_1+m_2\,,\qquad \nu=\frac{m_1m_2}{M^2}\,,\qquad \delta=\frac{m_1-m_2}{M}\,.
\label{eq:massvars}\end{equation}
The Newtonian source moments are
\begin{equation}\begin{aligned}
 \cI^L_N&=\sum_a m_a x_a^{\langle L\rangle}\,,\qquad
 L^i_N=\sum_a m_a(\bm x_a\times\bm v_a)^i\,,\\
 \cJ^L_N&=\sum_a m_a x_a^{\langle L-1}(\bm x_a\times\bm v_a)^{i_l\rangle}\,.
\end{aligned}\label{eq:binarymultipoles}\end{equation}
In the centre-of-mass frame, the moments needed reduce to
\begin{equation}
\begin{aligned}
  \cI_N^{ij}&=\nu M\,r^{\langle ij\rangle}\,,\qquad
  \cI_N^{ijk}=-\nu M\delta\,r^{\langle ijk\rangle}\,,\\
  \cI_N^{ijkl}&=\nu M(1-3\nu)\,r^{\langle ijkl\rangle}\,,\\
  \cJ_N^{ij}&=-\nu M\delta\,r^{\langle i}(\bm r\times\bm v)^{j\rangle}\,,\\
  \cJ_N^{ijk}&=\nu M(1-3\nu)\,r^{\langle ij}(\bm r\times\bm v)^{k\rangle}\,,\\
  L^i_N&=\nu M\,(\bm r\times\bm v)^i\,.
\end{aligned}
\label{eq:cmmoments}
\end{equation}
\paragraph*{From the action to the acceleration.}
After doubling the constituent worldlines,
$\bm x_a\to\bm x_{a,1},\bm x_{a,2}$, with Keldysh variables $\bm x_{a,\pm}$ as in Eq.~\eqref{eq:keldysh}, the \inin\ action $\Gamma[Q_1,Q_2]$ becomes a functional of the binary trajectories.
Together with the orbital action of Eq.~\eqref{eq:NZmatching}, it forms the \inin\ action of the binary,
\begin{equation}
    S_{\mathrm{bin}}[\bm x_{a,1},\bm x_{a,2}]=S_{\mathrm{orb}}[\bm x_{a,1}]-S_{\mathrm{orb}}[\bm x_{a,2}]+\Gamma[Q_1,Q_2]\,.
\label{eq:Sbin}
\end{equation}
Varying it with respect to $\bm x_{a,-}$ in the physical limit gives the Euler--Lagrange equations of $S_{\mathrm{orb}}$, with the generalised force~\cite{Galley:2009px,Galley:2012hx,Galley:2012qs},
\begin{equation}
 \bm F_a^{\mathrm{far}}(t)=\left[\frac{\delta\,\Gamma[Q_1,Q_2]}{\delta\bm x_{a,-}(t)}\right]_{\mathrm{PL}}\,,
\label{eq:eom}\end{equation}
where the physical limit sets $\bm x_{a,-}=0$ and $\bm x_{a,+}=\bm x_a$.
In the centre-of-mass frame, neglecting recoil effects, the source moments depend only on the relative variables $\bm r$ and $\bm v$, so that $\bm F_1^{\mathrm{far}}=-\bm F_2^{\mathrm{far}}\equiv\bm F^{\mathrm{far}}$ is the force conjugate to $\bm r$.
The boundary term $\Gamma^{\mathrm{bdy}}$ of Eq.~\eqref{eq:memsplit} does not enter the equations of motion. Hence, the trajectories, the impulse and the angle at fixed incoming angular momentum are unchanged.
The relative orbit therefore obeys
\begin{equation}
    \frac{\dd}{\dd t}\frac{\partial\cL_{\mathrm{orb}}}{\partial v^i}-\frac{\partial\cL_{\mathrm{orb}}}{\partial r^i}=F^{\mathrm{far}}_i\,,
\label{eq:EL}
\end{equation}
for which we use the harmonic-coordinate Lagrangian~\cite{Goldberger:2004jt},
\begin{equation}
\begin{aligned}
 \cL_{\mathrm{orb}}=\nu M\Big[&\frac{v^2}{2}+\frac{GM}{r}+(1-3\nu)\frac{v^4}{8}\\
 &+\frac{GM}{2r}\big((3+\nu)v^2+\nu\dot r^2\big)-\frac{G^2M^2}{2r^2}\Big]\,.
\end{aligned}
\label{eq:Lag1PN}
\end{equation}
Without the far-zone force, Eq.~\eqref{eq:EL} gives the conservative orbital acceleration $\bm a_{\mathrm{orb}}(\bm r,\bm v)$.
We write the relative equation of motion as
\begin{equation}
    \ddot{\bm r}=\bm a_{\mathrm{orb}}+\bm a^{\mathrm{far}}\,.
\label{eq:accelsplit}
\end{equation}
To evaluate the force, we note that each term of $\Gamma$ depends on $\bm r_-$ and $\bm v_-$ only through its root $Q_{s,-}$.
The matched moments are expressed in terms of $\bm r$ and $\bm v$ after eliminating accelerations with the conservative orbital equations.
The chain rule, with the velocity term integrated by parts, gives the components of the relative force,
\begin{equation}
    F^{\mathrm{far}}_i=\sum_{s}\bigg[\frac{\partial Q_{s}}{\partial r^i}\cdot\mathcal F_{s}
    -\frac{\dd}{\dd t}\Big(\frac{\partial Q_{s}}{\partial v^i}\cdot\mathcal F_{s}\Big)\bigg]\,,
\label{eq:chainforce}
\end{equation}
where $\mathcal F_{s}$ is the force conjugate to the root, Eq.~\eqref{eq:Kr}.
The sum runs over every root of the action.
At fixed $\bm r$ and $\bm v$, the force changes the acceleration by
\begin{equation}
    \frac{\partial^2\cL_{\mathrm{orb}}}{\partial v^i\partial v^j}\,a_{\mathrm{far}}^j=F_i^{\mathrm{far}}\,.
\label{eq:hessian}
\end{equation}
At Newtonian order the velocity Hessian is $\nu M\,\delta_{ij}$, and the corresponding acceleration is
\begin{equation}
    \bm a^{\mathrm{far}}_N=\frac{\bm F^{\mathrm{far}}}{\nu M}\,.
\end{equation}
At 1PN, inverting the velocity Hessian gives
\begin{equation}
\begin{aligned}
 \bm a^{\mathrm{far}}={}&\bm a^{\mathrm{far}}_N-(1-3\nu)\Big[\tfrac12v^2\,\bm a^{\mathrm{far}}_N+(\bm v\cdot\bm a^{\mathrm{far}}_N)\,\bm v\Big]\\
 &-\frac{GM}{r}\Big[(3+\nu)\,\bm a^{\mathrm{far}}_N+\nu\,(\hat{\bm r}\cdot\bm a^{\mathrm{far}}_N)\,\hat{\bm r}\Big]\,.
\end{aligned}
\label{eq:response}\end{equation}
The relative accelerations of the leading 6PN sectors are collected in the ancillary file \texttt{6pn\_accelerations.m}.
\subsection{Scattering orbits and angles}
\label{sec:matching}
\paragraph*{Kinematics.}
The full scattering angle is expanded as
\begin{equation}\begin{aligned}
 \frac\chi2&=\sum_{n\ge1}\frac{\chi_j^{(n)}(p_\infty)}{j^n}\,,
 &j&=\frac{J}{GM^2\nu}\,,\\
 p_\infty&=\sqrt{\gamma^2-1}\,,
 &\Gamma_{\mathrm{cm}}&=\frac{E_{\mathrm{cm}}}{M}=\sqrt{1+2\nu(\gamma-1)}\,,\\
 \widetilde\chi^{(n)}&=\Gamma_{\mathrm{cm}}^{n-1}\chi_j^{(n)}\,,
 &\gamma&=\frac{E_{\mathrm{cm}}^2-m_1^2-m_2^2}{2m_1m_2}\,,
\end{aligned}\label{eq:angleconventions}\end{equation}
where $E_{\mathrm{cm}}$ and $J$ are the incoming energy and angular momentum in the centre-of-mass frame, and $\gamma$ the relative Lorentz factor.
Here, $\chi$ is the relative scattering angle, between the incoming and outgoing values of the relative momentum, conjugate to $\bm r$,
\begin{equation}
    \bm p=\frac{\partial\cL_{\mathrm{orb}}}{\partial\bm v}\,.
\label{eq:relmomentum}
\end{equation}
With recoil neglected, it is also the deflection of each body in the centre-of-mass frame.
Through 1PN order, the asymptotic variables are related to the relative velocity at infinity $v_\infty$ and the impact parameter $b$ by
\begin{equation}
\begin{aligned}
 p_\infty=v_\infty\Big[1+\tfrac{1-2\nu}{2}v_\infty^2\Big]\,,\ \
 J=\nu M bv_\infty\Big[1+\tfrac{1-3\nu}{2}v_\infty^2\Big]\,.
\end{aligned}
\label{eq:kin1PN}
\end{equation}
The 6PN coefficients of $\widetilde\chi^{(n)}$ also receive contributions from the 5PN terms, through Eq.~\eqref{eq:kin1PN} and the factor $\Gamma_{\mathrm{cm}}^{n-1}$.
\paragraph*{The orbit.}
The relative orbit is expanded about the straight line $\bm r_0(t)$,
\begin{equation}
\begin{aligned}
    \bm r(t)&=\bm r_0(t)+\sum_{n\ge1}[\bm r]_{G^n}(t)+\delta\bm r(t)\,,\\
    \bm r_0(t)&=(b,v_\infty t,0)\,,\qquad
    \delta\bm r(t)=\sum_{n\ge2}[\delta\bm r]_{G^n}(t)\,,
\end{aligned}
\label{eq:orbitexp}
\end{equation}
where $[\bm r]_{G^n}$ are the $\cO(G^n)$ deflections of the conservative orbit, while 
$\delta\bm r$ is the displacement produced by the far-zone accelerations, at first order in them.
In this section, $[X]_{G^n}$ denotes the $\cO(G^n)$ part of a quantity $X$ on the orbit, including the factor $G^n$.
On $\bm r_0$, the Newtonian multipole moments of rank $l$ are polynomials in $t$ of degree at most $l$.
Their time derivatives beyond their rank vanish there and start at $\cO(G)$.
In every term of the failed-tail and memory forces, at least one moment is differentiated beyond its rank, so that these accelerations start at $\cO(G^3)$.
For these sectors the $\cO(G^3)$ acceleration has vanishing time integral,
\begin{equation}
    \int_{-\infty}^{+\infty}\!\dd t\,[\bm a^{\mathrm{far}}]_{G^3}=0\,.
\label{eq:zeroimpulse}
\end{equation}
Their first nonzero change of velocity therefore arises at $\cO(G^4)$, from the deflection of the orbit or from the orbital response.
\paragraph*{Total angle.}
The total scattering angle follows from the retarded far-zone acceleration, which perturbs the conservative orbit at fixed $v_\infty$ and $b$, that is at fixed incoming $E_{\mathrm{cm}}$ and $J$.
The contribution of these accelerations has two sources: the deflection of the orbit inside $\bm a^{\mathrm{far}}$, and the response of the orbit to the displacement produced by $\bm a^{\mathrm{far}}$.
At first order in $\bm a^{\mathrm{far}}$, both are captured by the linearised equation of motion,
\begin{equation}
 \delta\ddot{\bm r}=(\delta\bm r\cdot\partial_{\bm r}+\delta\bm v\cdot\partial_{\bm v})\,\bm a_{\mathrm{orb}}+\bm a^{\mathrm{far}}\,,
\label{eq:chitot}\end{equation}
with both accelerations evaluated on the conservative orbit.
With $\delta\bm r$ and $\delta\bm v$ vanishing in the far past, its solution is
\begin{equation}
\begin{aligned}
    \delta\bm v(t)&=\int_{-\infty}^{t}\!\dd t'\,\Big[(\delta\bm r\cdot\partial_{\bm r}+\delta\bm v\cdot\partial_{\bm v})\,\bm a_{\mathrm{orb}}+\bm a^{\mathrm{far}}\Big]\,,\\
    \delta\bm r(t)&=\int_{-\infty}^{t}\!\dd t'\,\delta\bm v(t')\,.
\end{aligned}
\label{eq:deltav}
\end{equation}
The change of the angle is the rotation of the outgoing relative velocity, which at infinity is parallel to the relative momentum~\eqref{eq:relmomentum},
\begin{equation}
\begin{aligned}
    \delta\chi^{\mathrm{tot}}&=\arctan\frac{\hat{\bm e}_\perp\cdot\delta\bm v(+\infty)}{v_\infty+\hat{\bm e}_\parallel\cdot\delta\bm v(+\infty)}\\
    &=\frac{\hat{\bm e}_\perp\cdot\delta\bm v(+\infty)}{v_\infty}\\
    &\quad-\frac{\big(\hat{\bm e}_\parallel\cdot\delta\bm v(+\infty)\big)\big(\hat{\bm e}_\perp\cdot\delta\bm v(+\infty)\big)}{v_\infty^2}+\cdots\,,
\end{aligned}
\label{eq:chitotangle}\end{equation}
where $v_\infty$ and $\hat{\bm e}_\parallel$ are the magnitude and direction of the conservative outgoing velocity, and $\hat{\bm e}_\perp$ is orthogonal to it and oriented towards increasing deflection.
We iterate Eq.~\eqref{eq:deltav} through $\cO(G^6)$, as in Ref.~\cite{Porto:2026fsd}, with the conservative orbit through its $\cO(G^3)$ deflection.
The integrands are elementary functions of $t$, and all integrals are evaluated in closed form.
\paragraph*{Conservative angle.}
The conservative angle follows instead from the real part of the \inout\ action~\eqref{eq:inout},
\begin{equation}
    \delta S^{\mathrm{cons}}=\operatorname{Re}S^{\mathrm{in\text{-}out}}[Q]\,.
\label{eq:Scons}
\end{equation}
We insert the binary moments in the \inout\ kernels of Sec.~\ref{sec:results} and evaluate them on the conservative orbit of Eq.~\eqref{eq:orbitexp}.
Each single sign function of the kernels becomes the Hilbert transform~\eqref{eq:Hilbert} in time.
The time integral then gives $\delta S^{\mathrm{cons}}$ as a function of $b$ and $v_\infty$, which we express in terms of $E_{\mathrm{cm}}$ and $J$ through Eq.~\eqref{eq:kin1PN}.
For a conservative orbit, the angle follows from the radial action~\cite{Kalin:2019inp},
\begin{equation}
    I_r=2\int_{r_{\min}}^{\infty}\!\dd r\,p_r\,,\qquad
    \pi+\chi=-\left.\frac{\partial I_r}{\partial J}\right|_{E_{\mathrm{cm}}}\,,
\label{eq:radialaction}
\end{equation}
with $p_r=\bm p\cdot\hat{\bm r}$ the radial momentum as a function of $r$, $E_{\mathrm{cm}}$ and $J$, and $r_{\min}$ the turning point.
At first order in the far-zone action, the change of $I_r$ at fixed $E_{\mathrm{cm}}$ and $J$ is the far-zone action evaluated on the conservative orbit~\cite{Kalin:2019inp,Porto:2024cwd,Porto:2026fsd},
\begin{equation}
    \delta I_r=\delta S^{\mathrm{cons}}\,,
\label{eq:dIr}
\end{equation}
so that
\begin{equation}
    \delta\chi^{\mathrm{cons}}=-\left.\frac{\partial\,\delta S^{\mathrm{cons}}}{\partial J}\right|_{E_{\mathrm{cm}}}\,.
\label{eq:chicons}\end{equation}
\paragraph*{Tail-like and memory-like terms.}
The kernels of $\delta S^{\mathrm{cons}}$ contain products of the sign functions of two insertions, Eq.~\eqref{eq:signPB}, and we call the third the middle insertion.
Its terms are either tail-like (T-like) or memory-like (M-like),
\begin{equation}
    \delta S^{\mathrm{cons}}=\delta S^{\mathrm{cons}}_{\mathrm{T\text{-}like}}+\delta S^{\mathrm{cons}}_{\mathrm{M\text{-}like}}\,.
\label{eq:TMsplit}
\end{equation}
In the tail-like terms, one insertion is evaluated on the free motion $\bm r_0$ and therefore carries zero frequency.
If it is the middle insertion, the nonlocal terms of Eq.~\eqref{eq:signPB} vanish and the product of the two sign functions reduces to its local part, $-1$, while otherwise the term vanishes, since the radiative kernel of that insertion vanishes at zero frequency~\cite{Porto:2024cwd}.
The tail-like terms are therefore fixed without any prescription.
At $\cO(G^4)$ all terms are tail-like.
In the memory-like terms, all three insertions are evaluated on the deflected orbit.
They first appear at $\cO(G^5)$, where all three insertions are of order $G$, and there the nonlocal terms of Eq.~\eqref{eq:signPB} contribute, so that the conservative angle depends on the prescription~\cite{Porto:2026fsd,Driesse:2026qiz}.
The failed tails, whose static insertion is $L$, contain only tail-like terms, while the memory sectors, the iterated radiation reaction and the quadratic-curvature coupling contain both classes.
In the local prescription of Ref.~\cite{Porto:2026fsd}, the product in the memory-like terms is replaced by its local part, $-1$.
The $\gamma$-3 prescription of Ref.~\cite{Driesse:2026qiz} treats the tail-like terms in the same way, while in the memory-like terms the radiative propagators are retarded and point towards the middle insertion, so that each product is $+1$~\cite{Porto:2026fsd}.
The two prescriptions therefore share the tail-like terms, and their memory-like terms differ by a sign.
\subsection{1PN corrections to the 5PN sectors}
\label{sec:iterations}
The 5PN failed tail $\{L,\cI_2,\cI_2\}$ and memory $\{\cI_2,\cI_2,\cI_2\}$ enter the 6PN result through their relative 1PN corrections.
We use the 1PN mass quadrupole~\cite{Galley:2012qs,Blanchet:2013haa}, with accelerations eliminated by the conservative orbital equations,
\begin{equation}\begin{aligned}
\cI^{ij}=\nu M\Big\{&r^{\langle ij\rangle}\Big[1+\tfrac{29}{42}(1-3\nu)v^2
-\tfrac{5-8\nu}{7}\tfrac{GM}{r}\Big]\\
&+\tfrac{11}{21}(1-3\nu)r^2v^{\langle ij\rangle}\\
&-\tfrac47(1-3\nu)(\bm r\cdot\bm v)r^{\langle i}v^{j\rangle}\Big\}\,,
\end{aligned}\label{eq:I1PN}\end{equation}
and the 1PN angular momentum,
\begin{equation}
\bm L=\nu M\,\bm r\times\bm v\left[1+\tfrac12(1-3\nu)v^2+(3+\nu)\frac{GM}{r}\right]\,.
\label{eq:L1PN}
\end{equation}
On the conservative orbit, $\bm L$ is conserved, and its magnitude is the incoming angular momentum $J$ of Eq.~\eqref{eq:kin1PN}.
Every insertion is expanded through 1PN, and each term is kept to first order in the 1PN corrections.
The time derivatives of the moments are reduced with the orbital equations~\eqref{eq:EL} without the far-zone force, and the force is converted with the response~\eqref{eq:response}.
The contributions of the two sectors to the scattering angles are given in the ancillary file \texttt{6pn\_scattering\_angles.m}.
\subsection{Iterated radiation reaction}
\label{sec:RR2}
\begin{figure}[t]
\centering
\includegraphics[width=0.72\columnwidth]{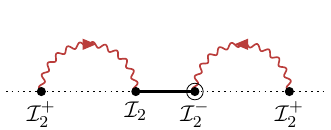}
\caption{The RR--RR term of the iterated radiation reaction.
The retarded field of one quadrupole (left) displaces the orbit, and the field of a second quadrupole (right) acts on the displaced moment.
The thick segment denotes the displacement of the orbit, which forms the middle insertion.}
\label{fig:rr2}
\end{figure}
The iterated radiation reaction contributes to the total and conservative scattering angles from $\cO(G^4)$.
\paragraph*{Forces and their iteration.}
The iteration starts from the radiation-reaction functional~\eqref{eq:RR}.
Its quadrupole term gives the Burke--Thorne acceleration~\cite{Burke:1970wx,Iyer:1993xi},
\begin{equation}
 a^i_{2.5}=-\frac{2G}{5}\,\cI_N^{(5)ij}\,r^j\,.
\label{eq:BT}
\end{equation}
The 3.5PN acceleration $\bm a_{3.5}$ contains the leading mass-octupole and current-quadrupole contributions~\cite{Iyer:1995rn,Galley:2012qs},
\begin{equation}
\begin{aligned}
 a^i_{\mathrm{oct}}&=-\frac{G\delta}{63}\,\cI_N^{(7)ijk}\,r^jr^k\,,\\
 a^i_{\mathrm{cq}}&=\frac{16G\delta}{45}\Big[\cJ_N^{(5)ik}(\bm r\times\bm v)^k
 +\big(\bm v\times\cJ_N^{(5)}\bm r\big)^i\\
 &\hspace{1.7cm}+\frac{\dd}{\dd t}\big(\bm r\times\cJ_N^{(5)}\bm r\big)^i\Big]\,,
\end{aligned}
\label{eq:a35pieces}\end{equation}
and the 1PN correction to the quadrupole term, obtained as in Sec.~\ref{sec:iterations}.
An inserted dissipative acceleration $\bm a'$ acts on a radiation-reaction acceleration $\bm a$ in two ways,
\begin{equation}\begin{aligned}
 \bm a[\bm a']&=\bm a_{\mathrm{RR}^2}[\bm a']+\bm a_{\mathrm{RR\text{--}RR}}[\bm a']\,,\\
 \bm a_{\mathrm{RR}^2}[\bm a']&=\Big[\bm a\big|_{\ddot{\bm r}\to\bm a_{\mathrm{orb}}+\bm a'}\Big]_{\bm a'}\,,\\
 \bm a_{\mathrm{RR\text{--}RR}}[\bm a']&=(\delta\bm r\cdot\partial_{\bm r}+\delta\bm v\cdot\partial_{\bm v})\,\bm a\,,\\
 \delta\ddot{\bm r}&=(\delta\bm r\cdot\partial_{\bm r}+\delta\bm v\cdot\partial_{\bm v})\,\bm a_{\mathrm{orb}}
 +\bm a'\,,
\end{aligned}
\label{eq:RR2ops}
\end{equation}
where the subscript $\bm a'$ selects the terms linear in $\bm a'$.
In the $\mathrm{RR}^2$ term, $\bm a'$ eliminates the higher time derivatives of the orbit inside $\bm a$, together with $\bm a_{\mathrm{orb}}$, while in the $\mathrm{RR\text{--}RR}$ term $\bm a$, reduced with the conservative orbital equations, is evaluated on the orbit displaced by $\bm a'$, shown in Fig.~\ref{fig:rr2}.
Their sum is the iterated radiation reaction, $2\mathrm{RR}=\mathrm{RR}^2+\mathrm{RR\text{--}RR}$.

The iteration of the Burke--Thorne acceleration with itself, $\bm a_{2.5}[\bm a_{2.5}]$, starts at 5PN~\cite{Almeida:2022jrv,Porto:2024cwd,Porto:2026fsd}.
At 6PN, the iteration also involves the 3.5PN force and the 1PN corrections to the 5PN iteration,
\begin{equation}
 \bm a^{2\mathrm{RR}}_{6\PN}=\bm a_{2.5}[\bm a_{3.5}]+\bm a_{3.5}[\bm a_{2.5}]
 +\big(\bm a_{2.5}[\bm a_{2.5}]\big)_{1\PN}\,.
\label{eq:RR2sum}
\end{equation}
In the last term, the 1PN corrections come from the orbit and from the response~\eqref{eq:response}.
The $\mathrm{RR}^2$ term of the iteration~\eqref{eq:RR2sum}, which depends only on $\bm r$ and $\bm v$, is given, together with the 2.5PN and 3.5PN accelerations, in the ancillary file \texttt{6pn\_accelerations.m}.
\paragraph*{Contribution to the total angle.}
At $\cO(G^4)$, the iteration~\eqref{eq:RR2sum} enters Eq.~\eqref{eq:chitot} as the acceleration $\bm a^{\mathrm{far}}$, and the $\cO(G)$ terms of the response~\eqref{eq:response} act on the 5PN iteration $\bm a_{2.5}[\bm a_{2.5}]$.
At $\cO(G^5)$ and $\cO(G^6)$, both terms of Eq.~\eqref{eq:RR2ops} still enter Eq.~\eqref{eq:chitot} as $\bm a^{\mathrm{far}}$~\cite{Porto:2026fsd}.
The $\mathrm{RR}^2$ term keeps its form, evaluated on the deflected orbit, while the $\mathrm{RR\text{--}RR}$ term of Eq.~\eqref{eq:RR2ops} becomes
\begin{equation}
\begin{aligned}
    \bm a_{\mathrm{RR\text{--}RR}}={}&(\delta\bm r\cdot\partial_{\bm r}+\delta\bm v\cdot\partial_{\bm v})\,(\bm a_{2.5}+\bm a_{3.5})\\
    &+\frac12\,(\delta\bm r\cdot\partial_{\bm r}+\delta\bm v\cdot\partial_{\bm v})^2\,\bm a_{\mathrm{orb}}\,,
\end{aligned}
\label{eq:RRRRforce}
\end{equation}
where $\delta\bm r$ is the displacement~\eqref{eq:RR2ops} produced by $\bm a'=\bm a_{2.5}+\bm a_{3.5}$.
The second term, quadratic in the displacement, starts at $\cO(G^5)$.
\paragraph*{Conservative action.}
The conservative contribution follows from the radiation-reaction part $S^{\mathrm{in\text{-}out}}_{\mathrm{RR}}$ of the \inout\ action~\eqref{eq:inout}, in which each kernel is multiplied by $\sgn\omega$, with the Feynman force
\begin{equation}
    \bm F^{\mathrm F}=\frac{\delta S^{\mathrm{in\text{-}out}}_{\mathrm{RR}}}{\delta\bm r}\,.
\label{eq:FFdef}
\end{equation}
Expanding about the conservative orbit $\bm r$ in the displacement $\delta\bm r$, as in Ref.~\cite{Porto:2026fsd}, gives
\begin{equation}
\begin{aligned}
    \Big[S_{\mathrm{orb}}+S^{\mathrm{in\text{-}out}}_{\mathrm{RR}}\Big]_{\bm r+\delta\bm r}
    ={}&\Big[S_{\mathrm{orb}}+S^{\mathrm{in\text{-}out}}_{\mathrm{RR}}\Big]_{\bm r}
    +\bm p\cdot\delta\bm r\,\Big|_{t\to-\infty}^{t\to+\infty}\\
    &+S^{\mathrm{cons}}_{2\mathrm{RR}}+\cdots\,.
\end{aligned}
\label{eq:RRexpansion}
\end{equation}
The first term is the value on the conservative orbit, the second the first variation of $S_{\mathrm{orb}}$, which is a boundary term since $\bm r$ solves Eq.~\eqref{eq:EL} without the far-zone force, and the ellipsis is of third order in $\bm F^{\mathrm F}$.
The first-variation term is removed by the Legendre transform to the radial action~\eqref{eq:dIr}~\cite{Porto:2026fsd}.
The conservative action is the second-order term,
\begin{equation}
    S^{\mathrm{cons}}_{2\mathrm{RR}}=\int\dd t\,\bm F^{\mathrm F}\cdot\delta\bm r+\frac12\,\delta^2S_{\mathrm{orb}}[\delta\bm r,\delta\bm r]\,,
\label{eq:RR2stationary}
\end{equation}
with
\begin{equation}
    \delta^2S_{\mathrm{orb}}[\delta\bm r,\delta\bm r]=\int\dd t\,(\delta\bm r\cdot\partial_{\bm r}+\delta\bm v\cdot\partial_{\bm v})^2\,\cL_{\mathrm{orb}}\,.
\label{eq:d2Sorb}
\end{equation}
The first term is the change of the radiation-reaction action when its source is displaced, and the second the quadratic change of the orbital action.
Its variation with $\delta\bm r$ fixed at the endpoints gives Eq.~\eqref{eq:EL} linearised about the conservative orbit, with force $\bm F^{\mathrm F}$, and on its solution
\begin{equation}
    S^{\mathrm{cons}}_{2\mathrm{RR}}=\frac12\int\dd t\,\bm F^{\mathrm F}\cdot\delta\bm r+\frac12\,\delta\bm r\cdot\delta\bm p\,\Big|_{t\to-\infty}^{t\to+\infty}\,,
\label{eq:RR2relation}
\end{equation}
with $\delta p_i=(\delta\bm r\cdot\partial_{\bm r}+\delta\bm v\cdot\partial_{\bm v})\,\partial\cL_{\mathrm{orb}}/\partial v^i$ the change of the relative momentum~\eqref{eq:relmomentum}.
\paragraph*{Conservative part at $\cO(G^4)$.}
At $\cO(G^4)$, the two force insertions are of order $G^2$, so the middle insertion is on $\bm r_0$, all terms are tail-like, and only the $G^0$ part of the orbital response is required.
We retain its relative 1PN correction.
The retarded $G^2$ force per unit reduced mass on the straight line $\bm r_0$ is
\begin{equation}
 \bm f=\Big[\bm a_{2.5}+\bm a_{3.5}+(1-3\nu)\Big(\tfrac12v^2\bm a_{2.5}+(\bm v\cdot\bm a_{2.5})\,\bm v\Big)\Big]_{G^2}\,.
\label{eq:RRnormforce}
\end{equation}
The last term removes the velocity-dependent 1PN part of the response~\eqref{eq:response}, which is contained in $\bm a_{3.5}$.
At this order, Eq.~\eqref{eq:FFdef} gives
\begin{equation}
    \bm F^{\mathrm F}=i\,\nu M\,\cH[\bm f]\,,
\label{eq:inoutforce}
\end{equation}
with $\cH$ the Hilbert transform of Eq.~\eqref{eq:Hilbert}.
With the time integral of the force and the corresponding change of velocity,
\begin{equation}
\begin{aligned}
    \bm V_F(t)&=\int_{-\infty}^{t}\bm f(t')\,\dd t'\,,\\
    \bm V_R&=\bm V_F-(1-3\nu)\Big(\tfrac12v^2\bm V_F+(\bm v\cdot\bm V_F)\,\bm v\Big)\,,
\end{aligned}
\label{eq:RRnormvariables}
\end{equation}
the displacement satisfies $\delta\dot{\bm r}=i\cH[\bm V_R]$.
Integrating Eq.~\eqref{eq:RR2relation} by parts, and using that the Hilbert transform preserves the inner product, gives
\begin{equation}
    S^{\mathrm{cons}}_{2\mathrm{RR}}=\frac{\nu M}{2}\int\dd t\,\bm V_F\cdot\bm V_R\,.
\label{eq:RRnormderiv}
\end{equation}
The boundary terms, including that of Eq.~\eqref{eq:RR2relation}, vanish because $\bm V_F(\pm\infty)=0$ and the displacement remains bounded.
Through 6PN, the result is
\begin{equation}
    S^{\mathrm{cons}}_{2\mathrm{RR}}
    =\frac{\pi\nu^3G^4M^5v_\infty^3}{b^3}\Big[\frac{106}{25}+\frac{6639-21249\nu}{700}\,v_\infty^2\Big]\,.
\label{eq:RRnorm}
\end{equation}
Its first term reproduces the 5PN result of Refs.~\cite{Porto:2024cwd,Porto:2026fsd}, and the second is its 6PN correction.
With the asymptotic variables~\eqref{eq:kin1PN} and Eq.~\eqref{eq:chicons}, it gives the conservative angle
\begin{equation}
 \widetilde\chi_{2\mathrm{RR}}^{(4),\mathrm{cons}}
 =\pi\nu^2p_\infty^6\Big[\frac{159}{25}+\Big(\frac{6561}{1400}-\frac{7407}{280}\,\nu\Big)p_\infty^2\Big]\,.
\label{eq:RRangle}
\end{equation}
\paragraph*{Conservative part at $\cO(G^5)$.}
The $\cO(G^5)$ part of Eq.~\eqref{eq:RR2stationary} is
\begin{equation}
\begin{aligned}
    \big[S^{\mathrm{cons}}_{2\mathrm{RR}}\big]_{G^5}={}&\int\dd t\,\Big([\bm F^{\mathrm F}]_{G^3}\cdot[\delta\bm r]_{G^2}
    +[\bm F^{\mathrm F}]_{G^2}\cdot[\delta\bm r]_{G^3}\Big)\\
    &+\frac12\,[\delta^2S_{\mathrm{orb}}]_{G}\big[[\delta\bm r]_{G^2},[\delta\bm r]_{G^2}\big]\\
    &+[\delta^2S_{\mathrm{orb}}]_{G^0}\big[[\delta\bm r]_{G^2},[\delta\bm r]_{G^3}\big]\,.
\end{aligned}
\label{eq:RRG5full}
\end{equation}
Since $[\delta\bm r]_{G^2}$ solves the linearised orbital equation with force $[\bm F^{\mathrm F}]_{G^2}$, the two terms with $[\delta\bm r]_{G^3}$ combine into the boundary term $[\delta\bm r]_{G^3}\cdot[\delta\bm p]_{G^2}$, which vanishes at both ends.
The action therefore requires the displacement only at $\cO(G^2)$, where it is the response to the force~\eqref{eq:RRnormforce},
\begin{equation}
\begin{aligned}
    \big[S^{\mathrm{cons}}_{2\mathrm{RR}}\big]_{G^5}={}&\int\dd t\,[\bm F^{\mathrm F}]_{G^3}\cdot[\delta\bm r]_{G^2}\\
    &+\frac12\,[\delta^2S_{\mathrm{orb}}]_{G}\big[[\delta\bm r]_{G^2},[\delta\bm r]_{G^2}\big]\,,
\end{aligned}
\label{eq:RRG5action}
\end{equation}
with $[\delta^2S_{\mathrm{orb}}]_G$ the $\cO(G)$ part of Eq.~\eqref{eq:d2Sorb} on the deflected orbit.
The additional power of $G$ enters either a radiating quadrupole of Fig.~\ref{fig:rr2}, and the term is tail-like, or the middle insertion, through the deflection of the orbit, and the term is memory-like.
Its memory-like terms are treated in the two prescriptions of Sec.~\ref{sec:matching}.
Through 6PN, the orbit and $\delta^2S_{\mathrm{orb}}$ are taken at 1PN.
\subsection{Quadratic-curvature coupling}
\label{sec:source_matching}
We now evaluate the contribution of the quadratic-curvature coupling~\eqref{eq:S2R} to the dynamics, whose \inin\ action is given by the diagrams of Fig.~\ref{fig:e2vertex}.
The coefficient $C^{abcd}$ is not fixed by the multipolar couplings of Eq.~\eqref{eq:Smult}.
\paragraph*{Matching to the constituents.}
We fix $C^{abcd}$ by comparing the gravitational Compton amplitudes computed from the constituents and from the multipolar action, in App.~\ref{app:E2onshell}, which gives
\begin{equation}
    \begin{aligned}
        \cL_{E^2}=\frac{\nu M(1-3\nu)}{168}\Big[
        &-19\big(r^iE_{ij}r^j\big)^2\\
        &+4r^2r^i(E^2)_{ij}r^j
        +4r^4\operatorname{tr}(E^2)\Big]\,,
    \end{aligned}
\label{eq:matchedE2}
\end{equation}
where $\nu M(1-3\nu)r^4=\sum_am_a|\bm x_a|^4$ in the Newtonian centre-of-mass frame.
Corrections proportional to $GM/r$ or $v^2$ enter the dynamics one PN order higher.
\paragraph*{Contribution to the dynamics.}
In the dynamics of the binary, both electric fields of Eq.~\eqref{eq:matchedE2} are the radiation-reaction field of its own quadrupole.
At leading order, this is the field of the Burke--Thorne acceleration~\eqref{eq:BT},
\begin{equation}
    E^{\mathrm{RR}}_{ij}=-\frac{2G}{5}\,\cI^{(5)}_{N,ij}\,.
\label{eq:ERR}
\end{equation}
Its magnetic curvature vanishes at the source, so only the electric operator contributes at this order.
Inserting Eq.~\eqref{eq:ERR} in Eq.~\eqref{eq:matchedE2} gives the coupling in terms of the orbit and the quadrupole,
\begin{equation}
\begin{aligned}
    \cL_{E^2}\big(\bm r,E^{\mathrm{RR}}\big)=&\frac{G^2\nu M(1-3\nu)}{1050}\Big[-19\big(r^i\cI^{(5)}_{N,ij}r^j\big)^2\\
        & +4r^2\,r^i\cI^{(5)}_{N,ik}\cI^{(5)}_{N,kj}r^j+4r^4\,\cI^{(5)}_{N,ij}\cI^{(5)}_{N,ij}\Big]\,,
\end{aligned}
\label{eq:E2binary}
\end{equation}
and the causal action of the binary follows by inserting $C^{abcd}(\bm r)$ and $\cI_N^{ij}$ in Eq.~\eqref{eq:E2causal}, which fixes the relative sign of the two roots.
Varying with respect to the orbit, through the coefficient and through the quadrupole, gives the acceleration
\begin{equation}
    \begin{aligned}
        a^k_{E^2}&=\frac{4G^2}{25\nu M}\,
        \frac{\partial C^{abcd}}{\partial r_k}\,
        \cI_{N,ab}^{(5)}\cI_{N,cd}^{(5)}-\frac{G}{5\nu M}\,
        \frac{\partial\cI_N^{ij}}{\partial r_k}\,
        \cI_{C,ij}^{(5)}\,,
    \end{aligned}
\label{eq:E2roots}
\end{equation}
with $\cI_C$ the induced quadrupole~\eqref{eq:E2induced}, evaluated with $\cI=\cI_N$.
On a scattering orbit, $\cI_N^{(5)}$ first becomes nonzero through the $\cO(G)$ acceleration, so $E^{\mathrm{RR}}=\cO(G^2)$.
The coefficient-root acceleration starts at $\cO(G^4)$.
The quadrupole-root acceleration starts at $\cO(G^3)$, but its time integral at this order vanishes.
Its first contribution to the angle requires a further power of $G$, from the deflection of the orbit or from the orbital response, so that both roots enter the angle from $\cO(G^4)$.
Through $\cO(G^6)$, the acceleration~\eqref{eq:E2roots} enters Eq.~\eqref{eq:chitot} as $\bm a^{\mathrm{far}}$.
\paragraph*{Conservative contribution.}
At $\cO(G^4)$, the conservative action~\eqref{eq:E2generalcons} requires the coefficient on the straight line $\bm r_0$ and each reaction field~\eqref{eq:ERR} at $\cO(G^2)$.
On $\bm r_0$, $C^{abcd}$ is a polynomial of degree four in $t$, while the zeroth and first time moments of $[\cI_N^{(5)}]_G$ vanish.
In the time domain, the two sign functions are Hilbert transforms~\eqref{eq:Hilbert} of the fields, and can therefore be removed under the time integral, giving
\begin{equation}
    \begin{aligned}
        S^{\mathrm{cons}}_{E^2}\big|_{G^4}
        &=-\int\dd t\,
        \cL_{E^2}\big(\bm r_0,[E^{\mathrm{RR}}]_{G^2}\big)\\
        &=-\frac{4G^2}{25}\int\dd t\,
        C^{abcd}(\bm r_0)
        \big[\cI_{N,ab}^{(5)}\big]_G
        \big[\cI_{N,cd}^{(5)}\big]_G\,.
    \end{aligned}
\label{eq:E2cons}
\end{equation}
At higher orders, the tail-like terms keep $C^{abcd}$ on $\bm r_0$ and expand the fields, while the memory-like terms contain the deflection of the orbit in $C^{abcd}$.
At $\cO(G^5)$, the memory-like part of Eq.~\eqref{eq:E2generalcons} comes from the $\cO(G)$ part of $C^{abcd}$, with both quadrupoles at $\cO(G)$,
\begin{equation}
\begin{aligned}
    S^{\mathrm{cons}}_{E^2}\big|_{\mathrm{M\text{-}like}}={}&-\frac{4G^2}{25}\int\!\frac{\dd\omega_1\dd\omega_2}{(2\pi)^2}\,
    \omega_1^5\,\omega_2^5\,\sgn(\omega_1)\,\sgn(\omega_2)\\
    &\times\big[C^{abcd}\big]_G(-\omega_1-\omega_2)\\
    &\times\big[\cI_{N,ab}\big]_G(\omega_1)\,\big[\cI_{N,cd}\big]_G(\omega_2)\,,
\end{aligned}
\label{eq:E2mem}
\end{equation}
where the local and $\gamma$-3 prescriptions replace the product of the two sign functions by $-1$ and $+1$, respectively.
The contributions of the coupling to the total and conservative angles are given in the ancillary file \texttt{6pn\_scattering\_angles.m}.

\section{Scattering angles through 6PN}
\label{sec:angle}
The scattering angle through 6PN receives contributions from the potential region, the tail sector, and the two-radiation sector,
\begin{equation}
    \chi=\chi_{\mathrm{pot}}+\chi_{\mathrm{T}}+\chi_{2\mathrm{R}}+\cdots\,.
    \label{eq:anglesplit}
\end{equation}
The ellipsis includes further $n\mathrm{R}$ radiation contributions that appear at higher PN orders.
We retain the integer-PN terms through 6PN and omit the half-integer PN terms in the total angle.
We give the two-radiation contributions to the total angle at 4PM, 5PM and 6PM, and to the conservative angle at 4PM and 5PM.
We then combine them with the potential and tail contributions at 4PM and 5PM, specifying the PN orders and powers of $\nu$ retained in each result.
A superscript \textit{tot} denotes the retarded angle at fixed incoming energy and angular momentum, and \textit{cons} the conservative angle in the specified prescription.
The $2\mathrm{R}$ contribution at 6PN is given by
\begin{equation}
    \begin{aligned}
        \widetilde\chi^{(n)}_{2\mathrm{R}}\big|_{6\PN}
        ={}&\widetilde\chi^{(n)}_{\mathrm{FT}}
        +\widetilde\chi^{(n)}_{\mathrm{M}}
        +\widetilde\chi^{(n)}_{1\PN}+\widetilde\chi^{(n)}_{2\mathrm{RR}}+\widetilde\chi^{(n)}_{E^2}\,,
    \end{aligned}
    \label{eq:assembly}
\end{equation}
where $\widetilde\chi^{(n)}_{\mathrm{FT}}$ denotes the 6PN failed-tail sectors,
$\widetilde\chi^{(n)}_{\mathrm{M}}$ the 6PN memory sectors,
$\widetilde\chi^{(n)}_{1\PN}$ the 1PN corrections to the 5PN failed tail and memory,
$\widetilde\chi^{(n)}_{2\mathrm{RR}}$ the iterated radiation reaction,
and $\widetilde\chi^{(n)}_{E^2}$ the quadratic-curvature coupling of the composite source.
The iterated radiation-reaction contribution splits into the $\mathrm{RR}^2$ and $\mathrm{RR\text{--}RR}$ terms of Eq.~\eqref{eq:RR2ops},
\begin{equation}
    \widetilde\chi^{(n)}_{2\mathrm{RR}}
    =\widetilde\chi^{(n)}_{\mathrm{RR}^2}
    +\widetilde\chi^{(n)}_{\mathrm{RR\text{--}RR}}\,.
    \label{eq:2RRdef}
\end{equation}
In the total angle at 6PM, $\widetilde\chi^{(6),\mathrm{tot}}_{2\mathrm{RR}}$ also contains the term of Eq.~\eqref{eq:chitotangle} quadratic in the first-order radiation-reaction change of velocity.
We further separate the two-radiation contribution into the tail-like and memory-like parts of Sec.~\ref{sec:matching},
\begin{equation}
    \widetilde\chi^{(n)}_{2\mathrm{R}}
    =\widetilde\chi^{(n)}_{\mathrm{T\text{-}like}}
    +\widetilde\chi^{(n)}_{\mathrm{M\text{-}like}}\,.
    \label{eq:regions}
\end{equation}

For the conservative angle, we give results in both the local prescription of Ref.~\cite{Porto:2026fsd} and the $\gamma$-3 prescription of Ref.~\cite{Driesse:2026qiz}.
The two prescriptions are described in Sec.~\ref{sec:matching}, and through the orders considered here they have the same tail-like part and opposite memory-like parts,
\begin{equation}
    \begin{aligned}
        \widetilde\chi^{(n),\mathrm{cons}}_{2\mathrm{R}}\big|_{\mathrm{local}}
        & =\widetilde\chi^{(n),\mathrm{cons}}_{\mathrm{T\text{-}like}}
        -\widetilde\chi^{(n),\mathrm{cons}}_{\mathrm{M\text{-}like}}\big|_{\gamma\text{-}3}\,,\\
        \widetilde\chi^{(n),\mathrm{cons}}_{2\mathrm{R}}\big|_{\gamma\text{-}3}
        & =\widetilde\chi^{(n),\mathrm{cons}}_{\mathrm{T\text{-}like}}
        +\widetilde\chi^{(n),\mathrm{cons}}_{\mathrm{M\text{-}like}}\big|_{\gamma\text{-}3}\,.
    \end{aligned}
    \label{eq:prescriptions}
\end{equation}
\subsection{Two-radiation contribution to the total angle}
\label{sec:2Rtotal}
The retarded forces of Sec.~\ref{sec:forces}, including the iterated radiation reaction, give the following two-radiation contributions through 6PN:
\begin{equation}
    \begin{aligned}
        \widetilde\chi^{(4),\mathrm{tot}}_{2\mathrm{R}}
        ={}&\frac{1131}{112}\,\pi\nu^2p_\infty^6-\pi p_\infty^8\Big[\frac{256999}{28800}\,\nu^2
        +\frac{200729}{57600}\,\nu^3\Big]\,,\\[2pt]
        \widetilde\chi^{(5),\mathrm{tot}}_{2\mathrm{R}}
        ={}&\nu^2p_\infty^5\Big[\frac{245888}{945}
        -\frac{63591872}{165375}\,p_\infty^2\Big]-\frac{4096}{105}\,\nu^3p_\infty^7\,,\\[2pt]
        \widetilde\chi^{(6),\mathrm{tot}}_{2\mathrm{R}}
        ={}&\pi\nu^2p_\infty^4\Big[\frac{8455667}{25200}
        -\frac{43738223}{75600}\,p_\infty^2\Big]\\ 
        & -\frac{312469}{11200}\,\pi\nu^3p_\infty^6\,.
    \end{aligned}
    \label{eq:total2R}
\end{equation}
The 5PN terms agree with the corresponding two-radiation contributions of Ref.~\cite{Porto:2026fsd}, while the 6PN coefficients
at 5PM and 6PM are new.
The sector contributions through 6PN are listed in App.~\ref{sec:app_scattering_tables}.
\subsection{Total angle}
\label{sec:totalangle}
Adding the potential and tail contributions to the two-radiation contribution gives the total angle, as in Eq.~\eqref{eq:anglesplit}.
\paragraph*{4PM.}
At 4PM, the total angle at 6PN, omitting the test-body term, is
\begin{equation}
    \begin{aligned}
        &\widetilde\chi^{(4),\mathrm{tot}}\big|_{6\PN}^{\nu^1,\nu^2,\nu^3}
        =\pi\nu p_\infty^8\Big[\frac{29201523}{33554432}\,\pi^2
        -\frac{5824797}{627200}\\
        &\qquad-\frac{27953}{3360}\log\frac{p_\infty}{2}
        -\frac{43077}{11200}\,\nu-\frac{1491}{400}\,\nu^2\Big]\,,
    \end{aligned}
    \label{eq:fulltotal4PM}
\end{equation}
and it reproduces the known 4PM angle~\cite{Bern:2021yeh,Dlapa:2021vgp,Dlapa:2022lmu}.
\paragraph*{5PM.}
At 5PM, the total angle also receives a contribution 
from the dissipative part of the tail.
The tail removes energy and angular momentum from the binary, starting at $\cO(G^4)$, where the losses follow from the multipole moments evaluated through $\cO(G)$.
The radiated energy and angular momentum due to the tail are
related to the losses as
\begin{equation}
    E_{\mathrm{rad}}=-\Delta M_{\mathrm{tail}}\,,\qquad
    J_{\mathrm{rad}}=-a_i\Delta L^i_{\mathrm{tail}}\,.
    \label{eq:tail_losses_signs}
\end{equation}
Here $\bm a$ is the unit vector along the incoming orbital angular momentum, with the projection defined in Eq.~\eqref{eq:Lproj}.
Evaluating the finite tail flux~\eqref{eq:tailcheck} gives
\begin{equation}
    \begin{aligned}
        &E_{\mathrm{rad}}\big|_{G^4}
        =\frac{M\nu^2p_\infty^6}{j^4\Gamma_{\mathrm{cm}}^5}
        \Big[\frac{3136}{45}
        +\Big(\frac{1216}{105}-\frac{704}{45}\,\nu\Big)p_\infty^2\\
        &\quad+\Big(\frac{117248}{1575}-\frac{1808}{225}\,\nu\Big)p_\infty^4
        +\cO(p_\infty^6)\Big]\,,\\[3pt]
        &J_{\mathrm{rad}}\big|_{G^4}
        =\frac{GM^2\nu^2p_\infty^4}{j^3\Gamma_{\mathrm{cm}}^4}
        \Big[\frac{448}{5}
        +\frac{17760-31552\nu}{315}\,p_\infty^2\\
        &\quad+\frac{-13736-9008\nu+26880\nu^2}{315}\,p_\infty^4
        +\cO(p_\infty^6)\Big]\,.
    \end{aligned}
    \label{eq:tail_losses_integrated}
\end{equation}
The losses agree through 6PN with the 4PM radiated energy of Ref.~\cite{Dlapa:2022lmu} and the radiated angular momentum of Ref.~\cite{Heissenberg:2025ocy}.
At first order in the dissipative force, their contribution to the angle is given by the linear response~\cite{Bini:2021gat},
\begin{equation}
    \delta\chi=-\frac12\Big(\frac{\partial\chi}{\partial E_{\mathrm{cm}}}\,E_{\mathrm{rad}}+\frac{\partial\chi}{\partial J}\,J_{\mathrm{rad}}\Big)\,,
\label{eq:linresponse}
\end{equation}
where the derivatives act on the conservative angle and are evaluated at the incoming energy and angular momentum.
The $\cO(G^4)$ tail losses therefore contribute to the 5PM angle through the 1PM coefficient $\chi_j^{(1),\mathrm{cons}}=(2\gamma^2-1)/p_\infty$.
Substituting the tail losses~\eqref{eq:tail_losses_integrated} into Eq.~\eqref{eq:linresponse}, we obtain
\begin{equation}
    \begin{aligned}
        \widetilde\chi^{(5),\mathrm{diss}}_{\mathrm{T}}
        ={}&\nu p_\infty^3\Big[\frac{3584}{45}+\frac{2496}{35}\,p_\infty^2
        +\frac{37664}{1575}\,p_\infty^4\Big]\\
        &-\nu^2p_\infty^5\Big[\frac{1216}{21}
        +\frac{168128}{1575}\,p_\infty^2\Big]
        +\frac{128}{3}\,\nu^3p_\infty^7\,.
    \end{aligned}
    \label{eq:disstail5PM}
\end{equation}
Its $\nu^1$ part agrees with the difference between the total and conservative angles at first self-force order~\cite{Driesse:2024xad,Driesse:2024feo,Driesse:2026qiz}.
Combining the dissipative tail~\eqref{eq:disstail5PM} and the two-radiation contribution~\eqref{eq:total2R} with the sum of the potential region of Ref.~\cite{Bern:2025wyd} and the conservative tail, Eq.~\eqref{eq:pottail5PM}, gives the following 5PN and 6PN terms at $\nu^2$ and $\nu^3$,
\begin{equation}
\begin{aligned}
    &\widetilde\chi^{(5),\mathrm{tot}}\big|^{\nu^2,\nu^3}\\
    &\quad=\nu^2p_\infty^5\Big[\frac{35493331}{21600}-\frac{224057}{1440}\,\pi^2+\frac{1408}{45}\log(2p_\infty)\Big]\\
    &\quad+\nu^2p_\infty^7\Big[\frac{56472579463}{21168000}-\frac{6230951}{20160}\,\pi^2\\
    &\qquad-\frac{6944}{225}\log(2p_\infty)\Big]-\frac{11008}{225}\,\nu^3p_\infty^7\,.
\end{aligned}
\label{eq:total5PM}
\end{equation}
At 5PN, the result agrees with Ref.~\cite{Porto:2026fsd}.
The 6PN $\nu^2$ coefficient is new, and the $\nu^3$ coefficient agrees with Ref.~\cite{Driesse:2024feo}.
\subsection{Two-radiation contribution to the conservative angle}
\label{sec:2Rcons}
\paragraph*{4PM.}
At $\mathcal{O}(G^4)$ all terms are tail-like, so the two prescriptions coincide.
The two-radiation part gives
\begin{equation}
    \begin{aligned}
        \widetilde\chi^{(4),\mathrm{cons}}_{2\mathrm{R}}
        ={}&\pi\nu^2p_\infty^6\Big[\frac{8919}{1400}-p_\infty^2\Big(\frac{1023607}{201600}
        -\frac{559}{2304}\,\nu\Big)\Big]\,.
    \end{aligned}
    \label{eq:consfarzone}
\end{equation}
The 5PN term agrees with Ref.~\cite{Porto:2026fsd}.
Subtracting it from the retarded contribution~\eqref{eq:total2R} gives
\begin{equation}
    \begin{aligned}
        &\big(\widetilde\chi^{(4),\mathrm{tot}}_{2\mathrm{R}}
        -\widetilde\chi^{(4),\mathrm{cons}}_{2\mathrm{R}}\big)\big|_{6\PN}\!=\!-\pi\nu^2p_\infty^8\Big[\frac{43077}{11200}
        +\frac{1491}{400}\,\nu\Big]\,.
    \end{aligned}
    \label{eq:closedcheck}
\end{equation}
This difference reproduces the dissipative two-radiation contribution to the 4PM angle of Ref.~\cite{Dlapa:2022lmu} at 6PN.
\paragraph*{5PM.}
At 5PM, the two conservative prescriptions give
\begin{equation}
    \begin{aligned}
        \widetilde\chi^{(5),\mathrm{cons}}_{2\mathrm{R}}\big|_{\mathrm{local}}
        ={}&\nu^2p_\infty^5\Big[\frac{138752}{675}
        -\frac{55523392}{165375}\,p_\infty^2\Big]\\
        &+\frac{82816}{1575}\,\nu^3p_\infty^7\,,\\
        \widetilde\chi^{(5),\mathrm{cons}}_{2\mathrm{R}}\big|_{\gamma\text{-}3}
        ={}&\nu^2p_\infty^5\Big[\frac{90112}{675}
        -\frac{538688}{165375}\,p_\infty^2\Big]\\
        &+\frac{82816}{1575}\,\nu^3p_\infty^7\,.
    \end{aligned}
    \label{eq:cons5PM2R}
\end{equation}
The local result agrees with Ref.~\cite{Porto:2026fsd} at 5PN and extends it to 6PN, while the $\gamma$-3 result agrees with Ref.~\cite{Driesse:2026qiz} through 6PN.
The tail-like part, common to both prescriptions, and the memory-like part in the $\gamma$-3 prescription are
\begin{equation}
    \begin{aligned}
        \widetilde\chi^{(5),\mathrm{cons}}_{\mathrm{T\text{-}like}}
        ={}&\nu^2p_\infty^5\Big[\frac{38144}{225}
        -\frac{622912}{3675}\,p_\infty^2\Big]\\
        &+\frac{82816}{1575}\,\nu^3p_\infty^7\,,\\
        \widetilde\chi^{(5),\mathrm{cons}}_{\mathrm{M\text{-}like}}\big|_{\gamma\text{-}3}
        ={}&\nu^2p_\infty^5\Big[-\frac{4864}{135}
        +\frac{27492352}{165375}\,p_\infty^2\Big]\,.
    \end{aligned}
    \label{eq:cons5PMregions}
\end{equation}
\subsection{Conservative angle}
\label{sec:consangle}
The potential region and conservative tail contributions are common to both prescriptions.
Their sum with Eq.~\eqref{eq:prescriptions} determines the conservative angle.
The conservative tail action of Eq.~\eqref{eq:tails}, evaluated on the binary, reads
\begin{equation}
    \begin{aligned}
        S^{\mathrm{cons}}_{\mathrm{T}}
        =G^2E_{\mathrm{cm}}\sum_L c_l\int\!\frac{\dd\omega}{2\pi}\,&\omega^{2l+2}
        \big|\cI_L(\omega)\big|_d^2\\
        &\times\Big[\frac{1}{2\eps}+K_l
        -\log\Big(\frac{\omega^2}{\bar\mu^2}\Big)\Big]\,,
    \end{aligned}
    \label{eq:tailonshell}
\end{equation}
where $|\cI_L|_d^2$ denotes the $d$-dimensional trace-free contraction.
The current moments contribute with the corresponding coefficients $c_l^J$ and $K_l^J$ of Eqs.~\eqref{eq:cl} and~\eqref{eq:Klclosed}.
The pole in $\eps$ multiplies the $\cO(\eps)$ terms of the moments, their contractions and the orbital equations, leaving finite contributions to the angle~\cite{Blumlein:2020znm,Bini:2020hmy,Blumlein:2021txe,Bini:2021gat}.
Through 6PN, the tail requires the mass quadrupole through 2PN, the mass octupole and current quadrupole through 1PN, and the mass hexadecapole and current octupole at Newtonian order.
Each moment enters with $l+1$ time derivatives, beyond its rank, and therefore starts at $\cO(G)$ on the scattering orbit, so the tail starts at $\cO(G^4)$.
The 4PM angle requires these moments through $\cO(G)$, while the 5PM angle also requires the $\cO(G^2)$ terms specified in App.~\ref{sec:apptail}.
The dimensional-regularisation inputs and the contributions of each tail sector are given in the ancillary file \texttt{6pn\_tail\_dimreg.m}.
\paragraph*{4PM.}
Mass polynomiality requires $\widetilde\chi^{(4),\mathrm{cons}}$ to be linear in $\nu$ at fixed $p_\infty$~\cite{Damour:2019lcq}.
The $\nu^2$ and $\nu^3$ terms therefore cancel between the three contributions of Eq.~\eqref{eq:anglesplit}.
At 6PN, the conservative tail evaluated with the $d$-dimensional multipoles of App.~\ref{sec:apptail} gives
\begin{equation}
    \left.\widetilde\chi^{(4),\mathrm{cons}}_{\mathrm{T}}\right|^{\nu^2,\nu^3}_{6\PN}
    =\pi p_\infty^8\left(\frac{1627687}{201600}\nu^2-\frac{45419}{11520}\nu^3\right)\,.
\label{eq:tail6PNcoeff}
\end{equation}
The poles and scale logarithms cancel among these tail sectors at $\nu^2$ and $\nu^3$.
The potential of Ref.~\cite{Blumlein:2021txj} contributes
\begin{equation}
    \widetilde\chi^{(4)}_{\mathrm{pot}}\Big|^{\nu^2,\nu^3}_{6\PN}
    =-\pi\nu^2p_\infty^8\Big(\frac{839}{280}-\frac{37}{10}\,\nu\Big)\,.
\label{eq:potlog}
\end{equation}
This includes the finite term from the pole-removing canonical transformation acting on the $\cO(\eps)$ part of the $\cO(G)$ Hamiltonian~\cite{Blumlein:2020pog}.
At fixed impact parameter, the potential contribution is linear in $\nu$~\cite{Bern:2021dqo}.
Its $\nu^2$ and $\nu^3$ terms arise from the logarithm accompanying the infrared pole when expressed through $b=GM\,j\,\Gamma_{\mathrm{cm}}/p_\infty$.
Together with Eq.~\eqref{eq:consfarzone}, these terms give
\begin{equation}
\begin{aligned}
    \widetilde\chi^{(4),\mathrm{cons}}\Big|^{\nu^2,\nu^3}_{6\PN}
    &=\Big(\widetilde\chi^{(4),\mathrm{cons}}_{2\mathrm{R}}+\widetilde\chi^{(4),\mathrm{cons}}_{\mathrm{T}}+\widetilde\chi^{(4)}_{\mathrm{pot}}\Big)\Big|^{\nu^2,\nu^3}_{6\PN}\\
    &=0\,.
\end{aligned}
\label{eq:polycheck}
\end{equation}
At $\nu^1$, the infrared pole of the potential cancels the ultraviolet pole of the tail.
Their sum reproduces the known conservative angle~\cite{Bern:2021yeh,Dlapa:2021vgp},
\begin{equation}
\begin{aligned}
    \widetilde\chi^{(4),\mathrm{cons}}_{\mathrm{pot+T}}\Big|^{\nu^1}_{6\PN}
    =\pi\nu p_\infty^8\Big[&\frac{29201523}{33554432}\pi^2-\frac{5824797}{627200}\\
    &-\frac{27953}{3360}\log\frac{p_\infty}{2}\Big]\,.
\end{aligned}
\label{eq:nu1check}
\end{equation}
The same agreement holds at 4PN and 5PN.
The cancellation implements mass polynomiality for the complete conservative 4PM angle.
\paragraph*{5PM.}
The infrared poles of the potential~\cite{Bern:2025wyd} cancel the ultraviolet poles of the conservative tail.
At 5PN and 6PN, the $\nu^2$ and $\nu^3$ terms of their finite sum are
\begin{equation}
    \begin{aligned}
        &\Big(\widetilde\chi^{(5)}_{\mathrm{pot}}
        +\widetilde\chi^{(5),\mathrm{cons}}_{\mathrm{T}}\Big)\Big|^{\nu^2,\nu^3}
        =\nu^2p_\infty^5\Big[\frac{217866437}{151200}\\
        &\quad-\frac{224057}{1440}\,\pi^2
        +\frac{1408}{45}\log(2p_\infty)
        +p_\infty^2\Big(\frac{66871979399}{21168000}\\
        &\quad-\frac{6230951}{20160}\,\pi^2
        -\frac{6944}{225}\log(2p_\infty)
        -\frac{82816}{1575}\,\nu\Big)\Big]\,.
    \end{aligned}
    \label{eq:pottail5PM}
\end{equation}
Adding Eq.~\eqref{eq:cons5PM2R} gives the corresponding 5PN and 6PN terms of both conservative angles,
\begin{equation}
    \begin{aligned}
        &\widetilde\chi^{(5),\mathrm{cons}}\big|_{\mathrm{local}}^{\nu^2,\nu^3}
        =\nu^2p_\infty^5\Big[\frac{1844051}{1120}\\
        &\quad-\frac{224057}{1440}\,\pi^2
        +\frac{1408}{45}\log(2p_\infty)
        +p_\infty^2\Big(\frac{19921661741}{7056000}\\
        &\quad-\frac{6230951}{20160}\,\pi^2
        -\frac{6944}{225}\log(2p_\infty)\Big)\Big]\,,\\
        &\widetilde\chi^{(5),\mathrm{cons}}\big|_{\gamma\text{-}3}^{\nu^2,\nu^3}
        =\nu^2p_\infty^5\Big[\frac{9522061}{6048}\\
        &\quad-\frac{224057}{1440}\,\pi^2
        +\frac{1408}{45}\log(2p_\infty)
        +p_\infty^2\Big(\frac{13360605467}{4233600}\\
        &\quad-\frac{6230951}{20160}\,\pi^2
        -\frac{6944}{225}\log(2p_\infty)\Big)\Big]\,.
    \end{aligned}
    \label{eq:fullcons5PM}
\end{equation}
The local result agrees with Ref.~\cite{Porto:2026fsd} at 5PN and extends it to 6PN.
The $\gamma$-3 result agrees with Ref.~\cite{Driesse:2026qiz} through 6PN.
In both prescriptions the $\nu^3$ terms cancel, consistently with mass polynomiality of the conservative 5PM angle.
At order $\nu$, where the two-radiation contribution vanishes, the sum of the potential and tail contributions reproduces Ref.~\cite{Driesse:2026qiz}.

\section{Conclusions}
\label{sec:conclusions}

We have derived the $\cO(G^2)$ far-zone action of a generic multipolar source at 6PN, in both the causal and the conservative prescriptions.
Its new ingredients are the five memory sectors, in which the waves of two multipoles interact and the resulting field acts on a third.
At generic frequencies their causal action is local in time.
On scattering orbits the memory action acquires a boundary term, nonlocal in time, in which the nonlinear memory of the waves emitted by the other two multipoles acts on the change of the mass multipoles.
We fixed it by the balance of angular momentum.
The tails and failed tails entering at 6PN reproduce the general-multipole results of Refs.~\cite{Almeida:2021xwn,Almeida:2026clf}.
The \inin\ one-point function that generates the action also gives the nonlinear waveforms.
We verified that the emitted fluxes of energy and angular momentum balance the losses of the \inin\ action in every sector.
The zero-frequency part of the angular-momentum flux, computed independently, reproduces the coefficients of the memory boundary terms and the nonlinear-memory contributions to the radiative moments~\cite{Blanchet:2013haa,Faye:2014fra}.

For the binary, the 6PN far-zone dynamics also requires the 1PN corrections to the 5PN failed tail and memory, the iterated 2.5PN and 3.5PN radiation reaction, and a coupling quadratic in the electric curvature of the composite source.
The covariant multipolar couplings do not generate it, and we determined its coefficient by matching the gravitational Compton amplitude of the multipoles to that of the constituents.

On scattering orbits, we have computed the two-radiation contributions to the total angle at 4PM, 5PM and 6PM and to the conservative angle at 4PM and 5PM through integer 6PN order.
We have also evaluated the conservative tail in dimensional regularisation and the dissipative tail entering the total 5PM angle.

At 4PM, our results reproduce the known angle~\cite{Bern:2021yeh,Dlapa:2021vgp,Dlapa:2022lmu} through 6PN.
At 5PM, the conservative angle agrees with Ref.~\cite{Porto:2026fsd} at 5PN in the local prescription and with Ref.~\cite{Driesse:2026qiz} through 6PN in the $\gamma$-3 one.
For the total angle, the 6PN $\nu^2$ coefficient is new, and the other terms agree with Refs.~\cite{Porto:2026fsd,Driesse:2024feo}.
At 6PM, the total two-radiation contribution agrees with Ref.~\cite{Porto:2026fsd} at 5PN and extends it to 6PN.
The potential and tail contributions are still needed for the complete 6PM angle.

The complete conservative 6PN dynamics also requires the near-zone potential beyond $\cO(G^4)$.
Since the far-zone action holds for a generic multipolar source, it also applies to spinning binaries once the spin contributions to the source moments are included.
The same construction can be used to compute the far-zone dynamics at 7PN, where higher-order nonlinear interactions, such as the tails of tails of tails, first enter at integer PN order, and where the recoil of the source, neglected here, contributes to the dynamics.

\begin{acknowledgments}
I would like to thank Stefano De Angelis, Manoj K. Mandal, Pierpaolo Mastrolia, Raj Patil, Matteo Pegorin, Jan Plefka, Rafael Porto, Massimiliano M. Riva, Radu Roiban, Sid Smith, and Jan Steinhoff for useful discussions and/or comments on the manuscript.
This research is supported by the Italian MIUR under contract 20223ANFHR (PRIN2022), by the ERC (NOTIMEFORCOSMO, 101126304),
and by the INFN initiatives \textit{Amplitudes} and \textit{TPPC}.
AI tools were used to assist the automation of computational codes. 
All scientific results were independently verified by the author, who also takes full responsibility for their interpretation.
\end{acknowledgments}
\appendix
\section{Quadrupolar angular projections of the memory flux}
\label{sec:angproj}

We compute the quadrupolar angular projection of the leading energy flux generated by a pair of multipoles.
This determines their contribution to the radiative quadrupole memory and the coefficients of the boundary terms of Sec.~\ref{sec:soft}.
We separate the angular dependence of the leading waveform~\eqref{eq:wavecheck}, whose energy flux is a sum over pairs of multipoles, by defining
\begin{equation}
    \begin{aligned}
        A^{(\cI_l)}_{ab}&=n_{L-2}\,\cI_{abL-2}\,,\\
        A^{(\cJ_l)}_{ab}&=n_{cL-2}\,\epsilon_{cd(a}\cJ_{b)dL-2}\,.
    \end{aligned}
\end{equation}
Using the TT projector $\Lambda_{ijab}$ of Eq.~\eqref{eq:LambdaTT} and the angular average at $d=3$~\cite{Thorne:1980ru,Damour:1990gj} gives
\begin{equation}
\begin{aligned}
  &\int\dd\Omega\,n_{\langle ij\rangle}\,(\Lambda A^{(Q_l)})_{ab}(\Lambda A^{(Q'_l)})_{ab}\\
  &\quad=4\pi\,\frac{(l-3)(l+4)(l+2)\,(l-2)!}{(2l+3)!!}\;\big(Q_{iL-1}Q'_{jL-1}\big)^{\mathrm{STF}}\,,\\
  &\int\dd\Omega\,n_{\langle ij\rangle}\,(\Lambda A^{(Q_l)})_{ab}(\Lambda A^{(Q'_{l+2})})_{ab}\\
  &\quad=2\pi\,\frac{l!\,(l+3)(l+4)}{(2l+5)!!}\;\big(Q_{L}Q'_{ijL}\big)^{\mathrm{STF}}\,,\\
  &\int\dd\Omega\,n_{\langle ij\rangle}\,(\Lambda A^{(\cJ_l)})_{ab}(\Lambda A^{(\cI_{l+1})})_{ab}\\
  &\quad=8\pi\,\frac{(l+3)\,(l-1)!}{(2l+3)!!}\;\big(\epsilon_{iab}\cJ_{aL-1}\cI_{bjL-1}\big)^{\mathrm{STF}}\,.
\end{aligned}
\label{eq:angmoments}
\end{equation}
Here $Q_l$ and $Q'_{l'}$ are both mass or both current moments.
For the pair $(\cI_l,\cJ_{l+1})$, the relation for $(\cJ_l,\cI_{l+1})$ holds with $\cI$ and $\cJ$ exchanged and the sign reversed.
No other pair has a quadrupolar component of even parity.
Inserting Eq.~\eqref{eq:angmoments} in Eqs.~\eqref{eq:quadrupolememory} and~\eqref{eq:bdymemory} gives the boundary terms in the form of Eq.~\eqref{eq:bdyfreq}.
In Eq.~\eqref{eq:bdyfreq}, $[Q_2Q_3]_{ij}$ denotes the bare STF contractions of Eq.~\eqref{eq:angmoments}, and $\varrho$ absorbs their angular coefficients.
With the first moment at frequency $\omega$ and the second at $-\omega$, as in Eqs.~\eqref{eq:bdyfreq} and~\eqref{eq:DCflux}, these coefficients are
\begin{equation}
\begin{aligned}
  \varrho[\cI_l,\cI_l]&=-\frac{2\,(l-3)(l+4)(l+2)\,(l-2)!}{(l!)^2\,(2l+3)!!}\,G^2\,,\\
  \varrho[\cI_l,\cI_{l+2}]&=\frac{2\,(l+3)(l+4)}{(l+2)!\,(2l+5)!!}\,G^2\,,\\
  \varrho[\cJ_l,\cJ_l]&=\Big(\frac{2l}{l+1}\Big)^2\varrho[\cI_l,\cI_l]\,,\\
  \varrho[\cJ_l,\cJ_{l+2}]&=\frac{2l}{l+1}\,\frac{2(l+2)}{l+3}\,\varrho[\cI_l,\cI_{l+2}]\,,\\
  \varrho[\cI_l,\cJ_{l+1}]&=-\frac{16\,(l+3)}{l\,l!\,(l+2)\,(2l+3)!!}\,iG^2\,,\\
  \varrho[\cJ_l,\cI_{l+1}]&=\frac{16\,(l+3)}{(l+1)\,(l+1)!\,(2l+3)!!}\,iG^2\,.
\end{aligned}
\label{eq:Rallellrest}
\end{equation}
The coefficients combine the angular integrals~\eqref{eq:angmoments} with the prefactors $1/l!$ and $2l/(l+1)$ of the waveform~\eqref{eq:wavecheck}, which fix the ratios between the mass and current entries, and with the normalisations of Eqs.~\eqref{eq:quadrupolememory} and~\eqref{eq:bdymemory}.
At $l=2$ and $l=3$ they give the coefficients of Eq.~\eqref{eq:SFL6PN}.
The factor $l-3$ makes the quadrupole-root boundary term of $\{\cI_2,\cI_3,\cI_3\}$ vanish.

\section{On-shell matching of the quadratic-curvature coupling}
\label{app:E2onshell}
\begin{figure*}[t]
    \centering
    \(\underset{\text{(a)}}{\vcenter{\hbox{\includegraphics[width=0.112\textwidth]{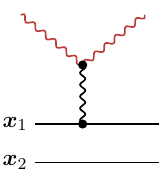}}}}
    \,+\,
    \underset{\text{(b)}}{\vcenter{\hbox{\includegraphics[width=0.112\textwidth]{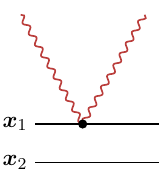}}}}
    \,+\,
    \underset{\text{(c)}}{\vcenter{\hbox{\includegraphics[width=0.112\textwidth]{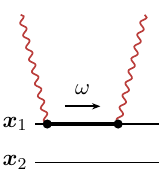}}}}
    \ =\
    \underset{\text{(d)}}{\vcenter{\hbox{\includegraphics[width=0.112\textwidth]{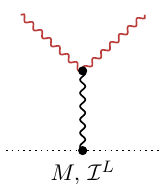}}}}
    \,+\,
    \underset{\text{(e)}}{\vcenter{\hbox{\includegraphics[width=0.112\textwidth]{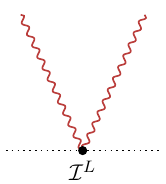}}}}
    \,+\,
    \underset{\text{(f)}}{\vcenter{\hbox{\includegraphics[width=0.112\textwidth]{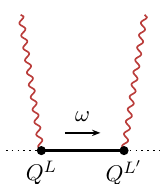}}}}
    \,+\,
    \underset{\text{(g)}}{\vcenter{\hbox{\includegraphics[width=0.112\textwidth]{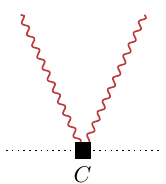}}}}\)
    \caption{
    Matching the constituent and multipolar Compton amplitudes at leading order in binding and constituent velocities.
    The graphs are static exchange, (a) and (d), double emission, (b) and (e), displacement, (c) and (f), and the quadratic-curvature contact (g).
    Red wavy lines denote external gravitons, black wavy lines static gravitons, solid lines constituent worldlines and dotted lines the source worldline.
    The thick segments in (c) and (f) denote the displacement of the constituent, carrying the frequency $\omega$ of one wave.
    For a constituent at rest in TT coordinates, (b) and (c) vanish.
    }
    \label{fig:onshellmatching}
\end{figure*}
We determine the Wilson tensor $C^{abcd}$ of Eq.~\eqref{eq:S2R} by matching elastic graviton scattering computed from the binary constituents and from the multipolar description.
At leading order in binding and constituent velocities, the constituent amplitude is a sum of free-particle amplitudes.
We therefore consider one particle of mass $m$ at position $\bm x$ relative to the reference worldline and sum over the constituents at the end.
The couplings linear in the curvature are fixed by one-graviton matching~\cite{Goldberger:2004jt,Goldberger:2009qd,Ross:2012fc}.
At second order, they contribute through both their two-graviton vertices and the displacement induced by one wave before the constituent couples to the other.
These contributions must be included in the amplitude before extracting the independent local contact, as shown in Fig.~\ref{fig:onshellmatching}.
We use the reduced STF moments of Eqs.~\eqref{eq:binarymultipoles} and~\eqref{eq:I1PN}, with accelerations eliminated by the orbital equations without the external field.
This choice fixes the operator basis, since field redefinitions can redistribute local terms between the multipolar couplings and $C^{abcd}$.
\paragraph*{Contact operator.}
For one constituent at rest, the contact term quadratic in the curvature and quartic in $\bm x$ is built from six rotational invariants, even under parity and time reversal,
\begin{equation}
    \begin{aligned}
        \cL_C=m\Big[&c_1\big(x^iE_{ij}x^j\big)^2+c_2x^2x^i(E^2)_{ij}x^j\\
        &+c_3x^4\operatorname{tr}(E^2)\Big]\\
        +m\Big[&c_1^B\big(x^iB_{ij}x^j\big)^2+c_2^Bx^2x^i(B^2)_{ij}x^j\\
        &+c_3^Bx^4\operatorname{tr}(B^2)\Big]\,,
    \end{aligned}
\label{eq:wilsonbasis}
\end{equation}
where $B_{ij}=\tfrac12\epsilon_{ipq}B_{q|pj}$ is the magnetic curvature of Eq.~\eqref{eq:EB}.
Mixed electric--magnetic structures are excluded by parity.
\paragraph*{Compton amplitude.}
In the rest frame of the source, an incoming graviton of momentum $k_1=\omega_1(1,\bm n_1)$ and polarisation label $\lambda_1$ scatters into an outgoing graviton of momentum $k_2=\omega_2(1,\bm n_2)$ and polarisation label $\lambda_2$.
With $\hat S=1+i\hat T$, the Compton amplitude is defined by
\begin{equation}
    \langle k_2,\lambda_2|\,i\hat T\,|k_1,\lambda_1\rangle=2\pi\delta(\omega_1-\omega_2)\,i\cM(\lambda_1,\lambda_2)\,.
\label{eq:comptondef}
\end{equation}
Since the source is static at this order, $\omega_1=\omega_2=\omega$, and the source absorbs the momentum $\bm q=\omega(\bm n_1-\bm n_2)$.
The waves have polarisations $\varepsilon_1$ and $\varepsilon_2^*$ and phases $e^{-ik_1\cdot x}$ and $e^{ik_2\cdot x}$, normalised as in Eq.~\eqref{eq:metricexp}.
We write the amplitude in terms of the gauge-invariant curvatures of the waves on the worldline,
\begin{equation}
    \begin{aligned}
        E_1&=\frac{\kappa}{2}\omega^2\varepsilon_1\,,\ \ 
        E_2=\frac{\kappa}{2}\omega^2\varepsilon^*_2\,,\ \
        (B_n)_{ij}=\epsilon_{ipq}n_n^p(E_n)_{qj}\,,
    \end{aligned}
\label{eq:E2wavecurvatures}
\end{equation}
and denote by $[\cM]_l$ its term of degree $l$ in $\bm x$.
The matching condition reads
\begin{equation}
    \cM^{\mathrm{near}}=\cM^{\mathrm{far}}\,,
\label{eq:E2matchcondition}
\end{equation}
with $\cM^{\mathrm{near}}$ the constituent amplitude
and $\cM^{\mathrm{far}}$ the multipolar one.
This is shown in Fig.~\ref{fig:onshellmatching}.
We impose this equality order by order in $\bm x$ for arbitrary wave directions and polarisations.
\paragraph*{Near zone.}
The constituent amplitude is
\begin{equation}
    \cM^{\mathrm{near}}=\cM^{\mathrm{near}}_{\mathrm{ex}}+\cM^{\mathrm{near}}_{\mathrm{de}}+\cM^{\mathrm{near}}_{\mathrm{dev}}\,,
\end{equation}
with $\cM^{\mathrm{near}}_{\mathrm{ex}}$ the static exchange, $\cM^{\mathrm{near}}_{\mathrm{de}}$ the double emission and $\cM^{\mathrm{near}}_{\mathrm{dev}}$ the displacement.
We choose TT coordinates for the external waves, so that $h_{00}=h_{0i}=0$.
The point-particle action~\eqref{eq:Spp} then gives neither a two-graviton vertex nor a linear force on a constituent initially at rest.
Hence, only $\cM^{\mathrm{near}}_{\mathrm{ex}}$ survives and gives
\begin{equation}
    \cM^{\mathrm{near}}=\cM^{\mathrm{near}}_{\mathrm{ex}}
    =\frac{2m}{\omega^2\bm q^2}\operatorname{tr}(E_1E_2)\,e^{i\bm q\cdot\bm x}\,.
\label{eq:nearamp}
\end{equation}
Expanding in $\bm x$, the STF terms keep the static pole, while the trace terms are polynomial in $\bm q$,
\begin{equation}
    \begin{aligned}
        \cM^{\mathrm{near}}={}&\frac{2m}{\omega^2}\operatorname{tr}(E_1E_2)
        \bigg[\sum_{l=0}^{4}\frac{i^l}{l!}\,\frac{q_Lx^{\langle L\rangle}}{\bm q^2}\\
        &-\frac{x^2}{6}-\frac{i}{10}x^2(\bm q\cdot\bm x)\\
        &+\frac{x^2(\bm q\cdot\bm x)^2}{28}
        -\frac{\bm q^2x^4}{280}\bigg]+\cO\big(x^5\big)\,.
    \end{aligned}
\label{eq:nearexp}
\end{equation}
\paragraph*{Far zone.}
The far-zone amplitude at leading order reads as
\begin{equation}
   \cM^{\mathrm{far}}=\cM^{\mathrm{far}}_{\mathrm{ex}}+\cM^{\mathrm{far}}_{\mathrm{de}}+\cM^{\mathrm{far}}_{\mathrm{dev}}+\cM_C\,,
\end{equation}
with $\cM^{\mathrm{far}}_{\mathrm{ex}}$ the static exchange, $\cM^{\mathrm{far}}_{\mathrm{de}}$ the double emission, $\cM^{\mathrm{far}}_{\mathrm{dev}}$ the displacement and $\cM_C$ the contact term.
In $\cM^{\mathrm{far}}_{\mathrm{ex}}$, the static field couples to the monopole and mass moments of Eq.~\eqref{eq:Smult},
\begin{equation}
    \cM^{\mathrm{far}}_{\mathrm{ex}}=\frac{2m}{\omega^2}\operatorname{tr}(E_1E_2)
    \sum_{l=0}^{4}\frac{i^l}{l!}\,\frac{q_Lx^{\langle L\rangle}}{\bm q^2}\,.
\label{eq:farex}
\end{equation}
For one constituent, the $l=1$ term is an auxiliary mass dipole, which cancels in the sum over the constituents.
The double emission $\cM^{\mathrm{far}}_{\mathrm{de}}$ follows by expanding the curvature, tetrad and covariant derivatives in the mass couplings to second order in the waves,
\begin{equation}
    \begin{aligned}
        \cM^{\mathrm{far}}_{\mathrm{de}}&=m\sum_{l=2}^{4}
        \frac{x^{\langle a_1\cdots a_l\rangle}}{l!}
        \big[\nabla_{a_3\cdots a_l}E_{a_1a_2}\big]_{12}\,,\\
        [E_{ab}]_{12}&=\frac{1}{\omega^2}\big(E_1E_2+E_2E_1\big)_{ab}\,,
    \end{aligned}
\label{eq:farde}
\end{equation}
where the bracket denotes the vertex bilinear in the two waves.
The multipolar position is measured in the local frame of the source.
It undergoes geodesic deviation even when the constituent stays at rest in TT coordinates.
Coupling the other wave to the displaced moments gives $\cM^{\mathrm{far}}_{\mathrm{dev}}$.
With $p_1=-p_2=\omega$, the linear couplings needed through quartic order in $\bm x$ are
\begin{equation}
    \begin{aligned}
        \frac{\cL_n^{(1)}}m={}&x^i(E_n)_{ij}x^j\Big[\tfrac12+\tfrac{ip_n}{6}\,\bm n_n\cdot\bm x\\
        &\qquad-\tfrac{\omega^2}{24}\big((\bm n_n\cdot\bm x)^2-\tfrac17x^2\big)\Big]\\
        &+(\bm x\times\bm v)^i(B_n)_{ij}x^j\Big[\tfrac23+\tfrac{ip_n}{6}\,\bm n_n\cdot\bm x\Big]\\
        &+\tfrac{ip_n}{12}\,\bm n_n\cdot(\bm x\times\bm v)\,x^i(B_n)_{ij}x^j\,.
    \end{aligned}
\label{eq:ttlinear}
\end{equation}
The terms proportional to $E_n$ contain the mass quadrupole, octupole and hexadecapole, and those proportional to $B_n$ contain the current quadrupole and octupole.
The terms linear in velocity must be retained when computing the force, even though the unperturbed constituent is at rest.
The resulting response is
\begin{equation}
    \begin{aligned}
        F^i_n&=\bigg[\frac{\partial\cL^{(1)}_n}{\partial x^i}
        +ip_n\frac{\partial\cL^{(1)}_n}{\partial v^i}\bigg]_{\bm v=0}\,,\\
        \delta\bm x_n&=-\frac{\bm F_n}{m\omega^2}\,,\\
        \cM^{\mathrm{far}}_{\mathrm{dev}}&=\bm F_2\cdot\delta\bm x_1=-\frac{\bm F_1\cdot\bm F_2}{m\omega^2}\,,
    \end{aligned}
\label{eq:fardev}
\end{equation}
where terms quadratic in the velocity, including those of the reduced 1PN quadrupole, do not contribute at $\bm v=0$.
For the binary, summing the two constituent responses includes both their relative motion and their common displacement.
The contact term $\cM_C$ is the part of Eq.~\eqref{eq:wilsonbasis} bilinear in the waves,
\begin{equation}
    \begin{aligned}
        \cM_C=m\Big[&2c_1\big(x^i(E_1)_{ij}x^j\big)\big(x^k(E_2)_{kl}x^l\big)\\
        &+c_2x^2x^i\big(E_1E_2+E_2E_1\big)_{ij}x^j\\
        &+2c_3x^4\operatorname{tr}(E_1E_2)\Big]\\
        +m\Big[&2c_1^B\big(x^i(B_1)_{ij}x^j\big)\big(x^k(B_2)_{kl}x^l\big)\\
        &+c_2^Bx^2x^i\big(B_1B_2+B_2B_1\big)_{ij}x^j\\
        &+2c_3^Bx^4\operatorname{tr}(B_1B_2)\Big]\,.
    \end{aligned}
\label{eq:farcontact}
\end{equation}
\paragraph*{Matching.}
The multipolar exchange~\eqref{eq:farex} reproduces the STF terms of Eq.~\eqref{eq:nearexp}, including the static pole.
The remaining comparison concerns the trace terms.
At quadratic order in $\bm x$, only the mass quadrupole contributes to the double emission and displacement,
\begin{equation}
    \begin{aligned}
        \big[\cM^{\mathrm{far}}_{\mathrm{de}}\big]_2&=\frac{m}{\omega^2}\Big[x^i(E_1E_2)_{ij}x^j-\tfrac13x^2\operatorname{tr}(E_1E_2)\Big]\,,\\
        \big[\cM^{\mathrm{far}}_{\mathrm{dev}}\big]_2&=-\frac{m}{\omega^2}\,x^i(E_1E_2)_{ij}x^j\,.
    \end{aligned}
\label{eq:farlow}
\end{equation}
Their sum reproduces the quadratic trace term of Eq.~\eqref{eq:nearexp}.
At cubic order, including the mass octupole and current quadrupole gives
\begin{equation}
    \big[\cM^{\mathrm{far}}_{\mathrm{de}}+\cM^{\mathrm{far}}_{\mathrm{dev}}\big]_3
    =-\frac{im}{5\omega^2}\,x^2(\bm q\cdot\bm x)\operatorname{tr}(E_1E_2)\,.
\label{eq:farcubic}
\end{equation}
This reproduces the cubic trace term, so the multipolar action accounts for the constituent amplitude through this order.
At quartic order, the contact supplies the remaining difference,
\begin{equation}
    \begin{aligned}
        \big[\cM^{\mathrm{far}}_{\mathrm{de}}+\cM^{\mathrm{far}}_{\mathrm{dev}}\big]_4
        ={}&\frac{2m}{\omega^2}\operatorname{tr}(E_1E_2)\Big[\frac{x^2(\bm q\cdot\bm x)^2}{28}\\
        &\qquad-\frac{\bm q^2x^4}{280}\Big]-\cM_C\,.
    \end{aligned}
\label{eq:E2quarticmatching}
\end{equation}
At this order, the double emission comes from the hexadecapole coupling.
The displacement pairs the quadrupole with the hexadecapole and with the current octupole, and the octupole and current quadrupole with each other.
Subtracting them from the quartic trace terms leaves the six structures of Eq.~\eqref{eq:farcontact}.
Equating their coefficients for arbitrary wave directions, polarisations and positions fixes
\begin{equation}
    \begin{aligned}
        (c_1,c_2,c_3)&=\tfrac{1}{168}(-19,4,4)\,,\\
        (c_1^B,c_2^B,c_3^B)&=\tfrac{1}{168}(-28,40,-4)\,.
    \end{aligned}
\label{eq:E2coefficients}
\end{equation}
Summing the electric couplings over the constituents gives the binary operator~\eqref{eq:matchedE2}.
\section{Dimensional-regularisation inputs for the tail}
\label{sec:apptail}
The tail action~\eqref{eq:tailonshell} has a pole in $\eps$, so its finite part contains the $\cO(\eps)$ terms of the multipole moments and of the orbital equations.
We therefore need the multipole moments in $d$ dimensions, expanded through $\cO(\eps)$:
the mass quadrupole through 2PN, the mass octupole and current quadrupole through 1PN, and the mass hexadecapole and current octupole at Newtonian order.
The multipole moments are integrals of the stress pseudo-tensor $\tau^{\mu\nu}$ of the matter and of the field in harmonic gauge, each element of the source taken at its retarded time.
We express the moments in the relative variables of Eq.~\eqref{eq:relcoords}, using the 1PN centre-of-mass relation in $d$ dimensions.

\paragraph*{Mass moments.}
The mass moments follow from the general-$d$ formula of Ref.~\cite{Amalberti:2023ohj},
\begin{equation}
\begin{aligned}
    \cI^L={}&\sum_{j\ge0}P_{l,j}\!\int\!\dd^dx\,\Big\{\big[1+4j\,\alpha_l(d+l+j-2)\big]\\
    &\times\partial_t^{2j}\big(\tau^{00}x^{\langle L\rangle}x^{2j}\big)\\
    &-2\alpha_l(d+l+2j-1)\,\partial_t^{2j+1}\big(\tau^{0a}x^ax^{\langle L\rangle}x^{2j}\big)\\
    &+\tfrac{1}{d-2}\big[1+2j(d-2)\alpha_l\big]\partial_t^{2j}\big(\tau^{aa}x^{\langle L\rangle}x^{2j}\big)\\
    &+\alpha_l\,\partial_t^{2j+2}\big(\tau^{ab}x^ax^bx^{\langle L\rangle}x^{2j}\big)\Big\}\,,
\end{aligned}
\label{eq:ILd}\end{equation}
where the trace-free projection is taken in $d$ dimensions and
\begin{equation}
\begin{aligned}
    \alpha_l&=\frac{d-1}{(d-2)(d+l-1)(d+l-2)}\,,\\
    P_{l,j}&=\frac{\Gamma(\tfrac d2+l)}{2^{2j}j!\,\Gamma(\tfrac d2+l+j)}\,.
\end{aligned}
\label{eq:ILdcoeff}\end{equation}
Expanding the retardation in time derivatives gives the sum over $j$: the $j$-th term carries $x^{2j}\partial_t^{2j}$ and is a relative $j$PN correction, so only the first few terms contribute at the orders considered here.

\paragraph*{Current moments.}
The current moments follow from the corresponding formula of Ref.~\cite{Amalberti:2023ohj},
\begin{equation}
\begin{aligned}
    \widetilde\cJ^{a|L}={}&{-}2\sum_{j\ge0}P_{l,j}\!\int\!\dd^dx\,\Big\{\Big[1+\tfrac{2j}{d+l-1}\Big]\partial_t^{2j}\big(\tau^{0a}x^{\langle L\rangle}x^{2j}\big)\\
    &-\tfrac{1}{d+l-1}\,\partial_t^{2j+1}\big(\tau^{ab}x^bx^{\langle L\rangle}x^{2j}\big)\Big\}\,,
\end{aligned}
\label{eq:JLd}\end{equation}
where the right-hand side is antisymmetrised in $a$ and $i_l$ and made trace-free.
In $d=3$, they reduce to the usual current moments,
\begin{equation}
    \widetilde\cJ_{i|L}=\epsilon_{ii_lj}\,\cJ_{jL-1}\,,\qquad \tfrac12\widetilde\cJ_{i|L}\widetilde\cJ_{i|L}=\cJ_L\cJ_L\,.
\label{eq:JLd3}\end{equation}
The multipolar action~\eqref{eq:Smult} contains instead the row-symmetric moments $\cJ^{c|L}$, which are $\tfrac{l}{l+1}$ times the symmetrisation of $\widetilde\cJ^{a|L}$ over $L$, as in Eq.~\eqref{eq:J3d} for $d=3$.
Two of them enter the tail through the contraction
\begin{equation}
    \cJ^{c|L}\cJ^{c|L}-\cJ^{c|iL-1}\cJ^{i|cL-1}=\tfrac12\widetilde\cJ_{c|L}\widetilde\cJ_{c|L}\,,
\label{eq:Jrowcontraction}\end{equation}
which holds for every $l$ and $d$.
With $c_l\to c_l^J$ and $K_l\to K_l^J$, it gives the current-moment form of Eq.~\eqref{eq:tailonshell}.

\paragraph*{Construction of the binary moments.}
To evaluate Eqs.~\eqref{eq:ILd} and~\eqref{eq:JLd}, we use the harmonic-gauge source~\cite{Blanchet:2013haa,Bini:2021gat},
\begin{equation}
    \tau^{\mu\nu}=|g|T^{\mu\nu}+\frac{\cN^{\mu\nu}}{16\pi G}\,.
\label{eq:taumunu}
\end{equation}
Here $T^{\mu\nu}$ is the stress-energy tensor of the point particles, and $\cN^{\mu\nu}$ collects the terms nonlinear in the gravitational field in the harmonic-coordinate Einstein equations.
The matter integrals reduce to particle sums, including the metric corrections generated by the companion.
The linearised metric is of order $G$, so the quadratic part of $\cN^{\mu\nu}$, divided by $G$ in Eq.~\eqref{eq:taumunu}, supplies the field contribution at $\cO(G)$.
We include the first retardation correction to the linearised metric.
The spatial integrals containing $\cN^{\mu\nu}$ reduce to scalar integrals evaluated by analytic continuation of the Riesz formula~\cite{Blanchet:2013haa},
\begin{equation}
\begin{aligned}
    \int\!\dd^dx\,r_1^\alpha r_2^\beta
    ={}&\pi^{d/2}r^{\alpha+\beta+d}\frac{\Gamma(\tfrac{\alpha+d}{2})\Gamma(\tfrac{\beta+d}{2})\Gamma(-\tfrac{\alpha+\beta+d}{2})}
    {\Gamma(-\tfrac\alpha2)\Gamma(-\tfrac\beta2)\Gamma(\tfrac{\alpha+\beta+2d}{2})}\,,
\end{aligned}
\label{eq:tailRiesz}
\end{equation}
where $r_a=|\bm x-\bm x_a|$.
Derivatives of the potentials include the distributional contact terms at the particle positions.

\paragraph*{Tail at $\cO(G^4)$.}
Each factor of Eq.~\eqref{eq:tailonshell} is a moment differentiated $l+1$ times, which on the straight line $\bm r_0$ of Eq.~\eqref{eq:orbitexp} starts at $\cO(G)$.
We compute the derivatives with the orbital equations,
\begin{equation}
    Q^{(l+1)}=\left(\bm v\cdot\partial_{\bm r}
    +\bm a_{\mathrm{orb}}\cdot\partial_{\bm v}\right)^{l+1}Q\,,
\label{eq:tailMomentDerivative}
\end{equation}
where $Q(\bm r,\bm v;d)$ is a mass or current moment and $\bm a_{\mathrm{orb}}$ the $d$-dimensional relative acceleration, which at $\cO(G)$ through 2PN follows from one potential-graviton exchange~\cite{Goldberger:2004jt,Gilmore:2008gq}.
We denote by $Q^{(l+1)}_{G^n}$ the $\cO(G^n)$ terms of Eq.~\eqref{eq:tailMomentDerivative} as functions of $\bm r$ and $\bm v$, before evaluation on the orbit.
Since Eq.~\eqref{eq:tailonshell} carries $G^2$, at $\cO(G^4)$ both factors are needed at $\cO(G)$, where the deflection of the orbit does not enter,
\begin{equation}
    [Q^{(l+1)}]_G=Q^{(l+1)}_G\big|_{\bm r_0}\,.
\label{eq:tailG1moment}
\end{equation}
We form the tensor contractions in $d$ dimensions and expand through $\cO(\eps)$.
The pole and the constant $K_l$ multiply the time integral of the contraction, while the frequency logarithm is finite and requires only the three-dimensional inputs.
The angle follows from Eq.~\eqref{eq:chicons}.

\paragraph*{Tail at $\cO(G^5)$.}
At $\cO(G^5)$, one factor is needed at $\cO(G^2)$, from the $\cO(G^2)$ terms on the straight line and the $\cO(G)$ terms on the deflected orbit,
\begin{equation}
    [Q^{(l+1)}]_{G^2}=Q^{(l+1)}_{G^2}\big|_{\bm r_0}+\big([\bm r]_G\cdot\partial_{\bm r}+[\bm v]_G\cdot\partial_{\bm v}\big)Q^{(l+1)}_{G}\big|_{\bm r_0}\,,
\label{eq:tailG2moment}
\end{equation}
with $[\bm r]_G$ and $[\bm v]_G$ the $\cO(G)$ deflection of Eq.~\eqref{eq:orbitexp}, computed with the $d$-dimensional acceleration.
Besides the $\cO(G)$ acceleration and moments, $Q^{(l+1)}_{G^2}$ contains the $\cO(G^2)$ terms of the moments and one insertion of the $\cO(G^2)$ acceleration.
Through 6PN, these are the static $\cO(G^2)$ term of the 2PN mass quadrupole,
\begin{equation}
    \cI^{ij}_{G^2}=\alpha_d\,\nu M\bigg[\frac{2(d-2)}{d-1}\,\frac{\Gamma(\frac d2-1)}{\pi^{d/2-1}}\,\frac{GM}{r^{d-2}}\bigg]^2 r^{\langle ij\rangle}\,,
\label{eq:staticG2quad}
\end{equation}
with
\begin{equation}
\begin{aligned}
    \alpha_d={}&\frac{N_0+\nu\,(d-6)N_1-3\nu^2(d-6)(d-4)(d-2)N_2}{12\,(d-6)(d-4)(d-2)^2(d+4)(d+6)}\,,\\
    N_0={}&-d^7+4d^6-232d^5+2332d^4-8728d^3\\
    &-2384d^2+80544d-120960\,,\\
    N_1={}&9d^6-31d^5+718d^4-3996d^3+7400d^2\\
    &+832d-12672\,,\\
    N_2={}&d^4-10d^3+60d^2-704d+1424\,,
\end{aligned}
\label{eq:alphad}
\end{equation}
and the relative acceleration at $\cO(G^2)$ through 2PN, which we derive from the $d$-dimensional near-zone Lagrangian in harmonic gauge~\cite{Gilmore:2008gq,Foffa:2011ub}.
\paragraph*{Checks.}
The 1PN mass quadrupole, Newtonian current quadrupole and 1PN relative acceleration agree through $\cO(G)$ and $\cO(\eps)$ with the $d$-dimensional results of Ref.~\cite{Bini:2021gat}.
At $d=3$, the mass quadrupole through 2PN and the mass octupole and current quadrupole through 1PN agree through $\cO(G)$ with Ref.~\cite{Amalberti:2024jaa}, and the 2PN acceleration reproduces the harmonic-coordinate result of Ref.~\cite{Blanchet:2013haa} through $\cO(G^2)$.
\section{Sector contributions to the scattering angles}
\label{sec:app_scattering_tables}
Table~\ref{tab:sectorangles} gives the sector contributions to the two-radiation part of the 4PM, 5PM and 6PM angles through 6PN.
The potential region and ordinary tail are kept separate.
\begin{table*}[p]
\centering\footnotesize\setlength{\tabcolsep}{4pt}
\renewcommand{\arraystretch}{1.2}
\begin{ruledtabular}
\begin{tabular}{lrrrrrr}
4PM & \multicolumn{3}{c}{Local} & \multicolumn{3}{c}{Retarded}\\
Sector & $\nu^2p_\infty^6$ & $\nu^2p_\infty^8$ & $\nu^3p_\infty^8$ & $\nu^2p_\infty^6$ & $\nu^2p_\infty^8$ & $\nu^3p_\infty^8$\\
\colrule
$\{L,\cI_3,\cI_3\}$ & $0$ & $130923/89600$ & $-130923/22400$ & $0$ & $43641/17920$ & $-43641/4480$ \\
$\{L,\cJ_2,\cJ_2\}$ & $0$ & $9/128$ & $-9/32$ & $0$ & $15/128$ & $-15/32$ \\
$\{L,\cI_2,\cJ_3\}$ & $0$ & $-993/2240$ & $2979/2240$ & $0$ & $-331/448$ & $993/448$ \\
$\{L,\cI_3,\cJ_2\}$ & $0$ & $369/11200$ & $-369/2800$ & $0$ & $123/2240$ & $-123/560$ \\
$\{\cI_2,\cI_2,\cI_4\}$ & $0$ & $-8771617/2822400$ & $8771617/940800$ & $0$ & $-14211289/2822400$ & $14211289/940800$ \\
$\{\cI_2,\cI_3,\cI_3\}$ & $0$ & $-216289/161280$ & $216289/40320$ & $0$ & $-391397/115200$ & $391397/28800$ \\
$\{\cI_2,\cJ_2,\cJ_2\}$ & $0$ & $-387/896$ & $387/224$ & $0$ & $-739/640$ & $739/160$ \\
$\{\cI_2,\cI_2,\cJ_3\}$ & $0$ & $-30099/22400$ & $90297/22400$ & $0$ & $-224803/67200$ & $224803/22400$ \\
$\{\cI_2,\cI_3,\cJ_2\}$ & $0$ & $-31603/22400$ & $31603/5600$ & $0$ & $-236981/67200$ & $236981/16800$ \\
$\{L,\cI_2,\cI_2\}$ & $69/80$ & $-969/896$ & $1359/448$ & $23/16$ & $-1615/896$ & $2265/448$ \\
$\{\cI_2,\cI_2,\cI_2\}$ & $-477/560$ & $-1101/980$ & $-19863/31360$ & $-529/280$ & $-295459/94080$ & $-7033/6272$ \\
$2\mathrm{RR}$ & $159/25$ & $6561/1400$ & $-7407/280$ & $211/20$ & $7939/560$ & $-23577/350$ \\
$E^2$ & $0$ & $-117/112$ & $351/112$ & $0$ & $-9987/2800$ & $29961/2800$ \\
\colrule
$\widetilde\chi^{(4)}_{2\mathrm{R}}$ & $8919/1400$ & $-1023607/201600$ & $559/2304$ & $1131/112$ & $-256999/28800$ & $-200729/57600$ \\
\end{tabular}
\end{ruledtabular}

\vspace{6pt}
\begin{ruledtabular}
\begin{tabular}{lrrrrrr}
5PM & \multicolumn{3}{c}{Local} & \multicolumn{3}{c}{Retarded}\\
Sector & $\nu^2p_\infty^5$ & $\nu^2p_\infty^7$ & $\nu^3p_\infty^7$ & $\nu^2p_\infty^5$ & $\nu^2p_\infty^7$ & $\nu^3p_\infty^7$\\
\colrule
$\{L,\cI_3,\cI_3\}$ & $0$ & $29312/525$ & $-117248/525$ & $0$ & $14656/175$ & $-58624/175$ \\
$\{L,\cJ_2,\cJ_2\}$ & $0$ & $512/225$ & $-2048/225$ & $0$ & $256/75$ & $-1024/75$ \\
$\{L,\cI_2,\cJ_3\}$ & $0$ & $-73984/4725$ & $73984/1575$ & $0$ & $-36992/1575$ & $36992/525$ \\
$\{L,\cI_3,\cJ_2\}$ & $0$ & $512/525$ & $-2048/525$ & $0$ & $256/175$ & $-1024/175$ \\
$\{\cI_2,\cI_2,\cI_4\}$ & $0$ & $-1136512/8575$ & $3409536/8575$ & $0$ & $-7788224/42875$ & $23364672/42875$ \\
$\{\cI_2,\cI_3,\cI_3\}$ & $0$ & $-235904/4725$ & $943616/4725$ & $0$ & $-8075584/70875$ & $32302336/70875$ \\
$\{\cI_2,\cJ_2,\cJ_2\}$ & $0$ & $-151808/11025$ & $607232/11025$ & $0$ & $-3338752/99225$ & $13355008/99225$ \\
$\{\cI_2,\cI_2,\cJ_3\}$ & $0$ & $-7654912/165375$ & $7654912/55125$ & $0$ & $-17017856/165375$ & $17017856/55125$ \\
$\{\cI_2,\cI_3,\cJ_2\}$ & $0$ & $-457984/11025$ & $1831936/11025$ & $0$ & $-1098688/11025$ & $4394752/11025$ \\
$\{L,\cI_2,\cI_2\}$ & $1088/45$ & $-173696/4725$ & $62464/525$ & $544/15$ & $-86848/1575$ & $31232/175$ \\
$\{\cI_2,\cI_2,\cI_2\}$ & $-3008/135$ & $-4580288/231525$ & $-2455552/77175$ & $-15136/315$ & $-10618624/231525$ & $-18054464/231525$ \\
$2\mathrm{RR}$ & $45824/225$ & $-134912/33075$ & $-149844736/165375$ & $257024/945$ & $1331968/4725$ & $-329821952/165375$ \\
$E^2$ & $0$ & $-8448/245$ & $25344/245$ & $0$ & $-604928/6125$ & $1814784/6125$ \\
\colrule
$\widetilde\chi^{(5)}_{2\mathrm{R}}$ & $138752/675$ & $-55523392/165375$ & $82816/1575$ & $245888/945$ & $-63591872/165375$ & $-4096/105$ \\
\end{tabular}
\end{ruledtabular}

\vspace{6pt}
\begin{ruledtabular}
\begin{tabular}{lrrr}
6PM & \multicolumn{3}{c}{Retarded}\\
Sector & $\nu^2p_\infty^4$ & $\nu^2p_\infty^6$ & $\nu^3p_\infty^6$\\
\colrule
$\{L,\cI_3,\cI_3\}$ & $0$ & $266099/1920$ & $-266099/480$ \\
$\{L,\cJ_2,\cJ_2\}$ & $0$ & $1421/288$ & $-1421/72$ \\
$\{L,\cI_2,\cJ_3\}$ & $0$ & $-5273/144$ & $5273/48$ \\
$\{L,\cI_3,\cJ_2\}$ & $0$ & $1411/720$ & $-1411/180$ \\
$\{\cI_2,\cI_2,\cI_4\}$ & $0$ & $-327569131/1058400$ & $327569131/352800$ \\
$\{\cI_2,\cI_3,\cI_3\}$ & $0$ & $-112019153/604800$ & $112019153/151200$ \\
$\{\cI_2,\cJ_2,\cJ_2\}$ & $0$ & $-293701/6048$ & $293701/1512$ \\
$\{\cI_2,\cI_2,\cJ_3\}$ & $0$ & $-3132257/20160$ & $3132257/6720$ \\
$\{\cI_2,\cI_3,\cJ_2\}$ & $0$ & $-706607/5040$ & $706607/1260$ \\
$\{L,\cI_2,\cI_2\}$ & $707/16$ & $-19283/288$ & $28103/96$ \\
$\{\cI_2,\cI_2,\cI_2\}$ & $-14891/252$ & $-1391527/47040$ & $-1913027/11760$ \\
$2\mathrm{RR}$ & $630803/1800$ & $1387993/3600$ & $-25108879/8400$ \\
$E^2$ & $0$ & $-288941/2100$ & $288941/700$ \\
\colrule
$\widetilde\chi^{(6)}_{2\mathrm{R}}$ & $8455667/25200$ & $-43738223/75600$ & $-312469/11200$ \\
\end{tabular}
\end{ruledtabular}

\caption{
Sector contributions to the two-radiation angles $\widetilde\chi^{(4)}_{2\mathrm{R}}$, $\widetilde\chi^{(5)}_{2\mathrm{R}}$ and $\widetilde\chi^{(6)}_{2\mathrm{R}}$ through 6PN.
At 4PM and 5PM they are given in the local prescription and with retarded propagators, and at 6PM with retarded propagators.
At 4PM, the local prescription coincides with the $\gamma$-3 one.
The 4PM and 6PM entries are in units of $\pi$.
The sectors $\{L,\cI_2,\cI_2\}$ and $\{\cI_2,\cI_2,\cI_2\}$ include their leading 5PN terms and 1PN corrections.
The $2\mathrm{RR}$ row includes both terms of Eq.~\eqref{eq:2RRdef}, and at 6PM also the quadratic term of Eq.~\eqref{eq:chitotangle}.
The last row of each block gives the sum of the sectors.
}
\label{tab:sectorangles}
\end{table*}
\clearpage

\bibliography{refs}

%apsrev4-2.bst 2019-01-14 (MD) hand-edited version of apsrev4-1.bst
%Control: key (0)
%Control: author (8) initials jnrlst
%Control: editor formatted (1) identically to author
%Control: production of article title (0) allowed
%Control: page (0) single
%Control: year (1) truncated
%Control: production of eprint (0) enabled
\begin{thebibliography}{181}%
\makeatletter
\providecommand \@ifxundefined [1]{%
 \@ifx{#1\undefined}
}%
\providecommand \@ifnum [1]{%
 \ifnum #1\expandafter \@firstoftwo
 \else \expandafter \@secondoftwo
 \fi
}%
\providecommand \@ifx [1]{%
 \ifx #1\expandafter \@firstoftwo
 \else \expandafter \@secondoftwo
 \fi
}%
\providecommand \natexlab [1]{#1}%
\providecommand \enquote  [1]{``#1''}%
\providecommand \bibnamefont  [1]{#1}%
\providecommand \bibfnamefont [1]{#1}%
\providecommand \citenamefont [1]{#1}%
\providecommand \href@noop [0]{\@secondoftwo}%
\providecommand \href [0]{\begingroup \@sanitize@url \@href}%
\providecommand \@href[1]{\@@startlink{#1}\@@href}%
\providecommand \@@href[1]{\endgroup#1\@@endlink}%
\providecommand \@sanitize@url [0]{\catcode `\\12\catcode `\$12\catcode `\&12\catcode `\#12\catcode `\^12\catcode `\_12\catcode `\%12\relax}%
\providecommand \@@startlink[1]{}%
\providecommand \@@endlink[0]{}%
\providecommand \url  [0]{\begingroup\@sanitize@url \@url }%
\providecommand \@url [1]{\endgroup\@href {#1}{\urlprefix }}%
\providecommand \urlprefix  [0]{URL }%
\providecommand \Eprint [0]{\href }%
\providecommand \doibase [0]{https://doi.org/}%
\providecommand \selectlanguage [0]{\@gobble}%
\providecommand \bibinfo  [0]{\@secondoftwo}%
\providecommand \bibfield  [0]{\@secondoftwo}%
\providecommand \translation [1]{[#1]}%
\providecommand \BibitemOpen [0]{}%
\providecommand \bibitemStop [0]{}%
\providecommand \bibitemNoStop [0]{.\EOS\space}%
\providecommand \EOS [0]{\spacefactor3000\relax}%
\providecommand \BibitemShut  [1]{\csname bibitem#1\endcsname}%
\let\auto@bib@innerbib\@empty
%</preamble>
\bibitem [{\citenamefont {Abbott}\ \emph {et~al.}(2016)\citenamefont {Abbott} \emph {et~al.}}]{LIGOScientific:2016aoc}%
  \BibitemOpen
  \bibfield  {author} {\bibinfo {author} {\bibfnamefont {B.~P.}\ \bibnamefont {Abbott}} \emph {et~al.} (\bibinfo {collaboration} {LIGO Scientific, Virgo}),\ }\bibfield  {title} {\bibinfo {title} {{Observation of Gravitational Waves from a Binary Black Hole Merger}},\ }\href {https://doi.org/10.1103/PhysRevLett.116.061102} {\bibfield  {journal} {\bibinfo  {journal} {Phys. Rev. Lett.}\ }\textbf {\bibinfo {volume} {116}},\ \bibinfo {pages} {061102} (\bibinfo {year} {2016})},\ \Eprint {https://arxiv.org/abs/1602.03837} {arXiv:1602.03837 [gr-qc]} \BibitemShut {NoStop}%
\bibitem [{\citenamefont {Abbott}\ \emph {et~al.}(2023)\citenamefont {Abbott} \emph {et~al.}}]{KAGRA:2021vkt}%
  \BibitemOpen
  \bibfield  {author} {\bibinfo {author} {\bibfnamefont {R.}~\bibnamefont {Abbott}} \emph {et~al.} (\bibinfo {collaboration} {KAGRA, VIRGO, LIGO Scientific}),\ }\bibfield  {title} {\bibinfo {title} {{GWTC-3: Compact Binary Coalescences Observed by LIGO and Virgo during the Second Part of the Third Observing Run}},\ }\href {https://doi.org/10.1103/PhysRevX.13.041039} {\bibfield  {journal} {\bibinfo  {journal} {Phys. Rev. X}\ }\textbf {\bibinfo {volume} {13}},\ \bibinfo {pages} {041039} (\bibinfo {year} {2023})},\ \Eprint {https://arxiv.org/abs/2111.03606} {arXiv:2111.03606 [gr-qc]} \BibitemShut {NoStop}%
\bibitem [{\citenamefont {Abac}\ \emph {et~al.}(2026)\citenamefont {Abac} \emph {et~al.}}]{LIGOScientific:2025slb}%
  \BibitemOpen
  \bibfield  {author} {\bibinfo {author} {\bibfnamefont {A.~G.}\ \bibnamefont {Abac}} \emph {et~al.} (\bibinfo {collaboration} {LIGO Scientific, VIRGO, KAGRA}),\ }\bibfield  {title} {\bibinfo {title} {{GWTC-4.0: Updating the Gravitational-Wave Transient Catalog with Observations from the First Part of the Fourth LIGO-Virgo-KAGRA Observing Run}},\ }\href {https://doi.org/10.3847/2041-8213/ae2c74} {\bibfield  {journal} {\bibinfo  {journal} {Astrophys. J. Lett.}\ }\textbf {\bibinfo {volume} {1004}},\ \bibinfo {pages} {L22} (\bibinfo {year} {2026})},\ \Eprint {https://arxiv.org/abs/2508.18082} {arXiv:2508.18082 [gr-qc]} \BibitemShut {NoStop}%
\bibitem [{\citenamefont {Punturo}\ \emph {et~al.}(2010)\citenamefont {Punturo} \emph {et~al.}}]{Punturo:2010zz}%
  \BibitemOpen
  \bibfield  {author} {\bibinfo {author} {\bibfnamefont {M.}~\bibnamefont {Punturo}} \emph {et~al.},\ }\bibfield  {title} {\bibinfo {title} {{The Einstein Telescope: A third-generation gravitational wave observatory}},\ }\href {https://doi.org/10.1088/0264-9381/27/19/194002} {\bibfield  {journal} {\bibinfo  {journal} {Class. Quant. Grav.}\ }\textbf {\bibinfo {volume} {27}},\ \bibinfo {pages} {194002} (\bibinfo {year} {2010})}\BibitemShut {NoStop}%
\bibitem [{\citenamefont {Reitze}\ \emph {et~al.}(2019)\citenamefont {Reitze} \emph {et~al.}}]{Reitze:2019iox}%
  \BibitemOpen
  \bibfield  {author} {\bibinfo {author} {\bibfnamefont {D.}~\bibnamefont {Reitze}} \emph {et~al.},\ }\bibfield  {title} {\bibinfo {title} {{Cosmic Explorer: The U.S. Contribution to Gravitational-Wave Astronomy beyond LIGO}},\ }\href@noop {} {\bibfield  {journal} {\bibinfo  {journal} {Bull. Am. Astron. Soc.}\ }\textbf {\bibinfo {volume} {51}},\ \bibinfo {pages} {035} (\bibinfo {year} {2019})},\ \Eprint {https://arxiv.org/abs/1907.04833} {arXiv:1907.04833 [astro-ph.IM]} \BibitemShut {NoStop}%
\bibitem [{\citenamefont {Amaro-Seoane}\ \emph {et~al.}(2017)\citenamefont {Amaro-Seoane} \emph {et~al.}}]{LISA:2017pwj}%
  \BibitemOpen
  \bibfield  {author} {\bibinfo {author} {\bibfnamefont {P.}~\bibnamefont {Amaro-Seoane}} \emph {et~al.} (\bibinfo {collaboration} {LISA}),\ }\href@noop {} {\bibinfo {title} {{Laser Interferometer Space Antenna}}} (\bibinfo {year} {2017}),\ \Eprint {https://arxiv.org/abs/1702.00786} {arXiv:1702.00786 [astro-ph.IM]} \BibitemShut {NoStop}%
\bibitem [{\citenamefont {P{\"u}rrer}\ and\ \citenamefont {Haster}(2020)}]{Purrer:2019jcp}%
  \BibitemOpen
  \bibfield  {author} {\bibinfo {author} {\bibfnamefont {M.}~\bibnamefont {P{\"u}rrer}}\ and\ \bibinfo {author} {\bibfnamefont {C.-J.}\ \bibnamefont {Haster}},\ }\bibfield  {title} {\bibinfo {title} {{Gravitational waveform accuracy requirements for future ground-based detectors}},\ }\href {https://doi.org/10.1103/PhysRevResearch.2.023151} {\bibfield  {journal} {\bibinfo  {journal} {Phys. Rev. Res.}\ }\textbf {\bibinfo {volume} {2}},\ \bibinfo {pages} {023151} (\bibinfo {year} {2020})},\ \Eprint {https://arxiv.org/abs/1912.10055} {arXiv:1912.10055 [gr-qc]} \BibitemShut {NoStop}%
\bibitem [{\citenamefont {Owen}\ \emph {et~al.}(2023)\citenamefont {Owen}, \citenamefont {Haster}, \citenamefont {Perkins}, \citenamefont {Cornish},\ and\ \citenamefont {Yunes}}]{Owen:2023mid}%
  \BibitemOpen
  \bibfield  {author} {\bibinfo {author} {\bibfnamefont {C.~B.}\ \bibnamefont {Owen}}, \bibinfo {author} {\bibfnamefont {C.-J.}\ \bibnamefont {Haster}}, \bibinfo {author} {\bibfnamefont {S.}~\bibnamefont {Perkins}}, \bibinfo {author} {\bibfnamefont {N.~J.}\ \bibnamefont {Cornish}},\ and\ \bibinfo {author} {\bibfnamefont {N.}~\bibnamefont {Yunes}},\ }\bibfield  {title} {\bibinfo {title} {{Waveform accuracy and systematic uncertainties in current gravitational wave observations}},\ }\href {https://doi.org/10.1103/PhysRevD.108.044018} {\bibfield  {journal} {\bibinfo  {journal} {Phys. Rev. D}\ }\textbf {\bibinfo {volume} {108}},\ \bibinfo {pages} {044018} (\bibinfo {year} {2023})},\ \Eprint {https://arxiv.org/abs/2301.11941} {arXiv:2301.11941 [gr-qc]} \BibitemShut {NoStop}%
\bibitem [{\citenamefont {Dhani}\ \emph {et~al.}(2025)\citenamefont {Dhani}, \citenamefont {V{\"o}lkel}, \citenamefont {Buonanno}, \citenamefont {Estelles}, \citenamefont {Gair}, \citenamefont {Pfeiffer}, \citenamefont {Pompili},\ and\ \citenamefont {Toubiana}}]{Dhani:2024jja}%
  \BibitemOpen
  \bibfield  {author} {\bibinfo {author} {\bibfnamefont {A.}~\bibnamefont {Dhani}}, \bibinfo {author} {\bibfnamefont {S.~H.}\ \bibnamefont {V{\"o}lkel}}, \bibinfo {author} {\bibfnamefont {A.}~\bibnamefont {Buonanno}}, \bibinfo {author} {\bibfnamefont {H.}~\bibnamefont {Estelles}}, \bibinfo {author} {\bibfnamefont {J.}~\bibnamefont {Gair}}, \bibinfo {author} {\bibfnamefont {H.~P.}\ \bibnamefont {Pfeiffer}}, \bibinfo {author} {\bibfnamefont {L.}~\bibnamefont {Pompili}},\ and\ \bibinfo {author} {\bibfnamefont {A.}~\bibnamefont {Toubiana}},\ }\bibfield  {title} {\bibinfo {title} {{Systematic Biases in Estimating the Properties of Black Holes Due to Inaccurate Gravitational-Wave Models}},\ }\href {https://doi.org/10.1103/5pks-qz6b} {\bibfield  {journal} {\bibinfo  {journal} {Phys. Rev. X}\ }\textbf {\bibinfo {volume} {15}},\ \bibinfo {pages} {031036} (\bibinfo {year} {2025})},\ \Eprint {https://arxiv.org/abs/2404.05811} {arXiv:2404.05811 [gr-qc]} \BibitemShut {NoStop}%
\bibitem [{\citenamefont {Einstein}\ \emph {et~al.}(1938)\citenamefont {Einstein}, \citenamefont {Infeld},\ and\ \citenamefont {Hoffmann}}]{Einstein:1938yz}%
  \BibitemOpen
  \bibfield  {author} {\bibinfo {author} {\bibfnamefont {A.}~\bibnamefont {Einstein}}, \bibinfo {author} {\bibfnamefont {L.}~\bibnamefont {Infeld}},\ and\ \bibinfo {author} {\bibfnamefont {B.}~\bibnamefont {Hoffmann}},\ }\bibfield  {title} {\bibinfo {title} {{The Gravitational equations and the problem of motion}},\ }\href {https://doi.org/10.2307/1968714} {\bibfield  {journal} {\bibinfo  {journal} {Annals Math.}\ }\textbf {\bibinfo {volume} {39}},\ \bibinfo {pages} {65} (\bibinfo {year} {1938})}\BibitemShut {NoStop}%
\bibitem [{\citenamefont {Blanchet}(2014)}]{Blanchet:2013haa}%
  \BibitemOpen
  \bibfield  {author} {\bibinfo {author} {\bibfnamefont {L.}~\bibnamefont {Blanchet}},\ }\bibfield  {title} {\bibinfo {title} {{Post-Newtonian Theory for Gravitational Waves}},\ }\href {https://doi.org/10.12942/lrr-2014-2} {\bibfield  {journal} {\bibinfo  {journal} {Living Rev. Rel.}\ }\textbf {\bibinfo {volume} {17}},\ \bibinfo {pages} {2} (\bibinfo {year} {2014})},\ \Eprint {https://arxiv.org/abs/1310.1528} {arXiv:1310.1528 [gr-qc]} \BibitemShut {NoStop}%
\bibitem [{\citenamefont {Porto}(2016{\natexlab{a}})}]{Porto:2016pyg}%
  \BibitemOpen
  \bibfield  {author} {\bibinfo {author} {\bibfnamefont {R.~A.}\ \bibnamefont {Porto}},\ }\bibfield  {title} {\bibinfo {title} {{The effective field theorist{\textquoteright}s approach to gravitational dynamics}},\ }\href {https://doi.org/10.1016/j.physrep.2016.04.003} {\bibfield  {journal} {\bibinfo  {journal} {Phys. Rept.}\ }\textbf {\bibinfo {volume} {633}},\ \bibinfo {pages} {1} (\bibinfo {year} {2016}{\natexlab{a}})},\ \Eprint {https://arxiv.org/abs/1601.04914} {arXiv:1601.04914 [hep-th]} \BibitemShut {NoStop}%
\bibitem [{\citenamefont {Sch{\"a}fer}\ and\ \citenamefont {Jaranowski}(2024)}]{Schafer:2018jfw}%
  \BibitemOpen
  \bibfield  {author} {\bibinfo {author} {\bibfnamefont {G.}~\bibnamefont {Sch{\"a}fer}}\ and\ \bibinfo {author} {\bibfnamefont {P.}~\bibnamefont {Jaranowski}},\ }\bibfield  {title} {\bibinfo {title} {{Hamiltonian formulation of general relativity and post-Newtonian dynamics of compact binaries}},\ }\href {https://doi.org/10.1007/s41114-024-00048-7} {\bibfield  {journal} {\bibinfo  {journal} {Living Rev. Rel.}\ }\textbf {\bibinfo {volume} {27}},\ \bibinfo {pages} {2} (\bibinfo {year} {2024})},\ \Eprint {https://arxiv.org/abs/1805.07240} {arXiv:1805.07240 [gr-qc]} \BibitemShut {NoStop}%
\bibitem [{\citenamefont {Levi}(2020)}]{Levi:2018nxp}%
  \BibitemOpen
  \bibfield  {author} {\bibinfo {author} {\bibfnamefont {M.}~\bibnamefont {Levi}},\ }\bibfield  {title} {\bibinfo {title} {{Effective Field Theories of Post-Newtonian Gravity: A comprehensive review}},\ }\href {https://doi.org/10.1088/1361-6633/ab12bc} {\bibfield  {journal} {\bibinfo  {journal} {Rept. Prog. Phys.}\ }\textbf {\bibinfo {volume} {83}},\ \bibinfo {pages} {075901} (\bibinfo {year} {2020})},\ \Eprint {https://arxiv.org/abs/1807.01699} {arXiv:1807.01699 [hep-th]} \BibitemShut {NoStop}%
\bibitem [{\citenamefont {Jaranowski}\ and\ \citenamefont {Schaefer}(1998)}]{Jaranowski:1997ky}%
  \BibitemOpen
  \bibfield  {author} {\bibinfo {author} {\bibfnamefont {P.}~\bibnamefont {Jaranowski}}\ and\ \bibinfo {author} {\bibfnamefont {G.}~\bibnamefont {Schaefer}},\ }\bibfield  {title} {\bibinfo {title} {{Third postNewtonian higher order ADM Hamilton dynamics for two-body point mass systems}},\ }\href {https://doi.org/10.1103/PhysRevD.57.7274} {\bibfield  {journal} {\bibinfo  {journal} {Phys. Rev. D}\ }\textbf {\bibinfo {volume} {57}},\ \bibinfo {pages} {7274} (\bibinfo {year} {1998})},\ \bibinfo {note} {[Erratum: Phys.Rev.D 63, 029902 (2001)]},\ \Eprint {https://arxiv.org/abs/gr-qc/9712075} {arXiv:gr-qc/9712075} \BibitemShut {NoStop}%
\bibitem [{\citenamefont {Damour}\ \emph {et~al.}(2014)\citenamefont {Damour}, \citenamefont {Jaranowski},\ and\ \citenamefont {Sch{\"a}fer}}]{Damour:2014jta}%
  \BibitemOpen
  \bibfield  {author} {\bibinfo {author} {\bibfnamefont {T.}~\bibnamefont {Damour}}, \bibinfo {author} {\bibfnamefont {P.}~\bibnamefont {Jaranowski}},\ and\ \bibinfo {author} {\bibfnamefont {G.}~\bibnamefont {Sch{\"a}fer}},\ }\bibfield  {title} {\bibinfo {title} {{Nonlocal-in-time action for the fourth post-Newtonian conservative dynamics of two-body systems}},\ }\href {https://doi.org/10.1103/PhysRevD.89.064058} {\bibfield  {journal} {\bibinfo  {journal} {Phys. Rev. D}\ }\textbf {\bibinfo {volume} {89}},\ \bibinfo {pages} {064058} (\bibinfo {year} {2014})},\ \Eprint {https://arxiv.org/abs/1401.4548} {arXiv:1401.4548 [gr-qc]} \BibitemShut {NoStop}%
\bibitem [{\citenamefont {Jaranowski}\ and\ \citenamefont {Sch{\"a}fer}(2015)}]{Jaranowski:2015lha}%
  \BibitemOpen
  \bibfield  {author} {\bibinfo {author} {\bibfnamefont {P.}~\bibnamefont {Jaranowski}}\ and\ \bibinfo {author} {\bibfnamefont {G.}~\bibnamefont {Sch{\"a}fer}},\ }\bibfield  {title} {\bibinfo {title} {{Derivation of local-in-time fourth post-Newtonian ADM Hamiltonian for spinless compact binaries}},\ }\href {https://doi.org/10.1103/PhysRevD.92.124043} {\bibfield  {journal} {\bibinfo  {journal} {Phys. Rev. D}\ }\textbf {\bibinfo {volume} {92}},\ \bibinfo {pages} {124043} (\bibinfo {year} {2015})},\ \Eprint {https://arxiv.org/abs/1508.01016} {arXiv:1508.01016 [gr-qc]} \BibitemShut {NoStop}%
\bibitem [{\citenamefont {Bernard}\ \emph {et~al.}(2016)\citenamefont {Bernard}, \citenamefont {Blanchet}, \citenamefont {Boh{\'e}}, \citenamefont {Faye},\ and\ \citenamefont {Marsat}}]{Bernard:2015njp}%
  \BibitemOpen
  \bibfield  {author} {\bibinfo {author} {\bibfnamefont {L.}~\bibnamefont {Bernard}}, \bibinfo {author} {\bibfnamefont {L.}~\bibnamefont {Blanchet}}, \bibinfo {author} {\bibfnamefont {A.}~\bibnamefont {Boh{\'e}}}, \bibinfo {author} {\bibfnamefont {G.}~\bibnamefont {Faye}},\ and\ \bibinfo {author} {\bibfnamefont {S.}~\bibnamefont {Marsat}},\ }\bibfield  {title} {\bibinfo {title} {{Fokker action of nonspinning compact binaries at the fourth post-Newtonian approximation}},\ }\href {https://doi.org/10.1103/PhysRevD.93.084037} {\bibfield  {journal} {\bibinfo  {journal} {Phys. Rev. D}\ }\textbf {\bibinfo {volume} {93}},\ \bibinfo {pages} {084037} (\bibinfo {year} {2016})},\ \Eprint {https://arxiv.org/abs/1512.02876} {arXiv:1512.02876 [gr-qc]} \BibitemShut {NoStop}%
\bibitem [{\citenamefont {Bernard}\ \emph {et~al.}(2017{\natexlab{a}})\citenamefont {Bernard}, \citenamefont {Blanchet}, \citenamefont {Boh{\'e}}, \citenamefont {Faye},\ and\ \citenamefont {Marsat}}]{Bernard:2016wrg}%
  \BibitemOpen
  \bibfield  {author} {\bibinfo {author} {\bibfnamefont {L.}~\bibnamefont {Bernard}}, \bibinfo {author} {\bibfnamefont {L.}~\bibnamefont {Blanchet}}, \bibinfo {author} {\bibfnamefont {A.}~\bibnamefont {Boh{\'e}}}, \bibinfo {author} {\bibfnamefont {G.}~\bibnamefont {Faye}},\ and\ \bibinfo {author} {\bibfnamefont {S.}~\bibnamefont {Marsat}},\ }\bibfield  {title} {\bibinfo {title} {{Energy and periastron advance of compact binaries on circular orbits at the fourth post-Newtonian order}},\ }\href {https://doi.org/10.1103/PhysRevD.95.044026} {\bibfield  {journal} {\bibinfo  {journal} {Phys. Rev. D}\ }\textbf {\bibinfo {volume} {95}},\ \bibinfo {pages} {044026} (\bibinfo {year} {2017}{\natexlab{a}})},\ \Eprint {https://arxiv.org/abs/1610.07934} {arXiv:1610.07934 [gr-qc]} \BibitemShut {NoStop}%
\bibitem [{\citenamefont {Damour}\ \emph {et~al.}(2016)\citenamefont {Damour}, \citenamefont {Jaranowski},\ and\ \citenamefont {Sch{\"a}fer}}]{Damour:2016abl}%
  \BibitemOpen
  \bibfield  {author} {\bibinfo {author} {\bibfnamefont {T.}~\bibnamefont {Damour}}, \bibinfo {author} {\bibfnamefont {P.}~\bibnamefont {Jaranowski}},\ and\ \bibinfo {author} {\bibfnamefont {G.}~\bibnamefont {Sch{\"a}fer}},\ }\bibfield  {title} {\bibinfo {title} {{Conservative dynamics of two-body systems at the fourth post-Newtonian approximation of general relativity}},\ }\href {https://doi.org/10.1103/PhysRevD.93.084014} {\bibfield  {journal} {\bibinfo  {journal} {Phys. Rev. D}\ }\textbf {\bibinfo {volume} {93}},\ \bibinfo {pages} {084014} (\bibinfo {year} {2016})},\ \Eprint {https://arxiv.org/abs/1601.01283} {arXiv:1601.01283 [gr-qc]} \BibitemShut {NoStop}%
\bibitem [{\citenamefont {Blanchet}\ \emph {et~al.}(2023{\natexlab{a}})\citenamefont {Blanchet}, \citenamefont {Faye}, \citenamefont {Henry}, \citenamefont {Larrouturou},\ and\ \citenamefont {Trestini}}]{Blanchet:2023sbv}%
  \BibitemOpen
  \bibfield  {author} {\bibinfo {author} {\bibfnamefont {L.}~\bibnamefont {Blanchet}}, \bibinfo {author} {\bibfnamefont {G.}~\bibnamefont {Faye}}, \bibinfo {author} {\bibfnamefont {Q.}~\bibnamefont {Henry}}, \bibinfo {author} {\bibfnamefont {F.}~\bibnamefont {Larrouturou}},\ and\ \bibinfo {author} {\bibfnamefont {D.}~\bibnamefont {Trestini}},\ }\bibfield  {title} {\bibinfo {title} {{Gravitational-wave flux and quadrupole modes from quasicircular nonspinning compact binaries to the fourth post-Newtonian order}},\ }\href {https://doi.org/10.1103/PhysRevD.108.064041} {\bibfield  {journal} {\bibinfo  {journal} {Phys. Rev. D}\ }\textbf {\bibinfo {volume} {108}},\ \bibinfo {pages} {064041} (\bibinfo {year} {2023}{\natexlab{a}})},\ \Eprint {https://arxiv.org/abs/2304.11186} {arXiv:2304.11186 [gr-qc]} \BibitemShut {NoStop}%
\bibitem [{\citenamefont {Blanchet}\ \emph {et~al.}(2023{\natexlab{b}})\citenamefont {Blanchet}, \citenamefont {Faye}, \citenamefont {Henry}, \citenamefont {Larrouturou},\ and\ \citenamefont {Trestini}}]{Blanchet:2023bwj}%
  \BibitemOpen
  \bibfield  {author} {\bibinfo {author} {\bibfnamefont {L.}~\bibnamefont {Blanchet}}, \bibinfo {author} {\bibfnamefont {G.}~\bibnamefont {Faye}}, \bibinfo {author} {\bibfnamefont {Q.}~\bibnamefont {Henry}}, \bibinfo {author} {\bibfnamefont {F.}~\bibnamefont {Larrouturou}},\ and\ \bibinfo {author} {\bibfnamefont {D.}~\bibnamefont {Trestini}},\ }\bibfield  {title} {\bibinfo {title} {{Gravitational-Wave Phasing of Quasicircular Compact Binary Systems to the Fourth-and-a-Half Post-Newtonian Order}},\ }\href {https://doi.org/10.1103/PhysRevLett.131.121402} {\bibfield  {journal} {\bibinfo  {journal} {Phys. Rev. Lett.}\ }\textbf {\bibinfo {volume} {131}},\ \bibinfo {pages} {121402} (\bibinfo {year} {2023}{\natexlab{b}})},\ \Eprint {https://arxiv.org/abs/2304.11185} {arXiv:2304.11185 [gr-qc]} \BibitemShut {NoStop}%
\bibitem [{\citenamefont {Damour}\ and\ \citenamefont {Jaranowski}(2017)}]{Damour:2017ced}%
  \BibitemOpen
  \bibfield  {author} {\bibinfo {author} {\bibfnamefont {T.}~\bibnamefont {Damour}}\ and\ \bibinfo {author} {\bibfnamefont {P.}~\bibnamefont {Jaranowski}},\ }\bibfield  {title} {\bibinfo {title} {{Four-loop static contribution to the gravitational interaction potential of two point masses}},\ }\href {https://doi.org/10.1103/PhysRevD.95.084005} {\bibfield  {journal} {\bibinfo  {journal} {Phys. Rev. D}\ }\textbf {\bibinfo {volume} {95}},\ \bibinfo {pages} {084005} (\bibinfo {year} {2017})},\ \Eprint {https://arxiv.org/abs/1701.02645} {arXiv:1701.02645 [gr-qc]} \BibitemShut {NoStop}%
\bibitem [{\citenamefont {Goldberger}\ and\ \citenamefont {Rothstein}(2006)}]{Goldberger:2004jt}%
  \BibitemOpen
  \bibfield  {author} {\bibinfo {author} {\bibfnamefont {W.~D.}\ \bibnamefont {Goldberger}}\ and\ \bibinfo {author} {\bibfnamefont {I.~Z.}\ \bibnamefont {Rothstein}},\ }\bibfield  {title} {\bibinfo {title} {{An Effective field theory of gravity for extended objects}},\ }\href {https://doi.org/10.1103/PhysRevD.73.104029} {\bibfield  {journal} {\bibinfo  {journal} {Phys. Rev. D}\ }\textbf {\bibinfo {volume} {73}},\ \bibinfo {pages} {104029} (\bibinfo {year} {2006})},\ \Eprint {https://arxiv.org/abs/hep-th/0409156} {arXiv:hep-th/0409156} \BibitemShut {NoStop}%
\bibitem [{\citenamefont {Gilmore}\ and\ \citenamefont {Ross}(2008)}]{Gilmore:2008gq}%
  \BibitemOpen
  \bibfield  {author} {\bibinfo {author} {\bibfnamefont {J.~B.}\ \bibnamefont {Gilmore}}\ and\ \bibinfo {author} {\bibfnamefont {A.}~\bibnamefont {Ross}},\ }\bibfield  {title} {\bibinfo {title} {Effective field theory calculation of second post-{Newtonian} binary dynamics},\ }\href {https://doi.org/10.1103/PhysRevD.78.124021} {\bibfield  {journal} {\bibinfo  {journal} {Phys. Rev. D}\ }\textbf {\bibinfo {volume} {78}},\ \bibinfo {pages} {124021} (\bibinfo {year} {2008})},\ \Eprint {https://arxiv.org/abs/0810.1328} {arXiv:0810.1328 [gr-qc]} \BibitemShut {NoStop}%
\bibitem [{\citenamefont {Foffa}\ and\ \citenamefont {Sturani}(2011)}]{Foffa:2011ub}%
  \BibitemOpen
  \bibfield  {author} {\bibinfo {author} {\bibfnamefont {S.}~\bibnamefont {Foffa}}\ and\ \bibinfo {author} {\bibfnamefont {R.}~\bibnamefont {Sturani}},\ }\bibfield  {title} {\bibinfo {title} {Effective field theory calculation of conservative binary dynamics at third post-{Newtonian} order},\ }\href {https://doi.org/10.1103/PhysRevD.84.044031} {\bibfield  {journal} {\bibinfo  {journal} {Phys. Rev. D}\ }\textbf {\bibinfo {volume} {84}},\ \bibinfo {pages} {044031} (\bibinfo {year} {2011})},\ \Eprint {https://arxiv.org/abs/1104.1122} {arXiv:1104.1122 [gr-qc]} \BibitemShut {NoStop}%
\bibitem [{\citenamefont {Porto}\ and\ \citenamefont {Rothstein}(2017)}]{Porto:2017dgs}%
  \BibitemOpen
  \bibfield  {author} {\bibinfo {author} {\bibfnamefont {R.~A.}\ \bibnamefont {Porto}}\ and\ \bibinfo {author} {\bibfnamefont {I.~Z.}\ \bibnamefont {Rothstein}},\ }\bibfield  {title} {\bibinfo {title} {{Apparent ambiguities in the post-Newtonian expansion for binary systems}},\ }\href {https://doi.org/10.1103/PhysRevD.96.024062} {\bibfield  {journal} {\bibinfo  {journal} {Phys. Rev. D}\ }\textbf {\bibinfo {volume} {96}},\ \bibinfo {pages} {024062} (\bibinfo {year} {2017})},\ \Eprint {https://arxiv.org/abs/1703.06433} {arXiv:1703.06433 [gr-qc]} \BibitemShut {NoStop}%
\bibitem [{\citenamefont {Kol}\ and\ \citenamefont {Shir}(2013)}]{Kol:2013ega}%
  \BibitemOpen
  \bibfield  {author} {\bibinfo {author} {\bibfnamefont {B.}~\bibnamefont {Kol}}\ and\ \bibinfo {author} {\bibfnamefont {R.}~\bibnamefont {Shir}},\ }\bibfield  {title} {\bibinfo {title} {{Classical 3-loop 2-body diagrams}},\ }\href {https://doi.org/10.1007/JHEP09(2013)069} {\bibfield  {journal} {\bibinfo  {journal} {JHEP}\ }\textbf {\bibinfo {volume} {09}},\ \bibinfo {pages} {069}},\ \Eprint {https://arxiv.org/abs/1306.3220} {arXiv:1306.3220 [hep-th]} \BibitemShut {NoStop}%
\bibitem [{\citenamefont {Foffa}\ and\ \citenamefont {Sturani}(2013{\natexlab{a}})}]{Foffa:2012rn}%
  \BibitemOpen
  \bibfield  {author} {\bibinfo {author} {\bibfnamefont {S.}~\bibnamefont {Foffa}}\ and\ \bibinfo {author} {\bibfnamefont {R.}~\bibnamefont {Sturani}},\ }\bibfield  {title} {\bibinfo {title} {{Dynamics of the gravitational two-body problem at fourth post-Newtonian order and at quadratic order in the Newton constant}},\ }\href {https://doi.org/10.1103/PhysRevD.87.064011} {\bibfield  {journal} {\bibinfo  {journal} {Phys. Rev. D}\ }\textbf {\bibinfo {volume} {87}},\ \bibinfo {pages} {064011} (\bibinfo {year} {2013}{\natexlab{a}})},\ \Eprint {https://arxiv.org/abs/1206.7087} {arXiv:1206.7087 [gr-qc]} \BibitemShut {NoStop}%
\bibitem [{\citenamefont {Foffa}\ and\ \citenamefont {Sturani}(2019)}]{Foffa:2019rdf}%
  \BibitemOpen
  \bibfield  {author} {\bibinfo {author} {\bibfnamefont {S.}~\bibnamefont {Foffa}}\ and\ \bibinfo {author} {\bibfnamefont {R.}~\bibnamefont {Sturani}},\ }\bibfield  {title} {\bibinfo {title} {{Conservative dynamics of binary systems to fourth Post-Newtonian order in the EFT approach I: Regularized Lagrangian}},\ }\href {https://doi.org/10.1103/PhysRevD.100.024047} {\bibfield  {journal} {\bibinfo  {journal} {Phys. Rev. D}\ }\textbf {\bibinfo {volume} {100}},\ \bibinfo {pages} {024047} (\bibinfo {year} {2019})},\ \Eprint {https://arxiv.org/abs/1903.05113} {arXiv:1903.05113 [gr-qc]} \BibitemShut {NoStop}%
\bibitem [{\citenamefont {Foffa}\ \emph {et~al.}(2017)\citenamefont {Foffa}, \citenamefont {Mastrolia}, \citenamefont {Sturani},\ and\ \citenamefont {Sturm}}]{Foffa:2016rgu}%
  \BibitemOpen
  \bibfield  {author} {\bibinfo {author} {\bibfnamefont {S.}~\bibnamefont {Foffa}}, \bibinfo {author} {\bibfnamefont {P.}~\bibnamefont {Mastrolia}}, \bibinfo {author} {\bibfnamefont {R.}~\bibnamefont {Sturani}},\ and\ \bibinfo {author} {\bibfnamefont {C.}~\bibnamefont {Sturm}},\ }\bibfield  {title} {\bibinfo {title} {{Effective field theory approach to the gravitational two-body dynamics, at fourth post-Newtonian order and quintic in the Newton constant}},\ }\href {https://doi.org/10.1103/PhysRevD.95.104009} {\bibfield  {journal} {\bibinfo  {journal} {Phys. Rev. D}\ }\textbf {\bibinfo {volume} {95}},\ \bibinfo {pages} {104009} (\bibinfo {year} {2017})},\ \Eprint {https://arxiv.org/abs/1612.00482} {arXiv:1612.00482 [gr-qc]} \BibitemShut {NoStop}%
\bibitem [{\citenamefont {Foffa}\ \emph {et~al.}(2019{\natexlab{a}})\citenamefont {Foffa}, \citenamefont {Porto}, \citenamefont {Rothstein},\ and\ \citenamefont {Sturani}}]{Foffa:2019yfl}%
  \BibitemOpen
  \bibfield  {author} {\bibinfo {author} {\bibfnamefont {S.}~\bibnamefont {Foffa}}, \bibinfo {author} {\bibfnamefont {R.~A.}\ \bibnamefont {Porto}}, \bibinfo {author} {\bibfnamefont {I.}~\bibnamefont {Rothstein}},\ and\ \bibinfo {author} {\bibfnamefont {R.}~\bibnamefont {Sturani}},\ }\bibfield  {title} {\bibinfo {title} {{Conservative dynamics of binary systems to fourth Post-Newtonian order in the EFT approach II: Renormalized Lagrangian}},\ }\href {https://doi.org/10.1103/PhysRevD.100.024048} {\bibfield  {journal} {\bibinfo  {journal} {Phys. Rev. D}\ }\textbf {\bibinfo {volume} {100}},\ \bibinfo {pages} {024048} (\bibinfo {year} {2019}{\natexlab{a}})},\ \Eprint {https://arxiv.org/abs/1903.05118} {arXiv:1903.05118 [gr-qc]} \BibitemShut {NoStop}%
\bibitem [{\citenamefont {Bl{\"u}mlein}\ \emph {et~al.}(2020{\natexlab{a}})\citenamefont {Bl{\"u}mlein}, \citenamefont {Maier}, \citenamefont {Marquard},\ and\ \citenamefont {Sch{\"a}fer}}]{Blumlein:2020pog}%
  \BibitemOpen
  \bibfield  {author} {\bibinfo {author} {\bibfnamefont {J.}~\bibnamefont {Bl{\"u}mlein}}, \bibinfo {author} {\bibfnamefont {A.}~\bibnamefont {Maier}}, \bibinfo {author} {\bibfnamefont {P.}~\bibnamefont {Marquard}},\ and\ \bibinfo {author} {\bibfnamefont {G.}~\bibnamefont {Sch{\"a}fer}},\ }\bibfield  {title} {\bibinfo {title} {{Fourth post-Newtonian Hamiltonian dynamics of two-body systems from an effective field theory approach}},\ }\href {https://doi.org/10.1016/j.nuclphysb.2020.115041} {\bibfield  {journal} {\bibinfo  {journal} {Nucl. Phys. B}\ }\textbf {\bibinfo {volume} {955}},\ \bibinfo {pages} {115041} (\bibinfo {year} {2020}{\natexlab{a}})},\ \Eprint {https://arxiv.org/abs/2003.01692} {arXiv:2003.01692 [gr-qc]} \BibitemShut {NoStop}%
\bibitem [{\citenamefont {Bl{\"u}mlein}\ \emph {et~al.}(2020{\natexlab{b}})\citenamefont {Bl{\"u}mlein}, \citenamefont {Maier}, \citenamefont {Marquard},\ and\ \citenamefont {Sch{\"a}fer}}]{Blumlein:2020znm}%
  \BibitemOpen
  \bibfield  {author} {\bibinfo {author} {\bibfnamefont {J.}~\bibnamefont {Bl{\"u}mlein}}, \bibinfo {author} {\bibfnamefont {A.}~\bibnamefont {Maier}}, \bibinfo {author} {\bibfnamefont {P.}~\bibnamefont {Marquard}},\ and\ \bibinfo {author} {\bibfnamefont {G.}~\bibnamefont {Sch{\"a}fer}},\ }\bibfield  {title} {\bibinfo {title} {{Testing binary dynamics in gravity at the sixth post-Newtonian level}},\ }\href {https://doi.org/10.1016/j.physletb.2020.135496} {\bibfield  {journal} {\bibinfo  {journal} {Phys. Lett. B}\ }\textbf {\bibinfo {volume} {807}},\ \bibinfo {pages} {135496} (\bibinfo {year} {2020}{\natexlab{b}})},\ \Eprint {https://arxiv.org/abs/2003.07145} {arXiv:2003.07145 [gr-qc]} \BibitemShut {NoStop}%
\bibitem [{\citenamefont {Goldberger}\ \emph {et~al.}(2014)\citenamefont {Goldberger}, \citenamefont {Ross},\ and\ \citenamefont {Rothstein}}]{Goldberger:2012kf}%
  \BibitemOpen
  \bibfield  {author} {\bibinfo {author} {\bibfnamefont {W.~D.}\ \bibnamefont {Goldberger}}, \bibinfo {author} {\bibfnamefont {A.}~\bibnamefont {Ross}},\ and\ \bibinfo {author} {\bibfnamefont {I.~Z.}\ \bibnamefont {Rothstein}},\ }\bibfield  {title} {\bibinfo {title} {{Black hole mass dynamics and renormalization group evolution}},\ }\href {https://doi.org/10.1103/PhysRevD.89.124033} {\bibfield  {journal} {\bibinfo  {journal} {Phys. Rev. D}\ }\textbf {\bibinfo {volume} {89}},\ \bibinfo {pages} {124033} (\bibinfo {year} {2014})},\ \Eprint {https://arxiv.org/abs/1211.6095} {arXiv:1211.6095 [hep-th]} \BibitemShut {NoStop}%
\bibitem [{\citenamefont {Galley}\ \emph {et~al.}(2016)\citenamefont {Galley}, \citenamefont {Leibovich}, \citenamefont {Porto},\ and\ \citenamefont {Ross}}]{Galley:2015kus}%
  \BibitemOpen
  \bibfield  {author} {\bibinfo {author} {\bibfnamefont {C.~R.}\ \bibnamefont {Galley}}, \bibinfo {author} {\bibfnamefont {A.~K.}\ \bibnamefont {Leibovich}}, \bibinfo {author} {\bibfnamefont {R.~A.}\ \bibnamefont {Porto}},\ and\ \bibinfo {author} {\bibfnamefont {A.}~\bibnamefont {Ross}},\ }\bibfield  {title} {\bibinfo {title} {{Tail effect in gravitational radiation reaction: Time nonlocality and renormalization group evolution}},\ }\href {https://doi.org/10.1103/PhysRevD.93.124010} {\bibfield  {journal} {\bibinfo  {journal} {Phys. Rev. D}\ }\textbf {\bibinfo {volume} {93}},\ \bibinfo {pages} {124010} (\bibinfo {year} {2016})},\ \Eprint {https://arxiv.org/abs/1511.07379} {arXiv:1511.07379 [gr-qc]} \BibitemShut {NoStop}%
\bibitem [{\citenamefont {Foffa}\ \emph {et~al.}(2019{\natexlab{b}})\citenamefont {Foffa}, \citenamefont {Mastrolia}, \citenamefont {Sturani}, \citenamefont {Sturm},\ and\ \citenamefont {Torres~Bobadilla}}]{Foffa:2019hrb}%
  \BibitemOpen
  \bibfield  {author} {\bibinfo {author} {\bibfnamefont {S.}~\bibnamefont {Foffa}}, \bibinfo {author} {\bibfnamefont {P.}~\bibnamefont {Mastrolia}}, \bibinfo {author} {\bibfnamefont {R.}~\bibnamefont {Sturani}}, \bibinfo {author} {\bibfnamefont {C.}~\bibnamefont {Sturm}},\ and\ \bibinfo {author} {\bibfnamefont {W.~J.}\ \bibnamefont {Torres~Bobadilla}},\ }\bibfield  {title} {\bibinfo {title} {{Static two-body potential at fifth post-Newtonian order}},\ }\href {https://doi.org/10.1103/PhysRevLett.122.241605} {\bibfield  {journal} {\bibinfo  {journal} {Phys. Rev. Lett.}\ }\textbf {\bibinfo {volume} {122}},\ \bibinfo {pages} {241605} (\bibinfo {year} {2019}{\natexlab{b}})},\ \Eprint {https://arxiv.org/abs/1902.10571} {arXiv:1902.10571 [gr-qc]} \BibitemShut {NoStop}%
\bibitem [{\citenamefont {Bl{\"u}mlein}\ \emph {et~al.}(2020{\natexlab{c}})\citenamefont {Bl{\"u}mlein}, \citenamefont {Maier},\ and\ \citenamefont {Marquard}}]{Blumlein:2019zku}%
  \BibitemOpen
  \bibfield  {author} {\bibinfo {author} {\bibfnamefont {J.}~\bibnamefont {Bl{\"u}mlein}}, \bibinfo {author} {\bibfnamefont {A.}~\bibnamefont {Maier}},\ and\ \bibinfo {author} {\bibfnamefont {P.}~\bibnamefont {Marquard}},\ }\bibfield  {title} {\bibinfo {title} {{Five-Loop Static Contribution to the Gravitational Interaction Potential of Two Point Masses}},\ }\href {https://doi.org/10.1016/j.physletb.2019.135100} {\bibfield  {journal} {\bibinfo  {journal} {Phys. Lett. B}\ }\textbf {\bibinfo {volume} {800}},\ \bibinfo {pages} {135100} (\bibinfo {year} {2020}{\natexlab{c}})},\ \Eprint {https://arxiv.org/abs/1902.11180} {arXiv:1902.11180 [gr-qc]} \BibitemShut {NoStop}%
\bibitem [{\citenamefont {Foffa}\ \emph {et~al.}(2021)\citenamefont {Foffa}, \citenamefont {Sturani},\ and\ \citenamefont {Torres~Bobadilla}}]{Foffa:2020nqe}%
  \BibitemOpen
  \bibfield  {author} {\bibinfo {author} {\bibfnamefont {S.}~\bibnamefont {Foffa}}, \bibinfo {author} {\bibfnamefont {R.}~\bibnamefont {Sturani}},\ and\ \bibinfo {author} {\bibfnamefont {W.~J.}\ \bibnamefont {Torres~Bobadilla}},\ }\bibfield  {title} {\bibinfo {title} {{Efficient resummation of high post-Newtonian contributions to the binding energy}},\ }\href {https://doi.org/10.1007/JHEP02(2021)165} {\bibfield  {journal} {\bibinfo  {journal} {JHEP}\ }\textbf {\bibinfo {volume} {02}},\ \bibinfo {pages} {165}},\ \Eprint {https://arxiv.org/abs/2010.13730} {arXiv:2010.13730 [gr-qc]} \BibitemShut {NoStop}%
\bibitem [{\citenamefont {Bl{\"u}mlein}\ \emph {et~al.}(2021{\natexlab{a}})\citenamefont {Bl{\"u}mlein}, \citenamefont {Maier}, \citenamefont {Marquard},\ and\ \citenamefont {Sch{\"a}fer}}]{Blumlein:2020pyo}%
  \BibitemOpen
  \bibfield  {author} {\bibinfo {author} {\bibfnamefont {J.}~\bibnamefont {Bl{\"u}mlein}}, \bibinfo {author} {\bibfnamefont {A.}~\bibnamefont {Maier}}, \bibinfo {author} {\bibfnamefont {P.}~\bibnamefont {Marquard}},\ and\ \bibinfo {author} {\bibfnamefont {G.}~\bibnamefont {Sch{\"a}fer}},\ }\bibfield  {title} {\bibinfo {title} {{The fifth-order post-Newtonian Hamiltonian dynamics of two-body systems from an effective field theory approach: potential contributions}},\ }\href {https://doi.org/10.1016/j.nuclphysb.2021.115352} {\bibfield  {journal} {\bibinfo  {journal} {Nucl. Phys. B}\ }\textbf {\bibinfo {volume} {965}},\ \bibinfo {pages} {115352} (\bibinfo {year} {2021}{\natexlab{a}})},\ \Eprint {https://arxiv.org/abs/2010.13672} {arXiv:2010.13672 [gr-qc]} \BibitemShut {NoStop}%
\bibitem [{\citenamefont {Bl{\"u}mlein}\ \emph {et~al.}(2022)\citenamefont {Bl{\"u}mlein}, \citenamefont {Maier}, \citenamefont {Marquard},\ and\ \citenamefont {Sch{\"a}fer}}]{Blumlein:2021txe}%
  \BibitemOpen
  \bibfield  {author} {\bibinfo {author} {\bibfnamefont {J.}~\bibnamefont {Bl{\"u}mlein}}, \bibinfo {author} {\bibfnamefont {A.}~\bibnamefont {Maier}}, \bibinfo {author} {\bibfnamefont {P.}~\bibnamefont {Marquard}},\ and\ \bibinfo {author} {\bibfnamefont {G.}~\bibnamefont {Sch{\"a}fer}},\ }\bibfield  {title} {\bibinfo {title} {{The fifth-order post-Newtonian Hamiltonian dynamics of two-body systems from an effective field theory approach}},\ }\href {https://doi.org/10.1016/j.nuclphysb.2022.115900} {\bibfield  {journal} {\bibinfo  {journal} {Nucl. Phys. B}\ }\textbf {\bibinfo {volume} {983}},\ \bibinfo {pages} {115900} (\bibinfo {year} {2022})},\ \bibinfo {note} {[Erratum: Nucl.Phys.B 985, 115991 (2022)]},\ \Eprint {https://arxiv.org/abs/2110.13822} {arXiv:2110.13822 [gr-qc]} \BibitemShut {NoStop}%
\bibitem [{\citenamefont {Porto}\ \emph {et~al.}(2025)\citenamefont {Porto}, \citenamefont {Riva},\ and\ \citenamefont {Yang}}]{Porto:2024cwd}%
  \BibitemOpen
  \bibfield  {author} {\bibinfo {author} {\bibfnamefont {R.~A.}\ \bibnamefont {Porto}}, \bibinfo {author} {\bibfnamefont {M.~M.}\ \bibnamefont {Riva}},\ and\ \bibinfo {author} {\bibfnamefont {Z.}~\bibnamefont {Yang}},\ }\bibfield  {title} {\bibinfo {title} {{Nonlinear gravitational radiation reaction: failed tail, memories {\&} squares}},\ }\href {https://doi.org/10.1007/JHEP04(2025)050} {\bibfield  {journal} {\bibinfo  {journal} {JHEP}\ }\textbf {\bibinfo {volume} {04}},\ \bibinfo {pages} {050}},\ \Eprint {https://arxiv.org/abs/2409.05860} {arXiv:2409.05860 [gr-qc]} \BibitemShut {NoStop}%
\bibitem [{\citenamefont {Bl{\"u}mlein}\ \emph {et~al.}(2021{\natexlab{b}})\citenamefont {Bl{\"u}mlein}, \citenamefont {Maier}, \citenamefont {Marquard},\ and\ \citenamefont {Sch{\"a}fer}}]{Blumlein:2021txj}%
  \BibitemOpen
  \bibfield  {author} {\bibinfo {author} {\bibfnamefont {J.}~\bibnamefont {Bl{\"u}mlein}}, \bibinfo {author} {\bibfnamefont {A.}~\bibnamefont {Maier}}, \bibinfo {author} {\bibfnamefont {P.}~\bibnamefont {Marquard}},\ and\ \bibinfo {author} {\bibfnamefont {G.}~\bibnamefont {Sch{\"a}fer}},\ }\bibfield  {title} {\bibinfo {title} {{The 6th post-Newtonian potential terms at $O(G_N^4)$}},\ }\href {https://doi.org/10.1016/j.physletb.2021.136260} {\bibfield  {journal} {\bibinfo  {journal} {Phys. Lett. B}\ }\textbf {\bibinfo {volume} {816}},\ \bibinfo {pages} {136260} (\bibinfo {year} {2021}{\natexlab{b}})},\ \Eprint {https://arxiv.org/abs/2101.08630} {arXiv:2101.08630 [gr-qc]} \BibitemShut {NoStop}%
\bibitem [{\citenamefont {Porto}(2006)}]{Porto:2005ac}%
  \BibitemOpen
  \bibfield  {author} {\bibinfo {author} {\bibfnamefont {R.~A.}\ \bibnamefont {Porto}},\ }\bibfield  {title} {\bibinfo {title} {{Post-Newtonian corrections to the motion of spinning bodies in NRGR}},\ }\href {https://doi.org/10.1103/PhysRevD.73.104031} {\bibfield  {journal} {\bibinfo  {journal} {Phys. Rev. D}\ }\textbf {\bibinfo {volume} {73}},\ \bibinfo {pages} {104031} (\bibinfo {year} {2006})},\ \Eprint {https://arxiv.org/abs/gr-qc/0511061} {arXiv:gr-qc/0511061} \BibitemShut {NoStop}%
\bibitem [{\citenamefont {Porto}\ and\ \citenamefont {Rothstein}(2006)}]{Porto:2006bt}%
  \BibitemOpen
  \bibfield  {author} {\bibinfo {author} {\bibfnamefont {R.~A.}\ \bibnamefont {Porto}}\ and\ \bibinfo {author} {\bibfnamefont {I.~Z.}\ \bibnamefont {Rothstein}},\ }\bibfield  {title} {\bibinfo {title} {{The Hyperfine Einstein-Infeld-Hoffmann potential}},\ }\href {https://doi.org/10.1103/PhysRevLett.97.021101} {\bibfield  {journal} {\bibinfo  {journal} {Phys. Rev. Lett.}\ }\textbf {\bibinfo {volume} {97}},\ \bibinfo {pages} {021101} (\bibinfo {year} {2006})},\ \Eprint {https://arxiv.org/abs/gr-qc/0604099} {arXiv:gr-qc/0604099} \BibitemShut {NoStop}%
\bibitem [{\citenamefont {Porto}\ and\ \citenamefont {Rothstein}(2008)}]{Porto:2008tb}%
  \BibitemOpen
  \bibfield  {author} {\bibinfo {author} {\bibfnamefont {R.~A.}\ \bibnamefont {Porto}}\ and\ \bibinfo {author} {\bibfnamefont {I.~Z.}\ \bibnamefont {Rothstein}},\ }\bibfield  {title} {\bibinfo {title} {{Spin(1)Spin(2) Effects in the Motion of Inspiralling Compact Binaries at Third Order in the Post-Newtonian Expansion}},\ }\href {https://doi.org/10.1103/PhysRevD.78.044012} {\bibfield  {journal} {\bibinfo  {journal} {Phys. Rev. D}\ }\textbf {\bibinfo {volume} {78}},\ \bibinfo {pages} {044012} (\bibinfo {year} {2008})},\ \bibinfo {note} {[Erratum: Phys.Rev.D 81, 029904 (2010)]},\ \Eprint {https://arxiv.org/abs/0802.0720} {arXiv:0802.0720 [gr-qc]} \BibitemShut {NoStop}%
\bibitem [{\citenamefont {Levi}(2010)}]{Levi:2008nh}%
  \BibitemOpen
  \bibfield  {author} {\bibinfo {author} {\bibfnamefont {M.}~\bibnamefont {Levi}},\ }\bibfield  {title} {\bibinfo {title} {{Next to Leading Order gravitational Spin1-Spin2 coupling with Kaluza-Klein reduction}},\ }\href {https://doi.org/10.1103/PhysRevD.82.064029} {\bibfield  {journal} {\bibinfo  {journal} {Phys. Rev. D}\ }\textbf {\bibinfo {volume} {82}},\ \bibinfo {pages} {064029} (\bibinfo {year} {2010})},\ \Eprint {https://arxiv.org/abs/0802.1508} {arXiv:0802.1508 [gr-qc]} \BibitemShut {NoStop}%
\bibitem [{\citenamefont {Levi}\ and\ \citenamefont {Steinhoff}(2015)}]{Levi:2015msa}%
  \BibitemOpen
  \bibfield  {author} {\bibinfo {author} {\bibfnamefont {M.}~\bibnamefont {Levi}}\ and\ \bibinfo {author} {\bibfnamefont {J.}~\bibnamefont {Steinhoff}},\ }\bibfield  {title} {\bibinfo {title} {{Spinning gravitating objects in the effective field theory in the post-Newtonian scheme}},\ }\href {https://doi.org/10.1007/JHEP09(2015)219} {\bibfield  {journal} {\bibinfo  {journal} {JHEP}\ }\textbf {\bibinfo {volume} {09}},\ \bibinfo {pages} {219}},\ \Eprint {https://arxiv.org/abs/1501.04956} {arXiv:1501.04956 [gr-qc]} \BibitemShut {NoStop}%
\bibitem [{\citenamefont {Levi}\ and\ \citenamefont {Steinhoff}(2021)}]{Levi:2016ofk}%
  \BibitemOpen
  \bibfield  {author} {\bibinfo {author} {\bibfnamefont {M.}~\bibnamefont {Levi}}\ and\ \bibinfo {author} {\bibfnamefont {J.}~\bibnamefont {Steinhoff}},\ }\bibfield  {title} {\bibinfo {title} {{Complete conservative dynamics for inspiralling compact binaries with spins at the fourth post-Newtonian order}},\ }\href {https://doi.org/10.1088/1475-7516/2021/09/029} {\bibfield  {journal} {\bibinfo  {journal} {JCAP}\ }\textbf {\bibinfo {volume} {09}},\ \bibinfo {pages} {029}},\ \Eprint {https://arxiv.org/abs/1607.04252} {arXiv:1607.04252 [gr-qc]} \BibitemShut {NoStop}%
\bibitem [{\citenamefont {Kim}\ \emph {et~al.}(2022)\citenamefont {Kim}, \citenamefont {Levi},\ and\ \citenamefont {Yin}}]{Kim:2021rfj}%
  \BibitemOpen
  \bibfield  {author} {\bibinfo {author} {\bibfnamefont {J.-W.}\ \bibnamefont {Kim}}, \bibinfo {author} {\bibfnamefont {M.}~\bibnamefont {Levi}},\ and\ \bibinfo {author} {\bibfnamefont {Z.}~\bibnamefont {Yin}},\ }\bibfield  {title} {\bibinfo {title} {{Quadratic-in-spin interactions at fifth post-Newtonian order probe new physics}},\ }\href {https://doi.org/10.1016/j.physletb.2022.137410} {\bibfield  {journal} {\bibinfo  {journal} {Phys. Lett. B}\ }\textbf {\bibinfo {volume} {834}},\ \bibinfo {pages} {137410} (\bibinfo {year} {2022})},\ \Eprint {https://arxiv.org/abs/2112.01509} {arXiv:2112.01509 [hep-th]} \BibitemShut {NoStop}%
\bibitem [{\citenamefont {Levi}\ \emph {et~al.}(2021{\natexlab{a}})\citenamefont {Levi}, \citenamefont {Mcleod},\ and\ \citenamefont {Von~Hippel}}]{Levi:2020uwu}%
  \BibitemOpen
  \bibfield  {author} {\bibinfo {author} {\bibfnamefont {M.}~\bibnamefont {Levi}}, \bibinfo {author} {\bibfnamefont {A.~J.}\ \bibnamefont {Mcleod}},\ and\ \bibinfo {author} {\bibfnamefont {M.}~\bibnamefont {Von~Hippel}},\ }\bibfield  {title} {\bibinfo {title} {{N$^{3}$LO gravitational quadratic-in-spin interactions at G$^{4}$}},\ }\href {https://doi.org/10.1007/JHEP07(2021)116} {\bibfield  {journal} {\bibinfo  {journal} {JHEP}\ }\textbf {\bibinfo {volume} {07}},\ \bibinfo {pages} {116}},\ \Eprint {https://arxiv.org/abs/2003.07890} {arXiv:2003.07890 [hep-th]} \BibitemShut {NoStop}%
\bibitem [{\citenamefont {Levi}\ \emph {et~al.}(2021{\natexlab{b}})\citenamefont {Levi}, \citenamefont {Mougiakakos},\ and\ \citenamefont {Vieira}}]{Levi:2019kgk}%
  \BibitemOpen
  \bibfield  {author} {\bibinfo {author} {\bibfnamefont {M.}~\bibnamefont {Levi}}, \bibinfo {author} {\bibfnamefont {S.}~\bibnamefont {Mougiakakos}},\ and\ \bibinfo {author} {\bibfnamefont {M.}~\bibnamefont {Vieira}},\ }\bibfield  {title} {\bibinfo {title} {{Gravitational cubic-in-spin interaction at the next-to-leading post-Newtonian order}},\ }\href {https://doi.org/10.1007/JHEP01(2021)036} {\bibfield  {journal} {\bibinfo  {journal} {JHEP}\ }\textbf {\bibinfo {volume} {01}},\ \bibinfo {pages} {036}},\ \Eprint {https://arxiv.org/abs/1912.06276} {arXiv:1912.06276 [hep-th]} \BibitemShut {NoStop}%
\bibitem [{\citenamefont {Kim}\ \emph {et~al.}(2023{\natexlab{a}})\citenamefont {Kim}, \citenamefont {Levi},\ and\ \citenamefont {Yin}}]{Kim:2022pou}%
  \BibitemOpen
  \bibfield  {author} {\bibinfo {author} {\bibfnamefont {J.-W.}\ \bibnamefont {Kim}}, \bibinfo {author} {\bibfnamefont {M.}~\bibnamefont {Levi}},\ and\ \bibinfo {author} {\bibfnamefont {Z.}~\bibnamefont {Yin}},\ }\bibfield  {title} {\bibinfo {title} {{N$^{3}$LO spin-orbit interaction via the EFT of spinning gravitating objects}},\ }\href {https://doi.org/10.1007/JHEP05(2023)184} {\bibfield  {journal} {\bibinfo  {journal} {JHEP}\ }\textbf {\bibinfo {volume} {05}},\ \bibinfo {pages} {184}},\ \Eprint {https://arxiv.org/abs/2208.14949} {arXiv:2208.14949 [hep-th]} \BibitemShut {NoStop}%
\bibitem [{\citenamefont {Kim}\ \emph {et~al.}(2023{\natexlab{b}})\citenamefont {Kim}, \citenamefont {Levi},\ and\ \citenamefont {Yin}}]{Kim:2022bwv}%
  \BibitemOpen
  \bibfield  {author} {\bibinfo {author} {\bibfnamefont {J.-W.}\ \bibnamefont {Kim}}, \bibinfo {author} {\bibfnamefont {M.}~\bibnamefont {Levi}},\ and\ \bibinfo {author} {\bibfnamefont {Z.}~\bibnamefont {Yin}},\ }\bibfield  {title} {\bibinfo {title} {{N$^{3}$LO quadratic-in-spin interactions for generic compact binaries}},\ }\href {https://doi.org/10.1007/JHEP03(2023)098} {\bibfield  {journal} {\bibinfo  {journal} {JHEP}\ }\textbf {\bibinfo {volume} {03}},\ \bibinfo {pages} {098}},\ \Eprint {https://arxiv.org/abs/2209.09235} {arXiv:2209.09235 [hep-th]} \BibitemShut {NoStop}%
\bibitem [{\citenamefont {Levi}\ \emph {et~al.}(2023)\citenamefont {Levi}, \citenamefont {Morales},\ and\ \citenamefont {Yin}}]{Levi:2022dqm}%
  \BibitemOpen
  \bibfield  {author} {\bibinfo {author} {\bibfnamefont {M.}~\bibnamefont {Levi}}, \bibinfo {author} {\bibfnamefont {R.}~\bibnamefont {Morales}},\ and\ \bibinfo {author} {\bibfnamefont {Z.}~\bibnamefont {Yin}},\ }\bibfield  {title} {\bibinfo {title} {{From the EFT of spinning gravitating objects to Poincar{\'e} and gauge invariance at the 4.5PN precision frontier}},\ }\href {https://doi.org/10.1007/JHEP09(2023)090} {\bibfield  {journal} {\bibinfo  {journal} {JHEP}\ }\textbf {\bibinfo {volume} {09}},\ \bibinfo {pages} {090}},\ \Eprint {https://arxiv.org/abs/2210.17538} {arXiv:2210.17538 [hep-th]} \BibitemShut {NoStop}%
\bibitem [{\citenamefont {Levi}\ and\ \citenamefont {Yin}(2023)}]{Levi:2022rrq}%
  \BibitemOpen
  \bibfield  {author} {\bibinfo {author} {\bibfnamefont {M.}~\bibnamefont {Levi}}\ and\ \bibinfo {author} {\bibfnamefont {Z.}~\bibnamefont {Yin}},\ }\bibfield  {title} {\bibinfo {title} {{Completing the fifth PN precision frontier via the EFT of spinning gravitating objects}},\ }\href {https://doi.org/10.1007/JHEP04(2023)079} {\bibfield  {journal} {\bibinfo  {journal} {JHEP}\ }\textbf {\bibinfo {volume} {04}},\ \bibinfo {pages} {079}},\ \Eprint {https://arxiv.org/abs/2211.14018} {arXiv:2211.14018 [hep-th]} \BibitemShut {NoStop}%
\bibitem [{\citenamefont {Mandal}\ \emph {et~al.}(2023{\natexlab{a}})\citenamefont {Mandal}, \citenamefont {Mastrolia}, \citenamefont {Patil},\ and\ \citenamefont {Steinhoff}}]{Mandal:2022nty}%
  \BibitemOpen
  \bibfield  {author} {\bibinfo {author} {\bibfnamefont {M.~K.}\ \bibnamefont {Mandal}}, \bibinfo {author} {\bibfnamefont {P.}~\bibnamefont {Mastrolia}}, \bibinfo {author} {\bibfnamefont {R.}~\bibnamefont {Patil}},\ and\ \bibinfo {author} {\bibfnamefont {J.}~\bibnamefont {Steinhoff}},\ }\bibfield  {title} {\bibinfo {title} {{Gravitational spin-orbit Hamiltonian at NNNLO in the post-Newtonian framework}},\ }\href {https://doi.org/10.1007/JHEP03(2023)130} {\bibfield  {journal} {\bibinfo  {journal} {JHEP}\ }\textbf {\bibinfo {volume} {03}},\ \bibinfo {pages} {130}},\ \Eprint {https://arxiv.org/abs/2209.00611} {arXiv:2209.00611 [hep-th]} \BibitemShut {NoStop}%
\bibitem [{\citenamefont {Mandal}\ \emph {et~al.}(2023{\natexlab{b}})\citenamefont {Mandal}, \citenamefont {Mastrolia}, \citenamefont {Patil},\ and\ \citenamefont {Steinhoff}}]{Mandal:2022ufb}%
  \BibitemOpen
  \bibfield  {author} {\bibinfo {author} {\bibfnamefont {M.~K.}\ \bibnamefont {Mandal}}, \bibinfo {author} {\bibfnamefont {P.}~\bibnamefont {Mastrolia}}, \bibinfo {author} {\bibfnamefont {R.}~\bibnamefont {Patil}},\ and\ \bibinfo {author} {\bibfnamefont {J.}~\bibnamefont {Steinhoff}},\ }\bibfield  {title} {\bibinfo {title} {{Gravitational quadratic-in-spin Hamiltonian at NNNLO in the post-Newtonian framework}},\ }\href {https://doi.org/10.1007/JHEP07(2023)128} {\bibfield  {journal} {\bibinfo  {journal} {JHEP}\ }\textbf {\bibinfo {volume} {07}},\ \bibinfo {pages} {128}},\ \Eprint {https://arxiv.org/abs/2210.09176} {arXiv:2210.09176 [hep-th]} \BibitemShut {NoStop}%
\bibitem [{\citenamefont {Goldberger}\ and\ \citenamefont {Ross}(2010)}]{Goldberger:2009qd}%
  \BibitemOpen
  \bibfield  {author} {\bibinfo {author} {\bibfnamefont {W.~D.}\ \bibnamefont {Goldberger}}\ and\ \bibinfo {author} {\bibfnamefont {A.}~\bibnamefont {Ross}},\ }\bibfield  {title} {\bibinfo {title} {{Gravitational radiative corrections from effective field theory}},\ }\href {https://doi.org/10.1103/PhysRevD.81.124015} {\bibfield  {journal} {\bibinfo  {journal} {Phys. Rev. D}\ }\textbf {\bibinfo {volume} {81}},\ \bibinfo {pages} {124015} (\bibinfo {year} {2010})},\ \Eprint {https://arxiv.org/abs/0912.4254} {arXiv:0912.4254 [gr-qc]} \BibitemShut {NoStop}%
\bibitem [{\citenamefont {Ross}(2012)}]{Ross:2012fc}%
  \BibitemOpen
  \bibfield  {author} {\bibinfo {author} {\bibfnamefont {A.}~\bibnamefont {Ross}},\ }\bibfield  {title} {\bibinfo {title} {{Multipole expansion at the level of the action}},\ }\href {https://doi.org/10.1103/PhysRevD.85.125033} {\bibfield  {journal} {\bibinfo  {journal} {Phys. Rev. D}\ }\textbf {\bibinfo {volume} {85}},\ \bibinfo {pages} {125033} (\bibinfo {year} {2012})},\ \Eprint {https://arxiv.org/abs/1202.4750} {arXiv:1202.4750 [gr-qc]} \BibitemShut {NoStop}%
\bibitem [{\citenamefont {Porto}\ \emph {et~al.}(2011)\citenamefont {Porto}, \citenamefont {Ross},\ and\ \citenamefont {Rothstein}}]{Porto:2010zg}%
  \BibitemOpen
  \bibfield  {author} {\bibinfo {author} {\bibfnamefont {R.~A.}\ \bibnamefont {Porto}}, \bibinfo {author} {\bibfnamefont {A.}~\bibnamefont {Ross}},\ and\ \bibinfo {author} {\bibfnamefont {I.~Z.}\ \bibnamefont {Rothstein}},\ }\bibfield  {title} {\bibinfo {title} {{Spin induced multipole moments for the gravitational wave flux from binary inspirals to third Post-Newtonian order}},\ }\href {https://doi.org/10.1088/1475-7516/2011/03/009} {\bibfield  {journal} {\bibinfo  {journal} {JCAP}\ }\textbf {\bibinfo {volume} {03}},\ \bibinfo {pages} {009}},\ \Eprint {https://arxiv.org/abs/1007.1312} {arXiv:1007.1312 [gr-qc]} \BibitemShut {NoStop}%
\bibitem [{\citenamefont {Maia}\ \emph {et~al.}(2017)\citenamefont {Maia}, \citenamefont {Galley}, \citenamefont {Leibovich},\ and\ \citenamefont {Porto}}]{Maia:2017gxn}%
  \BibitemOpen
  \bibfield  {author} {\bibinfo {author} {\bibfnamefont {N.~T.}\ \bibnamefont {Maia}}, \bibinfo {author} {\bibfnamefont {C.~R.}\ \bibnamefont {Galley}}, \bibinfo {author} {\bibfnamefont {A.~K.}\ \bibnamefont {Leibovich}},\ and\ \bibinfo {author} {\bibfnamefont {R.~A.}\ \bibnamefont {Porto}},\ }\bibfield  {title} {\bibinfo {title} {{Radiation reaction for spinning bodies in effective field theory I: Spin-orbit effects}},\ }\href {https://doi.org/10.1103/PhysRevD.96.084064} {\bibfield  {journal} {\bibinfo  {journal} {Phys. Rev. D}\ }\textbf {\bibinfo {volume} {96}},\ \bibinfo {pages} {084064} (\bibinfo {year} {2017})},\ \Eprint {https://arxiv.org/abs/1705.07934} {arXiv:1705.07934 [gr-qc]} \BibitemShut {NoStop}%
\bibitem [{\citenamefont {Cho}\ \emph {et~al.}(2021)\citenamefont {Cho}, \citenamefont {Pardo},\ and\ \citenamefont {Porto}}]{Cho:2021mqw}%
  \BibitemOpen
  \bibfield  {author} {\bibinfo {author} {\bibfnamefont {G.}~\bibnamefont {Cho}}, \bibinfo {author} {\bibfnamefont {B.}~\bibnamefont {Pardo}},\ and\ \bibinfo {author} {\bibfnamefont {R.~A.}\ \bibnamefont {Porto}},\ }\bibfield  {title} {\bibinfo {title} {{Gravitational radiation from inspiralling compact objects: Spin-spin effects completed at the next-to-leading post-Newtonian order}},\ }\href {https://doi.org/10.1103/PhysRevD.104.024037} {\bibfield  {journal} {\bibinfo  {journal} {Phys. Rev. D}\ }\textbf {\bibinfo {volume} {104}},\ \bibinfo {pages} {024037} (\bibinfo {year} {2021})},\ \Eprint {https://arxiv.org/abs/2103.14612} {arXiv:2103.14612 [gr-qc]} \BibitemShut {NoStop}%
\bibitem [{\citenamefont {Cho}\ \emph {et~al.}(2022)\citenamefont {Cho}, \citenamefont {Porto},\ and\ \citenamefont {Yang}}]{Cho:2022syn}%
  \BibitemOpen
  \bibfield  {author} {\bibinfo {author} {\bibfnamefont {G.}~\bibnamefont {Cho}}, \bibinfo {author} {\bibfnamefont {R.~A.}\ \bibnamefont {Porto}},\ and\ \bibinfo {author} {\bibfnamefont {Z.}~\bibnamefont {Yang}},\ }\bibfield  {title} {\bibinfo {title} {{Gravitational radiation from inspiralling compact objects: Spin effects to the fourth post-Newtonian order}},\ }\href {https://doi.org/10.1103/PhysRevD.106.L101501} {\bibfield  {journal} {\bibinfo  {journal} {Phys. Rev. D}\ }\textbf {\bibinfo {volume} {106}},\ \bibinfo {pages} {L101501} (\bibinfo {year} {2022})},\ \Eprint {https://arxiv.org/abs/2201.05138} {arXiv:2201.05138 [gr-qc]} \BibitemShut {NoStop}%
\bibitem [{\citenamefont {Amalberti}\ \emph {et~al.}(2024{\natexlab{a}})\citenamefont {Amalberti}, \citenamefont {Larrouturou},\ and\ \citenamefont {Yang}}]{Amalberti:2023ohj}%
  \BibitemOpen
  \bibfield  {author} {\bibinfo {author} {\bibfnamefont {L.}~\bibnamefont {Amalberti}}, \bibinfo {author} {\bibfnamefont {F.}~\bibnamefont {Larrouturou}},\ and\ \bibinfo {author} {\bibfnamefont {Z.}~\bibnamefont {Yang}},\ }\bibfield  {title} {\bibinfo {title} {{Multipole expansion at the level of the action in d-dimensions}},\ }\href {https://doi.org/10.1103/PhysRevD.109.104027} {\bibfield  {journal} {\bibinfo  {journal} {Phys. Rev. D}\ }\textbf {\bibinfo {volume} {109}},\ \bibinfo {pages} {104027} (\bibinfo {year} {2024}{\natexlab{a}})},\ \Eprint {https://arxiv.org/abs/2312.02868} {arXiv:2312.02868 [gr-qc]} \BibitemShut {NoStop}%
\bibitem [{\citenamefont {Amalberti}\ \emph {et~al.}(2024{\natexlab{b}})\citenamefont {Amalberti}, \citenamefont {Yang},\ and\ \citenamefont {Porto}}]{Amalberti:2024jaa}%
  \BibitemOpen
  \bibfield  {author} {\bibinfo {author} {\bibfnamefont {L.}~\bibnamefont {Amalberti}}, \bibinfo {author} {\bibfnamefont {Z.}~\bibnamefont {Yang}},\ and\ \bibinfo {author} {\bibfnamefont {R.~A.}\ \bibnamefont {Porto}},\ }\bibfield  {title} {\bibinfo {title} {{Gravitational radiation from inspiralling compact binaries to N3LO in the effective field theory approach}},\ }\href {https://doi.org/10.1103/PhysRevD.110.044046} {\bibfield  {journal} {\bibinfo  {journal} {Phys. Rev. D}\ }\textbf {\bibinfo {volume} {110}},\ \bibinfo {pages} {044046} (\bibinfo {year} {2024}{\natexlab{b}})},\ \Eprint {https://arxiv.org/abs/2406.03457} {arXiv:2406.03457 [gr-qc]} \BibitemShut {NoStop}%
\bibitem [{\citenamefont {Mandal}\ \emph {et~al.}(2025)\citenamefont {Mandal}, \citenamefont {Mastrolia}, \citenamefont {Patil},\ and\ \citenamefont {Steinhoff}}]{Mandal:2024iug}%
  \BibitemOpen
  \bibfield  {author} {\bibinfo {author} {\bibfnamefont {M.~K.}\ \bibnamefont {Mandal}}, \bibinfo {author} {\bibfnamefont {P.}~\bibnamefont {Mastrolia}}, \bibinfo {author} {\bibfnamefont {R.}~\bibnamefont {Patil}},\ and\ \bibinfo {author} {\bibfnamefont {J.}~\bibnamefont {Steinhoff}},\ }\bibfield  {title} {\bibinfo {title} {{Radiating Love: adiabatic tidal fluxes and modes up to next-to-next-to-leading post-Newtonian order}},\ }\href {https://doi.org/10.1007/JHEP05(2025)008} {\bibfield  {journal} {\bibinfo  {journal} {JHEP}\ }\textbf {\bibinfo {volume} {05}},\ \bibinfo {pages} {008}},\ \Eprint {https://arxiv.org/abs/2412.01706} {arXiv:2412.01706 [gr-qc]} \BibitemShut {NoStop}%
\bibitem [{\citenamefont {Goldberger}(2007)}]{Goldberger:2007hy}%
  \BibitemOpen
  \bibfield  {author} {\bibinfo {author} {\bibfnamefont {W.~D.}\ \bibnamefont {Goldberger}},\ }\bibfield  {title} {\bibinfo {title} {{Les Houches lectures on effective field theories and gravitational radiation}},\ }in\ \href@noop {} {\emph {\bibinfo {booktitle} {{Les Houches Summer School - Session 86: Particle Physics and Cosmology: The Fabric of Spacetime}}}}\ (\bibinfo {year} {2007})\ \Eprint {https://arxiv.org/abs/hep-ph/0701129} {arXiv:hep-ph/0701129} \BibitemShut {NoStop}%
\bibitem [{\citenamefont {Foffa}\ and\ \citenamefont {Sturani}(2014)}]{Foffa:2013qca}%
  \BibitemOpen
  \bibfield  {author} {\bibinfo {author} {\bibfnamefont {S.}~\bibnamefont {Foffa}}\ and\ \bibinfo {author} {\bibfnamefont {R.}~\bibnamefont {Sturani}},\ }\bibfield  {title} {\bibinfo {title} {{Effective field theory methods to model compact binaries}},\ }\href {https://doi.org/10.1088/0264-9381/31/4/043001} {\bibfield  {journal} {\bibinfo  {journal} {Class. Quant. Grav.}\ }\textbf {\bibinfo {volume} {31}},\ \bibinfo {pages} {043001} (\bibinfo {year} {2014})},\ \Eprint {https://arxiv.org/abs/1309.3474} {arXiv:1309.3474 [gr-qc]} \BibitemShut {NoStop}%
\bibitem [{\citenamefont {Goldberger}(2022)}]{Goldberger:2022rqf}%
  \BibitemOpen
  \bibfield  {author} {\bibinfo {author} {\bibfnamefont {W.~D.}\ \bibnamefont {Goldberger}},\ }\href {https://doi.org/10.1007/978-981-19-3079-9\_2-1} {\bibinfo {title} {{Effective Field Theory for Compact Binary Dynamics}}} (\bibinfo {year} {2022}),\ \Eprint {https://arxiv.org/abs/2212.06677} {arXiv:2212.06677 [hep-th]} \BibitemShut {NoStop}%
\bibitem [{\citenamefont {Bernard}\ \emph {et~al.}(2017{\natexlab{b}})\citenamefont {Bernard}, \citenamefont {Blanchet}, \citenamefont {Boh{\'e}}, \citenamefont {Faye},\ and\ \citenamefont {Marsat}}]{Bernard:2017bvn}%
  \BibitemOpen
  \bibfield  {author} {\bibinfo {author} {\bibfnamefont {L.}~\bibnamefont {Bernard}}, \bibinfo {author} {\bibfnamefont {L.}~\bibnamefont {Blanchet}}, \bibinfo {author} {\bibfnamefont {A.}~\bibnamefont {Boh{\'e}}}, \bibinfo {author} {\bibfnamefont {G.}~\bibnamefont {Faye}},\ and\ \bibinfo {author} {\bibfnamefont {S.}~\bibnamefont {Marsat}},\ }\bibfield  {title} {\bibinfo {title} {{Dimensional regularization of the IR divergences in the Fokker action of point-particle binaries at the fourth post-Newtonian order}},\ }\href {https://doi.org/10.1103/PhysRevD.96.104043} {\bibfield  {journal} {\bibinfo  {journal} {Phys. Rev. D}\ }\textbf {\bibinfo {volume} {96}},\ \bibinfo {pages} {104043} (\bibinfo {year} {2017}{\natexlab{b}})},\ \Eprint {https://arxiv.org/abs/1706.08480} {arXiv:1706.08480 [gr-qc]} \BibitemShut {NoStop}%
\bibitem [{\citenamefont {Marchand}\ \emph {et~al.}(2018)\citenamefont {Marchand}, \citenamefont {Bernard}, \citenamefont {Blanchet},\ and\ \citenamefont {Faye}}]{Marchand:2017pir}%
  \BibitemOpen
  \bibfield  {author} {\bibinfo {author} {\bibfnamefont {T.}~\bibnamefont {Marchand}}, \bibinfo {author} {\bibfnamefont {L.}~\bibnamefont {Bernard}}, \bibinfo {author} {\bibfnamefont {L.}~\bibnamefont {Blanchet}},\ and\ \bibinfo {author} {\bibfnamefont {G.}~\bibnamefont {Faye}},\ }\bibfield  {title} {\bibinfo {title} {{Ambiguity-Free Completion of the Equations of Motion of Compact Binary Systems at the Fourth Post-Newtonian Order}},\ }\href {https://doi.org/10.1103/PhysRevD.97.044023} {\bibfield  {journal} {\bibinfo  {journal} {Phys. Rev. D}\ }\textbf {\bibinfo {volume} {97}},\ \bibinfo {pages} {044023} (\bibinfo {year} {2018})},\ \Eprint {https://arxiv.org/abs/1707.09289} {arXiv:1707.09289 [gr-qc]} \BibitemShut {NoStop}%
\bibitem [{\citenamefont {Porto}\ and\ \citenamefont {Riva}(2026)}]{Porto:2026fsd}%
  \BibitemOpen
  \bibfield  {author} {\bibinfo {author} {\bibfnamefont {R.~A.}\ \bibnamefont {Porto}}\ and\ \bibinfo {author} {\bibfnamefont {M.~M.}\ \bibnamefont {Riva}},\ }\bibfield  {title} {\bibinfo {title} {{Black hole dynamics at fifth post-Newtonian order}},\ }\href {https://doi.org/10.1007/JHEP09(2026)075} {\bibfield  {journal} {\bibinfo  {journal} {JHEP}\ }\textbf {\bibinfo {volume} {09}},\ \bibinfo {pages} {075}},\ \Eprint {https://arxiv.org/abs/2604.09545} {arXiv:2604.09545 [gr-qc]} \BibitemShut {NoStop}%
\bibitem [{\citenamefont {Bini}\ \emph {et~al.}(2020{\natexlab{a}})\citenamefont {Bini}, \citenamefont {Damour},\ and\ \citenamefont {Geralico}}]{Bini:2020wpo}%
  \BibitemOpen
  \bibfield  {author} {\bibinfo {author} {\bibfnamefont {D.}~\bibnamefont {Bini}}, \bibinfo {author} {\bibfnamefont {T.}~\bibnamefont {Damour}},\ and\ \bibinfo {author} {\bibfnamefont {A.}~\bibnamefont {Geralico}},\ }\bibfield  {title} {\bibinfo {title} {{Binary dynamics at the fifth and fifth-and-a-half post-Newtonian orders}},\ }\href {https://doi.org/10.1103/PhysRevD.102.024062} {\bibfield  {journal} {\bibinfo  {journal} {Phys. Rev. D}\ }\textbf {\bibinfo {volume} {102}},\ \bibinfo {pages} {024062} (\bibinfo {year} {2020}{\natexlab{a}})},\ \Eprint {https://arxiv.org/abs/2003.11891} {arXiv:2003.11891 [gr-qc]} \BibitemShut {NoStop}%
\bibitem [{\citenamefont {Bini}\ \emph {et~al.}(2020{\natexlab{b}})\citenamefont {Bini}, \citenamefont {Damour},\ and\ \citenamefont {Geralico}}]{Bini:2020nsb}%
  \BibitemOpen
  \bibfield  {author} {\bibinfo {author} {\bibfnamefont {D.}~\bibnamefont {Bini}}, \bibinfo {author} {\bibfnamefont {T.}~\bibnamefont {Damour}},\ and\ \bibinfo {author} {\bibfnamefont {A.}~\bibnamefont {Geralico}},\ }\bibfield  {title} {\bibinfo {title} {{Sixth post-Newtonian local-in-time dynamics of binary systems}},\ }\href {https://doi.org/10.1103/PhysRevD.102.024061} {\bibfield  {journal} {\bibinfo  {journal} {Phys. Rev. D}\ }\textbf {\bibinfo {volume} {102}},\ \bibinfo {pages} {024061} (\bibinfo {year} {2020}{\natexlab{b}})},\ \Eprint {https://arxiv.org/abs/2004.05407} {arXiv:2004.05407 [gr-qc]} \BibitemShut {NoStop}%
\bibitem [{\citenamefont {Bini}\ \emph {et~al.}(2020{\natexlab{c}})\citenamefont {Bini}, \citenamefont {Damour},\ and\ \citenamefont {Geralico}}]{Bini:2020hmy}%
  \BibitemOpen
  \bibfield  {author} {\bibinfo {author} {\bibfnamefont {D.}~\bibnamefont {Bini}}, \bibinfo {author} {\bibfnamefont {T.}~\bibnamefont {Damour}},\ and\ \bibinfo {author} {\bibfnamefont {A.}~\bibnamefont {Geralico}},\ }\bibfield  {title} {\bibinfo {title} {{Sixth post-Newtonian nonlocal-in-time dynamics of binary systems}},\ }\href {https://doi.org/10.1103/PhysRevD.102.084047} {\bibfield  {journal} {\bibinfo  {journal} {Phys. Rev. D}\ }\textbf {\bibinfo {volume} {102}},\ \bibinfo {pages} {084047} (\bibinfo {year} {2020}{\natexlab{c}})},\ \Eprint {https://arxiv.org/abs/2007.11239} {arXiv:2007.11239 [gr-qc]} \BibitemShut {NoStop}%
\bibitem [{\citenamefont {Brunello}\ \emph {et~al.}(2026{\natexlab{a}})\citenamefont {Brunello}, \citenamefont {Mandal}, \citenamefont {Mastrolia}, \citenamefont {Patil}, \citenamefont {Pegorin}, \citenamefont {Ronca}, \citenamefont {Smith}, \citenamefont {Steinhoff},\ and\ \citenamefont {Torres~Bobadilla}}]{Brunello:2025gpf}%
  \BibitemOpen
  \bibfield  {author} {\bibinfo {author} {\bibfnamefont {G.}~\bibnamefont {Brunello}}, \bibinfo {author} {\bibfnamefont {M.~K.}\ \bibnamefont {Mandal}}, \bibinfo {author} {\bibfnamefont {P.}~\bibnamefont {Mastrolia}}, \bibinfo {author} {\bibfnamefont {R.}~\bibnamefont {Patil}}, \bibinfo {author} {\bibfnamefont {M.}~\bibnamefont {Pegorin}}, \bibinfo {author} {\bibfnamefont {J.}~\bibnamefont {Ronca}}, \bibinfo {author} {\bibfnamefont {S.}~\bibnamefont {Smith}}, \bibinfo {author} {\bibfnamefont {J.}~\bibnamefont {Steinhoff}},\ and\ \bibinfo {author} {\bibfnamefont {W.~J.}\ \bibnamefont {Torres~Bobadilla}},\ }\bibfield  {title} {\bibinfo {title} {{Six-Loop Gravitational Interactions at the Sixth Post-Newtonian Order}},\ }\href {https://doi.org/10.1103/c2r3-cqtf} {\bibfield  {journal} {\bibinfo  {journal} {Phys. Rev. Lett.}\ }\textbf {\bibinfo {volume} {137}},\ \bibinfo {pages} {111401} (\bibinfo {year} {2026}{\natexlab{a}})},\ \Eprint {https://arxiv.org/abs/2512.19498} {arXiv:2512.19498 [hep-th]} \BibitemShut
  {NoStop}%
\bibitem [{\citenamefont {Brunello}\ \emph {et~al.}(2026{\natexlab{b}})\citenamefont {Brunello}, \citenamefont {Mandal}, \citenamefont {Mastrolia}, \citenamefont {Patil}, \citenamefont {Pegorin}, \citenamefont {Smith},\ and\ \citenamefont {Steinhoff}}]{Brunello:2026anu}%
  \BibitemOpen
  \bibfield  {author} {\bibinfo {author} {\bibfnamefont {G.}~\bibnamefont {Brunello}}, \bibinfo {author} {\bibfnamefont {M.~K.}\ \bibnamefont {Mandal}}, \bibinfo {author} {\bibfnamefont {P.}~\bibnamefont {Mastrolia}}, \bibinfo {author} {\bibfnamefont {R.}~\bibnamefont {Patil}}, \bibinfo {author} {\bibfnamefont {M.}~\bibnamefont {Pegorin}}, \bibinfo {author} {\bibfnamefont {S.}~\bibnamefont {Smith}},\ and\ \bibinfo {author} {\bibfnamefont {J.}~\bibnamefont {Steinhoff}},\ }\href@noop {} {\bibinfo {title} {{All-order structure of static gravitational interactions and the seventh post-Newtonian potential}}} (\bibinfo {year} {2026}{\natexlab{b}}),\ \Eprint {https://arxiv.org/abs/2604.14134} {arXiv:2604.14134 [hep-th]} \BibitemShut {NoStop}%
\bibitem [{\citenamefont {Westpfahl}\ and\ \citenamefont {Goller}(1979)}]{Westpfahl:1979gu}%
  \BibitemOpen
  \bibfield  {author} {\bibinfo {author} {\bibfnamefont {K.}~\bibnamefont {Westpfahl}}\ and\ \bibinfo {author} {\bibfnamefont {M.}~\bibnamefont {Goller}},\ }\bibfield  {title} {\bibinfo {title} {{Gravitational scattering of two relativistic particles in postlinear approximation}},\ }\href {https://doi.org/10.1007/BF02817047} {\bibfield  {journal} {\bibinfo  {journal} {Lett. Nuovo Cim.}\ }\textbf {\bibinfo {volume} {26}},\ \bibinfo {pages} {573} (\bibinfo {year} {1979})}\BibitemShut {NoStop}%
\bibitem [{\citenamefont {Westpfahl}(1985)}]{Westpfahl:1985tsl}%
  \BibitemOpen
  \bibfield  {author} {\bibinfo {author} {\bibfnamefont {K.}~\bibnamefont {Westpfahl}},\ }\bibfield  {title} {\bibinfo {title} {{High-Speed Scattering of Charged and Uncharged Particles in General Relativity}},\ }\href {https://doi.org/10.1002/prop.2190330802} {\bibfield  {journal} {\bibinfo  {journal} {Fortsch. Phys.}\ }\textbf {\bibinfo {volume} {33}},\ \bibinfo {pages} {417} (\bibinfo {year} {1985})}\BibitemShut {NoStop}%
\bibitem [{\citenamefont {Damour}(2016)}]{Damour:2016gwp}%
  \BibitemOpen
  \bibfield  {author} {\bibinfo {author} {\bibfnamefont {T.}~\bibnamefont {Damour}},\ }\bibfield  {title} {\bibinfo {title} {{Gravitational scattering, post-Minkowskian approximation and Effective One-Body theory}},\ }\href {https://doi.org/10.1103/PhysRevD.94.104015} {\bibfield  {journal} {\bibinfo  {journal} {Phys. Rev. D}\ }\textbf {\bibinfo {volume} {94}},\ \bibinfo {pages} {104015} (\bibinfo {year} {2016})},\ \Eprint {https://arxiv.org/abs/1609.00354} {arXiv:1609.00354 [gr-qc]} \BibitemShut {NoStop}%
\bibitem [{\citenamefont {Damour}(2018)}]{Damour:2017zjx}%
  \BibitemOpen
  \bibfield  {author} {\bibinfo {author} {\bibfnamefont {T.}~\bibnamefont {Damour}},\ }\bibfield  {title} {\bibinfo {title} {{High-energy gravitational scattering and the general relativistic two-body problem}},\ }\href {https://doi.org/10.1103/PhysRevD.97.044038} {\bibfield  {journal} {\bibinfo  {journal} {Phys. Rev. D}\ }\textbf {\bibinfo {volume} {97}},\ \bibinfo {pages} {044038} (\bibinfo {year} {2018})},\ \Eprint {https://arxiv.org/abs/1710.10599} {arXiv:1710.10599 [gr-qc]} \BibitemShut {NoStop}%
\bibitem [{\citenamefont {Bjerrum-Bohr}\ \emph {et~al.}(2018)\citenamefont {Bjerrum-Bohr}, \citenamefont {Damgaard}, \citenamefont {Festuccia}, \citenamefont {Plant{\'e}},\ and\ \citenamefont {Vanhove}}]{Bjerrum-Bohr:2018xdl}%
  \BibitemOpen
  \bibfield  {author} {\bibinfo {author} {\bibfnamefont {N.~E.~J.}\ \bibnamefont {Bjerrum-Bohr}}, \bibinfo {author} {\bibfnamefont {P.~H.}\ \bibnamefont {Damgaard}}, \bibinfo {author} {\bibfnamefont {G.}~\bibnamefont {Festuccia}}, \bibinfo {author} {\bibfnamefont {L.}~\bibnamefont {Plant{\'e}}},\ and\ \bibinfo {author} {\bibfnamefont {P.}~\bibnamefont {Vanhove}},\ }\bibfield  {title} {\bibinfo {title} {{General Relativity from Scattering Amplitudes}},\ }\href {https://doi.org/10.1103/PhysRevLett.121.171601} {\bibfield  {journal} {\bibinfo  {journal} {Phys. Rev. Lett.}\ }\textbf {\bibinfo {volume} {121}},\ \bibinfo {pages} {171601} (\bibinfo {year} {2018})},\ \Eprint {https://arxiv.org/abs/1806.04920} {arXiv:1806.04920 [hep-th]} \BibitemShut {NoStop}%
\bibitem [{\citenamefont {Cheung}\ \emph {et~al.}(2018)\citenamefont {Cheung}, \citenamefont {Rothstein},\ and\ \citenamefont {Solon}}]{Cheung:2018wkq}%
  \BibitemOpen
  \bibfield  {author} {\bibinfo {author} {\bibfnamefont {C.}~\bibnamefont {Cheung}}, \bibinfo {author} {\bibfnamefont {I.~Z.}\ \bibnamefont {Rothstein}},\ and\ \bibinfo {author} {\bibfnamefont {M.~P.}\ \bibnamefont {Solon}},\ }\bibfield  {title} {\bibinfo {title} {{From Scattering Amplitudes to Classical Potentials in the Post-Minkowskian Expansion}},\ }\href {https://doi.org/10.1103/PhysRevLett.121.251101} {\bibfield  {journal} {\bibinfo  {journal} {Phys. Rev. Lett.}\ }\textbf {\bibinfo {volume} {121}},\ \bibinfo {pages} {251101} (\bibinfo {year} {2018})},\ \Eprint {https://arxiv.org/abs/1808.02489} {arXiv:1808.02489 [hep-th]} \BibitemShut {NoStop}%
\bibitem [{\citenamefont {Kosower}\ \emph {et~al.}(2019)\citenamefont {Kosower}, \citenamefont {Maybee},\ and\ \citenamefont {O'Connell}}]{Kosower:2018adc}%
  \BibitemOpen
  \bibfield  {author} {\bibinfo {author} {\bibfnamefont {D.~A.}\ \bibnamefont {Kosower}}, \bibinfo {author} {\bibfnamefont {B.}~\bibnamefont {Maybee}},\ and\ \bibinfo {author} {\bibfnamefont {D.}~\bibnamefont {O'Connell}},\ }\bibfield  {title} {\bibinfo {title} {{Amplitudes, Observables, and Classical Scattering}},\ }\href {https://doi.org/10.1007/JHEP02(2019)137} {\bibfield  {journal} {\bibinfo  {journal} {JHEP}\ }\textbf {\bibinfo {volume} {02}},\ \bibinfo {pages} {137}},\ \Eprint {https://arxiv.org/abs/1811.10950} {arXiv:1811.10950 [hep-th]} \BibitemShut {NoStop}%
\bibitem [{\citenamefont {K{\"a}lin}\ and\ \citenamefont {Porto}(2020{\natexlab{a}})}]{Kalin:2020mvi}%
  \BibitemOpen
  \bibfield  {author} {\bibinfo {author} {\bibfnamefont {G.}~\bibnamefont {K{\"a}lin}}\ and\ \bibinfo {author} {\bibfnamefont {R.~A.}\ \bibnamefont {Porto}},\ }\bibfield  {title} {\bibinfo {title} {{Post-Minkowskian Effective Field Theory for Conservative Binary Dynamics}},\ }\href {https://doi.org/10.1007/JHEP11(2020)106} {\bibfield  {journal} {\bibinfo  {journal} {JHEP}\ }\textbf {\bibinfo {volume} {11}},\ \bibinfo {pages} {106}},\ \Eprint {https://arxiv.org/abs/2006.01184} {arXiv:2006.01184 [hep-th]} \BibitemShut {NoStop}%
\bibitem [{\citenamefont {Mogull}\ \emph {et~al.}(2021)\citenamefont {Mogull}, \citenamefont {Plefka},\ and\ \citenamefont {Steinhoff}}]{Mogull:2020sak}%
  \BibitemOpen
  \bibfield  {author} {\bibinfo {author} {\bibfnamefont {G.}~\bibnamefont {Mogull}}, \bibinfo {author} {\bibfnamefont {J.}~\bibnamefont {Plefka}},\ and\ \bibinfo {author} {\bibfnamefont {J.}~\bibnamefont {Steinhoff}},\ }\bibfield  {title} {\bibinfo {title} {{Classical black hole scattering from a worldline quantum field theory}},\ }\href {https://doi.org/10.1007/JHEP02(2021)048} {\bibfield  {journal} {\bibinfo  {journal} {JHEP}\ }\textbf {\bibinfo {volume} {02}},\ \bibinfo {pages} {048}},\ \Eprint {https://arxiv.org/abs/2010.02865} {arXiv:2010.02865 [hep-th]} \BibitemShut {NoStop}%
\bibitem [{\citenamefont {K{\"a}lin}\ \emph {et~al.}(2023)\citenamefont {K{\"a}lin}, \citenamefont {Neef},\ and\ \citenamefont {Porto}}]{Kalin:2022hph}%
  \BibitemOpen
  \bibfield  {author} {\bibinfo {author} {\bibfnamefont {G.}~\bibnamefont {K{\"a}lin}}, \bibinfo {author} {\bibfnamefont {J.}~\bibnamefont {Neef}},\ and\ \bibinfo {author} {\bibfnamefont {R.~A.}\ \bibnamefont {Porto}},\ }\bibfield  {title} {\bibinfo {title} {{Radiation-reaction in the Effective Field Theory approach to Post-Minkowskian dynamics}},\ }\href {https://doi.org/10.1007/JHEP01(2023)140} {\bibfield  {journal} {\bibinfo  {journal} {JHEP}\ }\textbf {\bibinfo {volume} {01}},\ \bibinfo {pages} {140}},\ \Eprint {https://arxiv.org/abs/2207.00580} {arXiv:2207.00580 [hep-th]} \BibitemShut {NoStop}%
\bibitem [{\citenamefont {Jakobsen}\ \emph {et~al.}(2022)\citenamefont {Jakobsen}, \citenamefont {Mogull}, \citenamefont {Plefka},\ and\ \citenamefont {Sauer}}]{Jakobsen:2022psy}%
  \BibitemOpen
  \bibfield  {author} {\bibinfo {author} {\bibfnamefont {G.~U.}\ \bibnamefont {Jakobsen}}, \bibinfo {author} {\bibfnamefont {G.}~\bibnamefont {Mogull}}, \bibinfo {author} {\bibfnamefont {J.}~\bibnamefont {Plefka}},\ and\ \bibinfo {author} {\bibfnamefont {B.}~\bibnamefont {Sauer}},\ }\bibfield  {title} {\bibinfo {title} {{All things retarded: radiation-reaction in worldline quantum field theory}},\ }\href {https://doi.org/10.1007/JHEP10(2022)128} {\bibfield  {journal} {\bibinfo  {journal} {JHEP}\ }\textbf {\bibinfo {volume} {10}},\ \bibinfo {pages} {128}},\ \Eprint {https://arxiv.org/abs/2207.00569} {arXiv:2207.00569 [hep-th]} \BibitemShut {NoStop}%
\bibitem [{\citenamefont {Bern}\ \emph {et~al.}(2019{\natexlab{a}})\citenamefont {Bern}, \citenamefont {Cheung}, \citenamefont {Roiban}, \citenamefont {Shen}, \citenamefont {Solon},\ and\ \citenamefont {Zeng}}]{Bern:2019nnu}%
  \BibitemOpen
  \bibfield  {author} {\bibinfo {author} {\bibfnamefont {Z.}~\bibnamefont {Bern}}, \bibinfo {author} {\bibfnamefont {C.}~\bibnamefont {Cheung}}, \bibinfo {author} {\bibfnamefont {R.}~\bibnamefont {Roiban}}, \bibinfo {author} {\bibfnamefont {C.-H.}\ \bibnamefont {Shen}}, \bibinfo {author} {\bibfnamefont {M.~P.}\ \bibnamefont {Solon}},\ and\ \bibinfo {author} {\bibfnamefont {M.}~\bibnamefont {Zeng}},\ }\bibfield  {title} {\bibinfo {title} {{Scattering Amplitudes and the Conservative Hamiltonian for Binary Systems at Third Post-Minkowskian Order}},\ }\href {https://doi.org/10.1103/PhysRevLett.122.201603} {\bibfield  {journal} {\bibinfo  {journal} {Phys. Rev. Lett.}\ }\textbf {\bibinfo {volume} {122}},\ \bibinfo {pages} {201603} (\bibinfo {year} {2019}{\natexlab{a}})},\ \Eprint {https://arxiv.org/abs/1901.04424} {arXiv:1901.04424 [hep-th]} \BibitemShut {NoStop}%
\bibitem [{\citenamefont {Bern}\ \emph {et~al.}(2019{\natexlab{b}})\citenamefont {Bern}, \citenamefont {Cheung}, \citenamefont {Roiban}, \citenamefont {Shen}, \citenamefont {Solon},\ and\ \citenamefont {Zeng}}]{Bern:2019crd}%
  \BibitemOpen
  \bibfield  {author} {\bibinfo {author} {\bibfnamefont {Z.}~\bibnamefont {Bern}}, \bibinfo {author} {\bibfnamefont {C.}~\bibnamefont {Cheung}}, \bibinfo {author} {\bibfnamefont {R.}~\bibnamefont {Roiban}}, \bibinfo {author} {\bibfnamefont {C.-H.}\ \bibnamefont {Shen}}, \bibinfo {author} {\bibfnamefont {M.~P.}\ \bibnamefont {Solon}},\ and\ \bibinfo {author} {\bibfnamefont {M.}~\bibnamefont {Zeng}},\ }\bibfield  {title} {\bibinfo {title} {{Black Hole Binary Dynamics from the Double Copy and Effective Theory}},\ }\href {https://doi.org/10.1007/JHEP10(2019)206} {\bibfield  {journal} {\bibinfo  {journal} {JHEP}\ }\textbf {\bibinfo {volume} {10}},\ \bibinfo {pages} {206}},\ \Eprint {https://arxiv.org/abs/1908.01493} {arXiv:1908.01493 [hep-th]} \BibitemShut {NoStop}%
\bibitem [{\citenamefont {Cheung}\ and\ \citenamefont {Solon}(2020)}]{Cheung:2020gyp}%
  \BibitemOpen
  \bibfield  {author} {\bibinfo {author} {\bibfnamefont {C.}~\bibnamefont {Cheung}}\ and\ \bibinfo {author} {\bibfnamefont {M.~P.}\ \bibnamefont {Solon}},\ }\bibfield  {title} {\bibinfo {title} {{Classical gravitational scattering at $ \mathcal{O} $(G$^{3}$) from Feynman diagrams}},\ }\href {https://doi.org/10.1007/JHEP06(2020)144} {\bibfield  {journal} {\bibinfo  {journal} {JHEP}\ }\textbf {\bibinfo {volume} {06}},\ \bibinfo {pages} {144}},\ \Eprint {https://arxiv.org/abs/2003.08351} {arXiv:2003.08351 [hep-th]} \BibitemShut {NoStop}%
\bibitem [{\citenamefont {Parra-Martinez}\ \emph {et~al.}(2020)\citenamefont {Parra-Martinez}, \citenamefont {Ruf},\ and\ \citenamefont {Zeng}}]{Parra-Martinez:2020dzs}%
  \BibitemOpen
  \bibfield  {author} {\bibinfo {author} {\bibfnamefont {J.}~\bibnamefont {Parra-Martinez}}, \bibinfo {author} {\bibfnamefont {M.~S.}\ \bibnamefont {Ruf}},\ and\ \bibinfo {author} {\bibfnamefont {M.}~\bibnamefont {Zeng}},\ }\bibfield  {title} {\bibinfo {title} {{Extremal black hole scattering at $\mathcal{O}(G^3)$: graviton dominance, eikonal exponentiation, and differential equations}},\ }\href {https://doi.org/10.1007/JHEP11(2020)023} {\bibfield  {journal} {\bibinfo  {journal} {JHEP}\ }\textbf {\bibinfo {volume} {11}},\ \bibinfo {pages} {023}},\ \Eprint {https://arxiv.org/abs/2005.04236} {arXiv:2005.04236 [hep-th]} \BibitemShut {NoStop}%
\bibitem [{\citenamefont {K{\"a}lin}\ \emph {et~al.}(2020)\citenamefont {K{\"a}lin}, \citenamefont {Liu},\ and\ \citenamefont {Porto}}]{Kalin:2020fhe}%
  \BibitemOpen
  \bibfield  {author} {\bibinfo {author} {\bibfnamefont {G.}~\bibnamefont {K{\"a}lin}}, \bibinfo {author} {\bibfnamefont {Z.}~\bibnamefont {Liu}},\ and\ \bibinfo {author} {\bibfnamefont {R.~A.}\ \bibnamefont {Porto}},\ }\bibfield  {title} {\bibinfo {title} {{Conservative Dynamics of Binary Systems to Third Post-Minkowskian Order from the Effective Field Theory Approach}},\ }\href {https://doi.org/10.1103/PhysRevLett.125.261103} {\bibfield  {journal} {\bibinfo  {journal} {Phys. Rev. Lett.}\ }\textbf {\bibinfo {volume} {125}},\ \bibinfo {pages} {261103} (\bibinfo {year} {2020})},\ \Eprint {https://arxiv.org/abs/2007.04977} {arXiv:2007.04977 [hep-th]} \BibitemShut {NoStop}%
\bibitem [{\citenamefont {Bern}\ \emph {et~al.}(2021)\citenamefont {Bern}, \citenamefont {Parra-Martinez}, \citenamefont {Roiban}, \citenamefont {Ruf}, \citenamefont {Shen}, \citenamefont {Solon},\ and\ \citenamefont {Zeng}}]{Bern:2021dqo}%
  \BibitemOpen
  \bibfield  {author} {\bibinfo {author} {\bibfnamefont {Z.}~\bibnamefont {Bern}}, \bibinfo {author} {\bibfnamefont {J.}~\bibnamefont {Parra-Martinez}}, \bibinfo {author} {\bibfnamefont {R.}~\bibnamefont {Roiban}}, \bibinfo {author} {\bibfnamefont {M.~S.}\ \bibnamefont {Ruf}}, \bibinfo {author} {\bibfnamefont {C.-H.}\ \bibnamefont {Shen}}, \bibinfo {author} {\bibfnamefont {M.~P.}\ \bibnamefont {Solon}},\ and\ \bibinfo {author} {\bibfnamefont {M.}~\bibnamefont {Zeng}},\ }\bibfield  {title} {\bibinfo {title} {{Scattering Amplitudes and Conservative Binary Dynamics at ${\cal O}(G^4)$}},\ }\href {https://doi.org/10.1103/PhysRevLett.126.171601} {\bibfield  {journal} {\bibinfo  {journal} {Phys. Rev. Lett.}\ }\textbf {\bibinfo {volume} {126}},\ \bibinfo {pages} {171601} (\bibinfo {year} {2021})},\ \Eprint {https://arxiv.org/abs/2101.07254} {arXiv:2101.07254 [hep-th]} \BibitemShut {NoStop}%
\bibitem [{\citenamefont {Bern}\ \emph {et~al.}(2022{\natexlab{a}})\citenamefont {Bern}, \citenamefont {Parra-Martinez}, \citenamefont {Roiban}, \citenamefont {Ruf}, \citenamefont {Shen}, \citenamefont {Solon},\ and\ \citenamefont {Zeng}}]{Bern:2021yeh}%
  \BibitemOpen
  \bibfield  {author} {\bibinfo {author} {\bibfnamefont {Z.}~\bibnamefont {Bern}}, \bibinfo {author} {\bibfnamefont {J.}~\bibnamefont {Parra-Martinez}}, \bibinfo {author} {\bibfnamefont {R.}~\bibnamefont {Roiban}}, \bibinfo {author} {\bibfnamefont {M.~S.}\ \bibnamefont {Ruf}}, \bibinfo {author} {\bibfnamefont {C.-H.}\ \bibnamefont {Shen}}, \bibinfo {author} {\bibfnamefont {M.~P.}\ \bibnamefont {Solon}},\ and\ \bibinfo {author} {\bibfnamefont {M.}~\bibnamefont {Zeng}},\ }\bibfield  {title} {\bibinfo {title} {{Scattering Amplitudes, the Tail Effect, and Conservative Binary Dynamics at O(G4)}},\ }\href {https://doi.org/10.1103/PhysRevLett.128.161103} {\bibfield  {journal} {\bibinfo  {journal} {Phys. Rev. Lett.}\ }\textbf {\bibinfo {volume} {128}},\ \bibinfo {pages} {161103} (\bibinfo {year} {2022}{\natexlab{a}})},\ \Eprint {https://arxiv.org/abs/2112.10750} {arXiv:2112.10750 [hep-th]} \BibitemShut {NoStop}%
\bibitem [{\citenamefont {Bern}\ \emph {et~al.}(2022{\natexlab{b}})\citenamefont {Bern}, \citenamefont {Parra-Martinez}, \citenamefont {Roiban}, \citenamefont {Ruf}, \citenamefont {Shen}, \citenamefont {Solon},\ and\ \citenamefont {Zeng}}]{Bern:2022jvn}%
  \BibitemOpen
  \bibfield  {author} {\bibinfo {author} {\bibfnamefont {Z.}~\bibnamefont {Bern}}, \bibinfo {author} {\bibfnamefont {J.}~\bibnamefont {Parra-Martinez}}, \bibinfo {author} {\bibfnamefont {R.}~\bibnamefont {Roiban}}, \bibinfo {author} {\bibfnamefont {M.~S.}\ \bibnamefont {Ruf}}, \bibinfo {author} {\bibfnamefont {C.-H.}\ \bibnamefont {Shen}}, \bibinfo {author} {\bibfnamefont {M.~P.}\ \bibnamefont {Solon}},\ and\ \bibinfo {author} {\bibfnamefont {M.}~\bibnamefont {Zeng}},\ }\bibfield  {title} {\bibinfo {title} {{Scattering amplitudes and conservative dynamics at the fourth post-Minkowskian order}},\ }\href {https://doi.org/10.22323/1.416.0051} {\bibfield  {journal} {\bibinfo  {journal} {PoS}\ }\textbf {\bibinfo {volume} {LL2022}},\ \bibinfo {pages} {051} (\bibinfo {year} {2022}{\natexlab{b}})}\BibitemShut {NoStop}%
\bibitem [{\citenamefont {Dlapa}\ \emph {et~al.}(2022{\natexlab{a}})\citenamefont {Dlapa}, \citenamefont {K{\"a}lin}, \citenamefont {Liu},\ and\ \citenamefont {Porto}}]{Dlapa:2021npj}%
  \BibitemOpen
  \bibfield  {author} {\bibinfo {author} {\bibfnamefont {C.}~\bibnamefont {Dlapa}}, \bibinfo {author} {\bibfnamefont {G.}~\bibnamefont {K{\"a}lin}}, \bibinfo {author} {\bibfnamefont {Z.}~\bibnamefont {Liu}},\ and\ \bibinfo {author} {\bibfnamefont {R.~A.}\ \bibnamefont {Porto}},\ }\bibfield  {title} {\bibinfo {title} {{Dynamics of binary systems to fourth Post-Minkowskian order from the effective field theory approach}},\ }\href {https://doi.org/10.1016/j.physletb.2022.137203} {\bibfield  {journal} {\bibinfo  {journal} {Phys. Lett. B}\ }\textbf {\bibinfo {volume} {831}},\ \bibinfo {pages} {137203} (\bibinfo {year} {2022}{\natexlab{a}})},\ \Eprint {https://arxiv.org/abs/2106.08276} {arXiv:2106.08276 [hep-th]} \BibitemShut {NoStop}%
\bibitem [{\citenamefont {Dlapa}\ \emph {et~al.}(2022{\natexlab{b}})\citenamefont {Dlapa}, \citenamefont {K{\"a}lin}, \citenamefont {Liu},\ and\ \citenamefont {Porto}}]{Dlapa:2021vgp}%
  \BibitemOpen
  \bibfield  {author} {\bibinfo {author} {\bibfnamefont {C.}~\bibnamefont {Dlapa}}, \bibinfo {author} {\bibfnamefont {G.}~\bibnamefont {K{\"a}lin}}, \bibinfo {author} {\bibfnamefont {Z.}~\bibnamefont {Liu}},\ and\ \bibinfo {author} {\bibfnamefont {R.~A.}\ \bibnamefont {Porto}},\ }\bibfield  {title} {\bibinfo {title} {{Conservative Dynamics of Binary Systems at Fourth Post-Minkowskian Order in the Large-Eccentricity Expansion}},\ }\href {https://doi.org/10.1103/PhysRevLett.128.161104} {\bibfield  {journal} {\bibinfo  {journal} {Phys. Rev. Lett.}\ }\textbf {\bibinfo {volume} {128}},\ \bibinfo {pages} {161104} (\bibinfo {year} {2022}{\natexlab{b}})},\ \Eprint {https://arxiv.org/abs/2112.11296} {arXiv:2112.11296 [hep-th]} \BibitemShut {NoStop}%
\bibitem [{\citenamefont {Dlapa}\ \emph {et~al.}(2024)\citenamefont {Dlapa}, \citenamefont {K{\"a}lin}, \citenamefont {Liu},\ and\ \citenamefont {Porto}}]{Dlapa:2024cje}%
  \BibitemOpen
  \bibfield  {author} {\bibinfo {author} {\bibfnamefont {C.}~\bibnamefont {Dlapa}}, \bibinfo {author} {\bibfnamefont {G.}~\bibnamefont {K{\"a}lin}}, \bibinfo {author} {\bibfnamefont {Z.}~\bibnamefont {Liu}},\ and\ \bibinfo {author} {\bibfnamefont {R.~A.}\ \bibnamefont {Porto}},\ }\bibfield  {title} {\bibinfo {title} {{Local in Time Conservative Binary Dynamics at Fourth Post-Minkowskian Order}},\ }\href {https://doi.org/10.1103/PhysRevLett.132.221401} {\bibfield  {journal} {\bibinfo  {journal} {Phys. Rev. Lett.}\ }\textbf {\bibinfo {volume} {132}},\ \bibinfo {pages} {221401} (\bibinfo {year} {2024})},\ \Eprint {https://arxiv.org/abs/2403.04853} {arXiv:2403.04853 [hep-th]} \BibitemShut {NoStop}%
\bibitem [{\citenamefont {Damour}(2020{\natexlab{a}})}]{Damour:2020tta}%
  \BibitemOpen
  \bibfield  {author} {\bibinfo {author} {\bibfnamefont {T.}~\bibnamefont {Damour}},\ }\bibfield  {title} {\bibinfo {title} {{Radiative contribution to classical gravitational scattering at the third order in $G$}},\ }\href {https://doi.org/10.1103/PhysRevD.102.124008} {\bibfield  {journal} {\bibinfo  {journal} {Phys. Rev. D}\ }\textbf {\bibinfo {volume} {102}},\ \bibinfo {pages} {124008} (\bibinfo {year} {2020}{\natexlab{a}})},\ \Eprint {https://arxiv.org/abs/2010.01641} {arXiv:2010.01641 [gr-qc]} \BibitemShut {NoStop}%
\bibitem [{\citenamefont {Di~Vecchia}\ \emph {et~al.}(2021)\citenamefont {Di~Vecchia}, \citenamefont {Heissenberg}, \citenamefont {Russo},\ and\ \citenamefont {Veneziano}}]{DiVecchia:2021bdo}%
  \BibitemOpen
  \bibfield  {author} {\bibinfo {author} {\bibfnamefont {P.}~\bibnamefont {Di~Vecchia}}, \bibinfo {author} {\bibfnamefont {C.}~\bibnamefont {Heissenberg}}, \bibinfo {author} {\bibfnamefont {R.}~\bibnamefont {Russo}},\ and\ \bibinfo {author} {\bibfnamefont {G.}~\bibnamefont {Veneziano}},\ }\bibfield  {title} {\bibinfo {title} {{The eikonal approach to gravitational scattering and radiation at $ \mathcal{O} $(G$^{3}$)}},\ }\href {https://doi.org/10.1007/JHEP07(2021)169} {\bibfield  {journal} {\bibinfo  {journal} {JHEP}\ }\textbf {\bibinfo {volume} {07}},\ \bibinfo {pages} {169}},\ \Eprint {https://arxiv.org/abs/2104.03256} {arXiv:2104.03256 [hep-th]} \BibitemShut {NoStop}%
\bibitem [{\citenamefont {Herrmann}\ \emph {et~al.}(2021{\natexlab{a}})\citenamefont {Herrmann}, \citenamefont {Parra-Martinez}, \citenamefont {Ruf},\ and\ \citenamefont {Zeng}}]{Herrmann:2021tct}%
  \BibitemOpen
  \bibfield  {author} {\bibinfo {author} {\bibfnamefont {E.}~\bibnamefont {Herrmann}}, \bibinfo {author} {\bibfnamefont {J.}~\bibnamefont {Parra-Martinez}}, \bibinfo {author} {\bibfnamefont {M.~S.}\ \bibnamefont {Ruf}},\ and\ \bibinfo {author} {\bibfnamefont {M.}~\bibnamefont {Zeng}},\ }\bibfield  {title} {\bibinfo {title} {{Radiative classical gravitational observables at $ \mathcal{O} $(G$^{3}$) from scattering amplitudes}},\ }\href {https://doi.org/10.1007/JHEP10(2021)148} {\bibfield  {journal} {\bibinfo  {journal} {JHEP}\ }\textbf {\bibinfo {volume} {10}},\ \bibinfo {pages} {148}},\ \Eprint {https://arxiv.org/abs/2104.03957} {arXiv:2104.03957 [hep-th]} \BibitemShut {NoStop}%
\bibitem [{\citenamefont {Bini}\ \emph {et~al.}(2021)\citenamefont {Bini}, \citenamefont {Damour},\ and\ \citenamefont {Geralico}}]{Bini:2021gat}%
  \BibitemOpen
  \bibfield  {author} {\bibinfo {author} {\bibfnamefont {D.}~\bibnamefont {Bini}}, \bibinfo {author} {\bibfnamefont {T.}~\bibnamefont {Damour}},\ and\ \bibinfo {author} {\bibfnamefont {A.}~\bibnamefont {Geralico}},\ }\bibfield  {title} {\bibinfo {title} {{Radiative contributions to gravitational scattering}},\ }\href {https://doi.org/10.1103/PhysRevD.104.084031} {\bibfield  {journal} {\bibinfo  {journal} {Phys. Rev. D}\ }\textbf {\bibinfo {volume} {104}},\ \bibinfo {pages} {084031} (\bibinfo {year} {2021})},\ \Eprint {https://arxiv.org/abs/2107.08896} {arXiv:2107.08896 [gr-qc]} \BibitemShut {NoStop}%
\bibitem [{\citenamefont {Manohar}\ \emph {et~al.}(2022)\citenamefont {Manohar}, \citenamefont {Ridgway},\ and\ \citenamefont {Shen}}]{Manohar:2022dea}%
  \BibitemOpen
  \bibfield  {author} {\bibinfo {author} {\bibfnamefont {A.~V.}\ \bibnamefont {Manohar}}, \bibinfo {author} {\bibfnamefont {A.~K.}\ \bibnamefont {Ridgway}},\ and\ \bibinfo {author} {\bibfnamefont {C.-H.}\ \bibnamefont {Shen}},\ }\bibfield  {title} {\bibinfo {title} {{Radiated Angular Momentum and Dissipative Effects in Classical Scattering}},\ }\href {https://doi.org/10.1103/PhysRevLett.129.121601} {\bibfield  {journal} {\bibinfo  {journal} {Phys. Rev. Lett.}\ }\textbf {\bibinfo {volume} {129}},\ \bibinfo {pages} {121601} (\bibinfo {year} {2022})},\ \Eprint {https://arxiv.org/abs/2203.04283} {arXiv:2203.04283 [hep-th]} \BibitemShut {NoStop}%
\bibitem [{\citenamefont {Dlapa}\ \emph {et~al.}(2023)\citenamefont {Dlapa}, \citenamefont {K{\"a}lin}, \citenamefont {Liu}, \citenamefont {Neef},\ and\ \citenamefont {Porto}}]{Dlapa:2022lmu}%
  \BibitemOpen
  \bibfield  {author} {\bibinfo {author} {\bibfnamefont {C.}~\bibnamefont {Dlapa}}, \bibinfo {author} {\bibfnamefont {G.}~\bibnamefont {K{\"a}lin}}, \bibinfo {author} {\bibfnamefont {Z.}~\bibnamefont {Liu}}, \bibinfo {author} {\bibfnamefont {J.}~\bibnamefont {Neef}},\ and\ \bibinfo {author} {\bibfnamefont {R.~A.}\ \bibnamefont {Porto}},\ }\bibfield  {title} {\bibinfo {title} {{Radiation Reaction and Gravitational Waves at Fourth Post-Minkowskian Order}},\ }\href {https://doi.org/10.1103/PhysRevLett.130.101401} {\bibfield  {journal} {\bibinfo  {journal} {Phys. Rev. Lett.}\ }\textbf {\bibinfo {volume} {130}},\ \bibinfo {pages} {101401} (\bibinfo {year} {2023})},\ \Eprint {https://arxiv.org/abs/2210.05541} {arXiv:2210.05541 [hep-th]} \BibitemShut {NoStop}%
\bibitem [{\citenamefont {Jakobsen}\ \emph {et~al.}(2023{\natexlab{a}})\citenamefont {Jakobsen}, \citenamefont {Mogull}, \citenamefont {Plefka},\ and\ \citenamefont {Sauer}}]{Jakobsen:2023hig}%
  \BibitemOpen
  \bibfield  {author} {\bibinfo {author} {\bibfnamefont {G.~U.}\ \bibnamefont {Jakobsen}}, \bibinfo {author} {\bibfnamefont {G.}~\bibnamefont {Mogull}}, \bibinfo {author} {\bibfnamefont {J.}~\bibnamefont {Plefka}},\ and\ \bibinfo {author} {\bibfnamefont {B.}~\bibnamefont {Sauer}},\ }\bibfield  {title} {\bibinfo {title} {{Dissipative Scattering of Spinning Black Holes at Fourth Post-Minkowskian Order}},\ }\href {https://doi.org/10.1103/PhysRevLett.131.241402} {\bibfield  {journal} {\bibinfo  {journal} {Phys. Rev. Lett.}\ }\textbf {\bibinfo {volume} {131}},\ \bibinfo {pages} {241402} (\bibinfo {year} {2023}{\natexlab{a}})},\ \Eprint {https://arxiv.org/abs/2308.11514} {arXiv:2308.11514 [hep-th]} \BibitemShut {NoStop}%
\bibitem [{\citenamefont {Damgaard}\ \emph {et~al.}(2023)\citenamefont {Damgaard}, \citenamefont {Hansen}, \citenamefont {Plant{\'e}},\ and\ \citenamefont {Vanhove}}]{Damgaard:2023ttc}%
  \BibitemOpen
  \bibfield  {author} {\bibinfo {author} {\bibfnamefont {P.~H.}\ \bibnamefont {Damgaard}}, \bibinfo {author} {\bibfnamefont {E.~R.}\ \bibnamefont {Hansen}}, \bibinfo {author} {\bibfnamefont {L.}~\bibnamefont {Plant{\'e}}},\ and\ \bibinfo {author} {\bibfnamefont {P.}~\bibnamefont {Vanhove}},\ }\bibfield  {title} {\bibinfo {title} {{Classical observables from the exponential representation of the gravitational S-matrix}},\ }\href {https://doi.org/10.1007/JHEP09(2023)183} {\bibfield  {journal} {\bibinfo  {journal} {JHEP}\ }\textbf {\bibinfo {volume} {09}},\ \bibinfo {pages} {183}},\ \Eprint {https://arxiv.org/abs/2307.04746} {arXiv:2307.04746 [hep-th]} \BibitemShut {NoStop}%
\bibitem [{\citenamefont {Herrmann}\ \emph {et~al.}(2021{\natexlab{b}})\citenamefont {Herrmann}, \citenamefont {Parra-Martinez}, \citenamefont {Ruf},\ and\ \citenamefont {Zeng}}]{Herrmann:2021lqe}%
  \BibitemOpen
  \bibfield  {author} {\bibinfo {author} {\bibfnamefont {E.}~\bibnamefont {Herrmann}}, \bibinfo {author} {\bibfnamefont {J.}~\bibnamefont {Parra-Martinez}}, \bibinfo {author} {\bibfnamefont {M.~S.}\ \bibnamefont {Ruf}},\ and\ \bibinfo {author} {\bibfnamefont {M.}~\bibnamefont {Zeng}},\ }\bibfield  {title} {\bibinfo {title} {{Gravitational Bremsstrahlung from Reverse Unitarity}},\ }\href {https://doi.org/10.1103/PhysRevLett.126.201602} {\bibfield  {journal} {\bibinfo  {journal} {Phys. Rev. Lett.}\ }\textbf {\bibinfo {volume} {126}},\ \bibinfo {pages} {201602} (\bibinfo {year} {2021}{\natexlab{b}})},\ \Eprint {https://arxiv.org/abs/2101.07255} {arXiv:2101.07255 [hep-th]} \BibitemShut {NoStop}%
\bibitem [{\citenamefont {Riva}\ \emph {et~al.}(2023)\citenamefont {Riva}, \citenamefont {Vernizzi},\ and\ \citenamefont {Wong}}]{Riva:2023xxm}%
  \BibitemOpen
  \bibfield  {author} {\bibinfo {author} {\bibfnamefont {M.~M.}\ \bibnamefont {Riva}}, \bibinfo {author} {\bibfnamefont {F.}~\bibnamefont {Vernizzi}},\ and\ \bibinfo {author} {\bibfnamefont {L.~K.}\ \bibnamefont {Wong}},\ }\bibfield  {title} {\bibinfo {title} {{Angular momentum balance in gravitational two-body scattering: Flux, memory, and supertranslation invariance}},\ }\href {https://doi.org/10.1103/PhysRevD.108.104052} {\bibfield  {journal} {\bibinfo  {journal} {Phys. Rev. D}\ }\textbf {\bibinfo {volume} {108}},\ \bibinfo {pages} {104052} (\bibinfo {year} {2023})},\ \Eprint {https://arxiv.org/abs/2302.09065} {arXiv:2302.09065 [gr-qc]} \BibitemShut {NoStop}%
\bibitem [{\citenamefont {Jakobsen}\ and\ \citenamefont {Stoldt}(2026)}]{Jakobsen:2026ayw}%
  \BibitemOpen
  \bibfield  {author} {\bibinfo {author} {\bibfnamefont {G.~U.}\ \bibnamefont {Jakobsen}}\ and\ \bibinfo {author} {\bibfnamefont {K.}~\bibnamefont {Stoldt}},\ }\bibfield  {title} {\bibinfo {title} {{Runway to dissipation of angular momentum via worldline quantum field theory}},\ }\href {https://doi.org/10.1103/lx72-nz9r} {\bibfield  {journal} {\bibinfo  {journal} {Phys. Rev. D}\ }\textbf {\bibinfo {volume} {114}},\ \bibinfo {pages} {064020} (\bibinfo {year} {2026})},\ \Eprint {https://arxiv.org/abs/2605.12214} {arXiv:2605.12214 [hep-th]} \BibitemShut {NoStop}%
\bibitem [{\citenamefont {Heissenberg}(2025)}]{Heissenberg:2025ocy}%
  \BibitemOpen
  \bibfield  {author} {\bibinfo {author} {\bibfnamefont {C.}~\bibnamefont {Heissenberg}},\ }\bibfield  {title} {\bibinfo {title} {{Radiation-reaction and angular momentum loss at O(G4)}},\ }\href {https://doi.org/10.1103/xz14-87q7} {\bibfield  {journal} {\bibinfo  {journal} {Phys. Rev. D}\ }\textbf {\bibinfo {volume} {111}},\ \bibinfo {pages} {126012} (\bibinfo {year} {2025})},\ \Eprint {https://arxiv.org/abs/2501.02904} {arXiv:2501.02904 [hep-th]} \BibitemShut {NoStop}%
\bibitem [{\citenamefont {Georgoudis}\ \emph {et~al.}(2026)\citenamefont {Georgoudis}, \citenamefont {Heissenberg}, \citenamefont {Mogull},\ and\ \citenamefont {Russo}}]{Georgoudis:2026zzn}%
  \BibitemOpen
  \bibfield  {author} {\bibinfo {author} {\bibfnamefont {A.}~\bibnamefont {Georgoudis}}, \bibinfo {author} {\bibfnamefont {C.}~\bibnamefont {Heissenberg}}, \bibinfo {author} {\bibfnamefont {G.}~\bibnamefont {Mogull}},\ and\ \bibinfo {author} {\bibfnamefont {R.}~\bibnamefont {Russo}},\ }\href@noop {} {\bibinfo {title} {{Angular Momentum Loss at Fourth Post-Minkowskian Order}}} (\bibinfo {year} {2026}),\ \Eprint {https://arxiv.org/abs/2610.24072} {arXiv:2610.24072 [hep-th]} \BibitemShut {NoStop}%
\bibitem [{\citenamefont {Bohnenblust}\ \emph {et~al.}(2026{\natexlab{a}})\citenamefont {Bohnenblust}, \citenamefont {Jordan~Eriksen}, \citenamefont {Jakobsen}, \citenamefont {Stoldt},\ and\ \citenamefont {Yanovets}}]{Bohnenblust:2026ytj}%
  \BibitemOpen
  \bibfield  {author} {\bibinfo {author} {\bibfnamefont {L.}~\bibnamefont {Bohnenblust}}, \bibinfo {author} {\bibfnamefont {C.}~\bibnamefont {Jordan~Eriksen}}, \bibinfo {author} {\bibfnamefont {G.~U.}\ \bibnamefont {Jakobsen}}, \bibinfo {author} {\bibfnamefont {K.}~\bibnamefont {Stoldt}},\ and\ \bibinfo {author} {\bibfnamefont {A.}~\bibnamefont {Yanovets}},\ }\href@noop {} {\bibinfo {title} {{Mechanical angular momentum loss at $\mathcal{O}(G^4)$ from static and radiative gravitons}}} (\bibinfo {year} {2026}{\natexlab{a}}),\ \Eprint {https://arxiv.org/abs/2610.26822} {arXiv:2610.26822 [hep-th]} \BibitemShut {NoStop}%
\bibitem [{\citenamefont {Driesse}\ \emph {et~al.}(2024)\citenamefont {Driesse}, \citenamefont {Jakobsen}, \citenamefont {Mogull}, \citenamefont {Plefka}, \citenamefont {Sauer},\ and\ \citenamefont {Usovitsch}}]{Driesse:2024xad}%
  \BibitemOpen
  \bibfield  {author} {\bibinfo {author} {\bibfnamefont {M.}~\bibnamefont {Driesse}}, \bibinfo {author} {\bibfnamefont {G.~U.}\ \bibnamefont {Jakobsen}}, \bibinfo {author} {\bibfnamefont {G.}~\bibnamefont {Mogull}}, \bibinfo {author} {\bibfnamefont {J.}~\bibnamefont {Plefka}}, \bibinfo {author} {\bibfnamefont {B.}~\bibnamefont {Sauer}},\ and\ \bibinfo {author} {\bibfnamefont {J.}~\bibnamefont {Usovitsch}},\ }\bibfield  {title} {\bibinfo {title} {{Conservative Black Hole Scattering at Fifth Post-Minkowskian and First Self-Force Order}},\ }\href {https://doi.org/10.1103/PhysRevLett.132.241402} {\bibfield  {journal} {\bibinfo  {journal} {Phys. Rev. Lett.}\ }\textbf {\bibinfo {volume} {132}},\ \bibinfo {pages} {241402} (\bibinfo {year} {2024})},\ \Eprint {https://arxiv.org/abs/2403.07781} {arXiv:2403.07781 [hep-th]} \BibitemShut {NoStop}%
\bibitem [{\citenamefont {Bern}\ \emph {et~al.}(2024)\citenamefont {Bern}, \citenamefont {Herrmann}, \citenamefont {Roiban}, \citenamefont {Ruf}, \citenamefont {Smirnov}, \citenamefont {Smirnov},\ and\ \citenamefont {Zeng}}]{Bern:2024adl}%
  \BibitemOpen
  \bibfield  {author} {\bibinfo {author} {\bibfnamefont {Z.}~\bibnamefont {Bern}}, \bibinfo {author} {\bibfnamefont {E.}~\bibnamefont {Herrmann}}, \bibinfo {author} {\bibfnamefont {R.}~\bibnamefont {Roiban}}, \bibinfo {author} {\bibfnamefont {M.~S.}\ \bibnamefont {Ruf}}, \bibinfo {author} {\bibfnamefont {A.~V.}\ \bibnamefont {Smirnov}}, \bibinfo {author} {\bibfnamefont {V.~A.}\ \bibnamefont {Smirnov}},\ and\ \bibinfo {author} {\bibfnamefont {M.}~\bibnamefont {Zeng}},\ }\bibfield  {title} {\bibinfo {title} {{Amplitudes, supersymmetric black hole scattering at $ \mathcal{O}\left({G}^5\right) $, and loop integration}},\ }\href {https://doi.org/10.1007/JHEP10(2024)023} {\bibfield  {journal} {\bibinfo  {journal} {JHEP}\ }\textbf {\bibinfo {volume} {10}},\ \bibinfo {pages} {023}},\ \Eprint {https://arxiv.org/abs/2406.01554} {arXiv:2406.01554 [hep-th]} \BibitemShut {NoStop}%
\bibitem [{\citenamefont {Driesse}\ \emph {et~al.}(2025)\citenamefont {Driesse}, \citenamefont {Jakobsen}, \citenamefont {Klemm}, \citenamefont {Mogull}, \citenamefont {Nega}, \citenamefont {Plefka}, \citenamefont {Sauer},\ and\ \citenamefont {Usovitsch}}]{Driesse:2024feo}%
  \BibitemOpen
  \bibfield  {author} {\bibinfo {author} {\bibfnamefont {M.}~\bibnamefont {Driesse}}, \bibinfo {author} {\bibfnamefont {G.~U.}\ \bibnamefont {Jakobsen}}, \bibinfo {author} {\bibfnamefont {A.}~\bibnamefont {Klemm}}, \bibinfo {author} {\bibfnamefont {G.}~\bibnamefont {Mogull}}, \bibinfo {author} {\bibfnamefont {C.}~\bibnamefont {Nega}}, \bibinfo {author} {\bibfnamefont {J.}~\bibnamefont {Plefka}}, \bibinfo {author} {\bibfnamefont {B.}~\bibnamefont {Sauer}},\ and\ \bibinfo {author} {\bibfnamefont {J.}~\bibnamefont {Usovitsch}},\ }\bibfield  {title} {\bibinfo {title} {{Emergence of Calabi{\textendash}Yau manifolds in high-precision black-hole scattering}},\ }\href {https://doi.org/10.1038/s41586-025-08984-2} {\bibfield  {journal} {\bibinfo  {journal} {Nature}\ }\textbf {\bibinfo {volume} {641}},\ \bibinfo {pages} {603} (\bibinfo {year} {2025})},\ \Eprint {https://arxiv.org/abs/2411.11846} {arXiv:2411.11846 [hep-th]} \BibitemShut {NoStop}%
\bibitem [{\citenamefont {Dlapa}\ \emph {et~al.}(2025)\citenamefont {Dlapa}, \citenamefont {K{\"a}lin}, \citenamefont {Liu},\ and\ \citenamefont {Porto}}]{Dlapa:2025biy}%
  \BibitemOpen
  \bibfield  {author} {\bibinfo {author} {\bibfnamefont {C.}~\bibnamefont {Dlapa}}, \bibinfo {author} {\bibfnamefont {G.}~\bibnamefont {K{\"a}lin}}, \bibinfo {author} {\bibfnamefont {Z.}~\bibnamefont {Liu}},\ and\ \bibinfo {author} {\bibfnamefont {R.~A.}\ \bibnamefont {Porto}},\ }\bibfield  {title} {\bibinfo {title} {{Local-in-Time Conservative Binary Dynamics at Fifth Post-Minkowskian and First Self-Force Orders}},\ }\href {https://doi.org/10.1103/215k-27sj} {\bibfield  {journal} {\bibinfo  {journal} {Phys. Rev. Lett.}\ }\textbf {\bibinfo {volume} {135}},\ \bibinfo {pages} {251401} (\bibinfo {year} {2025})},\ \Eprint {https://arxiv.org/abs/2506.20665} {arXiv:2506.20665 [hep-th]} \BibitemShut {NoStop}%
\bibitem [{\citenamefont {Driesse}\ \emph {et~al.}(2026)\citenamefont {Driesse}, \citenamefont {Jakobsen}, \citenamefont {Mogull}, \citenamefont {Nega}, \citenamefont {Plefka}, \citenamefont {Sauer},\ and\ \citenamefont {Usovitsch}}]{Driesse:2026qiz}%
  \BibitemOpen
  \bibfield  {author} {\bibinfo {author} {\bibfnamefont {M.}~\bibnamefont {Driesse}}, \bibinfo {author} {\bibfnamefont {G.~U.}\ \bibnamefont {Jakobsen}}, \bibinfo {author} {\bibfnamefont {G.}~\bibnamefont {Mogull}}, \bibinfo {author} {\bibfnamefont {C.}~\bibnamefont {Nega}}, \bibinfo {author} {\bibfnamefont {J.}~\bibnamefont {Plefka}}, \bibinfo {author} {\bibfnamefont {B.}~\bibnamefont {Sauer}},\ and\ \bibinfo {author} {\bibfnamefont {J.}~\bibnamefont {Usovitsch}},\ }\bibfield  {title} {\bibinfo {title} {{Conservative Black Hole Scattering at Fifth Post-Minkowskian and Second Self-Force Order}},\ }\href {https://doi.org/10.1103/4dvk-nglx} {\bibfield  {journal} {\bibinfo  {journal} {Phys. Rev. Lett.}\ }\textbf {\bibinfo {volume} {137}},\ \bibinfo {pages} {081402} (\bibinfo {year} {2026})},\ \Eprint {https://arxiv.org/abs/2601.16256} {arXiv:2601.16256 [hep-th]} \BibitemShut {NoStop}%
\bibitem [{\citenamefont {Bern}\ \emph {et~al.}(2026)\citenamefont {Bern}, \citenamefont {Herrmann}, \citenamefont {Roiban}, \citenamefont {Ruf}, \citenamefont {Smirnov}, \citenamefont {Smith},\ and\ \citenamefont {Zeng}}]{Bern:2025wyd}%
  \BibitemOpen
  \bibfield  {author} {\bibinfo {author} {\bibfnamefont {Z.}~\bibnamefont {Bern}}, \bibinfo {author} {\bibfnamefont {E.}~\bibnamefont {Herrmann}}, \bibinfo {author} {\bibfnamefont {R.}~\bibnamefont {Roiban}}, \bibinfo {author} {\bibfnamefont {M.~S.}\ \bibnamefont {Ruf}}, \bibinfo {author} {\bibfnamefont {A.~V.}\ \bibnamefont {Smirnov}}, \bibinfo {author} {\bibfnamefont {S.}~\bibnamefont {Smith}},\ and\ \bibinfo {author} {\bibfnamefont {M.}~\bibnamefont {Zeng}},\ }\bibfield  {title} {\bibinfo {title} {{Scattering Amplitudes and Conservative Binary Dynamics at O(G5) without Self-Force Truncation}},\ }\href {https://doi.org/10.1103/zjcp-ssrw} {\bibfield  {journal} {\bibinfo  {journal} {Phys. Rev. Lett.}\ }\textbf {\bibinfo {volume} {137}},\ \bibinfo {pages} {021402} (\bibinfo {year} {2026})},\ \Eprint {https://arxiv.org/abs/2512.23654} {arXiv:2512.23654 [hep-th]} \BibitemShut {NoStop}%
\bibitem [{\citenamefont {Dlapa}\ \emph {et~al.}(2026)\citenamefont {Dlapa}, \citenamefont {K{\"a}lin}, \citenamefont {Liu},\ and\ \citenamefont {Porto}}]{Dlapa:2026oyq}%
  \BibitemOpen
  \bibfield  {author} {\bibinfo {author} {\bibfnamefont {C.}~\bibnamefont {Dlapa}}, \bibinfo {author} {\bibfnamefont {G.}~\bibnamefont {K{\"a}lin}}, \bibinfo {author} {\bibfnamefont {Z.}~\bibnamefont {Liu}},\ and\ \bibinfo {author} {\bibfnamefont {R.~A.}\ \bibnamefont {Porto}},\ }\bibfield  {title} {\bibinfo {title} {{Nonlocal-in-time tail effects in gravitational scattering to fifth post-Minkowskian and tenth self-force orders}},\ }\href {https://doi.org/10.1103/wkp4-vy6g} {\bibfield  {journal} {\bibinfo  {journal} {Phys. Rev. D}\ }\textbf {\bibinfo {volume} {114}},\ \bibinfo {pages} {024029} (\bibinfo {year} {2026})},\ \Eprint {https://arxiv.org/abs/2604.25916} {arXiv:2604.25916 [hep-th]} \BibitemShut {NoStop}%
\bibitem [{\citenamefont {Burke}(1971)}]{Burke:1970wx}%
  \BibitemOpen
  \bibfield  {author} {\bibinfo {author} {\bibfnamefont {W.~L.}\ \bibnamefont {Burke}},\ }\bibfield  {title} {\bibinfo {title} {{Gravitational radiation damping of slowly moving systems calculated using matched asymptotic expansions}},\ }\href {https://doi.org/10.1063/1.1665603} {\bibfield  {journal} {\bibinfo  {journal} {J. Math. Phys.}\ }\textbf {\bibinfo {volume} {12}},\ \bibinfo {pages} {401} (\bibinfo {year} {1971})}\BibitemShut {NoStop}%
\bibitem [{\citenamefont {Iyer}\ and\ \citenamefont {Will}(1993)}]{Iyer:1993xi}%
  \BibitemOpen
  \bibfield  {author} {\bibinfo {author} {\bibfnamefont {B.~R.}\ \bibnamefont {Iyer}}\ and\ \bibinfo {author} {\bibfnamefont {C.~M.}\ \bibnamefont {Will}},\ }\bibfield  {title} {\bibinfo {title} {{Post-Newtonian gravitational radiation reaction for two-body systems}},\ }\href {https://doi.org/10.1103/PhysRevLett.70.113} {\bibfield  {journal} {\bibinfo  {journal} {Phys. Rev. Lett.}\ }\textbf {\bibinfo {volume} {70}},\ \bibinfo {pages} {113} (\bibinfo {year} {1993})}\BibitemShut {NoStop}%
\bibitem [{\citenamefont {Iyer}\ and\ \citenamefont {Will}(1995)}]{Iyer:1995rn}%
  \BibitemOpen
  \bibfield  {author} {\bibinfo {author} {\bibfnamefont {B.~R.}\ \bibnamefont {Iyer}}\ and\ \bibinfo {author} {\bibfnamefont {C.~M.}\ \bibnamefont {Will}},\ }\bibfield  {title} {\bibinfo {title} {{Post-Newtonian gravitational radiation reaction for two-body systems: Nonspinning bodies}},\ }\href {https://doi.org/10.1103/PhysRevD.52.6882} {\bibfield  {journal} {\bibinfo  {journal} {Phys. Rev. D}\ }\textbf {\bibinfo {volume} {52}},\ \bibinfo {pages} {6882} (\bibinfo {year} {1995})}\BibitemShut {NoStop}%
\bibitem [{\citenamefont {Galley}\ and\ \citenamefont {Tiglio}(2009)}]{Galley:2009px}%
  \BibitemOpen
  \bibfield  {author} {\bibinfo {author} {\bibfnamefont {C.~R.}\ \bibnamefont {Galley}}\ and\ \bibinfo {author} {\bibfnamefont {M.}~\bibnamefont {Tiglio}},\ }\bibfield  {title} {\bibinfo {title} {{Radiation reaction and gravitational waves in the effective field theory approach}},\ }\href {https://doi.org/10.1103/PhysRevD.79.124027} {\bibfield  {journal} {\bibinfo  {journal} {Phys. Rev. D}\ }\textbf {\bibinfo {volume} {79}},\ \bibinfo {pages} {124027} (\bibinfo {year} {2009})},\ \Eprint {https://arxiv.org/abs/0903.1122} {arXiv:0903.1122 [gr-qc]} \BibitemShut {NoStop}%
\bibitem [{\citenamefont {Galley}\ and\ \citenamefont {Leibovich}(2012)}]{Galley:2012qs}%
  \BibitemOpen
  \bibfield  {author} {\bibinfo {author} {\bibfnamefont {C.~R.}\ \bibnamefont {Galley}}\ and\ \bibinfo {author} {\bibfnamefont {A.~K.}\ \bibnamefont {Leibovich}},\ }\bibfield  {title} {\bibinfo {title} {{Radiation reaction at 3.5 post-Newtonian order in effective field theory}},\ }\href {https://doi.org/10.1103/PhysRevD.86.044029} {\bibfield  {journal} {\bibinfo  {journal} {Phys. Rev. D}\ }\textbf {\bibinfo {volume} {86}},\ \bibinfo {pages} {044029} (\bibinfo {year} {2012})},\ \Eprint {https://arxiv.org/abs/1205.3842} {arXiv:1205.3842 [gr-qc]} \BibitemShut {NoStop}%
\bibitem [{\citenamefont {Leibovich}\ \emph {et~al.}(2020)\citenamefont {Leibovich}, \citenamefont {Maia}, \citenamefont {Rothstein},\ and\ \citenamefont {Yang}}]{Leibovich:2019cxo}%
  \BibitemOpen
  \bibfield  {author} {\bibinfo {author} {\bibfnamefont {A.~K.}\ \bibnamefont {Leibovich}}, \bibinfo {author} {\bibfnamefont {N.~T.}\ \bibnamefont {Maia}}, \bibinfo {author} {\bibfnamefont {I.~Z.}\ \bibnamefont {Rothstein}},\ and\ \bibinfo {author} {\bibfnamefont {Z.}~\bibnamefont {Yang}},\ }\bibfield  {title} {\bibinfo {title} {{Second post-Newtonian order radiative dynamics of inspiralling compact binaries in the Effective Field Theory approach}},\ }\href {https://doi.org/10.1103/PhysRevD.101.084058} {\bibfield  {journal} {\bibinfo  {journal} {Phys. Rev. D}\ }\textbf {\bibinfo {volume} {101}},\ \bibinfo {pages} {084058} (\bibinfo {year} {2020})},\ \Eprint {https://arxiv.org/abs/1912.12546} {arXiv:1912.12546 [gr-qc]} \BibitemShut {NoStop}%
\bibitem [{\citenamefont {Leibovich}\ \emph {et~al.}(2023)\citenamefont {Leibovich}, \citenamefont {Pardo},\ and\ \citenamefont {Yang}}]{Leibovich:2023xpg}%
  \BibitemOpen
  \bibfield  {author} {\bibinfo {author} {\bibfnamefont {A.~K.}\ \bibnamefont {Leibovich}}, \bibinfo {author} {\bibfnamefont {B.~A.}\ \bibnamefont {Pardo}},\ and\ \bibinfo {author} {\bibfnamefont {Z.}~\bibnamefont {Yang}},\ }\bibfield  {title} {\bibinfo {title} {{Radiation reaction for nonspinning bodies at 4.5PN in the effective field theory approach}},\ }\href {https://doi.org/10.1103/PhysRevD.108.024017} {\bibfield  {journal} {\bibinfo  {journal} {Phys. Rev. D}\ }\textbf {\bibinfo {volume} {108}},\ \bibinfo {pages} {024017} (\bibinfo {year} {2023})},\ \Eprint {https://arxiv.org/abs/2302.11016} {arXiv:2302.11016 [gr-qc]} \BibitemShut {NoStop}%
\bibitem [{\citenamefont {Blanchet}\ and\ \citenamefont {Damour}(1988)}]{Blanchet:1987wq}%
  \BibitemOpen
  \bibfield  {author} {\bibinfo {author} {\bibfnamefont {L.}~\bibnamefont {Blanchet}}\ and\ \bibinfo {author} {\bibfnamefont {T.}~\bibnamefont {Damour}},\ }\bibfield  {title} {\bibinfo {title} {{Tail Transported Temporal Correlations in the Dynamics of a Gravitating System}},\ }\href {https://doi.org/10.1103/PhysRevD.37.1410} {\bibfield  {journal} {\bibinfo  {journal} {Phys. Rev. D}\ }\textbf {\bibinfo {volume} {37}},\ \bibinfo {pages} {1410} (\bibinfo {year} {1988})}\BibitemShut {NoStop}%
\bibitem [{\citenamefont {Blanchet}\ and\ \citenamefont {Damour}(1992)}]{Blanchet:1992br}%
  \BibitemOpen
  \bibfield  {author} {\bibinfo {author} {\bibfnamefont {L.}~\bibnamefont {Blanchet}}\ and\ \bibinfo {author} {\bibfnamefont {T.}~\bibnamefont {Damour}},\ }\bibfield  {title} {\bibinfo {title} {{Hereditary effects in gravitational radiation}},\ }\href {https://doi.org/10.1103/PhysRevD.46.4304} {\bibfield  {journal} {\bibinfo  {journal} {Phys. Rev. D}\ }\textbf {\bibinfo {volume} {46}},\ \bibinfo {pages} {4304} (\bibinfo {year} {1992})}\BibitemShut {NoStop}%
\bibitem [{\citenamefont {Blanchet}(1998)}]{Blanchet:1997jj}%
  \BibitemOpen
  \bibfield  {author} {\bibinfo {author} {\bibfnamefont {L.}~\bibnamefont {Blanchet}},\ }\bibfield  {title} {\bibinfo {title} {{Gravitational wave tails of tails}},\ }\href {https://doi.org/10.1088/0264-9381/15/1/009} {\bibfield  {journal} {\bibinfo  {journal} {Class. Quant. Grav.}\ }\textbf {\bibinfo {volume} {15}},\ \bibinfo {pages} {113} (\bibinfo {year} {1998})},\ \bibinfo {note} {[Erratum: Class.Quant.Grav. 22, 3381 (2005)]},\ \Eprint {https://arxiv.org/abs/gr-qc/9710038} {arXiv:gr-qc/9710038} \BibitemShut {NoStop}%
\bibitem [{\citenamefont {Almeida}\ \emph {et~al.}(2023{\natexlab{a}})\citenamefont {Almeida}, \citenamefont {M{\"u}ller}, \citenamefont {Foffa},\ and\ \citenamefont {Sturani}}]{Almeida:2023yia}%
  \BibitemOpen
  \bibfield  {author} {\bibinfo {author} {\bibfnamefont {G.~L.}\ \bibnamefont {Almeida}}, \bibinfo {author} {\bibfnamefont {A.}~\bibnamefont {M{\"u}ller}}, \bibinfo {author} {\bibfnamefont {S.}~\bibnamefont {Foffa}},\ and\ \bibinfo {author} {\bibfnamefont {R.}~\bibnamefont {Sturani}},\ }\bibfield  {title} {\bibinfo {title} {{Conservative binary dynamics from gravitational tail emission processes}},\ }\href {https://doi.org/10.1103/PhysRevD.108.124010} {\bibfield  {journal} {\bibinfo  {journal} {Phys. Rev. D}\ }\textbf {\bibinfo {volume} {108}},\ \bibinfo {pages} {124010} (\bibinfo {year} {2023}{\natexlab{a}})},\ \bibinfo {note} {[Erratum: Phys.Rev.D 113, 089903 (2026)]},\ \Eprint {https://arxiv.org/abs/2307.05327} {arXiv:2307.05327 [gr-qc]} \BibitemShut {NoStop}%
\bibitem [{\citenamefont {Henry}\ and\ \citenamefont {Larrouturou}(2023)}]{Henry:2023sdy}%
  \BibitemOpen
  \bibfield  {author} {\bibinfo {author} {\bibfnamefont {Q.}~\bibnamefont {Henry}}\ and\ \bibinfo {author} {\bibfnamefont {F.}~\bibnamefont {Larrouturou}},\ }\bibfield  {title} {\bibinfo {title} {{Conservative tail and failed-tail effects at the fifth post-Newtonian order}},\ }\href {https://doi.org/10.1103/PhysRevD.108.084048} {\bibfield  {journal} {\bibinfo  {journal} {Phys. Rev. D}\ }\textbf {\bibinfo {volume} {108}},\ \bibinfo {pages} {084048} (\bibinfo {year} {2023})},\ \Eprint {https://arxiv.org/abs/2307.05860} {arXiv:2307.05860 [gr-qc]} \BibitemShut {NoStop}%
\bibitem [{\citenamefont {Almeida}\ \emph {et~al.}(2026)\citenamefont {Almeida}, \citenamefont {M{\"u}ller}, \citenamefont {Foffa},\ and\ \citenamefont {Sturani}}]{Almeida:2026clf}%
  \BibitemOpen
  \bibfield  {author} {\bibinfo {author} {\bibfnamefont {G.~L.}\ \bibnamefont {Almeida}}, \bibinfo {author} {\bibfnamefont {A.}~\bibnamefont {M{\"u}ller}}, \bibinfo {author} {\bibfnamefont {S.}~\bibnamefont {Foffa}},\ and\ \bibinfo {author} {\bibfnamefont {R.}~\bibnamefont {Sturani}},\ }\bibfield  {title} {\bibinfo {title} {{Angular momentum tail contributions to compact binary dynamics}},\ }\href {https://doi.org/10.1103/bkxs-xs4j} {\bibfield  {journal} {\bibinfo  {journal} {Phys. Rev. D}\ }\textbf {\bibinfo {volume} {114}},\ \bibinfo {pages} {064057} (\bibinfo {year} {2026})},\ \Eprint {https://arxiv.org/abs/2603.20096} {arXiv:2603.20096 [gr-qc]} \BibitemShut {NoStop}%
\bibitem [{\citenamefont {Christodoulou}(1991)}]{Christodoulou:1991cr}%
  \BibitemOpen
  \bibfield  {author} {\bibinfo {author} {\bibfnamefont {D.}~\bibnamefont {Christodoulou}},\ }\bibfield  {title} {\bibinfo {title} {{Nonlinear nature of gravitation and gravitational wave experiments}},\ }\href {https://doi.org/10.1103/PhysRevLett.67.1486} {\bibfield  {journal} {\bibinfo  {journal} {Phys. Rev. Lett.}\ }\textbf {\bibinfo {volume} {67}},\ \bibinfo {pages} {1486} (\bibinfo {year} {1991})}\BibitemShut {NoStop}%
\bibitem [{\citenamefont {Wiseman}\ and\ \citenamefont {Will}(1991)}]{Wiseman:1991ss}%
  \BibitemOpen
  \bibfield  {author} {\bibinfo {author} {\bibfnamefont {A.~G.}\ \bibnamefont {Wiseman}}\ and\ \bibinfo {author} {\bibfnamefont {C.~M.}\ \bibnamefont {Will}},\ }\bibfield  {title} {\bibinfo {title} {{Christodoulou's nonlinear gravitational wave memory: Evaluation in the quadrupole approximation}},\ }\href {https://doi.org/10.1103/PhysRevD.44.R2945} {\bibfield  {journal} {\bibinfo  {journal} {Phys. Rev. D}\ }\textbf {\bibinfo {volume} {44}},\ \bibinfo {pages} {R2945} (\bibinfo {year} {1991})}\BibitemShut {NoStop}%
\bibitem [{\citenamefont {Thorne}(1992)}]{Thorne:1992sdb}%
  \BibitemOpen
  \bibfield  {author} {\bibinfo {author} {\bibfnamefont {K.~S.}\ \bibnamefont {Thorne}},\ }\bibfield  {title} {\bibinfo {title} {{Gravitational-wave bursts with memory: The Christodoulou effect}},\ }\href {https://doi.org/10.1103/PhysRevD.45.520} {\bibfield  {journal} {\bibinfo  {journal} {Phys. Rev. D}\ }\textbf {\bibinfo {volume} {45}},\ \bibinfo {pages} {520} (\bibinfo {year} {1992})}\BibitemShut {NoStop}%
\bibitem [{\citenamefont {Favata}(2009)}]{Favata:2008yd}%
  \BibitemOpen
  \bibfield  {author} {\bibinfo {author} {\bibfnamefont {M.}~\bibnamefont {Favata}},\ }\bibfield  {title} {\bibinfo {title} {{Post-Newtonian corrections to the gravitational-wave memory for quasi-circular, inspiralling compact binaries}},\ }\href {https://doi.org/10.1103/PhysRevD.80.024002} {\bibfield  {journal} {\bibinfo  {journal} {Phys. Rev. D}\ }\textbf {\bibinfo {volume} {80}},\ \bibinfo {pages} {024002} (\bibinfo {year} {2009})},\ \Eprint {https://arxiv.org/abs/0812.0069} {arXiv:0812.0069 [gr-qc]} \BibitemShut {NoStop}%
\bibitem [{\citenamefont {Trestini}\ and\ \citenamefont {Blanchet}(2023)}]{Trestini:2023wwg}%
  \BibitemOpen
  \bibfield  {author} {\bibinfo {author} {\bibfnamefont {D.}~\bibnamefont {Trestini}}\ and\ \bibinfo {author} {\bibfnamefont {L.}~\bibnamefont {Blanchet}},\ }\bibfield  {title} {\bibinfo {title} {{Gravitational-wave tails of memory}},\ }\href {https://doi.org/10.1103/PhysRevD.107.104048} {\bibfield  {journal} {\bibinfo  {journal} {Phys. Rev. D}\ }\textbf {\bibinfo {volume} {107}},\ \bibinfo {pages} {104048} (\bibinfo {year} {2023})},\ \Eprint {https://arxiv.org/abs/2301.09395} {arXiv:2301.09395 [gr-qc]} \BibitemShut {NoStop}%
\bibitem [{\citenamefont {Schwinger}(1961)}]{Schwinger:1960qe}%
  \BibitemOpen
  \bibfield  {author} {\bibinfo {author} {\bibfnamefont {J.~S.}\ \bibnamefont {Schwinger}},\ }\bibfield  {title} {\bibinfo {title} {{Brownian motion of a quantum oscillator}},\ }\href {https://doi.org/10.1063/1.1703727} {\bibfield  {journal} {\bibinfo  {journal} {J. Math. Phys.}\ }\textbf {\bibinfo {volume} {2}},\ \bibinfo {pages} {407} (\bibinfo {year} {1961})}\BibitemShut {NoStop}%
\bibitem [{\citenamefont {Keldysh}(1965)}]{Keldysh:1964ud}%
  \BibitemOpen
  \bibfield  {author} {\bibinfo {author} {\bibfnamefont {L.~V.}\ \bibnamefont {Keldysh}},\ }\bibfield  {title} {\bibinfo {title} {{Diagram Technique for Nonequilibrium Processes}},\ }\href {https://doi.org/10.1142/9789811279461\_0007} {\bibfield  {journal} {\bibinfo  {journal} {Sov. Phys. JETP}\ }\textbf {\bibinfo {volume} {20}},\ \bibinfo {pages} {1018} (\bibinfo {year} {1965})}\BibitemShut {NoStop}%
\bibitem [{\citenamefont {Calzetta}\ and\ \citenamefont {Hu}(1987)}]{Calzetta:1986ey}%
  \BibitemOpen
  \bibfield  {author} {\bibinfo {author} {\bibfnamefont {E.}~\bibnamefont {Calzetta}}\ and\ \bibinfo {author} {\bibfnamefont {B.~L.}\ \bibnamefont {Hu}},\ }\bibfield  {title} {\bibinfo {title} {{Closed Time Path Functional Formalism in Curved Space-Time: Application to Cosmological Back Reaction Problems}},\ }\href {https://doi.org/10.1103/PhysRevD.35.495} {\bibfield  {journal} {\bibinfo  {journal} {Phys. Rev. D}\ }\textbf {\bibinfo {volume} {35}},\ \bibinfo {pages} {495} (\bibinfo {year} {1987})}\BibitemShut {NoStop}%
\bibitem [{\citenamefont {Galley}(2013)}]{Galley:2012hx}%
  \BibitemOpen
  \bibfield  {author} {\bibinfo {author} {\bibfnamefont {C.~R.}\ \bibnamefont {Galley}},\ }\bibfield  {title} {\bibinfo {title} {{Classical Mechanics of Nonconservative Systems}},\ }\href {https://doi.org/10.1103/PhysRevLett.110.174301} {\bibfield  {journal} {\bibinfo  {journal} {Phys. Rev. Lett.}\ }\textbf {\bibinfo {volume} {110}},\ \bibinfo {pages} {174301} (\bibinfo {year} {2013})},\ \Eprint {https://arxiv.org/abs/1210.2745} {arXiv:1210.2745 [gr-qc]} \BibitemShut {NoStop}%
\bibitem [{\citenamefont {Galley}\ \emph {et~al.}(2014)\citenamefont {Galley}, \citenamefont {Tsang},\ and\ \citenamefont {Stein}}]{Galley:2014wla}%
  \BibitemOpen
  \bibfield  {author} {\bibinfo {author} {\bibfnamefont {C.~R.}\ \bibnamefont {Galley}}, \bibinfo {author} {\bibfnamefont {D.}~\bibnamefont {Tsang}},\ and\ \bibinfo {author} {\bibfnamefont {L.~C.}\ \bibnamefont {Stein}},\ }\href@noop {} {\bibinfo {title} {{The principle of stationary nonconservative action for classical mechanics and field theories}}} (\bibinfo {year} {2014}),\ \Eprint {https://arxiv.org/abs/1412.3082} {arXiv:1412.3082 [math-ph]} \BibitemShut {NoStop}%
\bibitem [{\citenamefont {Foffa}\ and\ \citenamefont {Sturani}(2013{\natexlab{b}})}]{Foffa:2011np}%
  \BibitemOpen
  \bibfield  {author} {\bibinfo {author} {\bibfnamefont {S.}~\bibnamefont {Foffa}}\ and\ \bibinfo {author} {\bibfnamefont {R.}~\bibnamefont {Sturani}},\ }\bibfield  {title} {\bibinfo {title} {{Tail terms in gravitational radiation reaction via effective field theory}},\ }\href {https://doi.org/10.1103/PhysRevD.87.044056} {\bibfield  {journal} {\bibinfo  {journal} {Phys. Rev. D}\ }\textbf {\bibinfo {volume} {87}},\ \bibinfo {pages} {044056} (\bibinfo {year} {2013}{\natexlab{b}})},\ \Eprint {https://arxiv.org/abs/1111.5488} {arXiv:1111.5488 [gr-qc]} \BibitemShut {NoStop}%
\bibitem [{\citenamefont {Foffa}\ and\ \citenamefont {Sturani}(2020)}]{Foffa:2019eeb}%
  \BibitemOpen
  \bibfield  {author} {\bibinfo {author} {\bibfnamefont {S.}~\bibnamefont {Foffa}}\ and\ \bibinfo {author} {\bibfnamefont {R.}~\bibnamefont {Sturani}},\ }\bibfield  {title} {\bibinfo {title} {{Hereditary terms at next-to-leading order in two-body gravitational dynamics}},\ }\href {https://doi.org/10.1103/PhysRevD.101.064033} {\bibfield  {journal} {\bibinfo  {journal} {Phys. Rev. D}\ }\textbf {\bibinfo {volume} {101}},\ \bibinfo {pages} {064033} (\bibinfo {year} {2020})},\ \bibinfo {note} {[Erratum: Phys.Rev.D 103, 089901 (2021)]},\ \Eprint {https://arxiv.org/abs/1907.02869} {arXiv:1907.02869 [gr-qc]} \BibitemShut {NoStop}%
\bibitem [{\citenamefont {Brunello}(2022)}]{Brunello:2022zui}%
  \BibitemOpen
  \bibfield  {author} {\bibinfo {author} {\bibfnamefont {G.}~\bibnamefont {Brunello}},\ }\emph {\bibinfo {title} {{Effective Field Theory Approach to General Relativity and Feynman Diagrams for Coalescing Binary Systems}}},\ \href@noop {} {Master's thesis},\ \bibinfo  {school} {U. Padua, Dept. Phys. Astron.} (\bibinfo {year} {2022}),\ \Eprint {https://arxiv.org/abs/2211.01321} {arXiv:2211.01321 [hep-th]} \BibitemShut {NoStop}%
\bibitem [{\citenamefont {Almeida}\ \emph {et~al.}(2023{\natexlab{b}})\citenamefont {Almeida}, \citenamefont {Foffa},\ and\ \citenamefont {Sturani}}]{Almeida:2022jrv}%
  \BibitemOpen
  \bibfield  {author} {\bibinfo {author} {\bibfnamefont {G.~L.}\ \bibnamefont {Almeida}}, \bibinfo {author} {\bibfnamefont {S.}~\bibnamefont {Foffa}},\ and\ \bibinfo {author} {\bibfnamefont {R.}~\bibnamefont {Sturani}},\ }\bibfield  {title} {\bibinfo {title} {{Gravitational radiation contributions to the two-body scattering angle}},\ }\href {https://doi.org/10.1103/PhysRevD.107.024020} {\bibfield  {journal} {\bibinfo  {journal} {Phys. Rev. D}\ }\textbf {\bibinfo {volume} {107}},\ \bibinfo {pages} {024020} (\bibinfo {year} {2023}{\natexlab{b}})},\ \Eprint {https://arxiv.org/abs/2209.11594} {arXiv:2209.11594 [gr-qc]} \BibitemShut {NoStop}%
\bibitem [{\citenamefont {Almeida}\ \emph {et~al.}(2025)\citenamefont {Almeida}, \citenamefont {M{\"u}ller}, \citenamefont {Foffa},\ and\ \citenamefont {Sturani}}]{Almeida:2024lbv}%
  \BibitemOpen
  \bibfield  {author} {\bibinfo {author} {\bibfnamefont {G.~L.}\ \bibnamefont {Almeida}}, \bibinfo {author} {\bibfnamefont {A.}~\bibnamefont {M{\"u}ller}}, \bibinfo {author} {\bibfnamefont {S.}~\bibnamefont {Foffa}},\ and\ \bibinfo {author} {\bibfnamefont {R.}~\bibnamefont {Sturani}},\ }\bibfield  {title} {\bibinfo {title} {{Gravitational memory contributions to waveform and effective action}},\ }\href {https://doi.org/10.1103/1ncd-r238} {\bibfield  {journal} {\bibinfo  {journal} {Phys. Rev. D}\ }\textbf {\bibinfo {volume} {112}},\ \bibinfo {pages} {084077} (\bibinfo {year} {2025})},\ \Eprint {https://arxiv.org/abs/2410.10565} {arXiv:2410.10565 [gr-qc]} \BibitemShut {NoStop}%
\bibitem [{\citenamefont {Almeida}\ \emph {et~al.}(2021)\citenamefont {Almeida}, \citenamefont {Foffa},\ and\ \citenamefont {Sturani}}]{Almeida:2021xwn}%
  \BibitemOpen
  \bibfield  {author} {\bibinfo {author} {\bibfnamefont {G.~L.}\ \bibnamefont {Almeida}}, \bibinfo {author} {\bibfnamefont {S.}~\bibnamefont {Foffa}},\ and\ \bibinfo {author} {\bibfnamefont {R.}~\bibnamefont {Sturani}},\ }\bibfield  {title} {\bibinfo {title} {{Tail contributions to gravitational conservative dynamics}},\ }\href {https://doi.org/10.1103/PhysRevD.104.124075} {\bibfield  {journal} {\bibinfo  {journal} {Phys. Rev. D}\ }\textbf {\bibinfo {volume} {104}},\ \bibinfo {pages} {124075} (\bibinfo {year} {2021})},\ \Eprint {https://arxiv.org/abs/2110.14146} {arXiv:2110.14146 [gr-qc]} \BibitemShut {NoStop}%
\bibitem [{\citenamefont {Henry}\ \emph {et~al.}(2021)\citenamefont {Henry}, \citenamefont {Faye},\ and\ \citenamefont {Blanchet}}]{Henry:2021cek}%
  \BibitemOpen
  \bibfield  {author} {\bibinfo {author} {\bibfnamefont {Q.}~\bibnamefont {Henry}}, \bibinfo {author} {\bibfnamefont {G.}~\bibnamefont {Faye}},\ and\ \bibinfo {author} {\bibfnamefont {L.}~\bibnamefont {Blanchet}},\ }\bibfield  {title} {\bibinfo {title} {{The current-type quadrupole moment and gravitational-wave mode ({\ensuremath{\ell}}, m) = (2, 1) of compact binary systems at the third post-Newtonian order}},\ }\href {https://doi.org/10.1088/1361-6382/ac1850} {\bibfield  {journal} {\bibinfo  {journal} {Class. Quant. Grav.}\ }\textbf {\bibinfo {volume} {38}},\ \bibinfo {pages} {185004} (\bibinfo {year} {2021})},\ \Eprint {https://arxiv.org/abs/2105.10876} {arXiv:2105.10876 [gr-qc]} \BibitemShut {NoStop}%
\bibitem [{\citenamefont {Larrouturou}\ \emph {et~al.}(2022)\citenamefont {Larrouturou}, \citenamefont {Blanchet}, \citenamefont {Henry},\ and\ \citenamefont {Faye}}]{Larrouturou:2021gqo}%
  \BibitemOpen
  \bibfield  {author} {\bibinfo {author} {\bibfnamefont {F.}~\bibnamefont {Larrouturou}}, \bibinfo {author} {\bibfnamefont {L.}~\bibnamefont {Blanchet}}, \bibinfo {author} {\bibfnamefont {Q.}~\bibnamefont {Henry}},\ and\ \bibinfo {author} {\bibfnamefont {G.}~\bibnamefont {Faye}},\ }\bibfield  {title} {\bibinfo {title} {{The quadrupole moment of compact binaries to the fourth post-Newtonian order: II. Dimensional regularization and renormalization}},\ }\href {https://doi.org/10.1088/1361-6382/ac5ba0} {\bibfield  {journal} {\bibinfo  {journal} {Class. Quant. Grav.}\ }\textbf {\bibinfo {volume} {39}},\ \bibinfo {pages} {115008} (\bibinfo {year} {2022})},\ \Eprint {https://arxiv.org/abs/2110.02243} {arXiv:2110.02243 [gr-qc]} \BibitemShut {NoStop}%
\bibitem [{\citenamefont {Trestini}\ \emph {et~al.}(2023)\citenamefont {Trestini}, \citenamefont {Larrouturou},\ and\ \citenamefont {Blanchet}}]{Trestini:2022tot}%
  \BibitemOpen
  \bibfield  {author} {\bibinfo {author} {\bibfnamefont {D.}~\bibnamefont {Trestini}}, \bibinfo {author} {\bibfnamefont {F.}~\bibnamefont {Larrouturou}},\ and\ \bibinfo {author} {\bibfnamefont {L.}~\bibnamefont {Blanchet}},\ }\bibfield  {title} {\bibinfo {title} {{The quadrupole moment of compact binaries to the fourth post-Newtonian order: relating the harmonic and radiative metrics}},\ }\href {https://doi.org/10.1088/1361-6382/acb5de} {\bibfield  {journal} {\bibinfo  {journal} {Class. Quant. Grav.}\ }\textbf {\bibinfo {volume} {40}},\ \bibinfo {pages} {055006} (\bibinfo {year} {2023})},\ \Eprint {https://arxiv.org/abs/2209.02719} {arXiv:2209.02719 [gr-qc]} \BibitemShut {NoStop}%
\bibitem [{\citenamefont {Beneke}\ and\ \citenamefont {Smirnov}(1998)}]{Beneke:1997zp}%
  \BibitemOpen
  \bibfield  {author} {\bibinfo {author} {\bibfnamefont {M.}~\bibnamefont {Beneke}}\ and\ \bibinfo {author} {\bibfnamefont {V.~A.}\ \bibnamefont {Smirnov}},\ }\bibfield  {title} {\bibinfo {title} {{Asymptotic expansion of Feynman integrals near threshold}},\ }\href {https://doi.org/10.1016/S0550-3213(98)00138-2} {\bibfield  {journal} {\bibinfo  {journal} {Nucl. Phys. B}\ }\textbf {\bibinfo {volume} {522}},\ \bibinfo {pages} {321} (\bibinfo {year} {1998})},\ \Eprint {https://arxiv.org/abs/hep-ph/9711391} {arXiv:hep-ph/9711391} \BibitemShut {NoStop}%
\bibitem [{\citenamefont {Bini}\ \emph {et~al.}(2012)\citenamefont {Bini}, \citenamefont {Damour},\ and\ \citenamefont {Faye}}]{Bini:2012gu}%
  \BibitemOpen
  \bibfield  {author} {\bibinfo {author} {\bibfnamefont {D.}~\bibnamefont {Bini}}, \bibinfo {author} {\bibfnamefont {T.}~\bibnamefont {Damour}},\ and\ \bibinfo {author} {\bibfnamefont {G.}~\bibnamefont {Faye}},\ }\bibfield  {title} {\bibinfo {title} {{Effective action approach to higher-order relativistic tidal interactions in binary systems and their effective one body description}},\ }\href {https://doi.org/10.1103/PhysRevD.85.124034} {\bibfield  {journal} {\bibinfo  {journal} {Phys. Rev. D}\ }\textbf {\bibinfo {volume} {85}},\ \bibinfo {pages} {124034} (\bibinfo {year} {2012})},\ \Eprint {https://arxiv.org/abs/1202.3565} {arXiv:1202.3565 [gr-qc]} \BibitemShut {NoStop}%
\bibitem [{\citenamefont {Porto}(2016{\natexlab{b}})}]{Porto:2016zng}%
  \BibitemOpen
  \bibfield  {author} {\bibinfo {author} {\bibfnamefont {R.~A.}\ \bibnamefont {Porto}},\ }\bibfield  {title} {\bibinfo {title} {{The Tune of Love and the Nature(ness) of Spacetime}},\ }\href {https://doi.org/10.1002/prop.201600064} {\bibfield  {journal} {\bibinfo  {journal} {Fortsch. Phys.}\ }\textbf {\bibinfo {volume} {64}},\ \bibinfo {pages} {723} (\bibinfo {year} {2016}{\natexlab{b}})},\ \Eprint {https://arxiv.org/abs/1606.08895} {arXiv:1606.08895 [gr-qc]} \BibitemShut {NoStop}%
\bibitem [{\citenamefont {Bohnenblust}\ \emph {et~al.}(2026{\natexlab{b}})\citenamefont {Bohnenblust}, \citenamefont {Jordan~Eriksen}, \citenamefont {Hoogeveen}, \citenamefont {Jakobsen},\ and\ \citenamefont {Plefka}}]{Bohnenblust:2026ujk}%
  \BibitemOpen
  \bibfield  {author} {\bibinfo {author} {\bibfnamefont {L.}~\bibnamefont {Bohnenblust}}, \bibinfo {author} {\bibfnamefont {C.}~\bibnamefont {Jordan~Eriksen}}, \bibinfo {author} {\bibfnamefont {J.}~\bibnamefont {Hoogeveen}}, \bibinfo {author} {\bibfnamefont {G.~U.}\ \bibnamefont {Jakobsen}},\ and\ \bibinfo {author} {\bibfnamefont {J.}~\bibnamefont {Plefka}},\ }\bibfield  {title} {\bibinfo {title} {{Black hole response theory and its exact shockwave limit}},\ }\href {https://doi.org/10.1007/JHEP08(2026)064} {\bibfield  {journal} {\bibinfo  {journal} {JHEP}\ }\textbf {\bibinfo {volume} {08}},\ \bibinfo {pages} {064}},\ \Eprint {https://arxiv.org/abs/2604.22009} {arXiv:2604.22009 [hep-th]} \BibitemShut {NoStop}%
\bibitem [{\citenamefont {Berends}\ and\ \citenamefont {Giele}(1988)}]{Berends:1987me}%
  \BibitemOpen
  \bibfield  {author} {\bibinfo {author} {\bibfnamefont {F.~A.}\ \bibnamefont {Berends}}\ and\ \bibinfo {author} {\bibfnamefont {W.~T.}\ \bibnamefont {Giele}},\ }\bibfield  {title} {\bibinfo {title} {{Recursive Calculations for Processes with n Gluons}},\ }\href {https://doi.org/10.1016/0550-3213(88)90442-7} {\bibfield  {journal} {\bibinfo  {journal} {Nucl. Phys. B}\ }\textbf {\bibinfo {volume} {306}},\ \bibinfo {pages} {759} (\bibinfo {year} {1988})}\BibitemShut {NoStop}%
\bibitem [{\citenamefont {Jakobsen}\ \emph {et~al.}(2023{\natexlab{b}})\citenamefont {Jakobsen}, \citenamefont {Mogull}, \citenamefont {Plefka}, \citenamefont {Sauer},\ and\ \citenamefont {Xu}}]{Jakobsen:2023ndj}%
  \BibitemOpen
  \bibfield  {author} {\bibinfo {author} {\bibfnamefont {G.~U.}\ \bibnamefont {Jakobsen}}, \bibinfo {author} {\bibfnamefont {G.}~\bibnamefont {Mogull}}, \bibinfo {author} {\bibfnamefont {J.}~\bibnamefont {Plefka}}, \bibinfo {author} {\bibfnamefont {B.}~\bibnamefont {Sauer}},\ and\ \bibinfo {author} {\bibfnamefont {Y.}~\bibnamefont {Xu}},\ }\bibfield  {title} {\bibinfo {title} {{Conservative Scattering of Spinning Black Holes at Fourth Post-Minkowskian Order}},\ }\href {https://doi.org/10.1103/PhysRevLett.131.151401} {\bibfield  {journal} {\bibinfo  {journal} {Phys. Rev. Lett.}\ }\textbf {\bibinfo {volume} {131}},\ \bibinfo {pages} {151401} (\bibinfo {year} {2023}{\natexlab{b}})},\ \Eprint {https://arxiv.org/abs/2306.01714} {arXiv:2306.01714 [hep-th]} \BibitemShut {NoStop}%
\bibitem [{\citenamefont {Bautista}\ \emph {et~al.}(2026)\citenamefont {Bautista}, \citenamefont {Driesse}, \citenamefont {Haddad},\ and\ \citenamefont {Jakobsen}}]{Bautista:2026qse}%
  \BibitemOpen
  \bibfield  {author} {\bibinfo {author} {\bibfnamefont {Y.~F.}\ \bibnamefont {Bautista}}, \bibinfo {author} {\bibfnamefont {M.}~\bibnamefont {Driesse}}, \bibinfo {author} {\bibfnamefont {K.}~\bibnamefont {Haddad}},\ and\ \bibinfo {author} {\bibfnamefont {G.~U.}\ \bibnamefont {Jakobsen}},\ }\bibfield  {title} {\bibinfo {title} {{Gravitational wave scattering in spinless WQFT}},\ }\href {https://doi.org/10.1007/JHEP05(2026)252} {\bibfield  {journal} {\bibinfo  {journal} {JHEP}\ }\textbf {\bibinfo {volume} {05}},\ \bibinfo {pages} {252}},\ \Eprint {https://arxiv.org/abs/2602.06125} {arXiv:2602.06125 [hep-th]} \BibitemShut {NoStop}%
\bibitem [{\citenamefont {Brunello}\ \emph {et~al.}(2026{\natexlab{c}})\citenamefont {Brunello}, \citenamefont {Meo},\ and\ \citenamefont {Smith}}]{Brunello:2026rdk}%
  \BibitemOpen
  \bibfield  {author} {\bibinfo {author} {\bibfnamefont {G.}~\bibnamefont {Brunello}}, \bibinfo {author} {\bibfnamefont {M.}~\bibnamefont {Meo}},\ and\ \bibinfo {author} {\bibfnamefont {S.}~\bibnamefont {Smith}},\ }\bibfield  {title} {\bibinfo {title} {{Gravitational Compton scattering at the fourth post-Minkowskian order}},\ }\href {https://doi.org/10.1103/nbxl-68tz} {\bibfield  {journal} {\bibinfo  {journal} {Phys. Rev. D}\ }\textbf {\bibinfo {volume} {114}},\ \bibinfo {pages} {085022} (\bibinfo {year} {2026}{\natexlab{c}})},\ \Eprint {https://arxiv.org/abs/2606.28239} {arXiv:2606.28239 [hep-th]} \BibitemShut {NoStop}%
\bibitem [{\citenamefont {K{\"a}lin}\ and\ \citenamefont {Lim}(2026)}]{Kalin:2026txf}%
  \BibitemOpen
  \bibfield  {author} {\bibinfo {author} {\bibfnamefont {G.}~\bibnamefont {K{\"a}lin}}\ and\ \bibinfo {author} {\bibfnamefont {J.}~\bibnamefont {Lim}},\ }\href@noop {} {\bibinfo {title} {{Graph-Based Algorithms for Worldline EFT Integrands}}} (\bibinfo {year} {2026}),\ \Eprint {https://arxiv.org/abs/2610.35007} {arXiv:2610.35007 [hep-th]} \BibitemShut {NoStop}%
\bibitem [{\citenamefont {Marchand}\ \emph {et~al.}(2016)\citenamefont {Marchand}, \citenamefont {Blanchet},\ and\ \citenamefont {Faye}}]{Marchand:2016vox}%
  \BibitemOpen
  \bibfield  {author} {\bibinfo {author} {\bibfnamefont {T.}~\bibnamefont {Marchand}}, \bibinfo {author} {\bibfnamefont {L.}~\bibnamefont {Blanchet}},\ and\ \bibinfo {author} {\bibfnamefont {G.}~\bibnamefont {Faye}},\ }\bibfield  {title} {\bibinfo {title} {{Gravitational-wave tail effects to quartic non-linear order}},\ }\href {https://doi.org/10.1088/0264-9381/33/24/244003} {\bibfield  {journal} {\bibinfo  {journal} {Class. Quant. Grav.}\ }\textbf {\bibinfo {volume} {33}},\ \bibinfo {pages} {244003} (\bibinfo {year} {2016})},\ \Eprint {https://arxiv.org/abs/1607.07601} {arXiv:1607.07601 [gr-qc]} \BibitemShut {NoStop}%
\bibitem [{\citenamefont {Edison}\ and\ \citenamefont {Levi}(2024)}]{Edison:2023qvg}%
  \BibitemOpen
  \bibfield  {author} {\bibinfo {author} {\bibfnamefont {A.}~\bibnamefont {Edison}}\ and\ \bibinfo {author} {\bibfnamefont {M.}~\bibnamefont {Levi}},\ }\bibfield  {title} {\bibinfo {title} {{Higher-order tails and RG flows due to scattering of gravitational radiation from binary inspirals}},\ }\href {https://doi.org/10.1007/JHEP08(2024)161} {\bibfield  {journal} {\bibinfo  {journal} {JHEP}\ }\textbf {\bibinfo {volume} {08}},\ \bibinfo {pages} {161}},\ \Eprint {https://arxiv.org/abs/2310.20066} {arXiv:2310.20066 [hep-th]} \BibitemShut {NoStop}%
\bibitem [{\citenamefont {Vermaseren}(2000)}]{Vermaseren:2000nd}%
  \BibitemOpen
  \bibfield  {author} {\bibinfo {author} {\bibfnamefont {J.~A.~M.}\ \bibnamefont {Vermaseren}},\ }\href@noop {} {\bibinfo {title} {{New features of FORM}}} (\bibinfo {year} {2000}),\ \Eprint {https://arxiv.org/abs/math-ph/0010025} {arXiv:math-ph/0010025} \BibitemShut {NoStop}%
\bibitem [{\citenamefont {Ruijl}\ \emph {et~al.}(2017)\citenamefont {Ruijl}, \citenamefont {Ueda},\ and\ \citenamefont {Vermaseren}}]{Ruijl:2017dtg}%
  \BibitemOpen
  \bibfield  {author} {\bibinfo {author} {\bibfnamefont {B.}~\bibnamefont {Ruijl}}, \bibinfo {author} {\bibfnamefont {T.}~\bibnamefont {Ueda}},\ and\ \bibinfo {author} {\bibfnamefont {J.}~\bibnamefont {Vermaseren}},\ }\href@noop {} {\bibinfo {title} {{FORM version 4.2}}} (\bibinfo {year} {2017}),\ \Eprint {https://arxiv.org/abs/1707.06453} {arXiv:1707.06453 [hep-ph]} \BibitemShut {NoStop}%
\bibitem [{\citenamefont {Chetyrkin}\ and\ \citenamefont {Tkachov}(1981)}]{Chetyrkin:1981qh}%
  \BibitemOpen
  \bibfield  {author} {\bibinfo {author} {\bibfnamefont {K.~G.}\ \bibnamefont {Chetyrkin}}\ and\ \bibinfo {author} {\bibfnamefont {F.~V.}\ \bibnamefont {Tkachov}},\ }\bibfield  {title} {\bibinfo {title} {{Integration by parts: The algorithm to calculate $\beta$-functions in 4 loops}},\ }\href {https://doi.org/10.1016/0550-3213(81)90199-1} {\bibfield  {journal} {\bibinfo  {journal} {Nucl. Phys. B}\ }\textbf {\bibinfo {volume} {192}},\ \bibinfo {pages} {159} (\bibinfo {year} {1981})}\BibitemShut {NoStop}%
\bibitem [{\citenamefont {Laporta}(2000)}]{Laporta:2000dsw}%
  \BibitemOpen
  \bibfield  {author} {\bibinfo {author} {\bibfnamefont {S.}~\bibnamefont {Laporta}},\ }\bibfield  {title} {\bibinfo {title} {{High-precision calculation of multiloop Feynman integrals by difference equations}},\ }\href {https://doi.org/10.1142/S0217751X00002159} {\bibfield  {journal} {\bibinfo  {journal} {Int. J. Mod. Phys. A}\ }\textbf {\bibinfo {volume} {15}},\ \bibinfo {pages} {5087} (\bibinfo {year} {2000})},\ \Eprint {https://arxiv.org/abs/hep-ph/0102033} {arXiv:hep-ph/0102033} \BibitemShut {NoStop}%
\bibitem [{\citenamefont {Lee}(2012)}]{Lee:2012cn}%
  \BibitemOpen
  \bibfield  {author} {\bibinfo {author} {\bibfnamefont {R.~N.}\ \bibnamefont {Lee}},\ }\href@noop {} {\bibinfo {title} {{Presenting LiteRed: a tool for the Loop InTEgrals REDuction}}} (\bibinfo {year} {2012}),\ \Eprint {https://arxiv.org/abs/1212.2685} {arXiv:1212.2685 [hep-ph]} \BibitemShut {NoStop}%
\bibitem [{\citenamefont {Lee}(2014)}]{Lee:2013mka}%
  \BibitemOpen
  \bibfield  {author} {\bibinfo {author} {\bibfnamefont {R.~N.}\ \bibnamefont {Lee}},\ }\bibfield  {title} {\bibinfo {title} {{LiteRed 1.4: a powerful tool for reduction of multiloop integrals}},\ }\href {https://doi.org/10.1088/1742-6596/523/1/012059} {\bibfield  {journal} {\bibinfo  {journal} {J. Phys. Conf. Ser.}\ }\textbf {\bibinfo {volume} {523}},\ \bibinfo {pages} {012059} (\bibinfo {year} {2014})},\ \Eprint {https://arxiv.org/abs/1310.1145} {arXiv:1310.1145 [hep-ph]} \BibitemShut {NoStop}%
\bibitem [{\citenamefont {Blanchet}(1993)}]{Blanchet:1993ng}%
  \BibitemOpen
  \bibfield  {author} {\bibinfo {author} {\bibfnamefont {L.}~\bibnamefont {Blanchet}},\ }\bibfield  {title} {\bibinfo {title} {{Time asymmetric structure of gravitational radiation}},\ }\href {https://doi.org/10.1103/PhysRevD.47.4392} {\bibfield  {journal} {\bibinfo  {journal} {Phys. Rev. D}\ }\textbf {\bibinfo {volume} {47}},\ \bibinfo {pages} {4392} (\bibinfo {year} {1993})}\BibitemShut {NoStop}%
\bibitem [{\citenamefont {Thorne}(1980)}]{Thorne:1980ru}%
  \BibitemOpen
  \bibfield  {author} {\bibinfo {author} {\bibfnamefont {K.~S.}\ \bibnamefont {Thorne}},\ }\bibfield  {title} {\bibinfo {title} {{Multipole Expansions of Gravitational Radiation}},\ }\href {https://doi.org/10.1103/RevModPhys.52.299} {\bibfield  {journal} {\bibinfo  {journal} {Rev. Mod. Phys.}\ }\textbf {\bibinfo {volume} {52}},\ \bibinfo {pages} {299} (\bibinfo {year} {1980})}\BibitemShut {NoStop}%
\bibitem [{\citenamefont {Poincar{\'e}}(1910)}]{Poincare:1910}%
  \BibitemOpen
  \bibfield  {author} {\bibinfo {author} {\bibfnamefont {H.}~\bibnamefont {Poincar{\'e}}},\ }\href@noop {} {\emph {\bibinfo {title} {{Le{\c c}ons de m{\'e}canique c{\'e}leste}}}},\ Vol.~\bibinfo {volume} {3}\ (\bibinfo  {publisher} {Gauthier-Villars},\ \bibinfo {address} {Paris},\ \bibinfo {year} {1910})\BibitemShut {NoStop}%
\bibitem [{\citenamefont {Bertrand}(1923)}]{Bertrand:1923}%
  \BibitemOpen
  \bibfield  {author} {\bibinfo {author} {\bibfnamefont {G.}~\bibnamefont {Bertrand}},\ }\bibfield  {title} {\bibinfo {title} {{La th{\'e}orie des mar{\'e}es et les {\'e}quations int{\'e}grales}},\ }\href@noop {} {\bibfield  {journal} {\bibinfo  {journal} {Ann. Sci. \'Ec. Norm. Sup{\'e}r.}\ }\textbf {\bibinfo {volume} {40}},\ \bibinfo {pages} {151} (\bibinfo {year} {1923})}\BibitemShut {NoStop}%
\bibitem [{\citenamefont {Muskhelishvili}(1953)}]{Muskhelishvili:1953}%
  \BibitemOpen
  \bibfield  {author} {\bibinfo {author} {\bibfnamefont {N.~I.}\ \bibnamefont {Muskhelishvili}},\ }\href@noop {} {\emph {\bibinfo {title} {{Singular Integral Equations}}}}\ (\bibinfo  {publisher} {P. Noordhoff},\ \bibinfo {address} {Groningen},\ \bibinfo {year} {1953})\BibitemShut {NoStop}%
\bibitem [{\citenamefont {Di~Vecchia}\ \emph {et~al.}(2022)\citenamefont {Di~Vecchia}, \citenamefont {Heissenberg},\ and\ \citenamefont {Russo}}]{DiVecchia:2022owy}%
  \BibitemOpen
  \bibfield  {author} {\bibinfo {author} {\bibfnamefont {P.}~\bibnamefont {Di~Vecchia}}, \bibinfo {author} {\bibfnamefont {C.}~\bibnamefont {Heissenberg}},\ and\ \bibinfo {author} {\bibfnamefont {R.}~\bibnamefont {Russo}},\ }\bibfield  {title} {\bibinfo {title} {{Angular momentum of zero-frequency gravitons}},\ }\href {https://doi.org/10.1007/JHEP08(2022)172} {\bibfield  {journal} {\bibinfo  {journal} {JHEP}\ }\textbf {\bibinfo {volume} {08}},\ \bibinfo {pages} {172}},\ \Eprint {https://arxiv.org/abs/2203.11915} {arXiv:2203.11915 [hep-th]} \BibitemShut {NoStop}%
\bibitem [{\citenamefont {Bini}\ \emph {et~al.}(2023)\citenamefont {Bini}, \citenamefont {Damour},\ and\ \citenamefont {Geralico}}]{Bini:2022enm}%
  \BibitemOpen
  \bibfield  {author} {\bibinfo {author} {\bibfnamefont {D.}~\bibnamefont {Bini}}, \bibinfo {author} {\bibfnamefont {T.}~\bibnamefont {Damour}},\ and\ \bibinfo {author} {\bibfnamefont {A.}~\bibnamefont {Geralico}},\ }\bibfield  {title} {\bibinfo {title} {{Radiated momentum and radiation reaction in gravitational two-body scattering including time-asymmetric effects}},\ }\href {https://doi.org/10.1103/PhysRevD.107.024012} {\bibfield  {journal} {\bibinfo  {journal} {Phys. Rev. D}\ }\textbf {\bibinfo {volume} {107}},\ \bibinfo {pages} {024012} (\bibinfo {year} {2023})},\ \Eprint {https://arxiv.org/abs/2210.07165} {arXiv:2210.07165 [gr-qc]} \BibitemShut {NoStop}%
\bibitem [{\citenamefont {Faye}\ \emph {et~al.}(2015)\citenamefont {Faye}, \citenamefont {Blanchet},\ and\ \citenamefont {Iyer}}]{Faye:2014fra}%
  \BibitemOpen
  \bibfield  {author} {\bibinfo {author} {\bibfnamefont {G.}~\bibnamefont {Faye}}, \bibinfo {author} {\bibfnamefont {L.}~\bibnamefont {Blanchet}},\ and\ \bibinfo {author} {\bibfnamefont {B.~R.}\ \bibnamefont {Iyer}},\ }\bibfield  {title} {\bibinfo {title} {{Non-linear multipole interactions and gravitational-wave octupole modes for inspiralling compact binaries to third-and-a-half post-Newtonian order}},\ }\href {https://doi.org/10.1088/0264-9381/32/4/045016} {\bibfield  {journal} {\bibinfo  {journal} {Class. Quant. Grav.}\ }\textbf {\bibinfo {volume} {32}},\ \bibinfo {pages} {045016} (\bibinfo {year} {2015})},\ \Eprint {https://arxiv.org/abs/1409.3546} {arXiv:1409.3546 [gr-qc]} \BibitemShut {NoStop}%
\bibitem [{\citenamefont {K{\"a}lin}\ and\ \citenamefont {Porto}(2020{\natexlab{b}})}]{Kalin:2019inp}%
  \BibitemOpen
  \bibfield  {author} {\bibinfo {author} {\bibfnamefont {G.}~\bibnamefont {K{\"a}lin}}\ and\ \bibinfo {author} {\bibfnamefont {R.~A.}\ \bibnamefont {Porto}},\ }\bibfield  {title} {\bibinfo {title} {{From boundary data to bound states. Part II. Scattering angle to dynamical invariants (with twist)}},\ }\href {https://doi.org/10.1007/JHEP02(2020)120} {\bibfield  {journal} {\bibinfo  {journal} {JHEP}\ }\textbf {\bibinfo {volume} {02}},\ \bibinfo {pages} {120}},\ \Eprint {https://arxiv.org/abs/1911.09130} {arXiv:1911.09130 [hep-th]} \BibitemShut {NoStop}%
\bibitem [{\citenamefont {Damour}(2020{\natexlab{b}})}]{Damour:2019lcq}%
  \BibitemOpen
  \bibfield  {author} {\bibinfo {author} {\bibfnamefont {T.}~\bibnamefont {Damour}},\ }\bibfield  {title} {\bibinfo {title} {{Classical and quantum scattering in post-Minkowskian gravity}},\ }\href {https://doi.org/10.1103/PhysRevD.102.024060} {\bibfield  {journal} {\bibinfo  {journal} {Phys. Rev. D}\ }\textbf {\bibinfo {volume} {102}},\ \bibinfo {pages} {024060} (\bibinfo {year} {2020}{\natexlab{b}})},\ \Eprint {https://arxiv.org/abs/1912.02139} {arXiv:1912.02139 [gr-qc]} \BibitemShut {NoStop}%
\bibitem [{\citenamefont {Damour}\ and\ \citenamefont {Iyer}(1991)}]{Damour:1990gj}%
  \BibitemOpen
  \bibfield  {author} {\bibinfo {author} {\bibfnamefont {T.}~\bibnamefont {Damour}}\ and\ \bibinfo {author} {\bibfnamefont {B.~R.}\ \bibnamefont {Iyer}},\ }\bibfield  {title} {\bibinfo {title} {{Multipole analysis for electromagnetism and linearized gravity with irreducible cartesian tensors}},\ }\href {https://doi.org/10.1103/PhysRevD.43.3259} {\bibfield  {journal} {\bibinfo  {journal} {Phys. Rev. D}\ }\textbf {\bibinfo {volume} {43}},\ \bibinfo {pages} {3259} (\bibinfo {year} {1991})}\BibitemShut {NoStop}%
\end{thebibliography}%

\end{document}